\documentclass[a4paper,titlepage,12pt]{book}

\usepackage{amsmath,amssymb} 
\usepackage{fancyhdr} 
\usepackage{bbm}
\usepackage{epsfig}
\usepackage{multirow,rotating}

\def\be{\begin{equation}}
\def\ee{\end{equation}}
\def\bea{\begin{eqnarray}}
\def\eea{\end{eqnarray}}

\newcommand{\ft}[2]{{\textstyle\frac{#1}{#2}}}
\newcommand{\nn}{\nonumber}

\def\slashchar#1{\setbox0=\hbox{$#1$}           
   \dimen0=\wd0                                 
   \setbox1=\hbox{/} \dimen1=\wd1               
   \ifdim\dimen0>\dimen1                        
      \rlap{\hbox to \dimen0{\hfil/\hfil}}      
      #1                                        
   \else                                        
      \rlap{\hbox to \dimen1{\hfil$#1$\hfil}}   
      /                                         
   \fi}

\def\tr{\text{tr}}

\def\diag{\text{diag}}

\def\mua{\mu}
\def\mub{\nu}
\def\muc{\rho}
\def\mud{\lambda}
\def\mue{\sigma}
\def\muf{\tau}
\def\mug{\kappa}
\def\muh{\tau}
\def\mui{\chi}
\def\muj{\theta}
\def\muk{\xi}

\def\ma{m}
\def\mb{n}
\def\mc{p}
\def\md{q}
\def\me{r}

\def\Ja{M}
\def\Jb{N}
\def\Jc{P}
\def\Jd{Q}
\def\Je{R}
\def\Jf{S}
\def\Jg{T}
\def\Jh{U}
\def\Ji{V}
\def\Jj{W}
\def\Jk{X}
\def\Jl{Y}
\def\Jm{Z}
\def\Jn{A}
\def\Jo{B}
\def\Jp{C}
\def\Jq{D}
\def\Jr{E}
\def\ja{a}
\def\jb{b}
\def\jc{c}
\def\jd{d}
\def\je{e}
\def\jf{f}
\def\jg{g}
\def\jh{h}
\def\ji{i}
\def\jj{j}
\def\jk{k}
\def\jl{l}
\def\jm{m}
\def\jn{n}
\def\ka{\alpha}
\def\kb{\beta}
\def\kc{\gamma}
\def\kd{\delta}

\def\la{x}

\def\sa{\alpha}
\def\sb{\beta}
\def\sc{\gamma}

\newdimen\squaresize \squaresize=12pt
\newdimen\thickness \thickness=0.7pt

\def\square#1{\hbox{\vrule width \thickness
   \vbox to \squaresize{\hrule height \thickness\vss
      \hbox to \squaresize{\hss#1\hss}
   \vss\hrule height\thickness}
\unskip\vrule width \thickness} \kern-\thickness}

\def\cut#1{\hbox{\vrule width-1 \thickness
   \vbox to \squaresize{\hrule height-1 \thickness\vss
      \hbox to \squaresize{\hss#1\hss}
   \vss\hrule height-1\thickness}
\unskip\vrule width +4 \thickness} \kern-\thickness}

\def\vsquare#1{\vbox{\square{$#1$}}\kern-\thickness}

\def\young#1{
\vbox{\smallskip\offinterlineskip \halign{&\vsquare{##}\cr #1}}}

\newcommand{\tinyyoung}[1]{
\squaresize=7pt \thickness=0.4pt \mbox{\tiny\young{#1}}
\squaresize=12pt \thickness=0.7pt}

\def\Re{\text{Re}}
\def\Im{\text{Im}}

\def\hV{\hat {\cal V}}

\def\sMa{{\cal M}}
\def\sMb{{\cal N}}
\def\sMc{{\cal P}}
\def\sMd{{\cal Q}}
\def\sMe{{\cal R}}
\def\sMf{{\cal S}}
\def\sMg{{\cal T}}

\def\La{\Lambda}
\def\Lb{\Sigma}

\def\Le{\Xi}

\def\Le{\Upsilon}
\def\Ma{M}
\def\Mb{N}
\def\Mc{P}
\def\Md{Q}
\def\Me{R}
\def\Mf{S}
\def\Mg{T}
\def\Mh{U}
\def\Mi{V}
\def\jja{i}
\def\jjb{j}
\def\jjc{k}
\def\jjd{l}
\def\jje{m}
\def\jjf{n}
\def\jjg{p}
\def\jjh{q}
\def\jji{r}
\def\jjj{s}
\def\jjk{t}
\def\jjl{u}
\def\ya{m}
\def\yb{n}
\def\yc{o}
\def\yd{p}
\def\ye{q}
\def\yf{r}
\def\yg{s}
\def\yh{t}
\def\xa{a}
\def\xb{b}
\def\xc{c}

\def\aa{\alpha}
\def\ab{\beta}
\def\ac{\gamma}
\def\ad{\delta}
\def\aee{\epsilon}
\def\ca{x}

\def\Na{{\hat 0}}
\def\Nv{0}
\def\cMa{\sMa}
\def\cMb{\sMb}
\def\cMc{\sMc}
\def\cMd{\sMd}
\def\cMe{\sMe}
\def\cMf{\sMf}
\def\cMg{\sMg}
\def\LLa{\Lambda}
\def\LLb{\Gamma}
\def\LLc{\Sigma}
\def\LLd{\Psi}
\def\Le{\Delta}
\def\LLf{\Xi}

\def\cAa{{\cal A}}
\def\cAb{{\cal B}}
\def\cAc{{\cal C}}

\begin{document}

\begin{titlepage}

\thispagestyle{empty}

\begin{center}

\hfill {\tt hep-th/0702084}

\vskip 1.5cm 
\begin{center}
{\Large {\bf Gauged Supergravities in Various}} \\
{\Large {\bf Spacetime Dimensions}\footnote{
Based on the author's PhD thesis, defended on December 20, 2006.}}
\end{center}

\vskip 1.0cm

{\bf Martin Weidner\footnote{Email: {\tt mweidner@usc.edu},
New Adress: {\em Department of Economics,
University of Southern California, 3620 S. Vermont Ave. KAP 300,
Los Angeles, CA 90089, U.S.A.}}}

\vskip 22pt

{\em II. Institut f\"ur Theoretische Physik\\[-.6ex]
Universit\"at Hamburg\\[-.6ex]
Luruper Chaussee 149\\[-.6ex]
D-22761 Hamburg, Germany}

\vskip 1.5cm

\end{center}

\vskip 5pt

\begin{center} {\bf ABSTRACT}\\[3ex]

\begin{minipage}{13cm}
\small

In this review articel
we study the gaugings of extended supergravity theories in various space-time dimensions.
These theories describe the low-energy limit of non-trivial string compactifications.
For each theory under consideration
we review all possible gaugings that are compatible with supersymmetry.
They are parameterized by the so-called
embedding tensor which is a group theoretical object that has to satisfy
certain representation constraints.
This embedding tensor determines all couplings in the gauged theory
that are necessary to preserve gauge invariance and supersymmetry.
The concept of the embedding tensor and the general structure
of the gauged supergravities are explained in detail.
The methods are then applied
to the half-maximal ($N=4$) supergravities in $d=4$ and $d=5$
and to the maximal supergravities in $d=2$ and $d=7$.
Examples of particular gaugings are given.
Whenever possible, the higher-dimensional origin of these theories is identified
and it is shown how the compactification parameters
like fluxes and torsion are contained in the embedding tensor.

\end{minipage}
\end{center}
\noindent

%

\end{titlepage}

\newpage
\thispagestyle{empty}

\vspace*{\fill}
\noindent
{\bf\Large Acknowledgments}\par
\bigskip
\noindent

First of all I want to thank my supervisor Henning Samtleben for his support at anytime in the last three years,
all his valuable insights and 
for the friendly collaboration. I learned a lot in the course of this work and it was a highly fortunate decision to do this doctorate.
I am deeply indebted to Jan Louis for offering me the PhD position here in Hamburg and for his friendly advice and help in
various respects. I highly enjoyed the good working atmosphere in the string theory group.

I like to thank Reinhard~Lorenzen, Paolo~Merlatti and Mathias~de~Riese for bearing me as an office mate
and for their help in physical and non-physical questions.
I am very grateful to Tako~Mattik and Sakura~Sch\"afer-Nameki
for numerous discussions about string theory and for moral support.
I want to thank
  Iman~Benmachiche,
  David~Cerde\~no,
  Christoph~Ellmer, 
  Thomas~Grimm,
  Olaf~Hohm,
  Manuel~Hohmann, 
  Hans~Jockers,
  Anke~Knauf,
  Simon~K\"ors, 
  Jonas~Sch\"on,
  Bastiaan~Spanjaard,
  Silvia~Vaula,
and Mattias~Wohlfarth
for the nice atmosphere in the group and for helpful discussions.

As a member of the physics graduate school ``Future Developments in Particle Physics'' I enjoyed important financial and intellectual support
and I am deeply indebted to the DFG (the German Science Foundation) and to everybody who helped organizing and maintaining this
graduate school.

Last but not least I want to thank my parents for their constant support during all stages of my studies and for guaranteeing that
there is always a place were I feel at home.

\bigskip
\newpage
\pagenumbering{roman}
\thispagestyle{plain}

\tableofcontents
\cleardoublepage


\pagenumbering{arabic}

\chapter{Introduction} \label{ch:intro}

\section{String Theory and Supergravity}

One of the great challenges of modern physics is the unification of general relativity and quantum field theory.
On the one hand, the large scale structure of the universe is governed by gravitational interactions which 
are accurately described by Einstein's general relativity. On the other hand, quantum field theory is used
to explain the fundamental interactions at small distances. In particular the so-called standard model 
of particle physics gives a description of the strong and electroweak interactions of all known elementary particles
which has successfully passed many precision tests in collider experiments. However, this separation into large scale and
small scale domains is not universally applicable. The early universe and black holes are examples of situations where a 
quantum theory of gravity is needed. The situation is also unsatisfactory from a theoretical perspective since
the basic concepts of general relativity (coordinate independence) and quantum theory (uncertainty relation) 
seem incompatible. That is why standard approaches to a quantum theory of gravity are hampered by divergences
which prevent the theory from being predictive. To avoid these problems a new theoretical framework is necessary 
and one of the few possible candidates is string theory
\cite{Green:1981yb,GSW,Pol}.

In string theory the fundamental object is no longer a point particle but a one-dimensional string which can move and vibrate
in some target space, e.g.\ in Minkowski space. Elementary particles are identified as resonance modes of the string,
most of which have excitation energies far above the energy scale one can presently probe in experiments.
There is a fundamental constant of string theory that governs the scale of these massive string excitations.
This constant can be expressed as a string tension (string energy per unit length), as a string length or
directly as a mass, in which case it is typically of the order of the Planck mass. String theory has two main
appealing features: Firstly, one of the massless string excitations is a spin 2 particle that can be identified with the graviton,
i.e.\ with the exchange particle of the gravitational force that is necessary in every quantum theory of gravity. 
Secondly, the theory can be formulated as a conformal field theory on the two-dimensional world-sheet which is swept out by
the string while traversing the target space. For some examples of target spaces these conformal field theories are
well understood quantum field theories. In this sense string theory provides a consistent framework of quantum gravity.
However, the theory is even more ambitious, because in principle it aims to predict the complete particle spectrum and all interactions
of nature, i.e.\ to provide a ``theory of everything''.

There are different formulations of string theory which are related by duality transformations.
All these formulations need a ten-dimensional target space in order to be consistent
quantum theories\footnote{We are only considering supersymmetric string theories here and we neglect the subtlety
that heterotic strings partially ``live'' in 26 space-time dimensions.}. Since our observed world is four-dimensional
one needs to assume that six of these dimensions are compactified, i.e.\ are rolled up to such a small size that they
are practically unobservable. The number of consistent compactification schemes and thus of resulting four-dimensional
effective theories
is very large. At present, there is no criterion to single out one of these schemes as the one that is realized
in nature.

String theory on arbitrary curved target spaces is far from being fully understood. For many applications, however,
one can restrict to the low-energy limit of string theory which is supergravity. As mentioned above string theory is formulated
as a conformal field theory on the two-dimensional world-sheet. In contrast, supergravity is a field theory on 
the target space. 
Each massless string mode corresponds to a field in the supergravity, in particular the graviton corresponds
to the metric. Therefore, supergravity includes general relativity. 

A crucial ingredient for string theory and supergravity is supersymmetry.
Purely bosonic string theory suffers from various inconsistencies that are resolved in supersymmetric string theories.
This symmetry relates bosons and fermions of a theory. Its presence leads to various cancellations in quantum corrections.
Originally, supersymmetry was introduced as a global symmetry in field theory \cite{Golfand:1971iw,Wess:1973kz}.
When it is turned into a local symmetry, supergravity is obtained.
The gauge field of local supersymmetry is the gravitino. It carries spin $3/2$ and is the
super-partner of the graviton, i.e.\ of the
space-time metric. This approach to supergravity via the gauging of supersymmetry was found independently of string theory
\cite{Freedman:1976xh,Deser:1976eh,Freedman:1976py,Chamseddine:1976bf} and the relation between these
theories was only realized afterwards \cite{Schwarz:1982jn,Green:1982sw}.
Also independently of string theory and supergravity the concept of supersymmetry is very important.
One can, for example, cure some problems
of the standard model (large radiative corrections to the Higgs boson mass, hierarchy problem) within a supersymmetric extension
of the standard model, and it is hoped to discover supersymmetry at the next generation of particle colliders (LHC and ILC).
This discovery would be important from a string theory point of view because it would justify supersymmetry
as one of its basic assumptions.

Supergravity theories exist in all space-time dimensions $d\leq 11$ and can have different numbers of supersymmetry generators
(for a review see e.g.\ \cite{VanNieuwenhuizen:1981ae,deWit:2002vz,VanProeyen:2003zj} and references therein).
Having several of these generators means to have more independent supersymmetry transformations and more gravitini.
One then speaks of extended supergravity.
The maximal number of real supercharges is $Q=32$, independent of the dimension $d$.
The present article
is devoted to the study of maximal ($Q=32$) and half-maximal ($Q=16$) supergravities and
of their possible gaugings, as will be explained in the next section. Our analysis 
takes place at the level of classical field theory. The motivation for our considerations is always the string theory
origin of these theories, and it is string theory that should provide the correct quantum description.

\section{Gauged supergravity theories}

String theory compactifications from $D=10$ down to $d<10$ dimensions generically yield
at low energies gauged supergravity theories. For example,
the isometry group of the internal $(D-d)$-dimensional manifold usually shows up within
the gauge group of the effective $d$-dimensional theory.
An ungauged effective theory is obtained from compactifications of IIA or IIB string theory
if the internal manifold is locally flat, e.g.\ the ungauged maximal supergravities are obtained from torus reductions. 
Since we consider extended supergravities with a large number of supercharges,
these ungauged supergravities are unique as soon as the field content is specified\footnote{
For the maximal supergravities in $d<10$
there is only one ungauged theory. The half-maximal supergravities are specified by the number
of vector multiplets.}. Gaugings are the only known deformations of these theories that preserve supersymmetry\footnote{
The only known exceptions are the massive IIA supergravity \cite{Romans:1985tz} and a massive deformation of the six-dimensional
half-maximal supergravity \cite{Romans:1985tw}, see our comments in section \ref{sec:HalfMaxUngauged}.}.
Therefore, any more complicated compactification scheme that preserves a large number of supercharges ($Q \geq 16$) must yield
a gauging of the respective ungauged theory. This fact is our motivation to construct all possible gaugings that are
compatible with supersymmetry. As soon as this is achieved 
the compactification parameters such as fluxes (i.e.\ background values for the
field strengths of the $D=10$ tensor gauge fields), torsion, number of branes, etc.\
must be contained in the parameters of the general gauging. These more
general compactification schemes are of great interest because for example fluxes may give vacuum expectation values
to some of the numerous massless fields (``moduli'') that generically result from string theory compactifications.
In the ground state one may in particular find
supersymmetry breaking, a cosmological constant and masses for the scalar fields
(for a review we refer to \cite{Grana:2005jc}). These are requirements
for a phenomenologically viable effective theory.

Gauging a theory means to turn a global symmetry into a local one. In other words, the symmetry parameters which were previously constant
are allowed to have a space-time dependence in the gauged theory. As mentioned above supergravity itself can be obtained
by gauging global supersymmetry,
but we are now considering the gauging of ordinary bosonic symmetries. In order to preserve gauge invariance 
one needs to minimally couple vector fields $A_\mua$ to the symmetry generators, i.e.\ to replace partial derivatives by covariant
derivatives, schematically
\begin{align}
   \partial_\mua \; & \rightarrow \; D_\mua \, = \, \partial_\mua \,  + \, A_\mua \; .
   \label{MinCoup}
\end{align}
In addition to this replacement we will find various other couplings
to be necessary in the gauged theory in order to preserve gauge invariance and supersymmetry.
For extended supergravities the original global symmetry group is rather large and
there are various choices of subgroups that can consistently be gauged.
Gauge groups that result from flux compactifications of string theory are usually non-semi-simple, but rather have the form
of semi-direct products of various Abelian and non-Abelian factors. 

In this work we study $N=4$ (half-maximal) supergravities in four dimensions,
whose structure is fixed by the extended supersymmetry as soon as the number of vector multiplets is specified.
String compactifications of phenomenological relevance are mostly those that yield $N=2$ supersymmetry
in $d=4$, which is then spontaneously broken down to $N=1$ and eventually to $N=0$.
For the $N=4$ theories supersymmetry can be spontaneously broken as well and the theories can also
be truncated to theories with less supersymmetry. For example certain interesting $N=1$
K\"ahler potentials can be computed from the $N=4$ scalar potential
\cite{Derendinger:2004jn,Derendinger:2005ph,Derendinger:2006ed,Derendinger:2006hr}.
In addition to these four-dimensional theories we study gaugings of extended (maximal and half-maximal) supergravities
in various other space-time dimensions. These theories still have a string theory origin but are obviously less relevant from
a phenomenological point of view.

Nevertheless, there are good reasons to consider these extended supergravities.
Many aspects of string compactifications
are not yet fully understood and it is often useful to consider models that are more simple and more concise due to
the rigid structure of extended supergravity. For example non-geometric string compactifications can be better
understood in such a restricted context \cite{Hull:2004in,Dabholkar:2005ve,Hull:2006tp}.
Also the mathematical structure of these theories is interesting on its own. Maximal supersymmetry completely determines
the global symmetry group of the ungauged theory and exotic groups like the exceptional Lie groups ${\rm E}_n$
(and in $d=2$ the infinite dimensional affine Lie group ${\rm E}_9$) appear.
These global symmetry groups not only organize the structure of the ungauged
supergravity but also govern the possible gaugings.
Lie groups and their representation theory are therefore the
most important mathematical tools in this article.
In supergravities with less supercharges, group theory is still important,
but much more differential geometry is necessary, for example in the description of the scalar manifolds.
Nevertheless, the general lessons we learn from the extended supergravity theories
 (e.g.\ the form of the topological couplings, the possibility to derive duality equations from the Lagrangian, etc.)
can also be applied to theories with less supersymmetry, see for example \cite{deWit:2003ja} for the $d=3$ case.

A very different motivation to study maximal extended supergravities 
comes from the fact that string theory on particular target spaces
is believed to be dual to particular ordinary quantum field theories.
The prime example of this holographic principle is the AdS/CFT correspondence that relates IIB string theory
on an Anti-de Sitter background with four dimensional $N=4$ super-Yang-Mills theory\footnote{
We always denote by $N$ the number of supersymmetries, which is often referred to as ${\cal N}$.} \cite{Maldacena:1997re,Aharony:1999ti}.
From a supergravity perspective the fluctuations around the ${AdS_5\times S^5}$ background are described by
to the ${\rm SO}(6)$ gauged maximal
supergravity in $d=5$, which is obtained by a sphere reduction from ten dimensions and has a stable $AdS$ ground state
\cite{Gunaydin:1984qu}. 
Although the supergravity limit only accounts for a small subset of string states, it can be a very fruitful first approach to test
the duality conjecture. There are also more string backgrounds for which a holographic dual
is conjectured, all of which correspond to gaugings of extended supergravities.

\section{Outline of the paper}

We wish to construct the most general gaugings of extended supergravity theories such that supersymmetry
is preserved. To clarify the starting point of our construction we first introduce
the ungauged maximal and half-maximal supergravities in the next chapter. These theories are obtained from
torus reductions of eleven- and ten-dimensional supergravity. The general method of gauging these theories is then
presented in chapter \ref{ch:GenGauged}. The gaugings are parameterized by an embedding tensor, which is a tensor under 
the respective global symmetry group and subject to certain group theoretical constraints. The method of the embedding tensor
was first worked out for the three-dimensional maximal supergravities \cite{Nicolai:2000sc,Nicolai:2001sv} and subsequently
applied to extended supergravities in different dimensions \cite{deWit:2003ja,Nicolai:2001ac,dWST4,deWit:2004nw}.
We give a general account of this method and explain the tasks
and problems that have to be solved in its application. In particular, we describe the generic form of the general gauged 
Lagrangian.

The remaining chapters then demonstrate the implementation of this method to particular extended supergravities. 
The gaugings of four-dimensional half-maximal ($N=4$) supergravities are discussed in chapter \ref{ch:D4}. 
Since in $d=4$ vector fields can be dualized to vector fields there are subtleties in the description of the
general gauging. Already in the ungauged theory a symplectic frame needs to be chosen in order to give a Lagrangian
formulation of the theory. The global symmetry group is therefore only realized onshell. These problems can be resolved.
By using group theoretical methods we give a unified
description of all known gaugings, in particular of those originating from flux compactifications.
Also various new gaugings are found and we give the scalar potential and the Killing spinor equations for all of them, thus
laying the cornerstone for a future analysis of these theories. 
Closely related to our elaboration of these $d=4$ theories is the presentation of the gauged $d=5$ half-maximal
supergravities in chapter \ref{ch:D5}. We explicitly give the embedding of all five-dimensional gaugings into
the four-dimensional ones, which corresponds to a torus reduction from $d=5$ to $d=4$.

Chapter \ref{ch:D7} is devoted to the study of maximal supergravity in $d=7$.
In this case two-forms are dual to three-forms and the gauged theory combines all of them in
a tower of tensor gauge fields that transform under an intricate set of non-Abelian gauge transformations.
In this way we can present the general gauged theory and its supersymmetry rules. We then discuss particular 
gaugings, for example we find the ${\rm SO}(5)$, ${\rm CSO}(4,1)$ and ${\rm SO}(4)$ gaugings
that originate from (warped) sphere reductions from $D=11$, IIA and IIB supergravity, respectively.
In particular, the ${\rm SO}(4)$ gauging had not been worked out previously and gives rise to an important setup
for holography.

Finally, in chapter \ref{ch:D2} we apply the methods to study gaugings of $d=2$ maximal supergravity. 
The global symmetry group in $d=2$ is the affine Lie group ${\rm E}_{9(9)}$ which
in contrast to higher dimensions is  infinite dimensional.
This results in various technical and conceptual difficulties that have to be resolved in the description of these
gaugings. The parameters of the general gauging organize into one single tensor that transforms in the unique infinite
dimensional level one representation of ${\rm E}_{9(9)}$. 
In terms of this tensor the bosonic Lagrangian of the general gauging is given 
(except for the scalar potential)
and it is shown how the gaugings of the higher
dimensional maximal supergravities are incorporated in this tensor. We also find the ${\rm SO}(9)$ gauging
that originates from a warped sphere reduction of IIA supergravity.

Some of the results presented here
were already published previously~\cite{Samtleben:2005bp,Schon:2006kz}.

\chapter{Supergravity theories from dimensional reduction} \label{ch:ungauged}

In this chapter we explain how the maximal and half-maximal supergravities in dimension $d$ are obtained
from the unique $D=11$ supergravity and the minimal $D=10$ supergravity via dimensional reduction on a torus $T^{q}$,
$q=D-d$.
For simplicity we only consider bosonic fields and we focus our attention on how the respective global symmetry groups $G_0$
of the lower dimensional theories emerge. 
There is a vast literature dealing with the issues that are discussed in this chapter, 
and we do not try to give a comprehensive reference list here.
Overview articles for the supergravity theories are for example \cite{deWit:1998aq,deWit:2002vz,VanProeyen:2003zj} and
for the dimensional reduction of gravity and supergravity we refer to \cite{Coquereaux:1988ne,Cremmer:1997ct,Cremmer:1998px,Roest:2005zq}.

\section{Torus reduction of pure gravity}
\label{sec:TorusPure}

Let us first consider Einstein gravity on a $D$ dimensional manifold ${\cal M}_D$ with coordinates
$x^{\hat \mu}$, $\hat \mu=0 \ldots D-1$.
The metric $g_{\hat \mua \hat \mub}$ has Lorentzian signature $(-,+,+,\ldots,+)$
and its dynamic is described by the Einstein-Hilbert action
\begin{align}
   {\cal S}_{\text{EH}} &= \int d^D x \, {\cal L}_{\text{EH}} \; , &
   {\cal L}_{\text{EH}} &= \sqrt{-g} \left( R^{(D)} + {\cal L}_{\text{M}} \right)  \; ,
   \label{EHaction}   
\end{align}
where $g = \det(g_{\hat \mua \hat \mub})$, $R^{(D)}$ is the curvature scalar of
$g_{\hat \mua \hat \mub}$ and ${\cal L}_{\text{M}}$ describes additional matter, i.e.\ in the case of pure gravity
we have ${\cal L}_{\text{M}}=0$. The equations of motion are the Einstein equations
\begin{align}
   R^{\hat \mua \hat \mub} - \ft 1 2 \, R \, g^{\hat \mua \hat \mub} &\equiv G^{\hat \mua \hat \mub} 
      = T^{\hat \mua \hat \mub} 
      \equiv \frac 1 {\sqrt{-g}} \, \frac{\delta (\sqrt{-g} {\cal L}_{\text{M}})} {\delta g_{\hat \mua \hat \mub}} \; ,
   \label{EinsteinEq}      
\end{align}
where $R_{\hat \mua \hat \mub}$, $G_{\hat \mua \hat \mub}$ and $T_{\hat \mua \hat \mub}$ are the Ricci, Einstein and
energy-momentum tensor, respectively, and as usual indices are raised and lowered using the metric $g_{{\hat \mua \hat \mub}}$
and the inverse metric $g^{\hat \mua \hat \mub}$.

We want to dimensionally reduce this theory on a torus down to $d=D-q$ space-time dimensions,
i.e.\ we demand the $D$-dimensional manifold ${\cal M}_D$ to locally have the form ${\cal M}_D={\cal M}_d \times T^q$,
with ${\cal M}_d$ being a $d$ dimensional space-time manifold and $T^q$ being the $q$-dimensional torus.
We introduce coordinates $x^\mua$ on ${\cal M}_d$, $\mua=0\ldots d-1$, and coordinates $y^a$ on $T^q$,
$a=1\ldots q$, such that the metric on ${\cal M}_D$ can be written as\footnote{
In more geometric terms we only consider solutions to \eqref{EinsteinEq} that possess $q$
Killing vector fields
$\xi_a^{\hat \mua}$, $a=1\ldots q$, which shall be linearly independent at every point $x \in {\cal M}_D$.
In addition we demand the $\xi_a^{\hat \mua}$ to be mutually commuting. As a consequence
the manifold ${\cal M}_D$ is a principal bundle with structure group ${\rm U}(1)^q$
and base manifold ${\cal M}_d$ and is therefore locally of the form ${\cal M}_D={\cal M}_d \times T^q$.
One can then locally introduce coordinates
$(x^\mua,y^a)$ such that the Killing vector fields are given by
$\xi_a^{\hat \mua}=\partial x^{\hat\mua} / \partial y^a$, see e.g.\ \cite{Coquereaux:1988ne}.
Note that the Lie derivative in the direction $\xi_a^{\hat \mua}$ is then simply the partial
derivative wrt $y^a$.}
\begin{align}
   ds^2 &= g_{\hat \mua \hat \mub} \, dx^{\hat \mua} \, dx^{\hat \mub}  \nonumber \\
        &= \tilde g_{\mua\mub} \, dx^{\mua} \, dx^{\mub}
	   + \rho^{2/q} \, M_{ab} \, (dy^a + A_\mua^a dx^\mua) \, (dy^b + A_\mub^b dx^\mub)
   \label{RedMetric}
\end{align}
where $\tilde g_{\mua\mub}$, $A^a_\mua$, $\rho$ and $M_{ab}$ depend on $x^\mua$ but not on $y^a$.
The metric on ${\cal M}_d$ is $\tilde g_{\mua \mub}$ and the $A^a_\mua$ are the $n$ Kaluza-Klein
vector fields. The metric on $T^q$ has been split into the dilaton $\rho$ and the unimodular matrix
$M_{ab}$ (i.e.\ $\det M =1$). From a $d$-dimensional perspective these are
$q(q+1)/2$ scalar fields.

Plugging the Ansatz \eqref{RedMetric} into the Einstein-Hilbert action
\eqref{EHaction} yields the effective $d$-dimensional action
\begin{align}
   S_{\text{eff}} &= \int d^{d} x \, {\cal L_\text{eff}} \nonumber \\
    {\cal L}_{\text{eff}} &= e \rho R^{(d)}
              - \ft 1 4 \, e \rho^{1+2/q} \, M_{ab} A^a_{\mua \mub} A^{b \mua \mub} 
             - \ft 1 4 \, e \rho \, \tr(M^{-1} \partial_\mua M M^{-1} \partial^\mua M)
	   \nonumber \\ & \qquad  
	     + \frac{q-1} {q} \, e \, \rho^{-1} \, (\partial_\mua \rho) (\partial^\mua \rho) 
	     + e \rho {\cal L}_{\text{M}} \; ,
   \label{Seff1}	     
\end{align}
where $e=\sqrt{-\det \tilde g_{\mua\mub}}$ and
$A^a_{\mua\mub} = 2 \partial_{[\mua} A^a_{\mub]}$ are the Abelian field strengths
of the vector fields.
In order to find the usual Einstein-Hilbert term in the effective action one can perform
a Weyl-rescaling
of the metric, namely $\tilde g_{\mua\mub} \mapsto g_{\mua\mub} = \rho^{\alpha} \tilde g_{\mua\mub}$ with
$\alpha=-2/(d-2)$. Note that with a slight abuse of notation we now denote by $g_{\mua\mub}$
the lower dimensional metric.
The Weyl-rescaled effective Lagrangian reads
\begin{align}
    {\cal L}_{\text{eff}} &= e R^{(d)} 
             - \ft 1 4 e \rho^{[2/n+2/(d-2)]} M_{ab} A^a_{\mu \nu} A^{b \mu \nu} 
             - \ft 1 4 e \tr(M^{-1} \partial_\mu M M^{-1} \partial^\mu M)
	     \nonumber \\ & \qquad
	     +  \left( \frac{n-1} {n} - \frac{d-1}{d-2} \right)
	                 e (\rho^{-1} \partial_\mu \rho) (\rho^{-1} \partial^\mu \rho)
             + e \rho^{-d/(d-2)} {\cal L}_M \; .
   \label{Seff2}	     
\end{align}
In addition to the Einstein-Hilbert term we thus have kinetic terms for the Abelian vector fields and
for the scalars. We did not immediately incorporate the Weyl-rescaling into the Ansatz \eqref{RedMetric}
since in chapter \ref{ch:D2} we will deal with $d=2$, in which case a Weyl-rescaling is not possible.
We will then use the form \eqref{Seff1} of the effective action.

Let us now consider the symmetries of the effective actions \eqref{Seff1} and \eqref{Seff2}.
From the freedom of choosing arbitrary
coordinate systems on ${\cal M}_D$ there remains on the one hand the freedom to choose arbitrary coordinates on the space-time
${\cal M}_d$.
On the other hand for the internal manifold the only coordinate changes that are compatible with
the torus Ansatz are arbitrary changes of the origin and global linear transformations of the internal coordinates, i.e.\
\begin{align}
   y^a \,& \mapsto \, \lambda^{2/q} \, ( y^b + L^b(x) ) \Lambda_b{}^a \; ,   
   \label{ytrafo}
\end{align}
where $\lambda \in \mathbbm{R}$ is a constant rescaling factor, $\Lambda$ is a constant ${\rm SL}(q)$ matrix,
and $L \in \mathbbm{R}^n$ are $x$-dependent coordinate shifts. $L^a(x)$ describes the ${\rm U}(1)^q$ gauge symmetries of the
vector fields, i.e.\ $A_\mua^a \mapsto A_\mua^a + \partial_\mua L^a$.
$\lambda$ and $\Lambda$ act on the $d$-dimensional fields as
\begin{align}
    A_\mua^a &\mapsto A_\mua^b \Lambda_b{}^a \; , &
    M &\mapsto  \Lambda M \Lambda^T \; , &
   \rho &\mapsto \lambda \rho \; .
   \label{SymmAMr}
\end{align}
These are global ${\rm GL}(q)=\mathbbm{R}^+ \times {\rm SL}(q)$ transformations.
The vector fields transform in the vector representation of ${\rm SL}(q)$ while the scalars 
form an ${\rm SL}(q)/{\rm SO}(q)$ coset. To make this coset structure more transparent it is convenient to introduce
group valued representatives ${\cal V} \in {\rm SL}(q)$ via
\begin{align}
   M &= {\cal V} {\cal V}^T \; .
   \label{DefV1}
\end{align}
For given $M(x)$ the last equation only specifies ${\cal V}(x)$ up to arbitrary local ${\rm SO}(q)$ transformations from the right.
The global ${\rm SL}(q)$ transformations act linearly on ${\cal V}$ from the left, i.e.\ 
${\cal V}$ transforms as
\begin{align}
   {\cal V} \, & \mapsto \, \Lambda \, {\cal V} \, h(x)   \; , \qquad \Lambda \in {\rm SL}(q) \; , \quad h(x) \in {\rm SO}(q) \; .
   \label{SLnSymm}
\end{align}
The relation \eqref{DefV1} between ${\cal V}$ and $M$ is completely analogous to the relation between the vielbein and
the space-time metric. This is not merely accidental: considering the reduction Ansatz \eqref{RedMetric}
for the vielbein and not for the metric, one finds ${\cal V}$ to be a component of the
$D$-dimensional vielbein and
the local ${\rm SO}(q)$ symmetry then descends from the local Lorentz symmetry of the flat vielbein
indices.

In order to express the kinetic term in the Lagrangian in terms of ${\cal V}$ one introduces the scalar currents
\begin{align}
   P_\mua + Q_\mua &= {\cal V}^{-1} \partial_\mua {\cal V}  \; ,&
   P_\mua^T &= P_\mua \; , &
   Q_\mua^T &= - Q_\mua^T \; .
   \label{SLnPQ}
\end{align}
Note that $Q_\mua$ is ${\mathfrak so}(q)$ valued, i.e.\ it takes values in the compact part of ${\mathfrak sl}(q)$,
while $P_\mua$ takes values in the non-compact directions of ${\mathfrak sl}(q)$. 
Using these currents the kinetic term for $M$ can be written as
\begin{align}
   {\cal L}_{\text{kin}} &= - \ft 1 4 \, e \, \tr(M^{-1} \partial_\mu M M^{-1} \partial^\mu M) = - e \, \tr ( P_\mua P^\mua ) \; .
\end{align}
To summarize, we found that dimensional reduction of pure gravity on a torus $T^q$ yields a
$d$-dimensional theory which describes gravity coupled to $q$ Kaluza-Klein vector $A_\mua^a$, one dilaton $\rho$ and
scalars ${\cal V}$ that parameterize an ${\rm SL}(q)/{\rm SO}(q)$ coset. The global symmetry group is
${\rm GL}(q) = \mathbbm{R}^+ \times {\rm SL}(q)$.

\begin{figure}[h]
  \begin{center}
    \vspace{0.3cm}
    \epsfig{file=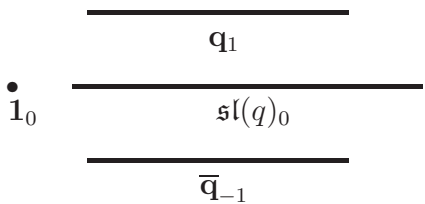,width=5.5cm}
    \caption{\label{Fig:D3}\small
                           Decomposition of $\mathfrak{sl}(q+1)$ under 
                           $\mathbbm{R}^+ \times {\rm SL}(q)$.
			   The subscripts denote the $\mathbbm{R}^+$ charges
			   which establish the vertical grading. At level $0$ one finds
			   the algebra of $\mathbbm{R}^+ \times {\rm SL}(q)$
			   while level $1$ and $-1$ contain an ${\rm SL}(q)$ vector and a
			   dual vector, respectively.
			   }
  \end{center}
\end{figure}

We now consider the particular case of $d=3$.
The Kaluza-Klein vector fields $A_\mua^a$ can then be dualized into scalars $A_a$ via the duality equation
\begin{align}
   \rho^{2+2/q} \, M_{ab} A^{b \mua \mub} &= \epsilon^{\mua\mub\muc} \partial_\muc A_a \; ,
   \label{DefDualA1}
\end{align}
where we use the covariant epsilon tensor, i.e.\  $\epsilon^{012}=e^{-1}$. 
The integrability condition for \eqref{DefDualA1} is given by the vector fields equation of motion
\begin{align}
   \partial_\mua ( e \rho^{1+2/q} \, M_{ab} A^{b \mua \mub} ) &= 0 \;.
\end{align}
Note that \eqref{DefDualA1} defines the scalars $A_a$ up to global shifts $A_a \mapsto A_a + \kappa_a$.
When formulating the theory without vector fields, i.e.\ entirely in terms of the metric and scalars,
these shift symmetries $\kappa_a$ become global symmetries, i.e.\ one expects ${\rm GL}(q) \ltimes \mathbbm{R}^q$
as global symmetry group. But a miraculous symmetry enhancement takes place and the complete global symmetry group
turns out to be $G_0={\rm SL}(q+1)$. 
Figure \ref{Fig:D3} shows the branching of
the Lie algebra of ${\rm SL}(q+1)$ under ${\rm GL}(q) = \mathbbm{R}^+ \times {\rm SL}(q)$.
For the resulting representations the dimensions are given as bold numbers and the subscripts denote the
charges under $\mathbbm{R}^+$.
The expected symmetry generators are ${\bf 1}_0$ (the generator of $\mathbbm{R}^+$), $\mathfrak{sl}(q)_0$ and 
${\bf  q}_{1}$ (the generators of the shift-symmetries $\kappa_a$). The symmetry enhancement yields the additional
generators $\overline {\bf q}_{-1}$, i.e.\ precisely those generators dual to the shift symmetries (in the
supergravity discussion we will find this to be a universal feature).

We now want to make the ${\rm SL}(q+1)$ symmetry explicit.
The scalars $\rho$, $A_a$ and ${\cal V}$ form an ${\rm SL}(q+1)/{\rm SO}(q+1)$
coset, the appropriate coset representative is defined as follows
\begin{align}
   \tilde {\cal V} &=  \begin{pmatrix} \rho^{-1} & 0 \\ \rho^{-1} A_a  & \rho^{1/q} {\cal V} \end{pmatrix}  \; .
   \label{SLn1tV}
\end{align}
Using the scalar current $\tilde P_\mua$ of $\tilde {\cal V}$, defined analogously to \eqref{SLnPQ}, 
the effective action takes the following compact form
\begin{align}
    {\cal L}_{\text{eff,d=3}} &= e \, R^{(3)}
             - e \, \tr(\tilde P_\mua \tilde P^\mua) + e \rho^{-3} {\cal L}_M \; .
   \label{Seff3}	     
\end{align}
The resulting equations of motion for $A_a$ are the integrability equations needed 
to reintroduce the vector fields $A^a_\mua$ via the duality equation \eqref{DefDualA1},
and by virtue of these duality equations all other equations of motion become equivalent to those derived from the previous Lagrangian
\eqref{Seff2}.

The ${\rm SL}(q+1)$ acts on $\tilde {\cal V}$ from the left, analogous to \eqref{SLnSymm},
the according ${\rm SL}(q+1)$ matrices read
\begin{align}
   \tilde \Lambda(\Lambda,\lambda) &=  \begin{pmatrix} \lambda^{-1/q} & 0 \\ 0  & \lambda^{1/q} \Lambda \end{pmatrix}  \; , &
   \tilde \Lambda(\kappa) &=  \begin{pmatrix} 0 & 0 \\ \kappa_a  & 0 \end{pmatrix}  \; , &
   \tilde \Lambda(\tau) &=  \begin{pmatrix} 0 & \tau^a \\ 0  & 0 \end{pmatrix} \; .
   \label{tlambda}
\end{align}
The transformations $\lambda$ and $\Lambda$ from \eqref{SymmAMr} correspond to $\tilde \Lambda(\Lambda,\lambda)$,
the shift symmetries $\kappa$ act via $\tilde \Lambda(\kappa)$, and the symmetry enhancement is described by
the additional ${\rm SL}(q+1)$ elements $\tilde \Lambda(\tau)$. 
Left action with $\tilde \Lambda(\tau)$ on the coset representative $\tilde {\cal V}$
destroys the block-form \eqref{SLn1tV},
and an appropriate ${\rm SO}(q+1)$ action is necessary to restore this
form. Therefore these new symmetry generators act highly nonlinear on the fields
$\rho$, $A_a$ and ${\cal V}$.

The pure gravity case we were just discussing already shows many universal features 
that we will re-encounter in
the following sections. In particular it is characteristic for maximal and half-maximal
supergravities that the scalars arrange in
the coset $G_0/H$, where $G_0$ is the global symmetry group and $H$ is its maximal compact
subgroup.
The formulation in terms of the coset representative ${\cal V}$ and the scalar currents
$P_\mua$ and $Q_\mua$ is used throughout the whole article.
Also the emergence of an enhanced symmetry group of the lower dimensional theory after
appropriate dualization of gauge fields is a characteristic that
will reappear in the following supergravity discussion. 
In the pure gravity case only vector gauge fields appear in the lower-dimensional theory, but
for the supergravities also higher rank $p$-form gauge fields are present and can be dualized. Symmetry
enhancement always takes place when the higher dimensional $p$-form fields give rise to scalar fields in the lower-dimensional theory.
We will make this explicit in the following section.

\section{Maximal supergravities from torus reductions} \label{sec:MaxTorus}

The unique supergravity theory in $D=11$ space-time dimensions contains as bosonic degrees
of freedom
the metric and a three-from gauge field $C_{\hat \mua \hat \mub \hat \muc}$ with
field strength
$G_{\hat \mua\hat\mub\hat\muc\hat\mud}=4 \partial_{[\hat\mua} C_{\hat\mub\hat\muc\hat\mud]}$
and gauge symmetry
$\delta C_{\hat\mua\hat\mub\hat\muc} = 3 \partial_{[\hat\mua} \Lambda_{\hat\mub\hat\muc]}$.
The bosonic part of the Lagrangian reads \cite{Cremmer:1978km}
\begin{align}
   {\cal L}_{D=11} &= \sqrt{-g} \left( R - 
     \ft 1 {12} \, G_{\hat\mua\hat\mub\hat\muc\hat\mud} \, G^{\hat\mua\hat\mub\hat\muc\hat\mud} 
     + \ft 2 {72^2} \, \epsilon^{\hat\mua\hat\mub\hat\muc\hat\mud\hat\mue\hat\muf\hat\mug\hat\muh\hat\mui\hat\muj\hat\muk} \,
       G_{\hat\mua\hat\mub\hat\muc\hat\mud} \, G_{\hat\mue\hat\muf\hat\mug\hat\muh}  \,
			              C_{\hat\mui\hat\muj\hat\muk}  \right) \; .
\end{align}
We dimensional reduce this theory on a torus $T^q$ down to $d=11-q$ dimensions, i.e.\
we make the Ansatz \eqref{RedMetric} for the metric and demand 
$C_{\hat\mua\hat\mub\hat\muc}$ to be constant along the torus coordinates $y^a$, i.e.\footnote{
The possibility to demand only the field strength to be constant along the internal
coordinates means to allow for a flux of the gauge field along the internal manifold.
These background fluxes yield gauged effective theories in $d$ dimensions.}
\begin{align}
   \frac{\partial} {\partial y^a} C_{\hat\mua\hat\mub\hat\muc} &= 0
   \label{RedAnsC}
\end{align}
In $d$ dimensions
the three-form then yields $q(q-1)(q-2)/6$ scalars $\chi_{[abc]}$, 
$q(q-1)/2$ vector gauge fields $B^{(1)}_{\mua [a b]}$,
$q$ two-form gauge fields $B^{(2)}_{\mua\mub a}$ and
one three-form gauge field $B^{(3)}_{\mua\mub\muc}$. The appropriate reduction Ansatz reads\footnote{
Under the projection $\pi:P \rightarrow M$ only vectors but not forms can be pushed
forward.}
\begin{align}
   \chi_{a b c} &= C_{a b c}  \; , &
   B^{(1)}_{\mua a b} &= \pi_\mua{}^{\hat \mua} \, C_{\hat\mua a b} \; , \nonumber \\               
   B^{(2)}_{\mua\mub a} &= \pi_\mua{}^{\hat \mua} \,\pi_\mub{}^{\hat \mub} \, 
                              C_{\hat\mua\hat\mub a} \; , &               
   B^{(3)}_{\mua\mub\muc} &= \pi_\mua{}^{\hat \mua} \, \pi_\mub{}^{\hat \mub} \,
                           \pi_\muc{}^{\hat \muc} \,  C_{\hat\mua\hat\mub\hat\muc} \; ,                
\end{align}
where
\begin{align}
   \pi_\mua{}^{\hat \mua} &= g_{\mua\mub} \, \frac{\partial x^\mub}{\partial x^{\hat \mub}} \,
                             g^{\hat \mub \hat \mua} \, .
\end{align}
If we identify $x^{\hat \mua}=(x^\mua,y^a)$ we have 
$\partial x^\mub / \partial x^{\hat \mub} = \delta^\mua_{\hat \mua}$ and thus find        
\begin{align}
   B^{(1)}_{\mua a b} &= C_{\mua a b} - A^c_{\mua} C_{a b c} \; , \nonumber \\
   B^{(2)}_{\mua\mub a} &= C_{\mua\mub a} - 2 A^b_{[\mua} C_{\mub] a b} 
                                          + A^b_{\mua} A^c_{\mub} C_{a b c} \; , \nonumber \\
   B^{(3)}_{\mua\mub\muc} &= C_{\mua\mub\muc} - 3 A^a_{[\mua} C_{\mub\muc]a} 
                                              + 3 A^a_{[\mua} A^b_{\mub} C_{\muc] a b}
					      - A^a_{\mua} A^b_{\mub} A^c_{\muc} C_{a b c} \; . &
   \label{Red3form}
\end{align}
The appearance of the Kaluza-Klein vector field $A^a_\mua$ ensures
that the forms $B^{(p)}$ do not transform under the gauge (coordinate)
transformations $L^a(x)$ that were introduced in \eqref{ytrafo}.
The forms $B^{(p)}$ and the scalars $\chi_{abc}$ transform under the torus ${\rm SL}(q)$
according to their index structure and are also charged under torus rescalings $\lambda$
under which also $\rho$ transform according to \eqref{SymmAMr}.
The field strengths of the forms $B^{(p)}$ are defined by\footnote{
This definition of the field strengths is motivated by dimensional reduction of the field strength
$G_{\hat\mua\hat\mub\hat\muc\hat\mud}$. Analogously to \eqref{Red3form} one has for example
$F^{(4)}_{\mua\mub\muc\mud} = G_{\mua\mub\muc\mud} + 4 A^a_{[\mua} G_{\mub\muc\mud]a} + \ldots$.
However, for $F^{(2)}_{\mua\mub a b}$ this would yield the natural definition
$2 \partial_{[\mua} B^{(1)}_{\mub] a b} +  A^c_{\mua\mub} \chi_{a b c}$ which we do not use
since otherwise scalar fields would appear in the definition of a field strengths.}
\begin{align}
    F^{(2)}_{\mua\mub a b} &= 2 \partial_{[\mua} B^{(1)}_{\mub] a b} \; , \nonumber \\
    F^{(3)}_{\mua\mub\muc a} &= 3 \partial_{[\mua} B^{(2)}_{\mub\muc] a}
                               + 3 A^b_{[\mua\mub} B^{(1)}_{\muc] a b}  \; ,	\nonumber \\		       
    F^{(4)}_{\mua\mub\muc\mud} &= 4 \partial_{[\mua} B^{(3)}_{\mub\muc\mud]}
                               + 6 A^a_{[\mua\mub} B^{(2)}_{\muc\mud] a}	\; .		       
\end{align}
The appropriate gauge transformations of the forms $B^{(p)}$ that leave these field strengths
invariant descend from those gauge transformations $\Lambda_{\hat\mua \hat \mub}$ of the
$D=11$ three-form that do not depend on the internal coordinates. But also a linear
dependence of $\Lambda_{\hat\mua \hat \mub}$ on the coordinates $y^a$ can be consistent with
the Ansatz \eqref{RedAnsC}, as long as it does not depend on
the space-time coordinates $x^\mua$.
Of these additional symmetries we are interested in the
particular case $\Lambda_{ab}= \kappa_{abc} \, y^c$, where $\kappa_{abc}$ has to be constant.
These three-from gauge transformations yield a global shift symmetry of
the scalars $\chi_{abc}$, but also act on the forms $B^{(p)}$ as follows
\begin{align}
   \delta \chi_{abc} &= \kappa_{abc} \; , &
   \delta B^{(1)}_{\mua a b} &= - \kappa_{abc} \, A_\mua^c \; , \nonumber \\
   \delta B^{(2)}_{\mua \mub b} &= A^b_{\mua} A^c_{\mub} \kappa_{a b c} \; , &
   \delta B^{(3)}_{\mua \mub \muc} &= - A^a_{\mua} A^b_{\mub} A^c_{\muc} \kappa_{a b c} \; .
   \label{ShiftChi}   
\end{align}
This is a global symmetry of the effective $d$-dimensional theory whose Lagrangian reads
\begin{align}
    {\cal L}_{\text{eff}} &= e R + {\cal L}^{(3)}_{\text{kin}} + {\cal L}^{(2)}_{\text{kin}}
                                 + {\cal L}^{(1)}_{\text{kin}} + {\cal L}^{(0)}_{\text{kin}}
   			         + e \rho^{-d/(d-2)} {\cal L}_{FFA}  \; ,
   \label{11dEffL}
\end{align}
where we have kinetic terms for the gauge fields and scalars
\begin{align}
   {\cal L}^{(3)}_{\text{kin}} &=  - \ft 1 {12} e \rho^{-1+6/(d-2)} F^{(4)}_{\mua\mub\muc\mud} F^{(4)\mua\mub\muc\mud} \; ,
   \nonumber \\
   {\cal L}^{(2)}_{\text{kin}} &= - \ft 1 {3} e \rho^{-1+4/(d-2)+2/q} M^{ab} F^{(3)}_{\mua\mub\muc a} F^{(3)\mua\mub\muc}_b  \; ,
   \nonumber \\
   {\cal L}^{(1)}_{\text{kin}} &= - \ft 1 4 e \rho^{18/(d-2)/q}
       \left( M_{ab} + 2 \rho^{(2-q)/q)} M^{ce} M^{df} \chi_{c d a} \chi_{e f b} \right) 
				  A^a_{\mua \mub} A^{b \mua \mub} 
  \nonumber \\ & \qquad \qquad 
  - e \rho^{-1+2/(d-2)+4/q} M^{ac} M^{bd} \chi_{abe} A_{\mua\mub}^e F^{(2)\mua\mub}_{c d}
  \nonumber \\ & \qquad \qquad \qquad \qquad
  - \ft 1 {2} e \rho^{-1+2/(d-2)+4/q} M^{ac} M^{bd} F^{(2)}_{\mua\mub a b} F^{(2)\mua\mub}_{c d} \; ,
  \nonumber \\ 
  {\cal L}^{(0)}_{\text{kin}} &= - \ft 1 4 e \tr(M^{-1} \partial_\mu M M^{-1} \partial^\mu M)
                                 - \ft {9}{(d-2)q} e (\rho^{-1} \partial_\mu \rho) (\rho^{-1} \partial^\mu \rho)
            \nonumber \\ & \qquad \qquad
              - \ft 1 {3} e \rho^{-1+6/q} M^{ad} M^{be} M^{cf}
                         (\partial_{\mua} \chi_{a b c}) (\partial^{\mua} \chi_{d e f}) \; ,
\end{align}
and ${\cal L}_{FFA}$ is a topological term that descends from the topological $GGC$-term in
$D=11$. The form of this term and also the further analysis depends on the particular
dimensions $d$ of the effective theory. In particular, the $p$-form gauge fields $B^{(p)}$
with field strengths $F^{(p+1)}$ can be dualized 
into $(d-p-2)$-form gauge fields $\tilde B^{(d-p-2)}$ with field strengths $\tilde F^{(d-p-1)}$.
The corresponding duality equation schematically reads
\begin{align}
   \tilde F^{(d-p-1)} &= \rho^x \, M \, * \, \left( F^{(p+1)} \, + \, \text{contributions from ${\cal L}_{FFA}$} \right) \; ,
   \label{DualityGen}   
\end{align}
where the asterisk denotes Hodge dualization and $\rho^x \, M$ indicates that some appropriate combination
of scalars is needed such that $F^{(d-p-1)}$ transforms dual to $F^{(p+1)}$ under ${\rm GL}(q)$. The duality 
equation is always such that the integrability equation is given by the equation of motion of the $p$-form.
A Lagrange formulation of the theory can then be given that contains $\tilde B^{(d-p-2)}$ instead of
$B^{(p)}$.
The ``standard'' formulation of the $d$-dimensional supergravity is obtained if those $p$-forms are dualized for
which $d-p-2 < p$, i.e.\ the rank of the gauge fields is minimized. In even dimensions there are $p$-form fields
with $d-p-2 = p$. Thus there is some freedom which of these $p$-form fields appear in the Lagrangian. For $d=4$ this
is the freedom of choosing a symplectic frame for the vector gauge fields ($p=1$).

\begin{figure}[tb]
\begin{minipage}{0.3\textwidth}
  \begin{center}
    \vspace{0.3cm}
    \epsfig{file=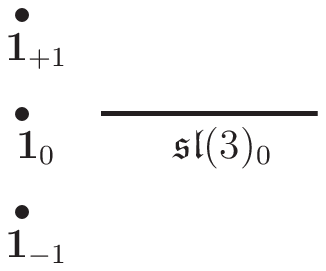,width=3cm}
  \end{center}
  \caption{\label{Fig:AlgD8}\small
                           Global symmetry in $d=8$: decomposition of $\mathfrak{sl}(2) \oplus \mathfrak{sl}(3)$ under 
                           $\mathbbm{R}^+ \times {\rm SL}(3)$.
			   }
\end{minipage}%
\hfill
\begin{minipage}{0.3\textwidth}
  \begin{center}
    \vspace{0.3cm}
    \epsfig{file=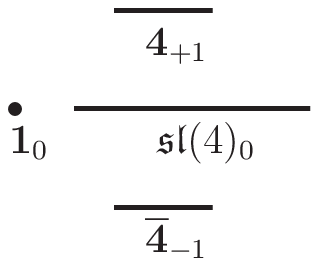,width=3cm}
  \end{center}
  \caption{\label{Fig:AlgD7}\small
                           Global symmetry in $d=7$: decomposition of $\mathfrak{sl}(5)$ under 
                           $\mathbbm{R}^+ \times {\rm SL}(4)$.
			   }
\end{minipage}%
\hfill
\begin{minipage}{0.3\textwidth}
  \begin{center}
    \vspace{0.3cm}
    \epsfig{file=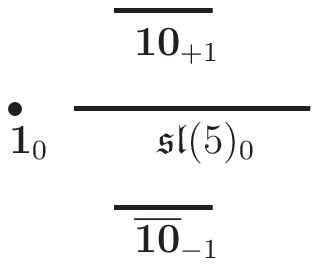,width=3cm}
  \end{center}
  \caption{\label{Fig:AlgD6}\small
                           Global symmetry in $d=6$: decomposition of $\mathfrak{so}(5,5)$
			   under $\mathbbm{R}^+ \times {\rm SL}(5)$.
			   }
\end{minipage}
\end{figure}

We are particularly interested in the
global onshell symmetry group of the effective theory.
From the torus reduction one expects an $\mathbbm{R}^+ \times {\rm SL}(q)$
symmetry group, where the $\mathbbm{R}^+$ factor corresponds to torus rescalings
$\lambda$.
For $d \leq 8$ the scalars $\chi_{abc}$ appear together with their shift symmetries
\eqref{ShiftChi}. Since the $\chi_{abc}$ are charged under torus rescalings $\lambda$
their shift-symmetries $\kappa_{abc}$ are as well, i.e.\ the action of $\kappa_{aba}$ does
not commute with the action of $\lambda$. In figures \ref{Fig:AlgD8} to \ref{Fig:AlgD6}
the symmetry generators for $3 \leq d \leq 6$ are depicted graphically. Again, the subscript at each
generator denotes its charge under torus rescalings and the vertical grading of the
generators corresponds to these charges. The generators of the
torus transformations are uncharged under $\lambda$ and denoted by
$\mathfrak{sl}(q)_0$, the generator of the torus rescalings itself is denoted
${\bf 1}_0$, and charge $+1$ is assigned to the shift symmetries $\kappa_{abc}$,
thus they are denoted ${\bf 1}_{+1}$, ${\bf 4}_{+1}$ etc. --- the number in
bold letters indicates their representation under ${\rm SL}(q)$. 

\begin{figure}[tb]
\begin{minipage}{0.3\textwidth}
  \begin{center}
    \vspace{0.3cm}
    \epsfig{file=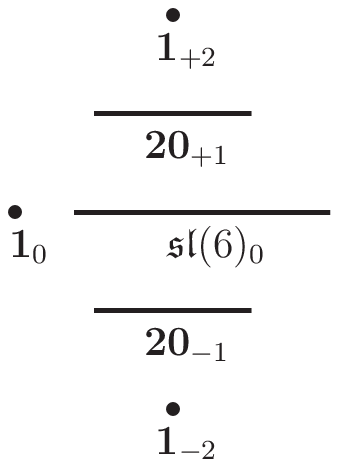,width=3.2cm}
  \end{center}
  \caption{\label{Fig:AlgD5}\small
                           Global symmetry in $d=5$: decomposition of $\mathfrak{e}(6)$ under 
                           $\mathbbm{R}^+ \times {\rm SL}(6)$.
			   }
\end{minipage}%
\hfill
\begin{minipage}{0.3\textwidth}
  \begin{center}
    \vspace{0.3cm}
    \epsfig{file=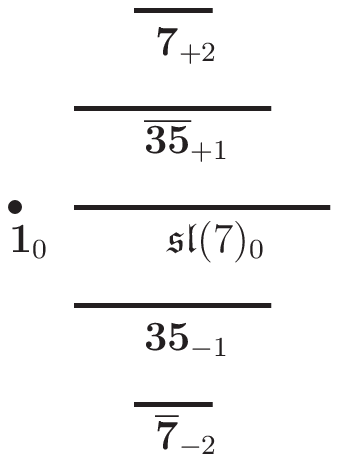,width=3.2cm}
  \end{center}
  \caption{\label{Fig:AlgD4}\small
                           Global symmetry in $d=4$: decomposition of $\mathfrak{e}(7)$ under 
                           $\mathbbm{R}^+ \times {\rm SL}(7)$.
			   }
\end{minipage}%
\hfill
\begin{minipage}{0.3\textwidth}
  \begin{center}
    \small
    \vspace{0.3cm}
    \epsfig{file=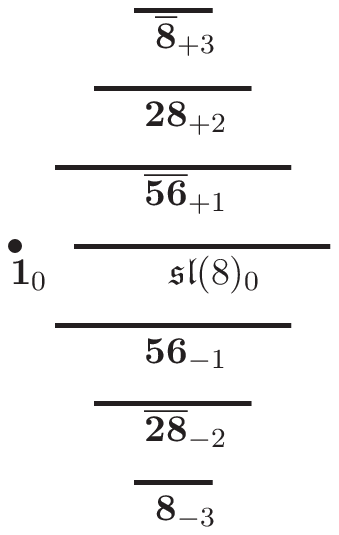,width=3.2cm}
  \end{center}
  \caption{\label{Fig:AlgD3}\small
                           Global symmetry in $d=3$: decomposition of $\mathfrak{e}(8)$ under 
                           $\mathbbm{R}^+ \times {\rm SL}(8)$.
			   }
\end{minipage}
\end{figure}

Similar to the above pure gravity case in $d=3$ the symmetry group becomes miraculously enhanced.
For each shift symmetry generator there also exists the dual
generator with negative charge under $\lambda$ and in the dual representation of
${\rm SL}(q)$. The global symmetry group $G_0$ of maximal supergravity turns out to be ${\rm SL}(2) \times {\rm SL}(3)$ for
$d=8$, ${\rm SL}(5)$ for $d=7$ and ${\rm SO}(5,5)$ for $d=6$. To prove this one
would have to show that the kinetic term of the scalars in \eqref{11dEffL} describes
the sigma model of the scalar cosets $G_0/H$ and that also the $p$-form gauge fields
arrange in representations of $G_0$ (after dualization) such that the field equations are $G_0$-invariant. 
In even dimensions there is the subtlety of self-duality, e.g.\ in $d=8$ the three-form $B^{(3)}$ forms an 
${\rm SL}(2)$ doublet together with its dual three-form, thus the whole global symmetry $G_0$
is not realized at the level of the Lagrangian but only at the level of the field equations.

For $d \leq 5$ additional scalars appear since according to \eqref{DualityGen} the forms $B^{(d-2)}$ can be dualized into
scalars.
As in the pure gravity case in $d=3$ these dual scalars also come equipped with
a shift symmetry. In figures \ref{Fig:AlgD5} to \ref{Fig:AlgD3} the generators of these
shift symmetries are denoted by ${\bf 1}_{2}$, ${\bf 7}_{2}$
and ${\bf 28}_{2}$. As before we also have the shift symmetries $\kappa_{abc}$, denoted by
${\bf 20}_{+1}$, $\overline{\bf 30}_{+1}$ and $\overline{\bf 56}_{+1}$ in the figures.
For $d=3$ also the Kaluza-Klein vector fields can be dualized into scalars
according to \eqref{DefDualA1}. 
Again, symmetry
enhancement takes place, i.e.\ for each shift symmetry generator also the dual symmetry
generator appears. This gives rise to the global symmetry group $G_0={\rm E}_{6(6)}$ in $d=5$, $G_0={\rm E}_{7(7)}$ in $d=4$,
and $G_0={\rm E}_{8(8)}$ in $d=3$.

\begin{figure}[tb]
\begin{minipage}{0.45\textwidth}
  \footnotesize
  \begin{center}
    \vspace{0.3cm}
    \epsfig{file=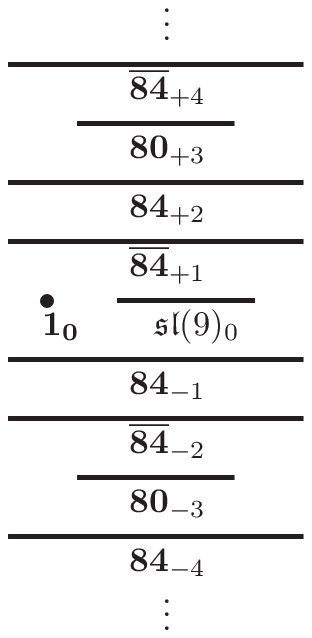,width=3cm}
  \end{center}
  \caption{\label{Fig:AlgD2a}\small
                           $d=2$, decomposition of $\mathfrak{e}(9)$ under ${\rm SL}(9)$.
			   }
\end{minipage}%
\hfill
\begin{minipage}{0.45\textwidth}
  \small
  \begin{center}
    \vspace{0.3cm}
    \epsfig{file=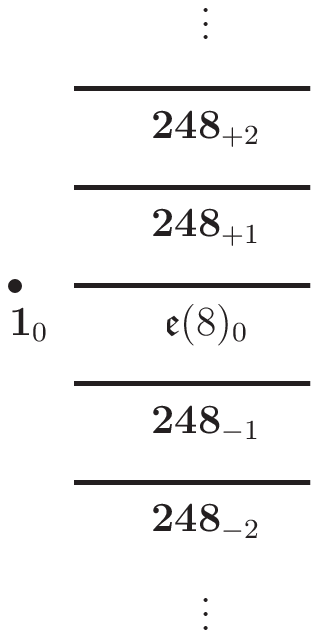,width=3cm}
  \end{center}
  \caption{\label{Fig:AlgD2b}\small
                           $d=2$, decomposition of $\mathfrak{e}(9)$ under $\mathbbm{R}^+ \times {\rm E}_{8(8)}$.
			   }
\end{minipage}
\end{figure}

For the case of $d=2$ we already mentioned that a Weyl-rescaling is not possible and thus the Lagrangian \eqref{11dEffL}
is not the appropriate starting point for the analysis. Nevertheless, the above discussion of the field content is
still applicable. From the three-form one only gets the scalars $\chi_{abc}$. The vectors and two-forms can be consistently set to
zero due to their field equations. The shift symmetries of the scalars are denoted ${\bf \overline{84}}_{+1}$ in figure
\ref{Fig:AlgD2a}, and again the corresponding dual symmetries ${\bf 84}_{-1}$ arise due to symmetry enhancement.
However, in $d=2$ scalars can be dualized to scalars and these new scalars can again be dualized, etc. This yields an infinite
tower of new scalars and thus of new shift symmetries. Accordingly, as depicted in figure \ref{Fig:AlgD2a},
an infinite symmetry enhancement
takes place. The global symmetry group is $G_0 = {\rm E}_{9(9)}$ which is the affine extension of ${\rm E}_{8(8)}$
\cite{Julia:1981wc}.
Thus, in contrast to higher dimensions
the on-shell symmetry group is infinite-dimensional in $d=2$. The symmetry algebra is an affine Lie algebra.

To understand why the affine extension of the $d=3$ symmetry group appears here we 
briefly consider the reduction of the $d=3$ maximal supergravity on a torus $T^1$ (i.e.\ on a circle).
This reduction yields the
${\rm E}_{8(8)}/{\rm SL}(16)$ coset of scalars in $d=2$. Onshell the dual scalars can be introduced, which transform
in the adjoint representation ${\bf 248}$ of ${\rm E}_{8(8)}$; these can be dualized again to find another ${\bf 248}$
scalars, etc. This gives an infinite stack of dual scalars and shift symmetries. Symmetry
enhancement yields also the dual symmetry generators, as shown in figure \ref{Fig:AlgD2b}. From this figure it is
quite intuitive that the loop group of ${\rm E}_{8(8)}$ appears as symmetry group in $d=2$.
The loop group also becomes centrally extended
to the affine extension ${\rm E}_{9(9)}$ of ${\rm E}_{8(8)}$.
Note that in figure \ref{Fig:AlgD2b} the charges that are indicated as subscripts correspond to the
($d=3 \rightarrow d=2$) torus rescalings $\lambda$.
These torus rescalings correspond to the generator $L_0$ of the Virasoro algebra
associated to ${\rm E}_{9(9)}$. In chapter \ref{ch:D2} we will come back to the $d=2$ theories and also give the ${\rm E}_{9(9)}$ symmetry
action explicitly. We will then also relate figures \ref{Fig:AlgD2a} and \ref{Fig:AlgD2b} by
explaining the appropriate embedding of the torus ${\rm GL}(9)$ into ${\rm E}_{9(9)}$.

\begin{table}[tb]
   \begin{center}
     \begin{tabular}{c|c@{\quad} c@{\quad} c@{\quad} c@{\quad} c@{\quad} c@{\quad} c@{\quad} c@{\quad} c}
        $d$ & \qquad $G_0$ & $H$ & $p=1$ & $p=2$ & $p=3$ & $p=4$ & $p=5$  \\ \hline
	      &&&&&&&& \\[-0.3cm]
        $8$ & ${\rm SL}(2) \times {\rm SL}(3)$ & ${\rm SO}(2) \times {\rm SO}(3)$ & 
	      ${\bf(2,3)}$ & ${\bf{(1,\overline 3)}}$ & ${\bf(2,1)}$ & ${\bf(1,3)}$ & ${\bf(2,\overline 3)}$  \\[0.08cm]
        $7$ & ${\rm SL}(5)$ & ${\rm SO}(5)$ & 
	          $\overline {\bf 10}$ & ${\bf 5}$ & $\overline {\bf 5}$ & ${\bf 10}$ & ${\bf 24}$ \\[0.08cm]
        $6$ & ${\rm SO}(5,5)$ & ${\rm SO}(5) \times {\rm SO}(5)$ & 
	         ${\bf 16_s}$ & ${\bf 10}$ & ${\bf 16_c}$ & ${\bf 45}$ & --- \\[0.08cm]
        $5$ & ${\rm E}_{6(6)}$ & ${\rm USp}(8)$ & ${\bf \overline{27}}$ & ${\bf 27}$ & ${\bf 78}$ & --- & --- \\[0.08cm]
        $4$ & ${\rm E}_{7(7)}$ & ${\rm SU}(8)$ & ${\bf 56}$ & ${\bf 133}$ & --- & --- & ---  \\[0.08cm]
        $3$ & ${\rm E}_{8(8)}$ & ${\rm SO}(16)$ & ${\bf 248}$ & --- & --- & --- & ---  \\[0.08cm]	
        $2$ & ${\rm E}_{9(9)}$ & $K({\rm E}_9)$ & --- & --- & --- & --- & --- 
     \end{tabular}
     \caption{\label{ListGH} \small For the maximal supergravities in $d$ dimensions the symmetry group $G_0$, its maximal
                                     compact subgroup $H$ and the representations of the $p$-form gauge fields ($p \leq 5$)
				     are listed.  For $d=8$ one also has a $6$-form that transforms as
				     ${\bf(3,1)} \oplus {\bf(1,8)}$. We also listed those $p$-forms (in $d=2$ scalars)
				     that can be introduced onshell via dualization, i.e.\ not all of the above fields carry independent
				     degrees of freedoms.
				      }
   \end{center}     
\end{table}

\begin{figure}[tb]
  \begin{center}
    \vspace{0.3cm}
    \epsfig{file=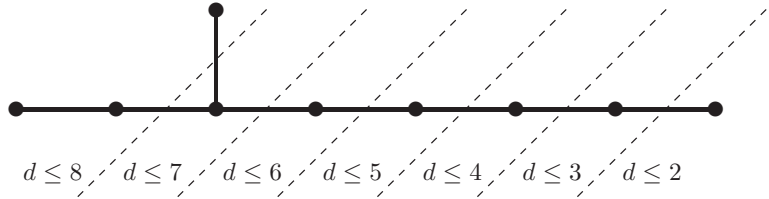,width=10cm}
  \caption{\label{Fig:Dynkin}\small
                           Dynkin diagram of $E_{11-d}$. For $d=8$ only the three knots on the left are present.
			   For every decrease in dimension one additional knot occurs, i.e.\ the rank increases by one.
			   Eventually, for $d=2$ one has $9$ knots and the above Dynkin diagram describes an affine Lie algebra.
			   }
  \end{center}
\end{figure}

In table \ref{ListGH} we summarize the symmetry groups $G_0$, the scalar cosets $G_0/H$ and
the representations of the $p$-form gauge fields for the maximal supergravities in $2\leq d \leq 8$.
The global symmetry group in dimensions
$2\leq d \leq 8$ turns out to be ${\rm E}_{q(q)}$, where $q=11-d$ is the dimension of the internal torus\footnote{We use the
common notation in denoting by ${\rm E}_q$ 
the complex Lie group (with rank $q$) and by ${\rm E}_{q(q)}$ the particular real form.
The number in brackets indicates the difference
between the number of compact and the number of non-compact generators of the real Lie algebra.
${\rm E}_{q(q)}$ is that real form with the maximal number of non-compact generators.}. The Dynkin diagrams
of the corresponding Lie algebras are depicted in figure \ref{Fig:Dynkin}. Note that the standard
notation for what we call $E_3$, $E_4$ and $E_5$ would be $A_1 \times A_2$, $A_4$ and $D_5$, but it is obviously very convenient to
depart from this in the present context.

All maximal supergravities that are obtained from $D=11$ supergravity are non-chiral,
i.e.\ there is an equal number of left- and right-handed supercharges in their supersymmetry algebra.
However, this distinction between left- and right-handed spinors only exists in $d=10$, $d=6$ and $d=2$. In all other dimensions
there are no Weyl-spinors and the maximal supergravities are unique, but also in $d=6$ and $d=2$ only the non-chiral
supergravities are of interest here, since for example in $d=6$ the chiral theories do not contain the metric in
their spectrum \cite{Townsend:1983xt,Hull:2000zn}. However, in ten dimensions the chiral IIB supergravity is as important
as the non-chiral IIA supergravity; each describes the low-energy limit of the corresponding string theory.

The bosonic Lagrangian of the IIA theory is given by \eqref{11dEffL} for $d=10$, the global symmetry group is only the
$\mathbbm{R}^+$ that corresponds to the circle rescalings. Note that in this case $B^{(1)}_{\mua ab}=0$ and $\chi_{abc}=0$,
thus the bosonic field content consists of the dilaton $\rho$, one Kaluza-Klein vector $A_\mua^a$, one two-form
$B^{(2)}_{\mua\mub a}$ ($a=1$) and one three-from $B^{(3)}_{\mua\mub\muc}$ --- of course, the corresponding dual forms can
also be introduced onshell.
In contrast, the IIB supergravity possesses a global ${\rm SL}(2)$ symmetry and its bosonic field
content consists of two scalars (the dilaton and the axion) that form an ${\rm SL}(2)/{\rm SO}(2)$ coset, two
two-form gauge fields that form a doublet under ${\rm SL}(2)$ and one self-dual four-form gauge field. 
We do not need the field equations of IIB supergravity here, but we want to mention that there is no complete
Lorentz invariant Lagrange formulation
of the theory since the self-duality condition of the four-form gauge fields always needs to be imposed as an extra constraint.

The existence of IIB supergravity and its ${\rm SL}(2)$ symmetry help to explain the
symmetry enhancement of the maximal lower dimensional supergravities, because all these theories
can also be obtained from torus reduction of IIB supergravity.
For $d=9$ the symmetry group of the maximal supergravity is
$G_0=\mathbbm{R}^+ \times {\rm SL}(2)$ and there is no symmetry enhancements, yet. Reduction
from $D=11$ explains $G_0$ as the symmetry group of the internal torus $T^2$,
and reduction from IIB supergravity explains it as the product of the ten-dimensional
${\rm SL}(2)$ symmetry and the $\mathbbm{R}^+$ (rescaling) symmetry of the internal circle.
But for $d=8$ one expects an $G_a=\mathbbm{R}^+ \times {\rm SL}(3)$ symmetry from $T^3$ reduction of $D=11$, and
an $G_b=\mathbbm{R}^+ \times {\rm SL}(2) \times {\rm SL}(2)$ symmetry from $T^2$ reduction of IIB. Neither
$G_a$ is a subgroup of $G_b$ nor vice versa, but they are both contained in the complete global symmetry group
$G_0={\rm SL}(2) \times {\rm SL}(3)$ and a careful analysis shows that $G_a$ and $G_b$ even generate this group.
The analysis for dimensions $d \leq 7$ is analogous.

So the miracle of symmetry enhancement is explained by the miracle of having different higher-dimensional ancestors
for the same effective theory. From a string theory perspective this is the miracle of $T$-duality 
($T$ refers to torus) which states
that IIA and IIB string theory are identical when compactified on a torus $T^q$, i.e.\ when the target manifold is of the
form ${\cal M}_d \times T^q$. Being identical means that they are just two different formulations of the same theory, and
this statements holds beyond the effective lower-dimensional supergravity, i.e.\ also for the whole tower of massive
string states. However, when the whole string theory is considered it turns out that the symmetry group is no longer
the real Lie group ${\rm E}_{11-d(11-d)}$, but only its discrete subgroup ${\rm E}_{11-d(11-d)}(\mathbbm{Z})$, which is referred to
as $U$-duality groups in a string theory context\footnote{
Note that the $T$-duality group ${\rm O}(10-d,10-d,\mathbbm{Z})$ is a subgroup of
${\rm E}_{11-d(11-d)}(\mathbbm{Z})$. $U$-duality combines $T$-duality with the $S$-duality of IIB supergravity.
From a low-energy perspective this $S$-duality is just the ${\rm SL}(2)$ symmetry of IIB supergravity, which in
string theory again is discretized to ${\rm SL}(2,\mathbbm{Z})$.}. One should keep in mind the duality origin of
these global symmetry groups, although here we will not pursue the string theory roots further.

\section{Half-maximal supergravities} \label{sec:HalfMaxUngauged}

We now want to depict the ungauged half-maximal supergravities, i.e.\ those with $Q=16$ real supercharges in their supersymmetry
algebra. Again we restrict the discussion to the non-chiral supergravities, thus avoiding subtleties in $d=6$ and
$d=2$ dimensions. Concerning the field content of half-maximal supergravity one does not have much freedom since
the bosonic and fermionic states have to arrange in multiplets of the supersymmetry
algebra. For the maximal $Q=32$ theories there is only one possible multiplet, the $Q=32$ gravity multiplet. 
For the half-maximal theories there are two types of multiplets: the $Q=16$ gravity and the $Q=16$ vector multiplet.
One gravity multiplet is always needed since it
contains the metric, but in addition there is the freedom of adding $m \in \mathbbm{N}$ vector multiplets.

The supersymmetry structure is still very rigid for $Q=16$. For the ungauged theory
there is only one way to consistently couple the vector multiplets to the gravity multiplet, i.e.\
the ungauged theory is completely determined when one specifies the dimension $d$ and the number
$m$ of vector multiplets. In dimensions $d \neq 6$ gaugings are the only known deformations of these half-maximal supergravities
that are compatible with supersymmetry. For $d=6$ one has the additional freedom to couple the vector fields
to the two-form gauge field of the gravitational multiplet via St\"uckelberg type couplings \cite{Romans:1985tw}.
We will introduce these type of couplings also for the gauged theories in all dimensions $d\geq4$,
but the difference in $d=6$ is that it can be switched on in addition to the gauging or without a gauging. 
This is analogous to IIA maximal supergravity in $d=10$, where also a massive deformation exists which is not a gauging
\cite{Romans:1985tz}.

The $Q=16$ supergravities in $d$ dimensions can be obtained from
torus reductions of $Q=16$ (i.e.\ $N=1$) supergravity in ten dimensions.
The $Q=16$, $d=10$ gravity multiplet contains as bosonic fields the metric, a scalar $\phi$ called the dilaton and
an antisymmetric two-form gauge field $B_{\mua\mub}$.
We label the vector multiplets by $i=1,\ldots,m$. Each vector multiplet contains only one vector gauge field $A^i_\mua$ 
as bosonic degrees of freedom. The bosonic Lagrangian in the Einstein frame reads
\cite{Chamseddine:1981ez,Bergshoeff:1981um,Chapline:1982ww}
\begin{align}
   {\cal L}_{D=10} &= \sqrt{-g} \left( R - \ft 9 8 (\phi^{-1} \partial_\mua \phi) (\phi^{-1} \partial^\mua \phi) 
                                         - \ft 3 2 \phi^{-4} {\cal G}_{\mua\mub\muc} {\cal G}^{\mua\mub\muc} 
					 - \ft 1 2 \phi^{-2} {\cal F}^i_{\mua\mub} {\cal F}^{i \, \mua\mub} \right) \; ,
   \label{Ln1d10}					 
\end{align}
where ${\cal F}^i_{\mua\mub}=2 \partial_{[\mua} A^i_{\mub]}$
and ${\cal G}_{\mua\mub\muc}= 3 \partial_{[\mua} B_{\mub\muc]} - A^i_{[\mua} F^i_{\mub\muc]}$ are the Abelian field strengths of the
vector-and two-form gauge fields. This Lagrangian is invariant under ${\rm SO}(m)$ rotations on the vector fields $A_\mua^i$
and under rescaling $\phi \mapsto \varphi \, \phi$,
$B_{\mua\mub} \mapsto \varphi^2 \, B_{\mua\mub}$ and $A^i_{\mua} \mapsto \varphi \, A^i_{\mua}$, where $\varphi \in \mathbbm{R}^+$.
Thus the global symmetry group is $G_0=\mathbbm{R}^+ \times {\rm SO}(m)$. On can deform the theory
by gauging a subgroup of $G_0$, using the $A_\mua^i$ as gauge fields. A particularly important example is the case
$m=496$ and a subgroup ${\rm E}_8 \times {\rm E}_8$ or ${\rm SO}(32)$ gauged. In these cases the appropriate deformation
of the Lagrangian \eqref{Ln1d10} describes the low energy limit of type I and heterotic string theory. But we continue
to consider the ungauged theory further.

When compactifying to $d=10-q$ dimensions the two form $B_{\mua\mub}$ yields one two-form, $q$ vector fields and
$q(q-1)/2$ scalars in the effective theory, while the vector fields $A_\mua^i$ yield $m$ vector fields and
$m \cdot q$ scalars. In total one thus obtains $n=m+q$ vector fields from the gauge fields of $D=10$. For $d\geq 4$ this is
also the number of vector multiplets one encounters in $d$ dimensions\footnote{
A linear combination of the Kaluza-Klein vector fields from the metric and of the vector fields from
the two-form $B_{\mua\mub}$ make up the vector fields in the $d$ dimensional $Q=16$ gravity multiplet.}. 
In other words, the $Q=16$ gravity multiplet in
$d+1$ dimensions always decomposes into one gravity and one vector multiplet in $d$ dimensions.
For $d=3$ an additional ``vector-multiplet'' appears since the dilaton from the metric 
can be dualized into a vector field and we then have $n=m+q+1$.

The analysis of the symmetry group $G_0$ of the effective theory is analogous
to the discussion in the last section, i.e.\ whenever a new scalar field appears 
it comes equipped with a shift symmetry and there is a symmetry enhancement by the generators dual to these
shift symmetries. This yields the global symmetry group
$G_0 =  \mathbbm{R}^+ \times {\rm SO}(q,n)$ for $5 \leq d \leq 9$. In $d=4$ also the
two-form can be dualized to a scalar and the symmetry group becomes enlarged to $G_0 = {\rm SL}(2) \times {\rm SL}(q,n)$.
Similarly, for $d=3$ the vector fields yield scalars via dualization such that the global symmetry group becomes
$G_0 = {\rm SO}(8,n)$,
and for $d=2$ the affine extension of the three-dimensional symmetry group is obtained.
As in the maximal case it turns out that the scalars always form a scalar coset $G_0/H$. The
respective maximal subgroups $H$ and the remaining bosonic fields are summarized in table \ref{ListGH2}.
Note that one obtains only $n \geq q$ (respectively $n\geq q+1$ for $d=3$) from torus reduction of $D=10$,
but there are $d$-dimensional theories for all numbers of vector multiplets $n \in \mathbbm{N}$.

\begin{table}[tb]
   \begin{center}
     \begin{tabular}{c|c@{\quad} c@{\quad} c@{\quad} c@{\quad} c@{\quad} c}
        $d$ & \qquad $G_0$ & $H$ & $p=1$ & $p=2$   \\ \hline
	      &&&&&& \\[-0.3cm]
        $8$ & $\mathbbm{R}^+ \times {\rm SO}(2,n)$ & ${\rm SO}(2) \times {\rm SO}(n)$ & 
	      ${\bf (2+n)_{+1/2}}$ & ${\bf{1}_{+1}}$   \\[0.08cm]
        $7$ & $\mathbbm{R}^+ \times {\rm SO}(3,n)$ & ${\rm SO}(3) \times {\rm SO}(n)$ & 
	          ${\bf (3+n)_{+1/2}}$ & ${\bf 1}_{+1}$ \\[0.08cm]
        $6$ & $\mathbbm{R}^+ \times {\rm SO}(4,n)$ & ${\rm SO}(4) \times {\rm SO}(n)$ & 
	         ${\bf (4+n)_{+1/2}}$ & ${\bf 1}_{+1} \oplus {\bf 1}_{-1}$  \\[0.08cm]
        $5$ & $\mathbbm{R}^+ \times {\rm SO}(5,n)$ & ${\rm SO}(5) \times {\rm SO}(n)$ 
               & ${\bf (5+n)_{+1/2} \oplus 1_{-1}}$ & ${\bf (5+n)_{-1/2} \oplus 1_{+1}}$ \\[0.08cm]
        $4$ & ${\rm SL}(2) \times {\rm SO}(6,n)$ & ${\rm SO}(6) \times {\rm SO}(n)$ 
	              & ${\bf (2,6+n)}$ & $\text{\bf adj}_{G_0}$
	                              \\[0.08cm]
        $3$ & ${\rm SO}(8,n)$ & ${\rm SO}(8) \times {\rm SO}(n)$ & $\text{\bf adj}_{G_0}$ & ---   \\[0.08cm]	
     \end{tabular}
     \caption{\label{ListGH2}{ \small For the $d$ dimensional half-maximal supergravities coupled to $n$ vector multiplets
                                      the symmetry group $G_0$, its maximal
                                     compact subgroup $H$ and the representations of the vector and two-form
				     gauge fields are listed. 
				     The subscripts at the representations denote the $\mathbbm{R}^+$ charges.
				     The $(d-2-p)$-forms always transform dual to the $p$-forms and
				     the $p=d-2$ forms always transform in the adjoint representation of $G_0$.
				     Note the respective dual forms can only be introduced onshell and only those fields
				     that appear in the ungauged Lagrangian carry degrees of freedom.
				     }}
   \end{center}     
\end{table}

\chapter[The general structure of gauged supergravities]{
The general structure of gauged supergravity theories} \label{ch:GenGauged}

In this chapter we start with some supersymmetric theory with global symmetry group $G_0$
and ask for the possible gaugings of this theory that are compatible
with supersymmetry, i.e.\ we demand the deformations of the theory not to break supersymmetry. Although the
answer to this question needs a case by case study, there exists a general technique to parameterize the deformations
via an embedding tensor~$\Theta$, which is a tensor under the global symmetry group $G_0$ and has to satisfy certain 
group theoretical constraints. Every single gauging breaks the global symmetry $G_0$ down to a local
gauge group $G \subset G_0$, but the set of all possible gaugings can be described in a $G_0$ covariant way by using $\Theta$.
This embedding tensor and the constraints it has to satisfy are introduced in the following section for an arbitrary
theory, and as far as possible we try to keep this generality in the remainder of this chapter. However, eventually we always
specialize to the maximal and half-maximal supergravities that were introduced in the last chapter.

\section{The embedding tensor} \label{sec:gen:emb}

We start from an ungauged supersymmetric theory with global symmetry group $G_0$.
The symmetry generators of the corresponding algebra $\mathfrak{g}_0$
are denotes $t_\sa$, $\sa=1,\ldots,\dim(\mathfrak{g}_0)$. They obey
\begin{align}
    [ t_\sa , t_\sb ] &= f_{\sa\sb}{}^\sc \, t_\sc \; ,
\end{align}
where $f_{\sa\sb}{}^\sc$ are the structure constants of $G_0$. 
Gauging the theory means to turn part of this global symmetry into a local one. In order to preserve gauge invariance
one needs to introduce minimal couplings of vector gauge fields, i.e.\
one replaces derivatives $\partial_\mua$ by covariant derivatives $D_\mua$.
The theory to start with contains vector fields $A_\mua^\Ma$ that transform in
some representation $V$ (indicated by the index $\Ma$) of the global symmetry group $G_0$.
These vector fields are ${\rm U}(1)$ gauge fields,
i.e.\ they do not only transform under $\mathfrak{g}_0$-transformations $L^\sa$, but also under
local gauge transformations $\Lambda^\Ma(x)$:
\begin{align}
   \delta_L A_\mua^\Ma &= - L^\sa \, t_{\sa\Mb}{}^\Ma \, A_\mua^\Mb \; , &
   \delta_\Lambda A_\mua^\Ma &= \partial_\mua \Lambda^\Ma \; .
\end{align}
In the covariant derivative of the
gauged theory these vector fields $A_\mua^\Ma$ need to be coupled to the $G_0$ symmetry generators $t_\sa$, i.e.~\cite{Nicolai:2000sc}
\begin{align}
   D_\mua &= \partial_\mua - \, g\, A_\mua^\Ma \, \Theta_\Ma{}^\sa \, t_\sa \; ,
   \label{GenCovDiv}
\end{align}
where $\Theta_\Ma{}^\sa$ is the so-called embedding tensor and $g \in \mathbbm{R}$ is the gauge coupling constant, which could
also be absorbed into $\Theta_\Ma{}^\sa$. The embedding tensor $\Theta_\Ma{}^\sa$ has to be real and appears
in \eqref{GenCovDiv} as a map
$\Theta \, : \, V \, \rightarrow \, \mathfrak{g}_0$. The image of this map defines the gauge group $G$ and 
the possible gauge transformations are parameterized by $\Lambda^\Ma(x)$. For example, a field $B_\Ma$ in the dual
representation $\overline V$ of the vector gauge fields transforms under $G$ as
\begin{align}
   \delta B_\Ma &= g \,\Lambda^\Mb \, \Theta_\Mb{}^\sa \, t_{\sa\Ma}{}^\Mc \, B_\Mc  
       = g \, \Lambda^\Mb \, X_{\Mb\Ma}{}^\Mc \, B_\Mc  \; . 
\end{align}
Here we introduced the gauge group generators $X_\Ma = \Theta_\Ma{}^\sa t_\sa$, which in the vector field
representation $X_{\Ma\Mb}{}^\Mc =  \Theta_\Ma{}^\sa \, t_{\sa\Mb}{}^\Mc$ take the role of
generalized structure constants for the gauge group $G$. Note that $X_{\Ma\Mb}{}^\Mc$
contains the whole information on $\Theta$ if the vector field representation is faithful.
The embedding tensor is not invariant under the global symmetry group $G_0$, but
to ensure the closure of the gauge group
and the gauge covariance of the following construction we demand $\Theta$ to be invariant under gauge transformations
$\delta = \Lambda^\Ma \delta_\Ma$, i.e.\
\begin{align}
   \delta_\Ma \Theta_\Mb{}^\sa &= g \Theta_\Ma{}^\sb 
               \left( t_{\sb\Mb}{}^\Mc \, \Theta_{\Mc}{}^\sa - f_{\sb\sc}{}^\sa \, \Theta_\Mb{}^\sc \right) = 0 \; .
   \label{QconGen1}
\end{align}
Equivalently one can demand the generators $X_{\Ma\Mb}{}^\Mc$ to be gauge invariant, and
the equation $\delta_\Ma X_{\Mb\Mc}{}^\Md = 0$ can be written as
\begin{align}
   [ X_\Ma , X_\Mb ] &= - X_{\Ma\Mb}{}^\Mc \, X_\Mc \; .
   \label{QconGen2}
\end{align}
This equation guarantees the closure of the gauge group and is the generalized Jacobi identity when evaluated in
the vector field representation. Note that the generators $X_{\Ma\Mb}{}^\Mc$
are generically not antisymmetric in the first two indices, and equation \eqref{QconGen2} only demands this antisymmetry
under projection with $X_\Mc$, that is with $\Theta_\Ma{}^\sa$. The two equivalent relations \eqref{QconGen1}
and \eqref{QconGen2} represent a quadratic constraint on $\Theta$. The embedding tensor has to satisfy this constraint
in order to describe a valid gauging.

In addition to this quadratic constraint a linear constraint on $\Theta$ is needed as well.
Eventually, it is supersymmetry which demands this linear constraint, but we will see in section \ref{sec:GenNonAVecTen}
that already the gauge invariance of the vector and tensor gauge field system yield at least parts
of this linear constraint. 
The embedding tensor transforms in the representation
$\overline V \otimes \mathfrak{g}_0 = \theta_1 \oplus \theta_2 \oplus \ldots \oplus \theta_n$,
where $\theta_i$, $i=1,\ldots,n$, are the irreducible components of the tensor product. The linear constraint
needs to be $G_0$ invariant. Thus, each irreducible component $\theta_i$ is either completely forbidden by the linear
constraint or not restricted at all\footnote{If two components $\theta_i$ and $\theta_j$ transform in the same $G_0$ representation,
a linear constraint of the form $\theta_i = \alpha \theta_j$, $\alpha \in \mathbbm{R}$, is possible as well. But one can
then form a linear combination $\theta'_i=\theta_i - \alpha \theta_j$ such that the linear constraint is again
of the form $\theta'_i = 0$. This happens, for example, for the maximal $d=8$ supergravities and for the half-maximal $d=5$
supergravities, see table \ref{LinCon1} and \ref{LinCon2}.},
i.e.\ there is a subset $S \subset \{ 1 , \ldots, n \}$ such that the linear constraint reads
\begin{align}
   \theta_i &= 0 \qquad \text{for all } i \in S \; .
   \label{LconGen}
\end{align}
This equation can be written as a projector equation $\mathbbm{P}_1 \Theta = 0$,
where $\mathbbm{P}_1$ projects onto those representation
in $\Theta$ that are forbidden. Similarly, the quadratic constraint can be written as $\mathbbm{P}_2 (\Theta \otimes \Theta) = 0$, where
$\mathbbm{P}_2$ projects on the appropriate representation in the symmetric tensor product of $\overline V \otimes \mathfrak{g}_0$.
One could
also imagine higher order constraints like $\mathbbm{P}_3(\Theta \otimes \Theta \otimes \Theta)=0$,
but it turns out that the linear and quadratic constraint are sufficient for the construction of the gauged theories.

For the maximal and half-maximal supergravities the global symmetry group $G_0$ and the representation $V$ of the vector fields 
were given in tables \ref{ListGH} and \ref{ListGH2}. For the known cases we collected the linear constraint 
in tables \ref{LinCon1} and \ref{LinCon2}. For the maximal theories a similar table was given in \cite{deWit:2002vt,deWit:2003hr}.
For the odd dimensions, i.e.\ $d=3$, $d=5$ and $d=7$, the maximal gauged theories were worked out explicitly
\cite{Nicolai:2001sv,deWit:2004nw,Samtleben:2005bp}, but via torus reduction one can infer the linear constraint for the 
even dimensions $d\geq 4$ as well --- in appendix \ref{app:DimRedEmb} this is explained in detail. 
By applying the methods of \cite{deWit:2005ub} one can also describe explicitly the general gaugings
of maximal $d=4$ supergravity \cite{dWST4}. The maximal theories for $d=8$ and $d=6$ were not yet worked out completely\footnote{
For the $d=8$ theories
there is a classification of the gaugings that does not use the embedding tensor but the Bianchi classification
of three-dimensional group manifolds \cite{Bergshoeff:2004vs}.
In this classification the possible gaugings are parameterized by a ${\bf 3}$ and a ${\bf 8}$ of ${\rm SL}(3)$,
which are only a subset of the complete embedding tensor.
We would thus expect that there are more general gaugings of $d=8$ maximal supergravity.}.
The maximal $d=2$ theory will be considered in chapter \ref{ch:D2}.
For the half-maximal theories we refer to \cite{Nicolai:2001ac,deWit:2003ja,Schon:2006kz} and to
chapters \ref{ch:D4} and \ref{ch:D5}.

\begin{table}[tb]
   \begin{center}
     \begin{tabular}{c|c@{\quad}l@{\quad$=$\quad}l@{\quad$\oplus$\quad}l}
        $d$ & $G_0$ & $\mathfrak{g}_0 \otimes \overline{V}$ & allowed & forbidden \\ \hline
	  \\[-0.3cm]
        $8$ & ${\rm SL}(2) \times {\rm SL}(3)$ & 
		   $\mathfrak{g}_0  \otimes {\bf(2,\overline{3})}$ &
		   ${\bf(2,\overline 3)} \oplus {\bf(2,6)}$ & 
		   ${\bf(2,\overline{3})} \oplus {\bf(2,\overline{15})} \oplus {\bf(4,\overline{3})}$
	  \\[0.08cm]
	$7$ & ${\rm SL}(5)$ & 
	               ${\bf 24} \otimes {\bf 10}$ &
	                ${\bf 15} \oplus \overline{\bf 40}$  &  ${\bf 10} \oplus {\bf 175}$ 
          \\[0.08cm]        
	$6$ & ${\rm SO}(5,5)$ & 
		     ${\bf 45} \otimes {\bf 16}_c$ & ${\bf 144}_s$  &  ${\bf 16}_c \oplus {\bf 560}_c$
          \\[0.08cm]        
	$5$ & ${\rm E}_{6(6)}$ & 
		  ${\bf 78} \otimes {\bf 27}$ & $\overline {\bf 351}$  &  ${\bf 27} \oplus {\bf 1728}$ 
          \\[0.08cm]        
	$4$ & ${\rm E}_{7(7)}$ & 
	               ${\bf 133} \otimes {\bf 56}$ &
	                ${\bf 912}$  &  ${\bf 56} \oplus {\bf 6480}$
          \\[0.08cm]        
	$3$ & ${\rm E}_{8(8)}$ & 
	                 ${\bf 248} \otimes {\bf 248}$ &
	                ${\bf 1}\oplus{\bf 3875}$  &  ${\bf 248} \oplus {\bf 27000} \oplus {\bf 30380}$ 			      
          \\[0.08cm]        
	$2$ & ${\rm E}_{9(9)}$ & 
	                 $\mathfrak{g}_0 \otimes \text{\bf basic}$ &
	                {\bf basic}  &  rest
     \end{tabular}
     \caption{\label{LinCon1}{ \small Decomposition of $\Theta$ for the $d$ dimensional maximal supergravities.
                                      The linear constraint only allows some of the irreducible components of $\Theta$. 
				      For $d=8$ we have $\mathfrak{g}_0={\bf(3,1)} \oplus {\bf(1,8)}$.
				      For $d=2$ the algebra is the affine extension of $\mathfrak{e}_{8(8)}$ and the vector fields transform
				      in the unique level one representation, called the basic representation 
				      --- see chapter \ref{ch:D2}.}}
   \end{center}     
\end{table}

\begin{table}[tb]
   \def\putX#1{  \setbox0=\hbox{#1}
                   \hspace{-5mm}\begin{minipage}{\wd0}\begin{center}\vspace{-1.5mm}#1\end{center}\end{minipage}\hspace{-1.5mm}}
   \begin{center}
     \begin{tabular}{c|c@{\,$\otimes$\,}c@{\;$=$\;}l@{\;\;$\oplus$\;\;}l}
        $d$ & $\mathfrak{g}_0$ & $\overline{V}$ & ~allowed & forbidden  \\ \hline
	 \\[-0.3cm]
	$5$ &    	$\Big({\bf 1}_0 \; \oplus \;$ \putX{$\tinyyoung{\cr\cr}_{\,0}$} $\Big)$&
	                $\Big({\bf 1}_{1} \; \oplus \;$ \putX{ \vspace{1mm}$\tinyyoung{\cr}_{\,-1/2}$}$\Big)$ & 
			~\putX{ \vspace{1mm}$\tinyyoung{\cr}_{\,-1/2}$}
			$\oplus\,$
			\putX{ \vspace{1mm}$\tinyyoung{\cr\cr}_{\,1}$}
			$\oplus\,$
			\putX{ \vspace{1mm}$\tinyyoung{\cr\cr\cr}_{\,-1/2}$} &
			${\bf 1}_1 \oplus\,$
			\putX{ \vspace{1mm}$\tinyyoung{\cr}_{\,-1/2}$}
			$\oplus\,$
			\putX{ \vspace{1mm}$\tinyyoung{&\cr\cr}_{\vspace{0.2cm}\hspace{-0.2cm}-1/2}$}			
        \\[0.55cm]
	$4$ &    	$\hspace{-0.2cm}\big[({\bf 3},\cdot)\oplus
	                ({\bf 1},$ \putX{ $\tinyyoung{\cr\cr}$ } $)\big]$ &	
	                $({\bf 2},$ \putX{ $\tinyyoung{\cr}$ } $)$ & 
			$({\bf 2},$ \putX{ $\tinyyoung{\cr}$ } $)$
			$\oplus$
			$({\bf 2},$ \putX{ $\tinyyoung{\cr\cr\cr}$ } $)$ &
			$({\bf 2},$ \putX{ $\tinyyoung{\cr}$ } $)$
			$\oplus$
			$\ldots$			
        \\[0.55cm]
	$3$ &           \putX{ $\tinyyoung{\cr \cr}$ }  &
	 		\putX{ $\tinyyoung{\cr \cr}$ }  &
                        {\bf 1}
 			$\;\oplus\;$
			\putX{ $\tinyyoung{&\cr}$ }
			$\;\oplus\;$
			\putX{ $\tinyyoung{\cr\cr\cr\cr}$ }  &  
			~\putX{ $\tinyyoung{\cr\cr}$ }
			$\;\oplus\;$
			\putX{ $\tinyyoung{&\cr&\cr}$ }
			$\;\oplus\;$
			\putX{ $\tinyyoung{&\cr\cr\cr}$ }
     \end{tabular}
     \caption{\label{LinCon2}{ \small Decomposition of $\Theta$ for the $d$ dimensional half-maximal supergravities.
                                      The linear constraint only allows some of the irreducible components of $\Theta$.
				      For $d=5$ the global symmetry group $G_0$ is $\mathbbm{R}^+ \times {\rm SO}(5,n)$,
				      and the $\mathbbm{R}^+$ charges are given as subscripts. For $d=4$ and
				      $d=3$ we have $G_0={\rm SL}(2)\times{\rm SO}(6,n)$ and $G_0={\rm SO}(8,n)$,
				      respectively, where $n$ is the number of vector multiplets. The Yang-tableaus
				      always refer to the respective ${\rm SO}$-group. }}
   \end{center}     
\end{table}

It should be mentioned that table \ref{LinCon1} and \ref{LinCon2} reflect our present knowledge of the methods
that can be used to work out the general gauged theories. It is not impossible that
a weaker linear constraints might suffice, if new methods are applied in the future. In this respect the
linear constraint is less robust than the quadratic one, which can immediately be traced back to gauge invariance and
closure of the gauge group.

We summarize this section. When describing the general gauging of a supersymmetric theory, the embedding tensor~$\Theta$ can be used to
parameterize the gauging. Any $\Theta$ that satisfies the appropriate linear constraint \eqref{LconGen} and the quadratic
constraint \eqref{QconGen1} describes a valid gauging and the construction of the gauged theory only requires these constraints
for consistency. When $\Theta$ is treated as a spurionic object, i.e.\ it transforms under the global symmetry group $G_0$,
one does formally preserve the $G_0$ symmetry in the gauged theory. This reflects the fact that the set of all possible gaugings
is $G_0$ invariant. But as soon as a particular gauging is considered, the embedding tensor that describes this
gauging breaks the $G_0$ invariance down to the gauge group $G \subset G_0$.

\section{Non-Abelian vector and tensor gauge fields} \label{sec:GenNonAVecTen}

In this section we mainly present the results of~\cite{deWit:2005hv} on the general form of vector/tensor gauge transformations 
in arbitrary space-time dimensions, but translated into a more convenient basis, see also the appendix of \cite{Samtleben:2005bp}.

\subsection{Gauge transformations and covariant field strengths} 
\label{subsec:gaugetrafo}

First, we want to introduce the gauge transformations and covariant field strengths for 
the $p$-form gauge fields that appear in gauged supergravity theories. 
Explicitly we will give all formulas for rank $p \leq 3$, but in principle the construction can be continued to arbitrary rank.
It will turn out that always the $(p+1)$-forms are needed to define a gauge invariant
field strengths for the $p$-forms.
In the next subsection we will explain how to truncate this tower of gauge fields to a finite subset without loosing gauge invariance.

In the ungauged theory we have (at least onshell)
vector gauge fields $A_\mua^\Ma$, two-form gauge fields $B_{\mua\mub\,I}$, three form-gauge fields $S^A_{\mua\mub\muc}$, etc.,
and all these fields come in possibly different representation of the global symmetry group $G_0$, indicated by the indices
$\Ma$, $I$ and $A$.  The Abelian field strengths of these tensor gauge fields take the form
\begin{align}
   {\cal F}^{0,\Ma}_{\mua\mub} &= 2 \, \partial_{[\mua} \, A^\Ma_{\mub]}
        \nonumber \\
   {\cal F}^0_{\mua\mub\muc\,I} &= 3 \, \partial_{[\mua} \, B_{\mub\muc] I} 
                     + 6 \, d_{I \Ma\Mb} \, A_{[\mua}^\Ma \, \partial_\mub \, A_{\muc]}^\Mb  \; ,
          \nonumber \\
   {\cal F}^{0,A}_{\mua\mub\muc\mud} &= 4 \, \partial_{[\mua}  \, S^A_{\mub\muc\mud]}
                                     - c^{AI}_\Ma \, \left( 12 \, B_{[\mua\mub\,I} \, \partial_{\muc} A^\Ma_{\mud]}
	              + 8 \, d_{I \Mb\Mc} \, A^\Ma_{[\mua} \, A^\Mb_\mub \, \partial_{\muc} \, A^\Mc_{\mud]} \right) \; ,
   \label{DefFFF}		      
\end{align}
where $d_{I\Ma\Mb}$ and $c^{AI}_\Ma$ are some appropriate $G_0$-invariant tensors.
To ensure invariance under the Abelian gauge transformations these tensors have to satisfy
\begin{align}
   d_{I[\Ma\Mb]} &= 0 \; , &   d_{I (\Ma\Mb} \, c^{AI}_{\Mc)} &= 0 \; .
   \label{ConstraintsDC}
\end{align}
The existence of $d_{I\Ma\Mb}=d_{I(\Ma\Mb)}$ means that the two-fold symmetric
tensor product of the representation of the vector-fields $A_\mua^\Ma$ contains the representation of the two-form fields
$B_{\mua\mub\,I}$. Similarly, the existence of $c^{AI}_\Ma$ means that the representation of $S_{\mua\mub\muc}^A$ is contained
in the tensor product of the representations of $A_\mua^\Ma$ and $B_{\mua\mub\,I}$.
Using table \ref{ListGH} one can easily check that these conditions are satisfied for the maximal
supergravities. The second equation in \eqref{ConstraintsDC} also holds since the three-fold symmetric tensor
product of the vector field-representation never contains the representation of ${\rm S_{\mua\mub\muc}^A}$.

We saw in chapter \ref{ch:ungauged} that in dimensional reduction of $D=11$ supergravity
additional terms $A \partial A$, etc., appear naturally in the field strength of the higher rank tensor fields.
From a lower dimensional perspective these terms are very important for
anomaly cancellation, and therefore always present. 
Using the relations \eqref{ConstraintsDC} one can show that under arbitrary variations
$\delta A_\mua^\Ma$, $\delta B_{\mua\mub\,I}$ and $\delta S^A_{\mua\mub\muc}$ the field strengths vary as
\begin{align}
   \delta\,{\cal F}_{\mua\mub}^{0,\Ma} &=  2\,\partial_{[\mua}\, (\Delta A_{\mub]}^{\Ma})  \; , \nonumber\\
   \delta\,{\cal F}^0_{\mua\mub\muc\,I} &= 3\, \partial_{[\mua} (\Delta B_{\mub\muc]\,I})
                        + 6 \, d_{I\Ma\Mb} \, {\cal F}_{[\mua\mub}^{0,\Ma} \, \Delta A_{\muc]}^{\Mb} \; , \nonumber \\
   \delta\,{\cal F}^{0,A}_{\mua\mub\muc\mud} &= 4\, \partial_{[\mua} (\Delta S^A_{\mua\mub\muc]})
                        - 6 \, c^{AI}_{\Ma} \, {\cal F}^{0,\Ma}_{[\mua\mub} \, \Delta B_{\muc\mud]\,I}
			+ 4 \, c^{AI}_{\Ma} \, {\cal F}^{0}_{[\mua\mub\muc\,I} \, \Delta A^\Ma_{\mud]} \; ,
   \label{VaryFFF}			
\end{align}
where we used the ``covariant variations''
\begin{align}
   \Delta A_\mua^\Ma &=  \delta A_\mua^\Ma \; , \nonumber \\
   \Delta B_{\mua\mub\,I} &= \delta B_{\mua\mub\,I} - 2 d_{I\Ma\Mb} \, A_{[\mua}^\Ma \, \delta A_{\mub]}^\Mb \; , \nonumber \\
   \Delta S_{\mua\mub\muc}^A &= \delta S_{\mua\mub\muc}^A - 3 \, c^{AI}_\Ma \, B_{[\mua\mub\,I} \, \delta A^\Ma_{\muc]} 
                                 - 2 \, c^{AI}_\Ma \, d_{I\Mb\Mc} \, A^\Ma_{[\mua} \, A^\Mb_\mub \, \delta A^\Mc_{\muc]} \; .
   \label{CovD1}				 
\end{align}
These covariant variations are very useful since they allow to express gauge transformations and 
variations of gauge invariant objects in a manifestly covariant form, i.e.\ without explicit appearance of gauge fields.

We now ask for the appropriate generalization of \eqref{DefFFF} in the gauged theory.
The gauge group generators $X_{\Ma\Mb}{}^\Mc$ were already introduced in the last section, and according to equation \eqref{QconGen2}
they take the role of generalized structure constants. Therefore, it would be natural to define the non-Abelian
field strength of the vector fields $A_\mua^\Ma$ as
\begin{align}
   {\cal F}^\Ma_{\mua\mub} &= 2 \, \partial_{[\mua} \, A^\Ma_{\mub]} + g X_{\Mb\Mc}{}^\Ma \, A_{[\mua}^\Mb \, A_{\mub]}^\Mc \;,
\end{align}
but under gauge transformations $\delta A_\mua^\Ma = D_\mua \Lambda^\Ma$ one finds this field strength to transform as
\begin{align}
   \delta {\cal F}^\Ma_{\mua\mub} &=    2 \, D_{[\mua} \delta A^\Ma_{\mub]}
                                \, = \, 2 \, D_{[\mua} D_{\mub]} \Lambda^\Ma
				\, = \, g \, {\cal F}^\Mb_{\mua\mub} \, X_{\Mb\Mc}{}^\Ma \, \Lambda^\Mc \nonumber \\
				  &= - g \, \Lambda^\Mb \, X_{\Mb\Mc}{}^\Ma \, {\cal F}^\Mc_{\mua\mub} 
				     + 2 \, g \, \Lambda^\Mb \, X_{(\Mb\Mc)}{}^\Ma \, {\cal F}^\Mc_{\mua\mub} 
   \label{Ftransforms}
\end{align}
where we used the Ricci identity $[D_\mua,D_\mub] = -g {\cal F}_{\mua\mub}^{\Ma} X_\Ma$, which is valid due to the quadratic
constraint, see also \cite{deWit:2004nw}. Here and in the following we use the covariant derivative as defined in \eqref{GenCovDiv}.
In the second line of equation \eqref{Ftransforms} the first term alone would describe the
correct covariant transformation of the field strength, but there is an unwanted second term
since the $X_{\Mb\Mc}{}^\Ma$ are typically not antisymmetric in the first two indices. Thus the field strengths
${\cal F}_{\mua\mub}^{\Ma}$ does not transform covariantly under gauge transformations.

This problem arises because the dimension of the gauge group $G$ can be smaller than the number of Abelian vector fields $A_\mua^\Ma$, 
and thus not all vector fields are really needed as gauge fields. For any particular gauge group one could split the
vector fields into the gauge fields and the remainder and treat them differently in the gauged theory. Those vector fields that
are neither used as gauge fields for $G$ nor are sterile under $G$ have to be absorbed into massive two-forms.
But an explicit split of the vector fields is not appropriate
for our purposes since we search for a general formulation valid for all allowed embedding tensors.
To solve this problem one introduces a covariant field strength of the vector fields that contains 
St\"uckelberg type couplings to the two-form gauge fields, i.e.\
\begin{align}
   {\cal H}^\Ma_{\mua\mub}  &= 2 \, \partial_{[\mua} \, A^\Ma_{\mub]} + g X_{\Mb\Mc}{}^\Ma \, A_{[\mua}^\Mb \, A_{\mub]}^\Mc
                               + g Z^{\Ma I} B_{\mua\mub\,I} \; .
   \label{covAH}
\end{align}
The tensor $Z^{\Ma I}$ should be such that the unwanted terms in \eqref{Ftransforms} can be absorbed into an
appropriate gauge transformation of the two-form gauge fields. Explicitly, $\delta B_{\mua\mub\,I}$ should
contain a term $(- 2 \, d_{I \Mb\Mc} \, \Lambda^\Mb \, {\cal F}^\Mc_{\mua\mub})$ and we need
\begin{align}
   X_{(\Ma\Mb)}{}^\Mc &= d_{I \Ma\Mb} Z^{\Mc I} \; .
   \label{DefZfromX}
\end{align}
This equation implicitly defines $Z^{\Ma I}$ as a linear function of the embedding tensor,
but it is also  a linear constraint on $\Theta$ itself, since $X_{(\Ma\Mb)}{}^\Mc$ not necessarily has the form \eqref{DefZfromX}.
For example for 
the maximal supergravity in $d=7$ this already yields the complete linear constraint\footnote{Probably the same
is true for all other dimensions $d \geq 4$, but we did not check this explicitly. However, the inverse statement, i.e.\
that the linear constraint on $\Theta$ implies equation \eqref{DefZfromX}, can easily be checked for $7 \geq d \geq 4$.
The point is that the $G_0$-tensors $X_{(\Ma\Mb)}{}^\Mc$ and $Z^{\Mc I}$ contain the allowed representations
both only once (or not at all for the ${\bf 15}$ in $d=7$) and $d_{I \Ma\Mb}$ is
injective (as a map from $Z^{* I}$ to $X_{(\Ma\Mb)}{}^{*}$), thus equation \eqref{DefZfromX} only fixes the factor between
these components of $X_{(\Ma\Mb)}{}^\Mc$ and $Z^{\Mc I}$.}. Note that the quadratic constraint \eqref{QconGen2} implies
\begin{align}
   X_{(\Ma\Mb)}{}^\Mc \, X_\Mc &= 0 \; , &
   \text{thus} &&
   Z^{\Ma I} X_\Ma &= 0 \; .
   \label{ZXrel}
\end{align}
Using this equation one can replace the field strength ${\cal F}^\Ma_{\mua\mub}$ in the Ricci identity
by the covariant derivative, i.e.\ we have
\begin{align}
   [D_\mua , D_\mub ] &= - g \, {\cal H}^\Ma_{\mua\mub} \, X_\Ma \; .
\end{align}

Continuing the analysis to higher rank gauge fields one finds that, analogous to equation \eqref{DefZfromX},
one needs a St\"uckelberg type couplings to the three-forms in the field strength of the two-forms,
and so on. Without explaining the details of the derivation we want to give the result.
The tensor $Y_{IA}$ that describes couplings to three-form
gauge fields in the covariant derivative of the two-from gauge fields is given by
\begin{align}
   X_{\Ma I}{}^J + 2 d_{I \Ma\Mb} \, Z^{\Mb J} &= c^{AJ}_{\Ma} \,  Y_{I A} \; .
   \label{DefGenY}
\end{align}
Again, this equation not only defines $Y_{I A}$ but also is a linear constraint on $\Theta$.
Note that equation \eqref{DefGenY} expresses the embedding
tensor in terms of $Z^{\Mb J}$ and $Y_{IA}$ if the representation of the two-form gauge fields is faithful.
From \eqref{DefGenY} and \eqref{ZXrel} we find the relations\footnote{To derive the second of relation in \eqref{YZGen1}
one starts from gauge invariance of $Z^{\Ma I}$, i.e.\ $\delta_\Ma Z^{\Ma I}=0$, and then applies \eqref{DefGenY} and \eqref{ZXrel}.}
\begin{align}
   c^{AJ}_{\Ma} \,  Y_{K A} \, Z^{\Ma I} - 2 d_{K \Ma\Mb} \, Z^{\Ma I} \, Z^{\Mb J} &= 0 \; , &
   Z^{\Ma I} \, Y_{I A} &= 0 \; .
   \label{YZGen1}
\end{align}

The covariant field strengths of respective gauge fields read
\begin{align}
   {\cal H}^\Ma_{\mua\mub}  &= 2 \, \partial_{[\mua} \, A^\Ma_{\mub]} + g X_{\Mb\Mc}{}^\Ma \, A_{[\mua}^\Mb \, A_{\mub]}^\Mc
                               + g Z^{\Ma I} B_{\mua\mub\,I} \; , \nonumber \\
   {\cal H}_{\mua\mub\mub\,I} &= 3\, D_{[\mua} B_{\mub\muc]\,I} 
                         + 6 d_{I \Ma\Mb}\,A_{[\mua}{}^{\Ma}(\partial_{\mub} A_{\muc]}{}^{\Mb}
		         + \ft13 \, g \, X_{[\Mc\Md]}{}^{\Mb} A_{\mub}{}^{\Mc}A_{\muc]}{}^{\Md}) 
                          + g \, Y_{I A} \, S_{\mua\mub\muc}^{A} \;, \nonumber \\
   {\cal H}^A_{\mua\mub\muc\mud} &= 4 \, D_{[\mua}  \, S^A_{\mub\muc\mud]}
                                     - c^{AI}_\Ma \, \Big( 6 \, B_{[\mua\mub\,I} \, {\cal H}^\Ma_{\muc\mud]}
				                           -3 g Z^{\Ma J} B_{[\mua\mub\,I} B_{\muc\mud]\,J}
	              + 8 \, d_{I \Mb\Mc} \, A^\Ma_{[\mua} \, A^\Mb_\mub \, \partial_{\muc} \, A^\Mc_{\mud]}
							   \nonumber \\ & \qquad \qquad \qquad \qquad \qquad
	      + 2 \, d_{I \Mb\Mc} \, X_{\Mc\Md}{}^\Me \, A^\Ma_{[\mua} \, A^\Mb_\mub \, A^\Mc_\muc \, A^\Md_{\mud]} 
	                   \Big)   + \text{four-form term.} & 
   \label{DefHHH}
\end{align}
The general variations of these field strengths read
\begin{align}
   \delta\,{\cal H}_{\mua\mub}^{\Ma} &=  2\,D_{[\mua}\, (\Delta A_{\mub]}^{\Ma}) 
                                      + g \, Z^{\Ma I} \, \Delta B_{\mua\mub\,I} \; , \nonumber\\
   \delta\,{\cal H}_{\mua\mub\muc\,I} &= 3\, D_{[\mua} (\Delta B_{\mub\muc]\,I})
                        + 6 \, d_{I\Ma\Mb} \, {\cal H}_{[\mua\mub}^{\Ma} \, \Delta A_{\muc]}^{\Mb} 
			 + g \, Y_{I A} \Delta S^A_{\mua\mub\muc}  \; , \nonumber \\
   \delta\,{\cal H}^A_{\mua\mub\muc\mud} &= 4\, D_{[\mua} (\Delta S^A_{\mub\muc\mud]})
                        - 6 \, c^{AI}_{\Ma} \, {\cal H}^\Ma_{[\mua\mub} \, \Delta B_{\muc\mud]\,I}
			+ 4 \, c^{AI}_{\Ma} \, {\cal H}_{[\mua\mub\muc\,I} \, \Delta A^\Ma_{\mud]} 
			  \nonumber \\ & \qquad \qquad \qquad \qquad \qquad \qquad \qquad \qquad + \text{four-form term} \; ,
   \label{VaryHHH}			
\end{align}
were we used the covariant variations defined in \eqref{CovD1}. 
In terms of these covariant variations the gauge transformations are given by
\begin{align}
   \Delta A_\mua^\Ma &= D_\mua \Lambda^\Ma - g Z^{\Ma I} \Sigma_{\mua\,I} \; , \nonumber \\
   \Delta B_{\mua\mub\,I} &= 2 D_{[\mua} \Sigma_{\mub]I} - 2 d_{I\Ma\Mb} \Lambda^\Ma {\cal H}^\Mb_{\mua\mub}
                             - g Y_{IA} \Phi^A_{\mua\mub} \; , \nonumber \\
   \Delta S_{\mua\mub\muc}^A &= 3 D_{[\mua} \Phi^A_{\mub\muc]} + 3 \, c^{AI}_\Ma \, {\cal H}^\Ma_{[\mua\mub} \, \Sigma_{\muc] I}
                                     + c^{AI}_\Ma \, \Lambda^\Ma \, {\cal H}_{\mua\mub\muc\,I} + \text{four-form gauge param.} \; ,
   \label{GenGauge}				     
\end{align}
where $\Lambda^\Ma(x)$, $\Sigma_{\mua\,I}(x)$ and $\Phi^A_{\mua\mub}(x)$ are the gauge parameters.
Plugging these gauge transformations into \eqref{VaryHHH} one finds that
the field strengths indeed transform covariantly, i.e.\ that
\begin{align}
   \delta {\cal H}^\Ma_{\mua\mub} &= - g \, \Lambda^\Mb \, X_{\Mb\Mc}{}^\Ma \, {\cal H}^\Mc_{\mua\mub}  \; , &
   \delta {\cal H}_{\mua\mub\muc\,I} &= g \, \Lambda^\Ma \, X_{\Ma I}{}^J \, {\cal H}_{\mua\mub\muc\,I}  \; .
   \label{TrafoCOV}   
\end{align}

For the field strength ${\cal H}^A_{\mua\mub\muc\mud}$ of the three-forms we did not
give the couplings to the four-form gauge fields,
but only with these couplings and with the appropriate gauge transformations of the four-forms this field strength
will transform covariantly.
Similarly, without four-form fields the gauge transformations \eqref{GenGauge} do not close on $S^A_{\mua\mub\muc}$,
but only on $A_\mua^\ma$ and $B_{\mua\mub}$. The corresponding algebra reads
\begin{align}
   [ \delta_{\Lambda_1} , \delta_{\Lambda_2} ] &=  \delta_{\tilde{\Lambda}}+\delta_{\tilde{\Xi}}+\delta_{\tilde{\Phi}}
   \nonumber\\
   [ \delta_{\Xi_1} , \delta_{\Xi_2} ] &= \delta_{{\Phi}}  \;,
   \label{GenGaAlg}
\end{align}
with
\begin{align}
   \tilde \Lambda^{\Ma} &= g {X_{\Mb\Mc}}^{\Ma}  \Lambda_{[1}^{\Mb}  \Lambda_{2]}^{\Mc} \;, \nonumber\\
   \tilde \Xi_{\mua\,I} &= d_{I\Ma\Mb} \, \left( \Lambda_1^{\Ma} D_\mua \Lambda_2^{\Mb} 
                                               - \Lambda_2^{\Ma} D_\mua \Lambda_1^{\Mb} \right) \;,\nonumber\\
   \tilde \Phi_{\mua\mub}^{A} &= 2 \, c^{AI}_\Mb \, d_{I\Ma\Mc}  
                            \,  {\cal H}_{\mua\mub}^{\Ma} \, \Lambda_{[1}^{\Mb} \, \Lambda_{2]}^{\Mc}  \;,\nonumber\\
   \Phi_{\mua\mub}^{A} &=  g \, c_\Ma^{A (I} \, Z^{|\Ma|J)}  \left( \Xi_{1\,\mua \, I} \, \Xi_{2\,\mub \, J} 
                                                       - \Xi_{2\,\mua \, I} \Xi_{1\,\mub \, J} \right)
    \;.
\end{align}
The quadratic constraint on $\Theta$ is crucial when checking these commutators.
Finally, we also give the modified Bianchi identities for the covariant field strengths
\begin{align}
   D_{[\mua} \, {\cal H}^\Ma_{\mub\muc]} &= \ft 1 3 \, g \, Z^{\Ma I} \, {\cal H}_{\mua\mub\muc\,I} \; , \nonumber \\
   D_{[\mua} \, {\cal H}_{\mub\muc\mud]\,I} &= \ft 3 2 \, d_{I\Ma\Mb} \, {\cal H}^\Ma_{[\mua\mub} \, {\cal H}^\Mb_{\muc\mud]}
                                               + \ft 1 4 \, g \, Y_{IA} \, {\cal H}^{A}_{\mua\mub\muc\mud} \; .
   \label{GenBianchi}					       
\end{align}
It is very convenient to use these identities when checking \eqref{TrafoCOV} and \eqref{GenGaAlg}.

\subsection{Truncations of the tower of $p$-form gauge fields} 

In the last section we found the
couplings to the $(p+1)$-forms necessary in the field strengths of the $p$-forms in order to ensure gauge invariance.
We now ask how this infinite tower of $p$-form gauge fields can be truncated to
a finite subsystem without loosing gauge invariance.
The answer to this question
comes from the fact that not all $(p+1)$-form gauge fields are really needed to make the field strength of the $p$-form
gauge fields invariant. For example, in the field strengths of the vector fields $A_\mua^\Ma$ the two-form fields
$B_{\mua\mub\,I}$ only enter
projected with $Z^{\Ma I}$. 
A finite gauge invariant set of gauge fields
is given by $\{ A_\mua^\Ma , \; Z^{\Ma I} B \, _{\mua\mub\,I} \}$.
Indeed, due to \eqref{YZGen1}
the three-forms $S^A_{\mua\mub\muc}$ drop out of the projected two-form field strength $Z^{\Ma I} \, {\cal H}_{\mua\mub\muc\,I}$.
Using \eqref{DefGenY} one can write this result without any reference to the three-from representation as
\begin{align}
    Z^{\Ma I} \left( X_{\Mb I}{}^J + 2 d_{I\Mb\Mc} \, Z^{\Mc J} \right) &= 0  \; , &&    
    \text{thus } \{ A_\mua^\Ma , \; Z^{\Ma I} \, B_{\mua\mub\,I} \} \text{ is closed.} 
   \label{AZB}
\end{align}
This is the truncation scheme used for the $d=4$ and $d=5$ maximal and half-maximal supergravities,
see \cite{deWit:2004nw,dWST4} for the maximal theories and chapter \ref{ch:D4} and \ref{ch:D5} for the half-maximal ones. 
For the higher-dimensional supergravities one finds that
the two-forms $B_{\mua\mub\,I}$ appear already unprojected in the ungauged theory, thus
a different truncation scheme is needed.

The three forms only enter projected with $Y_{IA}$ into the field strength of the two-form gauge fields.
We find $\{ A_\mua^\Ma , \; B_{\mua\mub\,I}, \; Y_{IA} S^A_{\mua\mub\muc} \}$ to be a set of gauge fields that is closed
under gauge transformations. The consistency condition for this is
\begin{align}
   Y_{IA} \left( X_{\Ma B}{}^A + c^{AJ}_\Ma \, Y_{JB} \right) &= 0 \; , &&
   \text{thus } \{ A_\mua^\Ma , \; B_{\mua\mub\,I}, \; Y_{IA} S^A_{\mua\mub\muc} \} \text{ is closed.} 
   \label{ABYS}
\end{align}
This condition is satisfied due to the quadratic constraint on $\Theta$. To prove \eqref{ABYS} one
starts with the gauge invariance of $Y_{IA}$, i.e.\ $\delta_\Ma Y_{IA}=0$ and then applies equations \eqref{DefGenY}
and \eqref{YZGen1}.

For the $d=3$ supergravities the vector fields are introduced as duals to the scalars, they thus transform in the adjoint
representation, i.e.\ in this case we have vector fields $A_\mua^\alpha$, an embedding tensor $\Theta_\alpha{}^\beta$ and
gauge group generators $X_{\alpha\beta}{}^\gamma = - \Theta_\alpha{}^\delta f_{\delta\beta}{}^\gamma$. In this case it turns out that
no higher rank gauge fields are needed since the $\Theta$-projected vector field $A^\alpha \, \Theta_\alpha{}^\beta$ are closed
under gauge transformations. The crucial relation for this is
\begin{align}
    X_{(\alpha \beta)}{}^\gamma \,  \Theta_\gamma{}^\delta &= 0 \; , &&
   \text{thus }  A^\alpha \, \Theta_\alpha{}^\beta \text{ is closed in $d=3$.}
   \label{OnlyA}
\end{align}
This condition is equivalent to the quadratic constraint in $d=3$ if the embedding tensor 
$\Theta_{\alpha\beta}$ is symmetric in $\alpha$ and $\beta$.\footnote{The Cartan-Killing form was used to lower the index $\beta$.}. 
The symmetry of $\Theta_{\alpha\beta}$ is always a consequence of the linear constraint in three-dimensions
\cite{Nicolai:2001sv,Nicolai:2001ac,deWit:2003ja}.

It thus depends on the particular theory which of the truncation schemes \eqref{AZB}, \eqref{ABYS} or \eqref{OnlyA} is used.
In each case only the corresponding
projected gauge transformations are present, i.e.\ only $\Lambda^\alpha \Theta_\alpha{}^\beta$ for $d=3$ and
in the higher dimensions $\{ \Lambda^\Ma , \; Z^{\Ma I} \, \Sigma_{\mua\,I} \}$ or
$\{ \Lambda^\Ma , \; \Sigma_{\mua\,I}, \; Y_{IA} \Phi^A_{\mua\mub} \}$.
For the $d=8$ maximal supergravity one also needs four-form fields and thus an
even larger set of gauge transformations, but the corresponding field strengths and gauge transformations were
not yet worked out in detail.

\begin{table}[tb]
   \begin{center}
     \begin{tabular}{l|@{\qquad}c@{\qquad}c@{\qquad}c@{\qquad}c@{\qquad}c@{\qquad}c@{\qquad}c}
        $d$      & $2$ & $3$ & $4$ & $5$ & $6$ & $7$ & $8$ \\ \hline
	$p_{\text{max}}$ & $0^*$ & $0$ & $1^*$ & $1$ & $2^*$ & $2$ & $3^*$
     \end{tabular}
     \caption{\label{MaxP}\small The highest rank $p_{\text{max}}$ of the tensor gauge fields that appears necessarily 
                          in the ungauged maximal supergravity in $d$ 
                          space-time dimensions. Scalars correspond to $p=0$, vector gauge fields to $p=1$, etc. 
			  The asterisk indicates self-duality of the $p_{\text{max}}$-form fields.}
   \end{center}   
\end{table}

For the maximal supergravities
we list in table \ref{MaxP} the maximal rank $p_{\text{max}}$ of forms that appear in the ungauged theory,
always referring to that formulation of the theory in which all forms have been dualized to smallest possible rank\footnote{
Typically different formulations in terms of dual $p$-forms also exist, and onshell one can always introduce all forms
up to rank $d-2$ via dualization, see table \ref{ListGH}.}. In the gauged theory only the
tensor gauge fields up to rank $p_{\text{max}}+1$ appear, and we saw in the above truncation schemes that
these $(p_{\text{max}}+1)$-form gauge fields are only introduced projected with some tensor $\Theta$, $Z$ or $Y$, while all other
gauge fields are introduced unprojected\footnote{In even dimensions there are subtleties since typically
only half of the $p_{\text{max}}$-form gauge fields appear in the ungauged Lagrangian. The others can only be introduced onshell
in the ungauged theory. In the gauged theory they also appear in the Lagrangian, but
like the $(p_{\text{max}}+1)$-forms only projected with some component of $\Theta$.}.
Thus for $\Theta \rightarrow 0$ these gauge fields decouple
and only the field content of the ungauged theory is left. Note also that the covariant field strengths \eqref{DefHHH}
become the ungauged field strengths \eqref{DefFFF} for $\Theta \rightarrow 0$.

\subsection{Topological terms in odd dimensions} 
\label{subsec:TopTermsOdd}

For all dimensions $d\geq 4$ the ungauged Lagrangian of maximal and half-maximal supergravity
contains a topological term and in this section we give
the appropriate generalization of this topological term in the gauged theory.
For simplicity, we restrict to odd dimensions. We also include the case $d=3$ for which a topological term
is present in the gauged theory but not in the ungauged one.
It turns out that gauge invariance already fixes the form of this term up to a factor\footnote{
In even dimensions the topological terms alone are not gauge invariant, instead there is a subtle interplay between these
terms and the kinetic terms of the gauge fields in the Lagrangian \cite{deWit:2005ub}.}.

We gave the general variations of the field strengths in \eqref{VaryHHH}. These variations have a much simpler 
form than the field strengths themselves, but we can infer the field strengths from their general variations via
integration. The same is true for the topological terms. In the following we therefore start with the presentation of
the general variation of the respective topological terms. We now go through the different cases.

\subsubsection*{d=3}

For $d=3$ we already explained that the vector fields $A_\mua^\alpha$ come in the adjoint representation,
i.e.\ we have to replace the indices $\Ma$, $\Mb$, etc. everywhere by indices $\alpha$, $\beta$, etc.
The embedding tensor then reads $\Theta_{\alpha}{}^\beta$ and we can use the Cartan-Killing form to raise- and lower
the algebra indices. The general variation of the topological term reads
\begin{align}
   \delta {\cal L}_{\text{top},d=3} &= \epsilon^{\mua\mub\muc} \, \Theta_{\alpha \beta} 
                                           \, (\delta A_\mua^\alpha) \, {\cal F}^\beta_{\mub\muc} 
					   + \text{total derivatives} \; .
\end{align}
This is the only possible Ansatz for the variation that yields covariant field equations.
This Ansatz has to pass two consistency checks.
Firstly, for gauge transformations $\delta A_\mua^\alpha = D_\mua \Lambda^\alpha$ this variation must yield a total derivative,
which is true due to the Jacobi identity $D_{[\mua} {\cal F}^\alpha_{\mub\muc]} = 0$.
Secondly, the variation must integrate up to a Lagrangian ${\cal L}_{\text{top},d=3}$.
If the linear constraint $\Theta_{[\alpha\beta]}=0$ and the quadratic constraint
\eqref{QconGen2} are satisfied\footnote{The quadratic constraint yields that $\Theta_{\alpha \beta} X_{\gamma\delta}{}^\beta$
is completely antisymmetric in $\alpha$, $\gamma$, $\delta$.}
the variation indeed integrates up to the topological term
\begin{align}
   {\cal L}_{\text{top},d=3} &= \epsilon^{\mua\mub\muc} \, \Theta_{\alpha \beta} \, A_\mua^\alpha \,
                                       \left( \partial_\mub A^\alpha_\muc 
     		                  + \ft 1 3 \, X_{\gamma\delta}{}^\beta \, A^\gamma_\mub \, A^\delta_\muc \right) \; .
\end{align}
This is the standard Chern-Simons term, but normally $\Theta_{\alpha\beta}$ is the Cartan Killing form and $X_{\alpha\beta}{}^\gamma$
are the structure constants, which need not to be the case here.

\subsubsection*{d=5}

In $d=5$ the vector gauge fields are dual to the two-form gauge fields, i.e.\ they transform in the dual representations
of $G_0$. We then have the index structure $B_{\mua\mub\,\Ma}$, $d_{\Ma\Mb\Mc}$, $Z^{\Ma\Mb}$, etc.
The general variation of the topological term reads \cite{deWit:2004nw}
\begin{align}
   \delta {\cal L}_{\text{top},d=5} &= \epsilon^{\mua\mub\muc\mud\mue} 
                                   \left[ \ft 1 3 \, g \, Z^{\Ma\Mb} \, (\Delta B_{\mua\mub \, \Ma}) \, {\cal H}_{\muc\mud\mue\,\Mb}
			     - (\Delta A_{\mua}^\Ma) \, d_{\Ma\Mb\Mc} \, {\cal H}^\Mb_{\mub\muc} \, {\cal H}^\Mc_{\mud\mue} \right] 
                   \nonumber \\ & \qquad \qquad \qquad
			  + \text{total derivatives}  \; .
\end{align}
The general Ansatz for $\delta {\cal L}_{\text{top},d=5}$ contains the two given terms
with an a priori arbitrary relative factor.
This factor is fixed since the variation must yield a total derivative for gauge transformations
\eqref{GenGauge}. This can easily be checked by using \eqref{GenBianchi}. However,
the additional constraints $d_{\Ma\Mb\Mc}=d_{(\Ma\Mb\Mc)}$ and $Z^{(\Ma\Mb)}=0$ are needed.
The complete symmetry of $d_{\Ma\Mb\Mc}$ is already necessary
in the ungauged theory to write down the appropriate ungauged topological term. The antisymmetry of $Z^{\Ma\Mb}$
is a consequence of the linear constraint on $\Theta$. With these two conditions and the quadratic constraint on $\Theta$ 
one can show that the above variation can be integrated up. The topological Lagrangian reads \cite{deWit:2004nw}
\begin{align}
   {\cal L}_{\text{top},d=5} &= \epsilon^{\mua\mub\muc\mud\mue} 
               \Big[  - \, \ft 4 3 \, d_{\Ma\Mb\Mc} \, A_\mua^\Ma \, \partial_\mub \, A_\muc^\Mb \, \partial_\mud \, A_\mue^\Mc
               		+   \ft 1 2 \, g \, Z^{\Ma\Mb} \, B_{\mua\mub\,\Ma} \, D_\muc \, B_{\mud\mue\,\Mb} 
		    \nonumber \\ & \qquad \qquad	
			- 2 \, g \, d_{\Ma\Mb\Mc} \, Z^{\Mc\Md} B_{\mua\mub\,\Md} A_\muc^\Ma 
		   \left( \partial_\mud \, A_\mue^\Mb \, + \, \ft 1 3 \, g \, X_{\Me\Mf}{}^\Mb \, A_\mud^\Me \, A_\mue^\Mf \right)
		    \nonumber \\ & \qquad \qquad
		        - 2 \, g \, d_{\Ma\Mb\Mc} \, X_{\Md\Me}{}^\Mc \, A_\mua^\Ma \, A_\mub^\Md \, A_\muc^\Me
         	\left( \partial_\mud \, A_\mue^\Mb \, + \, \ft 1 5 \, g \, X_{\Mf\Mg}{}^\Mb \, A_\mud^\Mf \, A_\mue^\Mg \right)
		               \Big]		\; .			
   \label{GenTopD5}
\end{align}
The first two terms already show that the symmetry of $d_{\Ma\Mb\Mc}$ and the antisymmetry of $Z^{\Ma\Mb}$
are needed in order that the variation
of the Lagrangian takes the above form. The first term already appears in the ungauged theory.

\subsubsection*{d=7}

In $d=7$ the two-form gauge fields are dual to three-form gauge fields and thus also transform in dual representations
of $G_0$. We therefore have three-forms $S^I_{\mua\mub\muc}$ and tensors $c^{IJ}_\Ma$, $Y_{IJ}$, etc.
The general variation of the topological term then reads \cite{Samtleben:2005bp}
\begin{align}
   \delta {\cal L}_{\text{top},d=7} &= - \ft 1 {18} \, \epsilon^{\mua\mub\muc\mud\mue\muf\mug}
                    \big[  Y_{IJ}  (\Delta S^I_{\mua\mub\muc})  {\cal H}^J_{\mud\mue\muf\mug}
		          + 6 \, c^{IJ}_\Ma \, (\Delta B_{\mua\mub\,I}) \, {\cal H}^\Ma_{\muc\mud} \, {\cal H}_{\mue\muf\mug\,J} 
		  \nonumber \\ & \qquad	\qquad  \qquad \qquad
			  + 2 \, c^{IJ}_\Ma \, (\Delta A_\mua^\Ma) \, {\cal H}_{\mub\muc\mud\,I} \, {\cal H}_{\mue\muf\mug\,J} \big]
		+ \text{total derivatives} \; .	  
\end{align}
This variation yields a total derivative under gauge transformations \eqref{GenGauge} and integrates up to a Lagrangian
${\cal L}_{\text{top},d=7}$ if $Y_{[IJ]}=0$ and $c^{(IJ)}_\Ma = 0$. For the maximal supergravities we give the complete topological
term in chapter \ref{ch:D7}. Here we restrict to the leading terms
\begin{align}
   {\cal L}_{\text{top},d=7} &= - \, \epsilon^{\mua\mub\muc\mud\mue\muf\mug}
                  \Big[  c_\Ma^{IJ} \,  B_{\mua\mub\, I} \, \partial_\muc \, A^\Ma_\mud \,
          \left( \partial_\mue \, B_{\muf\mug\,J} + 4 d_{J\Mb\Mc} \, A^\Mb_\mue \, \partial_\muf \, A^\Mc_\mug \right)
	  \nonumber \\ & \qquad \qquad \qquad
   - \ft 4 5 \, c_\Ma^{IJ} \, d_{I\Mb\Mc} \, d_{J\Md\Me} 
          \, A^\Ma_\mua \, A^\Mb_\mub \, A^\Md_\muc \, (\partial_\mud \, A^\Mc_\mue) \, (\partial_\muf A^\Me_\mug)
        \nonumber \\ & \qquad \qquad \qquad \qquad
    	+ \ft 1 9 \, g \, Y_{\Ma\Mb} \, S^\Ma_{\mua\mub\muc} \, D_\mud \, S^\Mb_{\mue\muf\mug} + \ldots \Big]	\; .			      
   \label{GenTopD7}	
\end{align}
The terms missing are of order $g^1$ or $g^2$, i.e.\ all terms of the ungauged theory are already given here.

\section{Preserving supersymmetry}
\label{sec:PresSUSY}

In this section we assume that the Lagrangian and the supersymmetry rules of the ungauged theory are known
and describe the modifications that have to be made in order to obtained the gauged theory. Note that minimal
substitution alone, i.e.\ replacement of all derivatives $\partial_\mua$ by covariant derivatives $D_\mua$, destroys
gauge invariance and supersymmetry. In the last section we already introduced the necessary covariant field strengths
and covariant topological terms that have to be introduced in order to restore gauge invariance. In order to restore
supersymmetry one introduces additional fermionic couplings and a scalar potential in the Lagrangian and also needs
to modify the supersymmetry rules of the fermions (the Killing spinor equations). These changes will be explained in the next
subsection. 

\subsection{Additional terms in Lagrangian and supersymmetry variations}

We saw that the bosonic fields of the maximal and half-maximal supergravities transform in some
representation of the global symmetry group $G_0$. In particular the scalars form the coset $G_0/H$ that
is described by a group element ${\cal V}$ subject to global $G_0$ transformations from the
left and local $H$ transformations from the right, i.e.~it transforms as
\begin{align}
   {\cal V} \, \mapsto \, \Lambda \, {\cal V} \, h(x) \; , \quad \Lambda \in G_0 \; , \quad h(x) \in H \; .
   \label{TrafoVGen}
\end{align}
See equation \eqref{SLnSymm} for the ${\rm SL}(q)/{\rm SO}(q)$ case,
and the following chapters for further examples. This particular description of the scalars is necessary since the
fermions also transform under local $H$-transformations, but not under $G_0$. Thus all couplings between 
$p$-form gauge fields and fermions have to be mediated by the scalar coset representative ${\cal V}$.

Let us first focus on the maximal supergravities, for which
the group $H$ coincides with the $R$-symmetry group $H_R$.
The latter is defined as the largest subgroup of the automorphism group of the supersymmetry algebra
that commutes with Lorentz transformations, i.e.\ it acts only on the internal indices of the supersymmetry generators
(not on their spinor indices) and leaves the supersymmetry algebra invariant. Every component of a super-multiplet thus transforms in
some representation of $H_R$, in particular the fermions. The gravity multiplet of maximal supergravity contains
the gravitini $\psi_\mua^\xa$ and matter fermions $\chi^\ya$, where the indices $\xa$ and $\ya$ refer to some
representation of $H=H_R$. In table \ref{ListFerm}  we listed the $R$-symmetry groups and the respective fermion representations
for dimensions $3 \leq d \leq 8$. For the even dimensions there always appears a representation $W$
together with its dual representation $\overline W$, which means that the corresponding fermions can be described by one complex Weyl spinor with representation
$W$ (its complex conjugate then carries $\overline W$). In odd dimensions one can always use (symplectic) Majorana spinors
that obey a (pseudo) reality condition. 

\begin{table}[tb]
   \small
   \begin{center}
     \begin{tabular}{c|| c | c@{\quad} c@{\quad} c | c@{\quad} c | c@{\quad}}
            & spinor &  &  \multicolumn{2}{c|}{representation under $H_R$}  & \multicolumn{2}{c|}{little group} &   \\[-0.05cm] 
        $d$ & type & $H_R$ & $\psi_\mua$ & $\chi$ & $\psi_\mua$ & $\chi$ & dof  \\ \hline &&&&&&& \\[-0.43cm]  \hline
	      &&&&&& \\[-0.3cm]
        $8$ & M,W & ${\rm U}(2)$ & ${\bf 2} \oplus {\bf 2}$ & ${\bf 2} \oplus \overline{\bf 2} \oplus {\bf 4} \oplus \overline{\bf 4}$ &
		   ${\bf 20}$ & ${\bf 4}$ & $80+48$                               \\[0.08cm]
        $7$ & S & ${\rm USp}(4)$ & ${\bf 4}$ & ${\bf 16}$ &
		   ${\bf 16}$ & ${\bf 4}$ & $64+64$                               \\[0.08cm]
        $6$ & SMW & ${\rm Usp}(4)\times{\rm Usp}(4)$ & ${\bf (4,1)} \oplus {\bf (1,4)}$ & ${\bf (4,5)} \oplus {\bf (5,4)}$ &
		   ${\bf 6}$ & ${\bf 4}$ & $48+80$                               \\[0.08cm]
        $5$ & S & ${\rm Usp}(8)$ & ${\bf 8}$ & ${\bf 48}$ &
		   ${\bf 4}$ & ${\bf 2}$ & $32+96$                               \\[0.08cm]
        $4$ & M,W & ${\rm SU}(8)$ & ${\bf 8} \oplus \overline {\bf 8}$ & ${\bf 56} \oplus \overline{\bf 56}$ &
		   ${\bf 1}$ & ${\bf 1}$ & $16+112$                               \\[0.08cm]
        $3$ & M & ${\rm SO}(16)$ & ${\bf 16}$ & ${\bf 128}$ &
		   --- & ${\bf 1}$ & $0+128$                               
     \end{tabular}
     \caption{\label{ListFerm}{ \small For the maximal supergravities in $d$ dimensions the $R$-symmetry groups
                                       and the corresponding representations of the gravitini $\psi_\mua$
				       and the matter spinors $\chi$ are listed. In addition the degrees of freedom
				       for each $\psi_\mua$ and $\chi$ are given, which corresponds to giving
				       the representation of these spinor under the respective little group ${\rm SO}(d-2)$.
				       The product of the dimensions of the $H_R$ and ${\rm SO}(d-2)$ representations  
				        yields the total degrees of freedom (dof) of $\psi_\mua$ and $\chi$,
				       which always sum up to 128.
				       In the second column the spinor types in the respective dimension
				       are given (M for Majorana, W for complex Weyl, 
				       S for symplectic Majorana and SMW for symplectic Majorana Weyl).
				       Note that $H_R$ coincides with (the complex covering group of) $H$ of table \ref{ListGH}.}} 
   \end{center}     
\end{table}

The Lagrangian of the gauged theory schematically takes the form
\begin{align}
   {\cal L} &= {\cal L}_0[\partial \rightarrow D,{\cal F}_0 \rightarrow {\cal H}]
              +{\cal L}_{\text{top}}
	      +{\cal L}_{\text{ferm.mass.}}
	      +{\cal L}_{\text{pot}} \; ,
   \label{GenL1}	      
\end{align}
where ${\cal L}_0$ is the ungauged Lagrangian without topological term, but including fermions.
All derivatives in ${\cal L}_0$ have to be replaced by covariant derivatives
and all $p$-form field strengths have to be replaced by covariant ones.
In addition one needs to add the respective gauge covariant topological term ${\cal L}_{\text{top}}$,
fermionic mass terms ${\cal L}_{\text{ferm.mass}}$
and a scalar potential ${\cal L}_{\text{pot}}$. In the last section we already gave ${\cal L}_{\text{top}}$, at least for the odd dimensions.
By fermionic mass terms we mean all bilinear couplings of the fermions that do not involve
$p$-form gauge fields or derivatives, i.e.\ schematically
\begin{align}
   e^{-1} \, {\cal L}_{\text{ferm.mass}} &= g \, A_{1\,\xa}{}^\xb \, \bar \psi_{\mua\,\xb} \, \Gamma^{\mua\mub} \, \psi^\xa_\mub
                         +  g \, A_{2\,\xa}{}^\ya \, \bar \chi_{\ya} \, \Gamma^{\mua} \, \psi_\mua^\xa
			 +  g \, A_{3\,\ya}{}^\yb \, \bar \chi_{\yb} \, \chi^\ya 
			   + \text{h.c.} \;   ,
   \label{GenFermMass}			   
\end{align}
where $A_1$, $A_2$ and $A_3$ are are some tensor that depend on scalar fields and linearly on the embedding tensor.
More precisely $A_1$, $A_2$ and $A_3$ are composed out of irreducible components of the $T$-tensor which we will introduce in
the next subsection. Note that no couplings of the form \eqref{GenFermMass} are present in the ungauged theory. In the gauged
theory these couplings are needed to cancel terms in the supersymmetry variations of the Lagrangian
that come from the new gauge field couplings.
But not all these new terms in are canceled in this way. One also needs to change the
Killing spinor equations as follows\footnote{For example, if we (schematically) write the
scalar kinetic terms as ${\cal L} \supset (D_\mua \phi) (D^\mua \phi)$ and the supersymmetry variations of the vector fields
as $\delta_\epsilon A_\mua = \bar \epsilon \psi_\mua + \bar \epsilon \Gamma_\mua \chi$, we find
in the variation of the Lagrangian terms of the form
$\delta_\epsilon {\cal L} \supset g (D_\mua \phi) \Theta (\bar \epsilon \psi_\mua + \bar \epsilon \Gamma_\mua \chi)$.
Those get canceled by terms from \eqref{GenFermMass} since
$\delta_\epsilon \psi_\mua \supset D_\mua \epsilon$ and $\delta_\epsilon \chi \subset D_\mua \phi \Gamma^\mua \epsilon$,
and by terms that follow when plugging \eqref{GenKilling} into the kinetic terms of the fermions
${\cal L}\supset \bar \psi_\mua \Gamma^{\mua\mub\muc} D_\mub \psi_\muc + \bar \chi \slashchar{D} \chi$.
More details are given in the following chapters for the concrete theories.}
\begin{align}
   \delta_\varepsilon \, \psi^\xa_\mua &= \text{ungauged terms} 
                                        + g \, A_{1\,\xb}{}^\xa \, \varepsilon^{\xb} \; , \nonumber \\
   \delta_\varepsilon \, \chi^\ya &= \text{ungauged terms} 
                                        + g \, A_{2\,\xa}{}^\ya \, \varepsilon^{\xa} \; ,
   \label{GenKilling}					
\end{align}
where $\varepsilon^{\xa}(x)$ is the parameter of supersymmetry transformations. 
Supersymmetry demands the same tensors $A_1$ and $A_2$ to appear here as in the Lagrangian.

Plugging the variations \eqref{GenKilling} into \eqref{GenFermMass} yields order $g^2$ terms in the variations of the Lagrangian.
In order to cancel those one needs a scalar potential of the form
\begin{align}
   e^{-1} \, {\cal L}_{\text{pot}} &= \, - \, g^2 \, V \, = \, 2 \, g^2 \, \left( A_{1\,\xa}{}^\xb \, \bar A_{1}{}^\xa{}_\xb  
                                     -  A_{2\,\xa}{}^\ya \, \bar A_{2}{}^{\xa}{}_\ya \right)				     
     \; ,
   \label{GenLpot}     
\end{align}
where the bar denotes complex conjugation.
This is a scalar potential since $A_1$ and $A_2$ depend on the scalar fields.
Note that we are not very explicit with our conventions here, but we assumed that complex conjugation lowers or highers the
indices $\xa$ and $\ya$. Of course, we will be much more concrete as soon as particular theories are discussed in the following
chapters. Supersymmetry then demands a quadratic constraint on $A_1$ and $A_2$ of the form\footnote{
This equation is obtained by considering terms of the
form $g^2 \bar \psi_\mua \Gamma^\mua \epsilon$ in the variation $\delta_\epsilon {\cal L}$.} \cite{D'Auria:2001kv}
\begin{align}
   A_{1\,\xa}{}^\xc \, \bar A_{1}{}^\xb{}_\xc
                -  A_{2\,\xa}{}^\ya \, \bar A_{2}{}^{\xb}{}_\ya &= - \, \frac 1 {2 r} \, \delta_\xa^\xb \, V \; ,
   \label{GenAA}		
\end{align}
where $r=\delta_\xa^\xa$ is the dimension of the gravitini representation. This constraint needs to be satisfied
as a consequence of the quadratic constraint on $\Theta$. Equation \eqref{GenAA} is sometimes denoted as generalized
Ward identity for extended supergravity.

According to table \ref{ListFerm} the fermionic degrees of freedom of the maximal supergravities
add up to $128$, and so do the
bosonic degrees of freedom in the ungauged theory. In order to preserve
supersymmetry, one is not allowed to alter the degrees of freedom. Nevertheless, as explained in the last section,
additional $(p_{\max}+1)$-forms are needed in
the gauged theory to get a gauge invariant field strength of the
$p_{\max}$-forms\footnote{And in even dimensions one also introduces those 
$p_{\max}$-forms (i.e.\ in $d=4$ vector fields)
in the Lagrangian that are normally only introduces onshell via dualization.}. 
According to \eqref{GenL1} these additional gauge fields do not get a 
kinetic term, but only appear via the St\"uckelberg type couplings
in the covariant field strengths and in the generalized topological term
and therefore do not yield additional degrees of freedom. Their field
equation will turn out to be a duality equation, which in odd dimensions
relates the $(p_{\max}+1)$-forms themselves to the $p_{\max}$-forms.
For even dimensions the construction is more subtle, for $d=4$ we again refer to \cite{deWit:2005ub}
and to chapter \ref{ch:D4}.

The construction of the gauged theory given in equations \eqref{GenL1} to \eqref{GenLpot} is not specific
for the maximal supergravities. The only thing that changes for supergravities with less supercharges is
that additional fermions from other multiplets are present. For example, for the half-maximal theories one still
has $\psi^\xa_\mua$ and $\chi^\ya$ from the gravity multiplet, but in addition one has matter fermions $\lambda^\ca$
from the $n$ vector multiplets. The indices $\xa$, $\ya$ and $\ca$ again indicate that these fields come in some representation
of $H$, but we now have $H=H_R \times {\rm SO}(n)$, i.e.\ $H$ is not identical with the $R$-symmetry group,
but contains it as a subgroup. The additional factor ${\rm SO}(n)$ refers to the transformations of the vector multiplets
into each other, i.e.\ $\lambda^\ca$ transforms as a vector under ${\rm SO}(n)$, while $\psi^\xa_\mua$ and $\chi^\ya$
are singlets under ${\rm SO}(n)$. In table \ref{ListFerm2} we summarize the representations of the fermions for
the half-maximal theories in $d=3,4,5$.\footnote{In section \ref{sec:HalfMaxUngauged} we explained that
from torus reduction of minimal supergravity in $d=10$ without vector multiplets one obtains the half-maximal theories in
$d=3,4,5$ with $n=8,6,5$ vector multiplets. According to table \ref{sec:HalfMaxUngauged} these theories all carry $64$ fermionic
degrees of freedom, i.e.\ half as much as the maximal theories.}

\begin{table}[tb]
   \small
   \begin{center}
     \begin{tabular}{c|| c | c@{\quad} c@{\quad} c@{\quad} c | c@{\quad}c@{\quad}c | c@{\quad}c@{\quad}c | c }
            & spinor &  &  \multicolumn{3}{c|}{under $H_R$} 
	              & \multicolumn{3}{c|}{under ${\rm SO}(n)$}  
		      & \multicolumn{3}{c|}{little group} &    \\[-0.05cm] 
        $d$ & type & $H_R$ & $\psi_\mua$ & $\chi$ & $\lambda$ & 
	                     $\psi_\mua$ & $\chi$ & $\lambda$ &
			     $\psi_\mua$ & $\chi$ & $\lambda$ & dof  \\ \hline &&&&&&&&&&& \\[-0.43cm]  \hline
	      &&&&&&&&&&&& \\[-0.3cm]
        $5$ & S & ${\rm Usp}(4)$ & ${\bf 4}$ & ${\bf 4}$ & ${\bf 4}$
	                         & ${\bf 1}$ & ${\bf 1}$ & ${\bf n}$
                                 & ${\bf 4}$ & ${\bf 2}$ & ${\bf 2}$  & 16+8+8n                          \\[0.08cm] 
       $4$ & M,W & ${\rm U}(4)$ & ${\bf 4} \oplus \overline{\bf 4}$ & ${\bf 4}\oplus \overline{\bf 4}$ & ${\bf 4}\oplus \overline{\bf 4}$
	                         & ${\bf 1}$ & ${\bf 1}$ & ${\bf n}$
                                 & ${\bf 1}$ & ${\bf 1}$ & ${\bf 1}$  & 8+8+8n           \\[0.08cm] 
        $3$ & M & ${\rm SO}(8)$ & ${\bf 8}_s$ & --- & ${\bf 8}_c$
	                         & ${\bf 1}$ & --- & ${\bf n}$
                                 & no dof & --- & ${\bf 1}$  & 0+0+8n
     \end{tabular}
     \caption{\label{ListFerm2}{ \small Analogous to table \ref{ListFerm}, but for
                                       the half-maximal supergravities in $d=3,4,5$ dimensions.
				       The $R$-symmetry groups
                                       and the corresponding representations of the gravitini $\psi_\mua$
				       and of the matter spinors $\chi$ and $\lambda$ are given. While $\psi_\mua$
				       and $\chi$ belong to the gravity multiplet, $\lambda$ belongs to the $n$
				       vector multiplets and thus transforms under the ${\rm SO}(n)$
				       that rotates these vector multiplets into each other.  } }
   \end{center}     
\end{table}

In the gauged theory
the supersymmetry rules for $\lambda$ have to be supplemented by a term  $g \, A_{2\,\xa}{}^\ca \, \varepsilon^{\xa}$ 
and the fermionic mass terms also contain all possible bilinear fermion coupling
that contain $\lambda^\ca$, in particular a term $g A_{2\,\xa}{}^\ca \bar \lambda_{\ca} \Gamma^{\mua} \psi_\mua^\xa$. Equation
\eqref{GenAA} then has to be modified as follows
\begin{align}
     A_{1\,\xa}{}^\xc \, \bar A_{1}{}^\xb{}_\xc
  -  A_{2\,\xa}{}^\ya \, \bar A_{2}{}^{\xb}{}_\ya
  -  A_{2\,\xa}{}^\ca \, \bar A_{2}{}^{\xb}{}_\ca		
		 &= - \, \frac 1 {2 r} \, \delta_\xa^\xb \, V \; ,
\end{align}
and this equation again has to be a consequence of the quadratic constraint on $\Theta$.

\subsection{The $T$-tensor}

In the last subsection we introduced tensors $A_1$, $A_2$ and $A_3$ to write down gravitational mass terms
for the fermions. These tensors transform under the maximal compact subgroup $H$ of $G_0$ and
have to be defined out of the embedding tensor~$\Theta$ which is a tensor under the global symmetry group $G_0$ itself. 
The object that relates representations of $G_0$ and $H$ is the scalar coset representative ${\cal V}$ which according
to \eqref{TrafoVGen} transforms under both groups. Since ${\cal V}$ is a group element of $G_0$ it has a natural
action ${\cal R}_{\cal V}$ on every $G_0$ representation. For example, if $G_0$ is some matrix group
(i.e.\ ${\rm SL}(q)$, ${\rm SO}(q,p)$ or ${\rm Sp}(q)$), then the natural action
on a vector $v$ is given by right multiplication, i.e.\ ${\cal R}_{\cal V}[v] = v \, {\cal V}$.

When acting with ${\cal V}$ on the embedding tensor~$\Theta$ one obtains the so-called $T$-tensor
\begin{align}
   T \, &\equiv  \, {\cal R}_{\cal V} \, [ \Theta ] \; .
   \label{DefTGen}
\end{align}
The $T$-tensor contains all the information on $\Theta$,
but it is scalar dependent and transforms under $H$, not under $G_0$. Every $G_0$-irreducible component of
$\Theta$ branches into one or more $H$-irreducible component of $T$, schematically
\begin{align}
   \Theta &= \theta_1 \oplus \theta_2 \oplus \ldots  &
   & \overset{H}{\rightarrow}   &
   T &= ( t_{11} \oplus t_{12} \oplus \ldots ) \oplus (t_{21} \oplus t_{22} \oplus \ldots) \oplus \ldots \; .
   \label{GenSplitT}
\end{align}
The irreducible components $t_{ij}$ of the $T$-tensor are used to build up the fermionic mass tensors
$A_1$, $A_2$ and $A_3$. This has first been observed for the maximal $d=4$ supergravity \cite{deWit:1982ig}.
When concrete examples are being discussed in the next chapters we will explicitly give the relations between $\Theta$, $T$ and
the $A$'s.

\begin{table}[tb]
     \begin{tabular}{c|c@{\qquad}c@{\qquad}c@{\qquad}c}
     $d$ & $G_0$ & $H$ & $\Theta$ & $T$ \\ \hline
     7 & ${\rm SL}(5)$ & ${\rm Usp}(4)$ & ${\bf 15}\oplus\overline{\bf 40}$ & ${\bf 1}\oplus{\bf 5}\oplus{\bf 14}\oplus{\bf 35}$ \\
     6 & ${\rm SO}(5,5)$ & ${\rm USp}(4)\times{\rm USp}(4)$ & ${\bf 144}_s$ & ${\bf (4,4)}\oplus{\bf (4,16)}\oplus{\bf (16,4)}$ \\
     5 & ${\rm E}_6$ & ${\rm USp}(8)$ & $\overline{\bf 351}$ & ${\bf 36}\oplus{\bf 315}$ \\
     4 & ${\rm E}_7$ & ${\rm SU}(8)$ & ${\bf 912}$ & ${\bf 36}\oplus \overline{\bf 36}\oplus{\bf 420}\oplus \overline{\bf 420}$ \\
     3 & ${\rm E}_8$ & ${\rm SO}(16)$ & ${\bf 1}\oplus{\bf 3875}$ & ${\bf 1}\oplus{\bf 135}\oplus{\bf 1820}\oplus{\bf 1920}$ 
     \end{tabular}
     \caption{\label{tab:TtenMax} \small  For the maximal supergravities in $d$ dimensions the $H$-irreducible components 
                                of the $T$-tensor are given. For convenience we again list the global symmetry groups $G_0$,
				its maximal compact subgroups $H$ and the irreducible components of the respective
				embedding tensor.}
     \vspace{2cm}
     \begin{tabular}{c|ccc@{\qquad}ccc}
     $d$ & $\psi_\mua \otimes \psi_\mub$ && $A_1$ & $\psi_\mua \otimes \chi$ && $A_2$  \\ \hline
     7 & $({\bf 4}\otimes{\bf 4})_{\text{antisym}}$ & $\supset$ & ${\bf 1}\oplus{\bf 5}$
       & ${\bf 4}\otimes{\bf 16}$ & $\supset$ & ${\bf 5}\oplus{\bf 14}\oplus{\bf 35}$  
     \\
     6 & ${\bf (4,1)}\otimes{\bf (1,4)}$ & $=$ & ${\bf (4,4)}$
       & ${\bf (4,1)}\otimes{\bf (5,4)}$ & $=$ & ${\bf (4,4)}\oplus{\bf (16,4)}$  
     \\
     5 & $({\bf 8}\otimes{\bf 8})_{\text{sym}}$ & $=$ & ${\bf 36}$
       & ${\bf 8}\otimes{\bf 48}$ & $\supset$ & ${\bf 315}$  
     \\
     4 & $({\bf 8}\otimes{\bf 8})_{\text{sym}}$ & $=$ & ${\bf 36}$
       & ${\bf 8}\otimes\overline{\bf 56}$ & $\supset$ & ${\bf 420}$  
     \\
     3 & $({\bf 16}\otimes{\bf 16})_{\text{sym}}$ & $=$ & ${\bf 1} \oplus {\bf 135}$
       & ${\bf 16}\otimes{\bf 128}$ & $\supset$ & ${\bf 1920}$  
     \end{tabular}
     \\[0.3cm]     
     \begin{tabular}{c|ccc}
     $d$ & $\chi \otimes \chi$ && $A_3$  \\ \hline
     7 & $({\bf 16}\otimes{\bf 16})_{\text{antisym}}$ & $\supset$ &  ${\bf 1}\oplus{\bf 5}\oplus{\bf 14}\oplus{\bf 35}$  \\
     6 & $({\bf (4,5)}\otimes{\bf (5,4)})_{\text{sym}}$ & $\supset$ &  ${\bf (4,4)}\oplus{\bf (4,16)}\oplus{\bf (16,4)}$  \\
     5 & $({\bf 48}\otimes{\bf 48})_{\text{sym}}$ & $\supset$ &  ${\bf 36}\oplus{\bf 315}$  \\
     4 & $({\bf 56}\otimes{\bf 56})_{\text{sym}}$ & $\supset$ &  ${\bf 420}$ \\
     3 & $({\bf 128}\otimes{\bf 128})_{\text{sym}}$ & $\supset$ &  ${\bf 1820}$ 
     \end{tabular}
     \caption{\label{tab:A1A2A3Max} \small For the maximal supergravities in $d$ dimensions it is listed
              which components of the $T$-tensor contribute to the fermionic mass tensors $A_1$, $A_2$ and $A_3$.
	      These tensors have to be composed out of $H$-representations that appear in
	      (appropriately (anti-) symmetrized) fermionic bilinears, as listed in
	      the table. The subset symbol $\supset$ is used if not all of the possible representations appear
	      (because they are not present in the $T$-tensor).
	      In even dimensions also the respective dual representations are present, e.g.\ in
	      $d=4$ we also have $(\overline{\bf 8}\otimes\overline{\bf 8})_{\text{sym}}=\overline{\bf 36}$
	      in $A_1$.
	      }
\end{table}

In table \ref{tab:TtenMax} and \ref{tab:A1A2A3Max} we list
the irreducible components of the $T$-tensor and of the fermionic mass matrices
for the maximal supergravities in dimensions $3\leq d \leq 7$. Comparing the two tables shows that
every component of the $T$-tensor appears somewhere in $A_1$, $A_2$ or $A_3$, i.e.\ all components are used
in the fermionic couplings. This however is a special feature of the maximal supergravities.
In general, not all components are used, as we will see for the half-maximal supergravities in the following chapters.

The description of the $T$-tensors completes our general discussion of gauged supergravity theories. 
In this chapter we first introduced the
embedding tensor~$\Theta$. This tensor parameterizes the minimal couplings of vector fields to symmetry generators in
the covariant derivative.
We then showed which additional changes in the Lagrangian and in the supersymmetry rules are necessary in order
to preserve gauge invariance and supersymmetry. All these couplings are parameterized in terms of $\Theta$.
We also introduced the linear and quadratic constraints that $\Theta$ has to satisfy in order to describe a valid gauging.
In the following chapters these general methods are applied to concrete examples.

\chapter{The $N=4$ supergravities in $d=4$} \label{ch:D4}

In this chapter we present the universal Lagrangian and the Killing spinor equations of 
the general gauged $N=4$ supergravities in four dimensions.
For an even number of spacetime dimensions there are subtleties that seem to hamper the universal description of the gauged
theory.
In particular, in four dimensions the global symmetry group $G_0$ of a supergravity theory is generically only realized on-shell
since it involves duality rotations between the electric and magnetic vector fields \cite{Gaillard:1981rj,deWit:2001pz}.
Only together the electric vector fields that appear in the ungauged Lagrangian and 
the magnetic vector fields that are introduced on-shell form a representation under $G_0$.
Thus, in a $G_0$ invariant formulation of the gaugings
the magnetic vector fields appear in the covariant derivative and therefore in the Lagrangian.
These issues were resolved in \cite{deWit:2005ub}, where for a general four-dimensional theory
it was explained
how to consistently couple electric and magnetic vector gauge fields together with two-form tensor gauge fields
in order to describe gaugings of a generic subgroup of $G_0$.
Here we apply these results to the case of gauged $N=4$ supergravities,
reviewing the work of \cite{Schon:2006kz}.
Very good lecture notes on the subject also exist already \cite{Derendinger:2006jb}.

Examples of $N=4$ supergravities in four dimensions are already known for more than twenty years
\cite{Das:1977uy, Cremmer:1977tc,Cremmer:1977tt,Freedman:1978ra,Gates:1982ct,Gates:1982db,Gates:1982an,
deRoo:1984gd,deRoo:1985jh,deRoo:1986yw,Bergshoeff:1985ms}.
From a string theory perspective these half-maximal supergravities can,
for example, result from orientifold compactifications of IIB supergravity
\cite{Frey:2002hf,Kachru:2002he}.
In this picture
parts of the embedding tensor correspond to fluxes or additional branes on the background
\cite{D'Auria:2002tc,D'Auria:2003jk,Angelantonj:2003rq,Angelantonj:2003up,Berg:2003ri},
but not all the known gaugings could so far be identified in this way.
Lower $N$ theories can be obtained by truncation of the $N=4$ supergravities.
For example certain relevant $N=1$
K\"ahler potentials can be computed from the $N=4$ scalar potential
\cite{Derendinger:2004jn,Derendinger:2005ph,Derendinger:2006ed,Derendinger:2006hr}.
It would also be interesting to find a gauged $N=4$ supergravity that possesses a de Sitter ground state,
since this is not the case for the theories investigated so far
\cite{deRoo:2003rm,deRoo:2006ms}.

\section{Embedding tensor and gauge fields}
\label{sec:D4con}

\subsection{Linear and quadratic constraint}

The global symmetry group of ungauged $d=4$ half-maximal supergravity
is $G_0={\rm SL}(2)\times {\rm SO}(6,n)$,
where $n$ denotes the number of vector multiplets.
We use indices $\aa=1,2$ and $\Ma=1,\ldots,6+n$ to label vector representations of ${\rm SL}(2)$
and ${\rm SO}(6,n)$. The generators of $G_0$ are $t_{\aa\ab}=t_{(\aa\ab)}$ and $t_{\Ma\Mb}=t_{[\Ma\Mb]}$.
In the respective vector representation they read
\begin{align}
   {(t_{\Ma\Mb})_\Mc}^\Md &= \delta^\Md_{[\Ma} \eta_{\Mb]\Mc} \; , &
   {(t_{\aa\ab})_\ac}^\ad &= \delta^\ad_{(\aa} \epsilon_{\ab)\ac} \; ,
\end{align}
where $\eta_{\Ma\Mb}$ is the ${\rm SO}(6,n)$ metric and $\epsilon_{\aa\ab}$ is the ${\rm SL}(2)$ invariant
Levi-Civita tensor.

The electric and magnetic vector fields together transform as a doublet under ${\rm SL}(2)$ and
a vector under ${\rm SO}(6,n)$, i.e.\ we have vector fields $A_\mua^{\Ma\aa}$.
The covariant derivative \eqref{GenCovDiv} takes the form
\begin{align}
   D_\mua &= \partial_\mua - g \, A_\mua{}^{\Ma\aa} \, {\Theta_{\Ma\aa}}^{\Mb\Mc} \, t_{\Mb\Mc}
                           - g \, A_\mua{}^{\Ma\aa} \, {\Theta_{\Ma\aa}}^{\ab\ac} \, t_{\ab\ac} \; .
   \label{D4CovDivTh}
\end{align}
It was already said that the gauge coupling constant $g$ could be absorbed into
the embedding tensor and is just used for convenience to keep track of the order of deformation.
The two components ${\Theta_{\Ma\aa}}^{\Mb\Mc}$ and ${\Theta_{\Ma\aa}}^{\ab\ac}$ further decompose into irreducible
representations of $G_0$. According to table \ref{LinCon2} the linear constraint only allows for two of these
irreducible components to be non-zero for a consistent gauging. These two components
are described by the tensors $\xi_{\aa\Ma}$ and $f_{\aa\Ma\Mb\Mc}=f_{\aa[\Ma\Mb\Mc]}$.
Both are doublets under ${\rm SL}(2)$, but $\xi_{\aa\Ma}$ is vector under ${\rm SO}(6,n)$ while
$f_{\aa\Ma\Mb\Mc}$ transforms as a thee-fold antisymmetric tensor. These tensors constitute the embedding
tensor as follows
\begin{align}
   \Theta_{\Ma\aa}{}^{\Mb\Mc} &= {f_{\aa\Ma}}^{\Mb\Mc} + \frac 1 2 \delta_\Ma^{[\Mb} \xi^{\Mc]}_\aa \; , &
   \Theta_{\Ma\aa}{}^{\ab\ac} &= \frac 1 2 \xi_{\ad\Ma} \epsilon^{\ad(\ab} \delta^{\ac)}_\aa \; .
   \label{D4ThetaFXI}
\end{align}
Working out the quadratic constraint \eqref{QconGen1} on $\Theta$ in terms of $\xi_{\aa\Ma}$ and $f_{\aa\Ma\Mb\Mc}$
yields the following set of constraints
\begin{align}
   \xi_\aa^\Ma \xi_{\ab\Ma} &= 0 \, , \nonumber \\[1ex]
   \xi^\Mc_{(\aa}  f_{\ab)\Mc\Ma\Mb} &= 0 \, , \nonumber \\[1ex]
   3 f_{\aa\Me[\Ma\Mb} {f_{\ab\Mc\Md]}}^\Me + 2 \xi_{(\aa[\Ma} f_{\ab)\Mb\Mc\Md]} &= 0 \; , \nonumber \\[1ex]
   \epsilon^{\aa\ab} \left( \xi_{\aa}^\Mc f_{\ab\Mc\Ma\Mb} + \xi_{\aa\Ma} \xi_{\ab\Mb} \right) &= 0 \, , \nonumber \\
   \epsilon^{\aa\ab} \left( f_{\aa\Ma\Mb\Me} {f_{\ab\Mc\Md}}^\Me - \xi^\Me_\aa f_{\ab\Me[\Ma[\Mc} \eta_{\Md]\Mb]}
       - \xi_{\aa[\Ma} f_{\Mb][\Mc\Md]\ab} + \xi_{\aa[\Mc} f_{\Md][\Ma\Mb]\ab} \right) &= 0 \, .
   \label{QConD4}
\end{align}
To summarize, gaugings are parameterized by the tensors
$\xi_{\aa\Ma}$ and $f_{\aa\Ma\Mb\Mc}$ that have to obey the constraints \eqref{QConD4}.
For any particular gauging these are constant tensors (their entries are fixed real numbers), but in the
construction of the general gauged theory they are treated as spurionic objects that transform under $G_0$.
The $G_0$ invariance is thus formally retained. This is possible because the constraints \eqref{QConD4} are
$G_0$ invariant, i.e.\ for a solution of \eqref{QConD4} a $G_0$ transformation yields another solution. Different
solutions that are related in this way describe equivalent gauged theories.

\subsection{Choice of symplectic frame}

It is convenient to define a composite index for the vector fields by
$A_\mua{}^\sMa = A_\mua{}^{\Ma\aa}$. On the linear space of vector fields there is 
a symplectic form $\Omega_{\sMa\sMb}$ defined by 
\begin{align}
   \Omega_{\sMa\sMb} \, &= \, \Omega_{\Ma\aa \, \Mb\ab} \, \equiv \, \eta_{\Ma\Mb} \epsilon_{\aa\ab} \, , &
   \Omega^{\sMa\sMb} \, &= \, \Omega^{\Ma\aa \, \Mb\ab} \, \equiv \, \eta^{\Ma\Mb} \epsilon^{\aa\ab} \, .
\end{align}
The existence of this symplectic form is a general feature of four-dimensional
gauge theory. Every decomposition $A_\mua^{\sMa}=(A_\mua^\La,A_{\mua\,\La})$ 
such that
\begin{align}
   \Omega_{\sMa\sMb} &=
   \begin{pmatrix} \Omega^{\La\Lb} & \Omega^{\La}{}_{\Lb} \\ 
                   \Omega_{\La}{}^{\Lb} & \Omega_{\La\Lb} \end{pmatrix}
		   =   
   \begin{pmatrix} 0 & \mathbbm{1} \\ - \mathbbm{1} & 0 \end{pmatrix} 
   \label{SplitElMag}
\end{align}
provides a consistent split into an equal number of electric $A_\mua^\La$ and magnetic
$A_{\mua\,\La}$ vector fields. That means the ungauged theory can be formulated such
that the electric fields $A_\mua^\La$ appear in the Lagrangian while their dual 
magnetic fields $A_{\mua\,\La}$ are only introduced onshell. Such a decomposition is
called a symplectic frame.
The symplectic group ${\rm Sp}(12+2n)$ is the group of linear transformations
that preserve $\Omega_{\sMa\sMb}$. Every two symplectic frames are related by
a symplectic rotation.

The gauge group generators in the vector field representation take the form
\begin{align}
   {X_{\sMa\sMb}}^\sMc &= {X_{\Ma\aa \, \Mb \ab}}^{\Mc\ac}  
                       =   \Theta_{\Ma\aa}{}^{\Md\Me} \, (t_{\Md\Me})_\Mb{}^\Mc \, \delta_\ab^\ac
		         + \Theta_{\Ma\aa}{}^{\ad\aee} \, (t_{\ad\aee})_\ab{}^\ac \, \delta_\Mb^\Mc
   \nonumber \\ &= -  \delta_\ab^\ac \, {f_{\aa\Ma\Mb}}^\Mc
      + \frac 1 2 \left( \delta_\Ma^\Mc \, \delta_\ab^\ac \, \xi_{\aa\Mb} -  \delta_\Mb^\Mc \, \delta_\aa^\ac \, \xi_{\ab\Ma} 
      - \delta_\ab^\ac \, \eta_{\Ma\Mb} \, \xi_{\aa}^\Mc
      + \epsilon_{\aa\ab} \, \delta_\Mb^\Mc \, \xi_{\ad\Ma} \, \epsilon^{\ad\ac} \right) \; .
  \label{DefXD4}      
\end{align}
These generators satisfy
\begin{align}
   {X_{\sMa[\sMb}}^\sMd \Omega_{\sMc]\sMd} &= 0  \; , &
   {X_{(\sMa\sMb}}^\sMd \Omega_{\sMc)\sMd} &= 0  \; .
   \label{D4XRel}
\end{align}
The first of these equations states that the symplectic form $\Omega_{\sMa\sMb}$
is invariant under gauge transformations. In fact, it is even invariant under
$G_0$ transformations, i.e.\ we have the following embedding of groups
$G \subset G_0 \subset {\rm Sp}(12+2n)$, where $G$ denotes the gauge group.
The second relation in \eqref{D4XRel}
was found in \cite{deWit:2005ub} to be the universal way of expressing the linear constraint
in four dimensions. This equation was used to work out the decomposition \eqref{D4ThetaFXI}
of the embedding tensor into its irreducible components $\xi_{\aa\Ma}$ and $f_{\aa\Ma\Mb\Mc}$.

In the following section we use a particular symplectic frame to give the Lagrangian
of the general gauging. The ${\rm SL}(2)$ doublet is decomposed as $\aa=(+,-)$ such that
$\epsilon_{+-}=\epsilon^{+-}=1$. We then use $A_\mua^{\Ma+}$ as electric and
$A_\mua^{\Ma-}$ as magnetic vector fields. This decomposition obviously satisfies
\eqref{SplitElMag}. If we only consider those symmetry transformations that do not
mix electric and magnetic vector fields the global symmetry group $G_0$ is
broken down to ${\rm SO}(1,1)\times{\rm SO}(6,n)$. Only this reduced group is realized
as symmetry group of the ungauged Lagrangian (and 
of the gauged Lagrangian when considering $\xi_{\aa\Ma}$ and $f_{\aa\Ma\Mb\Mc}$ as
spurionic objects). Note that the gauge group $G$ need not be contained in this reduced
offshell symmetry group. 

In order to illustrate the meaning of the quadratic constraints \eqref{QConD4}
we first consider the case of purely electric gaugings for the particular symplectic
frame just chosen. Purely electric gaugings are those for which only the electric vector
fields appear in the covariant derivative \eqref{D4CovDivTh}.
In this case we have $\xi_{\aa\Ma}=0$ and $f_{-\Ma\Mb\Mc}=0$.
We then find
${f_{+\Ma\Mb}}^\Mc = f_{+\Ma\Mb\Md} \, \eta^{\Md\Mc}$ to be the structure constants of
the gauge group and the constraint \eqref{QConD4} simplifies to the Jacobi identity
\begin{align}
   f_{+\Me[\Ma\Mb} {f_{+\Mc\Md]}}^\Me &= 0 \, .
   \label{JacobiFP}
\end{align}
The complete quadratic constraint \eqref{QConD4} can be viewed a generalization of this Jacobi
for more general gaugings.
Note that the ${\rm SO}(6,n)$ metric $\eta_{\Ma\Mb}$ is used in \eqref{JacobiFP} to contract the indices
in \eqref{JacobiFP},
while in the ordinary Jacobi identity the Cartan Killing form occurs.  Also the indices
$\Ma,\Mb,\ldots$ run over $6+n$ values while the gauge group might be of smaller dimension.
These issues will be discussed in section \ref{sec:D4examples}.

\subsection{Vector and tensor gauge fields}

The ungauged $N=4$ supergravity contains the metric, electric vector fields and scalars
as bosonic fields in the Lagrangian.
The dual magnetic vectors and two-form gauge fields are only introduced on-shell.
The latter come in the adjoint representation of $G_0$ and since $G_0$ has two factors
there are also two kinds of
two-form gauge fields, namely $B_{\mua\mub}^{\Ma\Mb}=B_{\mua\mub}^{[\Ma\Mb]}$ and 
$B_{\mua\mub}^{\aa\ab}=B_{\mua\mub}^{(\aa\ab)}=(B_{\mua\mub}^{++},B_{\mua\mub}^{+-},B_{\mua\mub}^{--})$.
For the general description of the gauged theory all these fields appear as free fields in the Lagrangian
\cite{deWit:2005ub}.
For the magnetic vectors this is necessary because they can appear as gauge fields
in the covariant derivative while the two-forms in turn are required in order to
consistently couple the vector fields.
Neither of these newly introduced gauge fields gets equipped with a kinetic term
and via their first order equations of motion they eventually turn out to be dual to the electric vector fields ${A_\mua}^{\Ma+}$
and to the scalars, respectively. Thus the number of degrees of freedom remains unchanged as compared to the ungauged theory.

We want to give the gauge invariant field strengths and the gauge transformations of
the vector and two-form gauge fields by applying the general formulas of
section \ref{sec:GenNonAVecTen}.
In these general formulas we used the tensors $d_{I\Ma\Mb}$ and $Z^{\Ma I}$ 
that now are given as follows
\begin{align}
   d_{I\Ma\Mb} \quad &\widehat= \quad
   \left\{ \begin{array}{ll}  
       d^{\aa\ab}{}_{\Mc\ac\Md\ad} &= - \eta_{\Mc\Md} \, \delta^{(\aa}_{(\ac} \, \delta^{\ab)}_{\ad)} 
        \\[0.2cm]
       d^{\Ma\Mb}{}_{\Mc\ac\Md\ad} &= \epsilon_{\ac\ad} \, \delta^{[\Ma}_{[\Mc} \, \delta^{\Mb]}_{\Md]} 
            \end{array} \right. \quad ,
   \nonumber \\[0.2cm]
   Z^{\Ma I} \quad &\widehat= \quad
   \left\{ \begin{array}{ll}  
       Z^{\Ma\aa\,\ab\ac} &= \eta^{\Ma\Md} \, \epsilon^{\aa\ad} \, \Theta_{\Md\ad}{}^{\ab\ac}  
       \\[0.2cm]
       Z^{\Ma\aa\,\Mb\Mc} &= \eta^{\Ma\Md} \, \epsilon^{\aa\ad} \, \Theta_{\Md\ad}{}^{\Mb\Mc}  
           \end{array} \right. \quad .
\end{align}
From equation \eqref{DefHHH} we then find
the following covariant field strengths\footnote{
Note that the indices $+$ and $-$ on the vector fields and on their field strengths
distinguish the electric ones from the magnetic ones and thus do not indicate complex self-dual combinations
of the field strengths as
is common in the literature. We hope note to confuse the reader with that notation.}
\begin{align}
   {\cal H}_{\mua\mub}^{\Ma+} &= 2 \partial_{[\mua} {A_{\mub]}}^{\Ma+} 
            - g \, \hat f{}_{\aa\Mb\Mc}{}^\Ma {A_{[\mua}}^{\Mb\aa} {A_{\mub]}}^{\Mc+} 
	       \nonumber \\ & \qquad \qquad
	    + \frac g 2 \, {{\Theta_{-}}^{\Ma}}_{\Mb\Mc} B_{\mua\mub}^{\Mb\Mc}
	    + \frac g 2 \, {\xi_+}^\Ma B_{\mua\mub}^{++} + \frac g 2 {\xi_{-}}^\Ma B_{\mua\mub}^{+-} \; ,
   \nonumber \\	    
   {\cal H}_{\mua\mub}^{\Ma-} &= 2 \partial_{[\mua} {A_{\mub]}}^{\Ma-} 
            - g \, \hat f{}_{\aa\Mb\Mc}{}^\Ma {A_{[\mua}}^{\Mb\aa} {A_{\mub]}}^{\Mc-} 
	       \nonumber \\ & \qquad \qquad
	    - \frac g 2 \, {{\Theta_{+}}^{\Ma}}_{\Mb\Mc} B_{\mua\mub}^{\Mb\Mc}
	    + \frac g 2 \, {\xi_-}^\Ma B_{\mua\mub}^{--} + \frac g 2 {\xi_{+}}^\Ma B_{\mua\mub}^{+-} \; ,
    \nonumber \\
    {\cal H}^{\Ma\Mb}_{\mua\mub\muc} &= 3 \, \partial^{\phantom{\nu}}_{[\mua} B^{\Ma\Mb}_{\mub\muc]} 
                 + 6 \, \epsilon_{\aa\ab} \, A^{\aa[\Ma}_{[\mua} \, \partial^{\phantom{\nu}}_{\mub} A^{\Mb]\ab}_{\muc]} + {\cal O}(g)\; ,
    \nonumber \\ 		       
    {\cal H}^{\aa\ab}_{\mua\mub\muc} &= 3 \, \partial^{\phantom{\nu}}_{[\mua} B^{\aa\ab}_{\mub\muc]}
                  + 6 \, \eta_{\Ma\Mb} \, A^{\Ma(\aa}_{[\mua} \, \partial^{\phantom{\nu}}_\mub
		  A^{\ab)\Mb}_{\muc]} + {\cal O}(g) \; .
   \label{FieldStrD4}	    
\end{align}
Only the electric field strength ${\cal H}_{\mua\mub}^{\Ma+}$ enters the Lagrangian,
but the magnetic and the two-form field strengths appear in the equations of motion.
For our purposes it is sufficient to know the two-form field strengths up to terms
of order $g$.

It is useful to define the following combinations of the electric field strengths
\begin{align}
   {{\cal G}_{\mua\mub}}^{\Ma+} &\equiv {{\cal H}_{\mua\mub}}^{\Ma+} \; , \nonumber \\
   {{\cal G}_{\mua\mub}}^{\Ma-} &\equiv 
          e^{-1} \, \eta^{\Ma\Mb} \, \epsilon_{\mua\mub\muc\mud} \, 
	    \frac{ \partial {\cal L}_{\text{kin}} } { \partial {\cal H}_{\muc\mud}^{\Mb+} }
	  \nonumber \\ &
          \; = \; - \ft 1 2 \,  \epsilon_{\mua\mub\muc\mud} \, \Im(\tau) M^{\Ma\Mb} \eta_{\Mb\Mc} {\cal H}^{\Mc+\,\muc\mud}
	     - \Re(\tau) {\cal H}^{\Ma+} _{\mua\mub} \; .
   \label{DefG}	     
\end{align}
We give the Lagrangian ${\cal L}_{\text{kin}}$ only in the next section, but we want to anticipate that 
in the ungauged theory (i.e.\ in the limit $g\rightarrow 0$) the equations of motion for the electric vector fields take the form
$\partial_{[\mua} {{\cal G}_{\mub\muc]}}^{\Ma-} = 0$.
The  magnetic vector fields can then be introduced via
${\cal H}^{\Ma-}_{\mua\mub} = {\cal G}^{\Ma-}_{\mua\mub}$.
Thus ${\cal G}^{\Ma\aa}=({\cal G}^{\Ma+},{\cal G}^{\Ma-})$ and ${\cal H}^{\Ma\aa}$ are
on-shell identical.

The existence of ${\cal G}^{\Ma\aa}$ results in the $d=4$ subtlety that in the
general gauge transformations \eqref{GenGauge}
we have to replace ${\cal H}^{\Ma\aa}$
by ${\cal G}^{\Ma\aa}$ in order to find a formulation
entirely in terms of electric vector fields in the limit $g \rightarrow 0$. 
Thus, the gauge transformations of the vector and tensor
gauge fields read
\begin{align}
   \delta A_{\mua}^{\Ma+} &= D_\mua \Lambda^{\Ma+} 
	    - \frac g 2 \, {{\Theta_{-}}^{\Ma}}_{\Mb\Mc} \Xi_{\mua}^{\Mb\Mc}
	    - \frac g 2 \, {\xi_+}^\Ma \Xi_{\mua}^{++} - \frac g 2 {\xi_{-}}^\Ma \Xi_{\mua}^{+-} \; ,
   \nonumber \\	    
   \delta A_{\mua}^{\Ma-} &= D_\mua \Lambda^{\Ma-} 
	    + \frac g 2 \, {{\Theta_{+}}^{\Ma}}_{\Mb\Mc} \Xi_{\mua}^{\Mb\Mc}
	    - \frac g 2 \, {\xi_-}^\Ma \Xi_{\mua}^{--} - \frac g 2 {\xi_{+}}^\Ma \Xi_{\mua}^{+-} \; ,
   \nonumber \\	    
    \Delta B^{\Ma\Mb}_{\mua\mub} &= 2 D_{[\mua} \Xi^{\Ma\Mb}_{\mub]} 
                       - 2 \epsilon_{\aa\ab} \Lambda^{\aa[\Ma} \, {\cal G}^{\Mb]\ab}_{\mua\mub} \; ,
    \nonumber \\ 		       
    \Delta B^{\aa\ab}_{\mua\mub} &= 2 D_{[\mua} \Xi^{\aa\ab}_{\mub]}
                       + 2 \eta_{\Ma\Mb} \Lambda^{\Ma(\aa} \, {\cal G}^{\ab)\Mb}_{\mua\mub} \; ,
    \label{D4GaugeTrafo}
\end{align}
where the gauge parameters are $\Lambda^{\Ma\aa}$,
$\Xi^{\Ma\Mb}_\mua=\Xi^{[\Ma\Mb]}_\mua$ and $\Xi^{\aa\ab}_\mua=\Xi^{(\aa\ab)}_\mua$,
and we used  the covariant variations \eqref{CovD1} of the two-form gauge fields 
\begin{align}
    \Delta B^{\Ma\Mb}_{\mua\mub} &= \delta B^{\Ma\Mb}_{\mua\mub} 
                       - 2 \epsilon_{\aa\ab} A^{\aa[\Ma}_{[\mua} \, \delta A^{\Mb]\ab}_{\mub]} \; ,
    \nonumber \\ 		       
    \Delta B^{\aa\ab}_{\mua\mub} &= \delta B^{\aa\ab}_{\mua\mub}
                       + 2 \eta_{\Ma\Mb} A^{\Ma(\aa}_{[\mua} \, \delta A^{\ab)\Mb}_{\mub]} \; .
    \label{CovDB}		       
\end{align}
In the Lagrangian the two-form gauge fields only appear projected with
$\Theta_{\Ma\aa}{}^{\ab\ac}$ and $\Theta_{\Ma\aa}{}^{\Ma\Mb}$, respectively.
On the two-forms the gauge transformations \eqref{D4GaugeTrafo} only close 
under this projection. The gauge algebra is a special case of \eqref{GenGaAlg}.

\section{Lagrangian and field equations}
\label{sec:D4lag}

The $N=4$ gravity multiplet contains as bosonic degrees of freedom the metric,
six massless vectors and two real massless scalars.
The scalar fields constitute an ${\rm SL}(2)/{\rm SO}(2)$ coset\footnote{
In the literature the symmetry group is usually denoted by ${\rm SU}(1,1)$,
however, we prefer to treat it as ${\rm SL}(2)$ 
which is of course the same group but with different conventions concerning
its fundamental representation.}.
This coset can 
equivalently be described by a complex number $\tau$ with $\Im(\tau)>0$ or
by a symmetric positive definite matrix $M_{\aa\ab} \in {\rm SL}(2)$.
The relation between these two descriptions is given by
\begin{align}
   M_{\aa\ab} &= \frac 1 {\Im(\tau)} \begin{pmatrix} |\tau|^2 & \Re(\tau) \\ \Re(\tau) & 1  \end{pmatrix}\;, &
   M^{\aa\ab} &= \frac 1 {\Im(\tau)} \begin{pmatrix} 1 & -\Re(\tau) \\ -\Re(\tau) & |\tau|^2 \end{pmatrix}\;, 
\end{align}
where $M^{\aa\ab}$ is the inverse of $M_{\aa\ab}$.
The ${\rm SL}(2)$ symmetry action on $M_{\aa\ab}$
\begin{align}
   M &\rightarrow g M g^T \;,&
   g &= \begin{pmatrix} a & b \\ c & d \end{pmatrix} \; \in {\rm SL}(2) \; ,
\end{align}
acts on $\tau$ as a M\"obius transformation $\tau \rightarrow (a \tau + b)/(c \tau + d)$.

We couple the gravity multiplet to $n$ vector multiplets, each containing one
vector and six real scalars. The scalars of the vector multiplets arrange
in the coset ${\rm SO}(6,n)/{\rm SO}(6) \times {\rm SO}(n)$
which is described by coset representatives
${{\cal V}_\Ma}^{\xa}$ and ${{\cal V}_\Ma}^{\ya}$
where $\ya=1,\ldots,6$ and $\xa=1,\ldots,n$ denote ${\rm SO}(6)$ and ${\rm SO}(n)$
vector indices, respectively.
The matrix ${\cal V}=({{\cal V}_\Ma}^{\ya},\,{{\cal V}_\Ma}^{\xa})$ 
is an element of ${\rm SO}(6,n)$, i.e.\
\begin{align}
   \eta_{\Ma\Mb} &= - {{\cal V}_\Ma}^{\ya} {{\cal V}_\Mb}^{\ya} + {{\cal V}_\Ma}^\xa {{\cal V}_\Mb}^\xa \; ,
   \label{DefEta}
\end{align}
where $\eta_{\Ma\Mb}=\diag(-1, -1, -1, -1, -1, -1, +1, \ldots, +1)$ is the ${\rm SO}(6,n)$ metric.
Global ${\rm SO}(6,n)$ transformations act on ${\cal V}$ from the left while local ${\rm SO}(6)  \times {\rm SO}(n)$ transformations
act from the right
\begin{align}
   {\cal V} \, &\rightarrow \, g {\cal V} h(x) \; , &&
   g \in {\rm SO}(6,n), \quad h(x) \in {\rm SO}(6)  \times {\rm SO}(n) \;.
   \label{CosetSO6n}
\end{align}
Analogous to $M_{\aa\ab}$ this coset space may be parameterized by
a symmetric positive definite scalar metric $M={\cal V}{\cal V}^T$,
explicitly given by
\begin{align}
   M_{\Ma\Mb} &= {{\cal V}_\Ma}^\xa {{\cal V}_\Mb}^\xa + {{\cal V}_\Ma}^{\ya} {{\cal V}_\Mb}^{\ya} \; .
   \label{DefMVV}
\end{align}
Its inverse is denoted by $M^{\Ma\Mb}$. Note that each of the matrices $M_{\Ma\Mb}$, ${{\cal V}_\Ma}^{\ya}$
and ${{\cal V}_\Ma}^{\xa}$ alone already parameterizes the ${\rm SO}(6,n)$ part of the scalar coset.

In order to give the scalar potential below
we also need to define the scalar dependent completely antisymmetric tensor
\begin{align}
   M_{\Ma\Mb\Mc\Md\Me\Mf} &= \epsilon_{\ya\yb\yc\yd\ye\yf} \,
                             {{\cal V}_\Ma}^{\ya} {{\cal V}_\Mb}^{\yb} {{\cal V}_\Mc}^{\yc} 
                             {{\cal V}_\Md}^{\yd} {{\cal V}_\Me}^{\ye} {{\cal V}_\Mf}^{\yf}  \; .
   \label{DefM6}			     
\end{align}

In addition to ${\Theta_{\Ma\aa}}^{\Mb\Mc}$ and ${\Theta_{\Ma\aa}}^{\ab\ac}$ defined in
\eqref{D4ThetaFXI} the following combination of $f_{\aa\Ma\Mb\Mc}$ and $\xi_{\aa\Ma}$
appears regularly
\begin{align}
   {\hat f}_{\aa\Ma\Mb\Mc} &= f_{\aa\Ma\Mb\Mc} - \xi_{\aa[\Ma}  \, \eta_{\Mc]\Mb} - \, \ft 3 2 \, \xi_{\aa\Mb} \eta_{\Ma\Mc} \; .
   \label{DefThetaF}
\end{align}

We can now present the bosonic Lagrangian of the general gauged theory\footnote{
Our space-time metric has signature $(-,+,+,+)$ and the Levi-Civita is a proper space-time tensor, i.e.\
$\epsilon^{0123}=e^{-1}$, $\epsilon_{0123}=- e$.}
\begin{align}
   {\cal L}_{\text{bos}} &= {\cal L}_{\text{kin}} + {\cal L}_{\text{top}} + {\cal L}_{\text{pot}} \; .
   \label{LagBosD4}
\end{align}
It consists of a kinetic term
\begin{align} 
   e^{-1} {\cal L}_{\text{kin}} & =  \ft 1 2 \, R 
                    + \ft 1 {16} \, (D_\mua M_{\Ma\Mb}) (D^\mua M^{\Ma\Mb})
     - \, \frac 1 {4 \, \Im(\tau)^2}  (D_\mua \tau) (D^\mua \tau^*) 
     \nonumber \\[1ex] 
     &  \qquad
     - \, \ft 1 4 \, \Im(\tau) \, M_{\Ma\Mb} {\cal H}_{\mua\mub}{}^{\Ma+} {\cal H}^{\mua\mub\Mb+}
               + \, \ft 1 8 \, \Re(\tau) \, \eta_{\Ma\Mb} \, \epsilon^{\mua\mub\muc\mud} 
    {{\cal H}_{\mua\mub}}^{\Ma+} {{\cal H}_{\muc\mud}}^{\Mb+}   \;,
\end{align}
a topological term for the vector and tensor gauge fields \cite{deWit:2005ub}
\begin{align}
   e^{-1} {\cal L}_{\text{top}} &= - \, \frac g 2 \, \epsilon^{\mua\mub\muc\mud}  \nonumber \\ &
          \bigg\{ 
          \xi_{+\Ma} \eta_{\Mb\Mc} A_\mua^{\Ma-} A_\mub^{\Mb+} \partial_\muc A_\mud^{\Mc+} 
     - \left( \hat f_{-\Ma\Mb\Mc} + 2 \, \xi_{-\Mb} \eta_{\Ma\Mc} \right) A_\mua^{\Ma-} A_\mub^{\Mb+} \partial_\muc A_\mud^{\Mc-} 
	\nonumber \\ & \,       
     - \, \frac g 4 \, \hat f{}_{\aa\Ma\Mb\Me} \hat f{}_{\ab\Mc\Md}{}^\Me A_\mua^{\Ma\aa} A_\mub^{\Mb+} A_\muc^{\Mc\ab} A_\mud^{\Md-} 
     + \, \frac g {16} \, \Theta_{+\Ma\Mb\Mc} {{\Theta_{-}}^{\Ma}}_{\Md\Me} B_{\mua\mub}^{\Mb\Mc} B_{\muc\mud}^{\Md\Me} 
	\nonumber \\ & \,  
       - \ft 1 4 \left( \Theta_{-\Ma\Mb\Mc} B_{\mua\mub}^{\Mb\Mc} 
		                     + \xi_{-\Ma} B_{\mua\mub}^{+-} + \xi_{+\Ma} B_{\mua\mub}^{++} \right)
	 \big( 2 \partial_\muc A_\mud^{\Ma-} - g {\hat f}{}_{\aa\Md\Me}{}^\Ma A_\muc^{\Md\aa} A_\mud^{\Me-} \big)
	  \bigg\} \, ,
\end{align}
and a scalar potential
\begin{align}
   e^{-1} {\cal L}_{\text{pot}} &= - g^2 V   
     \nonumber \\ & = - \frac{g^2} {16} \bigg\{
      f_{\aa\Ma\Mb\Mc} f_{\ab\Md\Me\Mf} M^{\aa\ab} \Big[
           \ft 1 3 \, M^{\Ma\Md} M^{\Mb\Me} M^{\Mc\Mf} 
        + ( \ft 2 3 \, \eta^{\Ma\Md} -  M^{\Ma\Md} ) \eta^{\Mb\Me} \eta^{\Mc\Mf}   \Big]
	\nonumber \\ & \qquad \qquad
	        - \ft 4 9 \, f_{\aa\Ma\Mb\Mc} f_{\ab\Md\Me\Mf} \epsilon^{\aa\ab} M^{\Ma\Mb\Mc\Md\Me\Mf}
                + 3 \, \xi_\aa^\Ma \xi_\ab^\Mb  M^{\aa\ab} M_{\Ma\Mb}  \bigg\} \; .
   \label{VD4}		
\end{align}
The action of the covariant derivative \eqref{D4CovDivTh} explicitly reads for the scalar fields
\begin{align}
   D_\mua M_{\aa\ab} &= \partial_\mua M_{\aa\ab} + g A_\mua^{\Ma\ac} \xi_{(\aa\Ma} M_{\ab)\ac}
        - g A_\mua^{\Ma\ad} \xi_{\epsilon\Ma} \epsilon_{\ad(\aa} \epsilon^{\epsilon\ac} M_{\ab)\ac} \; , \nonumber \\
   D_\mua M_{\Ma\Mb} &= \partial_\mua M_{\Ma\Mb} + 2 g A_\mua{}^{\Mc\aa} {\Theta_{\aa\Mc(\Ma}}^{\Md} M_{\Mb)\Md}   \; .
\end{align}
Note that $\Im(\tau)^{-2} (D_\mua \tau) (D^\mua \tau^*) = - \ft 1 2 (D_\mua M_{\aa\ab}) (D^\mua M^{\aa\ab})$,
i.e.\ the kinetic term for $\tau$ can equivalently be expressed in terms of $M_{\aa\ab}$.

Under general variations of the vector and two-form gauge fields 
the Lagrangian varies as
\begin{align}
   e^{-1} \delta & {\cal L}_{\text{bos}} = \ft 1 8 g \left( \Theta_{-\Ma\Mb\Mc} \Delta B^{\Mb\Mc}_{\mua\mub}
                                    + \xi_{-\Ma} \Delta B_{\mua\mub}^{+-} + \xi_{+\Ma} \Delta B_{\mua\mub}^{++} \right)
		 \epsilon^{\mua\mub\muc\mud}  \left( {\cal H}_{\muc\mud}^{\Ma-} - {\cal G}_{\muc\mud}^{\Ma-} \right)
		     \nonumber \\ & 
		   + \ft 1 2 (\delta A_\mua^{\Ma+}) \left( g \, \xi_{\ab\Ma} M_{+\ac} D^\mua M^{\ab\ac}
		                + \frac g 2 \, {\Theta_{+\Ma\Mc}}^{\Mb} M_{\Mb\Md} D^\mua M^{\Md\Mc} 
				  - \epsilon^{\mua\mub\muc\mud} \eta_{\Ma\Mb} \, D_\mub {\cal G}^{\Mb-}_{\muc\mud}
				  \right)
		     \nonumber \\ & 
		   + \ft 1 2 (\delta A_\mua^{\Ma-}) \left( g \, \xi_{\ab\Ma} M_{-\ac} D^\mua M^{\ab\ac}
                          + \frac g 2 \, {\Theta_{-\Ma\Mc}}^{\Mb} M_{\Mb\Md} D^\mua M^{\Md\Mc} 
			  + \epsilon^{\mua\mub\muc\mud} \eta_{\Ma\Mb} \, D_\mub {\cal G}^{\Mb+}_{\muc\mud}  \right)
          \nonumber \\ & + \, \text{total derivatives,}
   \label{varyL}			  
\end{align}
where we used the covariant variations \eqref{CovDB}.
Plugging the gauge transformations \eqref{D4GaugeTrafo} into these general variations 
one finds the Lagrangian to transform into a total derivative, i.e.\ the action is gauge invariant
\cite{deWit:2005ub}.

Equation \eqref{varyL} encodes the gauge field equations of motion of the theory.
Variation of the two-form gauge fields
yields a projected version of the duality equation ${\cal H}^{\Ma-}_{\mua\mub} = {\cal G}^{\Ma-}_{\mua\mub}$
between electric and magnetic vector fields. 
From varying the electric vector fields one obtains a field equation for the electric vectors themselves which contains scalar
currents as source terms. Finally, the variation of the magnetic vectors gives a duality equation between scalars and
two-form gauge fields. To make this transparent one needs the modified Bianchi identity for ${\cal H}^{\Ma+}_{\mua\mub}$
which reads
\begin{align}
   D^{\phantom{\Ma}}_{[\mua} {\cal H}^{\Ma+}_{\mub\muc]} 
      &= \frac g 6 \left( {\Theta_-}{}^{\Ma}{}_{\Mc\Md} {\cal H}^{\Mc\Md}_{\mua\mub\muc}
       +{\xi_+}^\Ma {\cal H}_{\mua\mub\muc}^{++} + {\xi_{-}}^\Ma {\cal H}_{\mua\mub\muc}^{+-} \right) \; .
\end{align}

Thus we find that the tensors $f_{\aa\Ma\Mb\Mc}$ and $\xi_{\aa\Ma}$ do not only specify the gauge group but also
organize the couplings of the various fields. They determine which vector gauge fields appear in the covariant derivatives,
how the field strengths have to be modified, which magnetic vector fields and which two-form gauge fields enter the Lagrangian
and how they become dual to electric vector fields and scalars via their equation of motion.
Consistency of the entire construction
crucially depends on the quadratic constraints \eqref{QConD4}.

In principle one should also give the fermionic contributions to the Lagrangian and check supersymmetry to verify
that \eqref{LagBosD4} really describes the bosonic part of a supergravity theory.
We have obtained the results by applying the general method of covariantly coupling electric and magnetic vector gauge fields in 
a gauged theory \cite{deWit:2005ub} to the particular case of $N=4$ supergravity.
This fixes the bosonic Lagrangian up to the scalar potential. The latter is also strongly restricted by gauge invariance,
only those terms that appear in \eqref{VD4} are allowed. We obtained the pre-factors between the various terms
by matching the scalar potential with the one known from half-maximal supergravity in three spacetime dimensions \cite{deWit:2003ja},
see appendix \ref{app:D3}. The general theory then was compared with various special cases that were already
worked out elsewhere \cite{deRoo:1985jh,Bergshoeff:1985ms,
deRoo:2003rm,D'Auria:2002tc,D'Auria:2003jk,Angelantonj:2003rq,
Angelantonj:2003up,Wagemans:1990mv,Villadoro:2004ci,Kaloper:1999yr},
see section \ref{sec:D4examples}.

A symplectic rotation of the vector fields yields a different Lagrangian 
which describes the same theory at the level of the equations of motion. All possible
Lagrangians of gauged $N=4$ supergravity are thus parameterized by
$\xi_{\aa\Ma}$, $f_{\aa\Ma\Mb\Mc}$ and an element of ${\rm Sp}(12+2n)$.
It can be shown that as a consequence of the constraints \eqref{QConD4}
one can perform for every gauging a symplectic rotation such that a
purely electric gauging is obtained \cite{deWit:2005ub}\footnote{In the maximal supersymmetric theory,
i.e.\ for $N=8$, this statement can even be reversed,
i.e.\ every gauging that is purely
electric in some symplectic frame is consistent,
i.e.\ solves the quadratic constraints for the embedding tensor \cite{dWST4}.
This is different in $N=4$ where a nontrivial quadratic constraint remains
also for purely electric gaugings.}. In other words, 
for every particular gauging there exists a natural symplectic
frame such that no magnetic vector fields
and no two-form fields are necessary in the Lagrangian. However, this natural
symplectic frame is only defined implicitly in terms of $\xi_{\aa\Ma}$
and $f_{\aa\Ma\Mb\Mc}$.
In order to have the general gauged Lagrangian explicitly 
parameterized by $\xi_{\aa\Ma}$ and $f_{\aa\Ma\Mb\Mc}$ one needs the above construction
with magnetic vectors and two-forms.

\section{Killing spinor equations}
\label{sec:D4kill}

So far we have only considered bosonic fields and we do not intend to give the entire fermionic Lagrangian
nor the complete supersymmetry action. They can e.g.\ be found in the paper of Bergshoeff, Koh and Sezgin \cite{Bergshoeff:1985ms}
for purely electric gaugings when only $f_{+\Ma\Mb\Mc}$ is non-zero, and we have chosen most of our conventions
to agree with their work in this special case\footnote{
The structure constants $f_{\Ma\Mb\Mc}$ in \cite{Bergshoeff:1985ms} equal minus $f_{+\Ma\Mb\Mc}$.}.
In particular all terms of order $g^0$, 
i.e.\ terms of the ungauged theory, can be found there.

Our aim in this section is to give the
Killing spinor equations of the general gauged theory,
i.e.\ the variations of the gravitini and of the spin $1/2$ fermions under supersymmetry.
Those are required for example when studying BPS solutions or when analyzing the supersymmetry breaking or preserving
of particular ground states.

All the fermions carry a representation of  $H={\rm SO}(2) \times {\rm SO}(6) \times {\rm SO}(n)$
which is the maximal compact subgroup of $G_0$.
Instead of ${\rm SO}(6)$ we work with its covering group ${\rm SU}(4)$ in the following.
The gravity multiplet contains four gravitini $\psi_\mua^\jja$ and four spin $1/2$ fermions $\chi^\jja$ and
in the $n$ vector multiplet there are $4n$ spin 1/2 fermions $\lambda^{\xa\jja}$, where
$\jja=1,\ldots, 4$ and $\xa=1,\ldots,n$ are vector indices of ${\rm SU}(4)$ and ${\rm SO}(n)$.
The ${\rm SO}(2)={\rm U}(1)$ acts on the fermions as a multiplication with a complex phase
$\exp(i q \lambda(x))$, where the charges $q$ are given in table \ref{FermRepD4}.

\begin{table}[tb]
   \begin{center}
     \begin{tabular}{r|ccc}
            & ${\rm SO}(2)$ charges & ${\rm SU}(4)$ rep. & ${\rm SO}(n)$ rep. 
\\ \hline && \\[-0.4cm]
        gravitini $\psi_\mua^\jja$ & $- \, \ft 1 2$ & ${\bf 4}$ & ${\bf 1}$ \\[0.2cm]
	spin $1/2$ fermions $\chi^\jja$~ & $+ \, \ft 3 2$ & ${\bf 4}$ & ${\bf 1}$ \\[0.2cm]
	spin $1/2$ fermions $\lambda^{\xa\jja}$ & $+ \, \ft 1 2$ & ${\bf 4}$ & ${\bf n}$
     \end{tabular}
     \caption{\label{FermRepD4}{ \small $H$-representations of the fermions of $d=4$, $N=4$ supergravity}}
   \end{center}     
\end{table}

As usual we use gamma-matrices with
\begin{align}
   \{ \Gamma_\mua, \Gamma_\mub \} &= 2 \eta_{\mua\mub} \; , &
   (\Gamma_\mua)^\dag &= \eta^{\mua\mub} \Gamma_{\mub} \; , &
   \Gamma_5 &= i \Gamma_0 \Gamma_1 \Gamma_2 \Gamma_3 \; .
\end{align}
All our fermions are chiral. We choose $\psi_\mua^\jja$ and $\lambda^{\xa\jja}$ to be right-handed while $\chi^\jja$ is left-handed,
that is
\begin{align}
   \Gamma_5 \psi_\mua^\jja &= + \psi_\mua^\jja \; , &
   \Gamma_5 \chi^\jja &= - \chi^\jja \; , &
   \Gamma_5 \lambda^{\xa\jja} &= + \lambda^{\xa\jja} \; . &
\end{align}
Vector indices of ${\rm SU}(4)$ are raised and lowered by complex conjugation, i.e.\ for an ordinary ${\rm SU}(4)$ vector
$v_\jja = (v^\jja)^*$.
However, for fermions we need the matrix $B=i \Gamma_5 \Gamma_2$
to define $\phi_\jja = B (\phi^\jja)^*$. This ensures
that $\phi_\jja$ transforms as a Dirac spinor when $\phi^\jja$ does. The complex conjugate of a chiral spinor has
opposite chirality, e.g.\ $\chi_\jja=B (\chi^\jja)^*$ is right-handed\footnote{
Right-handed spinors can be described by Weyl-spinors $\phi^A$, and left-handed ones then turn to conjugate Weyl-spinors
$\phi_{\dot A}$. Here $A$ and $\dot A$ are (conjugate) ${\rm SL}(2,\mathbb{C})$ vector indices.
In the chiral representation of the Gamma-matrices
\begin{align*}
  \Gamma_\mu&=\left(\begin{array}{ccc} 0 & \sigma^\mu \\
          \sigma_{\mu} & 0 \end{array}\right) \; , &
  \Gamma_5&=\left(\begin{array}{ccc} \mathbbm{1} & 0 \\
           0 & - \mathbbm{1} \end{array}\right) \; , &
   B &= i \Gamma_5 \Gamma_2 =  \left(\begin{array}{ccc} 0 & \epsilon \\
          - \epsilon & 0 \end{array}\right) \; ,
\end{align*}
where $\epsilon$ is the two-dimensional epsilon-tensor and $\sigma_\mua=(\mathbbm{1},\vec \sigma)$,
$\sigma^\mua=\eta^{\mua\mub} \sigma_{\mub}=(-\mathbbm{1},\vec \sigma)$ contains the Pauli matrices,
we find right-handed spinors to have the form $\phi=(\phi^A,0)^T$ while left-handed ones look like $\phi=(0,\phi_{\dot A})^T$.
Thus we have $\chi^\jja=(0,\chi^\jja_{\dot A})^T$ and its complex conjugate is given by $\chi_\jja=(\chi_\jja^A,0)^T$ where
the Weyl-spinors are related by $\chi_\jja^A=\epsilon^{AB} (\chi^\jja_{\dot B})^*$.}.
For $\bar \phi{}_\jja = (\phi^\jja)^\dag \Gamma_0$ we define the complex conjugate by
$\bar \phi{}^\jja = (\bar \phi{}_\jja)^* B$ which yields
$\bar \phi{}_\jja \chi^\jja = \bar \chi{}^\jja \phi_\jja = (\bar \phi{}^\jja \chi_\jja)^* = (\bar \chi{}_\jja \phi^\jja)^*$.

An ${\rm SO}(6)$ vector $v^\ya$ can alternatively be described by an antisymmetric tensor
$v^{\jja\jjb}=v^{[\jja\jjb]}$ subject to the pseudo-reality constraint
\begin{align}
   v_{\jja\jjb} &= (v^{\jja\jjb})^* = \frac 1 2 \epsilon_{\jja\jjb\jjc\jjd} v^{\jjc\jjd} \; .
\end{align}
We normalize the map $v^\ya \mapsto v^{\jja\jjb}$ such that the scalar product becomes
\begin{align}
   v^\ya w^\ya &= \frac 1 2 \epsilon_{\jja\jjb\jjc\jjd}  v^{\jja\jjb} w^{\jjc\jjd} \; .
\end{align}   
We can thus rewrite the coset representative ${{\cal V}_\Ma}^{\ya}$ as ${{\cal V}_\Ma}^{\jja\jjb}$
such that the equations \eqref{DefEta} and \eqref{DefM6} become
\begin{align}
   \eta_{\Ma\Mb} &= - \frac 1 2 \epsilon_{\jja\jjb\jjc\jjd} {{\cal V}_\Ma}^{\jja\jjb} {{\cal V}_\Mb}^{\jjc\jjd} 
                     + {{\cal V}_\Ma}^\xa {{\cal V}_\Mb}^\xa \; ,
   \nonumber \\
   M_{\Ma\Mb\Mc\Md\Me\Mf}  
         &= -  \, 2 \, i \, \epsilon_{\jja\jjb\jjg\jjj} \, \epsilon_{\jjc\jjd\jjh\jjk} \, \epsilon_{\jje\jjf\jji\jjl} \,
	         {{\cal V}_{[\Ma}}^{\jja\jjb} {{\cal V}_\Mb}^{\jjc\jjd} {{\cal V}_\Mc}^{\jje\jjf} 
                             {{\cal V}_\Md}^{\jjg\jjh} {{\cal V}_\Me}^{\jji\jjj} {{\cal V}_{\Mf]}}^{\jjk\jjl}  \; .
\end{align}
The scalar matrices ${{\cal V}_\Ma}^{\jja\jjb}$ and ${{\cal V}_\Ma}^{\xa}$ can be used to translate from ${\rm SO}(6,n)$
representations
under which the vector and tensor gauge fields transform into ${\rm SO}(6) \times {\rm SO}(n)$ representations carried by the
fermions. They are thus crucial when we want to couple fermions. For the same reason it is necessary to introduce
an ${\rm SL}(2)$ coset representative, namely a complex ${\rm SL}(2)$ vector ${\cal V}_{\aa}$ which satisfies
\begin{align}
   M_{\aa\ab} &= \Re( {\cal V}_{\aa} ({\cal V}_{\ab})^* ) \; .
\end{align}
Under ${\rm SO}(2)$ ${\cal V}_{\aa}$ carries charge $+1$ while its complex conjugate carries charge $-1$.\footnote{
The complex scalars $\phi$ and $\psi$ in \cite{Bergshoeff:1985ms} translate into our notation as
${\cal V}_+ = \psi$, ${\cal V}_- = i \phi$ and $\psi/\phi=i \tau^*$.}

In section \ref{sec:PresSUSY} we already described the modifications that are necessary in the fermionic sector when
gauging the theory. Those fermionic mass terms that involve the gravitini read
in our particular case
\begin{align}
   e^{-1} {\cal L}_{\text{f.mass}} \, &=
                           \, \ft 1 3 \, g \, A_1^{\jja\jjb} \, \bar \psi{}_{\mua\jja} \, \Gamma^{\mua\mub} \, \psi_{\mub\jjb}
                               - \ft 1 3 \, i \, g \, A_2^{\jja\jjb} \, \bar \psi{}_{\mua\jja} \, \Gamma^{\mua} \, \chi_{\jjb}
			       + i g \, {A_{2\,\xa\jja}}^\jjb \, \bar \psi{}^\jja_{\mua}  \, \Gamma^{\mua} \, \lambda^\xa_{\jjb}
			       + \text{h.c.} \; ,
\end{align}
where $A_1^{\jja\jjb}=A_1^{(\jja\jjb)}$, $A_2^{\jja\jjb}$ and ${A_{2\,\xa\jja}}^\jjb$ are the fermion shift matrices
which depend on the scalar fields.

Also the supersymmetry transformations of the fermions have to be endowed with corrections of order $g^1$, namely
\begin{align}
   \delta \psi_\mua^\jja &= 2 D_\mua \epsilon^\jja
             + \ft 1 4 \, i \, ({\cal V}_\aa)^* {{\cal V}_\Ma}^{\jja\jjb} \, {\cal G}^{\Ma\aa}_{\mub\muc}  
	                              \Gamma^{\mub\muc} \Gamma_\mua \epsilon_\jjb
	      - \ft 2 3 \, g \, A_1^{\jja\jjb} \Gamma_\mua \epsilon_\jjb \, , \nonumber \\
   \delta \chi^\jja &= i \, \epsilon^{\aa\ab} {\cal V}_\aa (D_\mua {\cal V}_\ab) \Gamma^\mua \epsilon^\jja
                + \ft 1 2 \, i \, {\cal V}_\aa {{\cal V}_\Ma}^{\jja\jjb} \, {\cal G}^{\Ma\aa}_{\mua\mub} \Gamma^{\mua\mub} \epsilon_\jjb 
		     - \ft 4 3 \, i \, g  \, A_2^{\jjb\jja} \epsilon_\jjb \, , \nonumber \\
   \delta \lambda_\xa^{\jja} &= 2 i \, {{\cal V}_{\xa}}^\Ma ( D_\mua {{\cal V}_\Ma}^{\jja\jjb} ) \Gamma^\mua \epsilon_{\jjb} 
		    - \ft 1 4 \, {\cal V}_\aa {\cal V}_{\Ma\xa} \, {\cal G}^{\Ma\aa}_{\mua\mub} \Gamma^{\mua\mub} \epsilon^\jja
                               + 2 \, i \, g \, A_{2\,\xa\jjb}{}^{\jja}  \, \epsilon^\jjb    \; ,
   \label{varyF}			       
\end{align}
where the same matrices $A_1$ and $A_2$ appear as in the Lagrangian. 
There are also higher order fermion terms in the supersymmetry rules, but those do not get corrections in the gauged theory.
We wrote the vector field contribution to the
fermion variations in an ${\rm SL}(2)$ covariant way. Using the definition \eqref{DefG} one finds
\begin{align}
   i \, {\cal V}_\aa {{\cal V}_\Ma}^{\jja\jjb} {\cal G}^{\Ma\aa}_{\mua\mub} \Gamma^{\mua\mub}
      &= ({\cal V}_-{}^*)^{-1} \, {{\cal V}_\Ma}^{\jja\jjb} \left( {\cal H}^{\Ma+}_{\mua\mub} 
                      + \ft 1 2 \, i \, \epsilon_{\mua\mub\muc\mud} {\cal H}^{\Ma+\,\muc\mud} \right) \Gamma^{\mua\mub}
     \nonumber \\ &		      
       = ({\cal V}_-{}^*)^{-1} \, {{\cal V}_\Ma}^{\jja\jjb}  {\cal H}^{\Ma+}_{\mua\mub} \Gamma^{\mua\mub} ( 1 - \Gamma_5 )  
       \, , \nonumber \\
   i \, {\cal V}_\aa {{\cal V}_\Ma}^{\xa} {\cal G}^{\Ma\aa}_{\mua\mub} \Gamma^{\mua\mub}
      &= ({\cal V}_-{}^*)^{-1} \, {{\cal V}_\Ma}^{\xa} \left( {\cal H}^{\Ma+}_{\mua\mub} 
                      - \ft 1 2 \, i \, \epsilon_{\mua\mub\muc\mud} {\cal H}^{\Ma+\,\muc\mud} \right) \Gamma^{\mua\mub}
     \nonumber \\ &		      
       = ({\cal V}_-{}^*)^{-1} \, {{\cal V}_\Ma}^{\xa}  {\cal H}^{\Ma+}_{\mua\mub} \Gamma^{\mua\mub} ( 1 + \Gamma_5 )  \; .
\end{align}
Explicitly, the fermion shift matrices are given by
\begin{align}
   A_1^{\jja\jjb} &= \epsilon^{\aa\ab} ({\cal V}_\aa)^* 
                   {{\cal V}_{[\jjc\jjd]}}^\Ma {{\cal V}_\Mb}^{[\jja\jjc]} {{\cal V}_\Mc}^{[\jjb\jjd]}  {f_{\ab\Ma}}^{\Mb\Mc}  \; ,
		   \nonumber \\
   A_2^{\jja\jjb} &= \epsilon^{\aa\ab} {\cal V}_\aa
                    {{\cal V}_{[\jjc\jjd]}}^\Ma {{\cal V}_\Mb}^{[\jja\jjc]} {{\cal V}_\Mc}^{[\jjb\jjd]}  {f_{\ab\Ma}}^{\Mb\Mc} 
		   + \frac 3 2 \epsilon^{\aa\ab} {\cal V}_\aa {{\cal V}_\Ma}^{\jja\jjb} {\xi_\ab}^\Ma  \; ,
		    \nonumber \\
   {A_{2\, \xa\jja}}^\jjb &= \epsilon^{\aa\ab} {\cal V}_\aa
                     {{\cal V}_\Ma}^{\xa} {{\cal V}^\Mb}_{[\jja\jjc]} {{\cal V}_\Mc}^{[\jjb\jjc]} {f_{\ab\Ma\Mb}}^\Mc 
		    - \frac 1 4 \delta_\jja^\jjb \epsilon^{\aa\ab} {\cal V}_\aa {{\cal V}_\xa}^{\Ma} \xi_{\ab\Ma}  \; .
\end{align}
In order that the Lagrangian is supersymmetric these matrices have to obey equation \eqref{GenAA},
which now reads
\begin{align}
    \ft 1 3 \, A_1^{\jja\jjc} \, {\bar A}_{1\,\jjb\jjc}  - \, \ft 1 9 \, A_2^{\jja\jjc} \, {\bar A}_{2\,\jjb\jjc}
               - \, \ft 1 2 \, {A_{2\, \xa\jjb}}^\jjc \, {\bar A}_{2\, \xa}{}^\jja{}_\jjc \,  &= \, - \, \ft 1 4 \, \delta^\jja_\jjb \, V \;,
    \label{WardD4}	       
\end{align}
where the scalar potential $V$ appears on the right hand side.
Equation \eqref{WardD4} is indeed satisfied as a consequence of the quadratic constraints \eqref{QConD4}.

If we have chosen $f_{\aa\Ma\Mb\Mc}$ and $\xi_{\aa\Ma}$ such that the scalar potential possesses an extremal point
one may wonder whether the associated ground state conserves some supersymmetry, i.e.\ whether
$\epsilon^\jja$ exists such the fermion variations \eqref{varyF} vanish in the ground state.
The usual Ansatz is $\epsilon^\jja=q^\jja \, \xi$, where $q^\jja$ is just an ${\rm SU}(4)$ vector
while $\xi$ is a right-handed Killing spinor of AdS ($V<0$) or Minkowski ($V=0$) space, i.e.\footnote{
Consistency of the AdS Killing spinor equation can be checked by using
$R_{\mua\mub\muc\mud}=- \ft 2 3 g^2 V g_{\mua[\muc} g_{\mud]\mub}$, $\Gamma_{[\mua} B \Gamma_{\mub]}^* B^* = - \Gamma_{\mua\mub}$
and $[D_\mua,D_\mub] \xi=-\ft 1 4 {R_{\mua\mub}}^{\muc\mud} \Gamma_{\muc\mud} \xi$.}
\begin{align}
   D_\mua \xi &= g \sqrt{- \ft 1 {12} V} \,  \Gamma_\mua B \xi^* \;.
\end{align}
The Killing spinor equations $\delta \psi^\jja = 0$, $\delta \chi^\jja = 0$ and  $\delta \lambda^{\xa\jja} = 0$ then take the form
\begin{align}
   A_1^{\jja\jjb} q_\jjb &= \sqrt{- \ft 3 4 V} \, q^\jja \; , &
   q_\jjb A_2^{\jjb\jja} &= 0 \; , &
   {A_{2\xa\jjb}}^\jja q^\jjb &= 0 \; .
   \label{KillingD4}
\end{align}
Due to \eqref{WardD4} the first equation of \eqref{KillingD4} already implies the other two.

\section{Examples}
\label{sec:D4examples}

In this section we give examples of tensors $f_{\aa\Ma\Mb\Mc}$ and $\xi_{\aa\Ma}$ that solve the constraints \eqref{QConD4},
therewith giving examples of gauged $N=4$ supergravities. 
In particular, we show how the embedding tensor contains the ${\rm SU}(1,1)$ phases that were introduced by de Roo and Wagemans
to find ground states with non-vanishing cosmological constant \cite{deRoo:1985jh,deRoo:1986yw,Wagemans:1990mv}.
Note that the possibility of these ${\rm SU}(1,1)$ phases was already discussed in \cite{Gates:1983ha}.
Similarly, the parameters that correspond to three-form fluxes in compactifications
from IIB supergravity \cite{D'Auria:2002tc,D'Auria:2003jk,Angelantonj:2003rq,Angelantonj:2003up} are identified.

\subsection{Purely electric gaugings}

In the particular symplectic frame we have chosen -- the one in which the electric and
magnetic vector fields each form a vector under ${\rm SO}(6,n)$ -- the purely electric gaugings are those
for which $f_{-\Ma\Mb\Mc}=0$ and $\xi_{\aa\Ma}=0$, thus only $f_{+\Ma\Mb\Mc}$ is non-vanishing.
This is the class of theories that were constructed by Bergshoeff, Koh and Sezgin \cite{Bergshoeff:1985ms}.
As mentioned above the quadratic constraint in this case simplifies to
the Jacobi identity \eqref{JacobiFP}, which may alternatively be written as
\begin{align}
   {f_{+\Me[\Ma}}^\Md {f_{+\Mb\Mc]}}^\Me &= 0 \, .
   \label{JacobiFP2}
\end{align}
This is a constraint on ${f_{+\Ma\Mb}}^\Mc=f_{+\Ma\Mb\Md} \eta^{\Md\Mc}$ only, but in addition  
the linear constraint $f_{+\Ma\Mb\Mc}=f_{+[\Ma\Mb\Mc]}$ has to be satisfied, such that 
the ${\rm SO}(6,n)$ metric $\eta_{\Ma\Mb}$ enters
non-trivially into this system of constraints. The dimension of the gauge group can at most be $6+n$, which is obvious in
the case that we consider here ($\Ma=1,\ldots,6+n$), but which is also the general limit for arbitrary gaugings.

We first consider semi-simple gaugings. Let ${f_{ab}}^c$ be the structure constants of a semi-simple gauge group $G$,
where $a,b,c=1 \ldots \dim(G)$, $\dim(G)\leq 6+n$, then $\eta_{ab}={f_{ac}}^d {f_{bd}}^c$  is the Cartan-Killing form and
we can choose a basis such that it becomes diagonal, i.e.\
\begin{align}
  \eta_{ab} &= \diag(\,\underbrace{1, \dots,}_{p}\underbrace{-1,\dots}_{q} \,) \;.
\end{align}
We can only realize the gauge group $G$ if we can embed its Lie algebra $\mathfrak{g}_0=\{v^a\}$ into the vector space 
of electric vector fields such that the preimage of $\eta_{\Ma\Mb}$ agrees with $\eta_{ab}$ up to a factor. This
puts a restriction on the signature of $\eta_{ab}$, namely either $p \leq 6$, $q \leq n$ (case 1) or $p \leq n$, $q \leq 6$
(case 2). To make the embedding explicit we define the index $\hat M$ with range $\hat M = 1 \ldots p , 7 \ldots 6+q$
(case 1) or $\hat M = 1 \ldots q , 7 \ldots 6+p$ (case 2). We then have $(\eta_{\hat M \hat N}) = \pm (\eta_{ab})$ and
we can define
\begin{align}
   ( f_{+\hat M \hat N \hat P} ) &= ( f_{abc} ) \; , \qquad \text{all other entries of $f_{+\Ma\Mb\Mc}$ zero,}
\end{align}
where $f_{abc} = {f_{ab}}^d \eta_{dc}$. Since $G$ is semi-simple $f_{abc}$ is completely antisymmetric and thus
$f_{+\Ma\Mb\Mc}$ satisfies the linear and the quadratic constraint. For $n \leq 6$ the possible simple groups
that can appear as factors in $G$ are
${\rm SU}(2)$, ${\rm SO}(2,1)$, ${\rm SO}(3,1)$, ${\rm SL}(3)$, ${\rm SU}(2,1)$, ${\rm SO}(4,1)$ and ${\rm SO}(3,2)$.
For larger $n$ we then find ${\rm SU}(3)$, ${\rm SO}(5)$, ${\rm G}_{2(2)}$, ${\rm SL}(4)$, ${\rm SU}(3,1)$,
${\rm SO}(5,1)$, etc.

Apart from these semi-simple gaugings there are various non-semi-simple gaugings that satisfy \eqref{JacobiFP2}.
Of those we only want to give an example. We can choose three mutual orthogonal lightlike vectors $a_\Ma$, $b_\Ma$ and $c_\Ma$
and define $f_{+\Ma\Mb\Mc}$ to be the volume form on their span, i.e.\
\begin{align}
   f_{+\Ma\Mb\Mc} &= a_{[\Ma} b_\Mb c_{\Mc]} \; .
\end{align}
The vectors have to be linearly independent in order that $f_{+\Ma\Mb\Mc}$ is non-vanishing.
The quadratic constraint is then satisfied trivially since it contains $\eta_{\Ma\Mb}$ which is vanishing on the domain
of $f_{+\Ma\Mb\Mc}$. The gauge group turns out to be $G={\rm U}(1)^3$. We can generalize this construction by choosing
$f_{+\Ma\Mb\Mc}$ to be any three-form that has as domain a lightlike subspace of the vector space $\{v^\Ma\}$.
All corresponding gauge groups are Abelian.

None of the purely electric gaugings can have a ground state 
with non-vanishing cosmological constant
since the scalar potential \eqref{VD4} in this case is proportional to
$M^{++}=\Im(\tau)^{-1}$. Therefore de Roo and Wagemans introduced a further deformation of the theory \cite{deRoo:1985jh}.
Starting from a semi-simple gauging as presented
above they introduced a phase for every simple group factor as additional parameters in the description of the gauging.
In the next subsection we will explain the relation of these phases to our parameters $f_{\aa\Ma\Mb\Mc}$
and show how these theories fit into our framework.

\subsection{The phases of de Roo and Wagemans}

We now allow for $f_{+\Ma\Mb\Mc}$ and $f_{-\Ma\Mb\Mc}$ to be non-zero but keep $\xi_\aa^\Ma=0$. The
quadratic constraint \eqref{QConD4} then reads
\begin{align}
   f_{\aa\Me[\Ma\Mb} {f_{\ab\Mc\Md]}}^\Me &= 0 \; ,\qquad\qquad
   \epsilon^{\aa\ab} f_{\aa\Ma\Mb\Me} {f_{\ab\Mc\Md}}^\Me = 0 \, .
   \label{QconRW}
\end{align}
To find solutions we start from the situation of the last subsection, i.e.\ we assume to have some structure constants
$f_{\Ma\Mb\Mc}=f_{[\Ma\Mb\Mc]}$ that satisfy the Jacobi-identity ${f_{\Me[\Ma}}^\Md {f_{\Mb\Mc]}}^\Me = 0$. In
addition we assume to have a decomposition of the vector space $\{v^\Ma\}$ into $K$ mutual orthogonal subspaces with projectors
$\mathbbm{P}_{i\Ma}{}^\Mb$, $i=1 \ldots K$, i.e.\ such that for a general vector $v_\Ma$ we have
\begin{align}
   v_\Ma &= \sum_{i=1}^K \, \mathbbm{P}_{i\Ma}{}^\Mb v_\Mb \; , &
   \eta^{\Ma\Mc} \, \mathbbm{P}_{i\Ma}{}^\Mb \, \mathbbm{P}_{j\Mc}{}^\Md &= 0 \qquad \text{for } i \neq j \; .
\end{align}
Furthermore this decomposition shall be such that the three-form $f_{\Ma\Mb\Mc}$ does not mix between the subspaces, i.e.\
it decomposes into a sum of three-forms on each subspace
\begin{align}
   f_{\Ma\Mb\Mc} &= \sum_{i=1}^K \, f^{(i)}_{\Ma\Mb\Mc} \; , &
   f^{(i)}_{\Ma\Mb\Mc} &= \mathbbm{P}_{i\Ma}{}^\Md \, \mathbbm{P}_{i\Mb}{}^\Me \, \mathbbm{P}_{i\Mc}{}^\Mf \, f_{\Md\Me\Mf} \; .
\end{align}
This implies that the gauge group splits into $K$ factors $G=G^{(1)} \times G^{(2)} \times \ldots \times G^{(K)}$ with
$f^{(i)}_{\Ma\Mb\Mc}$ being the structure constant of the $i$-th factor, each of them satisfying the above Jacobi-identity separately.
Solutions of the constraint \eqref{QconRW} are then given by
\begin{align}
   f_{\aa\Ma\Mb\Mc} &= \sum_{i=1}^K \, w^{(i)}_\aa  \, f^{(i)}_{\Ma\Mb\Mc} \; , &
   w^{(i)}_\aa &= ( w^{(i)}_+ , \, w^{(i)}_- ) = ( \cos \alpha_i ,  \, \sin \alpha_i ) ,
   \label{GaugingRW}
\end{align}
where the $w^{(i)}_\aa$ could be arbitrary ${\rm SL}(2)$ vectors which we could restrict to have unit length without
loss of generality. The $\alpha_i \in \mathbbm{R}$, $i=1\ldots K$, are the de Roo-Wagemans-phases
first introduces in \cite{deRoo:1985jh}.
In the following we use the abbreviations $c_i = \cos \alpha_i$, $s_i = \sin \alpha_i$.
If $K=1$ we find $f_{+\Ma\Mb\Mc}$ and $f_{-\Ma\Mb\Mc}$ to be proportional.
This case is equivalent to the purely electric gaugings of the last subsection
since one always finds an ${\rm SL}(2)$ transformation such that $w^{(1)}_\aa$ becomes $(1,0)$.

For a semi-simple gauging as described in the last subsection there is a natural decomposition of $\{v^\Ma\}$
into mutual orthogonal subspaces and $K$ equals the number of simple factors in $G$. But the above construction
also applies for non-semi-simple gaugings.

We have mentioned above that every consistent gauging is purely electric in a particular symplectic frame. 
Considering a concrete gauging it is therefore natural to formulate the theory in this particular frame,
and also the two-form gauge fields then disappear from the Lagrangian.
For those gaugings defined by \eqref{GaugingRW} we may perform the symplectic transformation
\begin{align}
   \tilde A{}^{\Ma+}_\mua &= \sum_{i=1}^K \, c_i \, \mathbbm{P}_i{}^{\Ma}{}_\Mb \, A_\mua^{\Mb+} 
                            +\sum_{i=1}^K \, s_i \, \mathbbm{P}_i{}^{\Ma}{}_\Mb \, A_\mua^{\Mb-} \nonumber \, , \\
   \tilde A{}^{\Ma-}_\mua &= - \sum_{i=1}^K \, s_i \, \mathbbm{P}_i{}^{\Ma}{}_\Mb \, A_\mua^{\Mb+} 
                            +\sum_{i=1}^K \, c_i \, \mathbbm{P}_i{}^{\Ma}{}_\Mb \, A_\mua^{\Mb-} \, ,
\end{align}
such that the covariant derivative depends exclusively on $\tilde A{}_\mua^{\Ma+}$
\begin{align}
    D_\mua      &= \nabla_\mua 
            - g \, \tilde A{}_\mua{}^{\Ma+} {f_{\Ma}}^{\Mb\Mc} t_{\Mb\Mc} \; .
\end{align}
Note that the new electric vector fields $\tilde A{}_\mua^{\Ma+}$ do not form a vector under ${\rm SO}(6,n)$, but transform
into $\tilde A{}_\mua^{\Ma-}$ under this group. The Lagrangian in the new symplectic frame reads
\begin{align}
   e^{-1} {\cal L} &= \, \ft 1 2 \, R 
                    + \, \ft 1 8 \, (D_\mua M_{\Ma\Mb}) (D^\mua M^{\Ma\Mb})
     - \frac 1 {4 \, \Im(\tau)^2}  (D_\mua \tau) (D^\mua \tau^*) 
     \nonumber \\[1ex]      &  \qquad
     - \, \ft 1 4 \, {\cal I}_{\Ma\Mb} {\tilde {\cal F}}{}_{\mua\mub}{}^{\Ma+} \tilde {\cal F}{}^{\mua\mub\Mb+}
               - \, \ft 1 8 \, {\cal R}_{\Ma\Mb}  \, \epsilon^{\mua\mub\muc\mud} 
    {\tilde {\cal F}_{\mua\mub}}^{\Ma+} {\tilde {\cal F}_{\muc\mud}}^{\Mb+}  
    - g^2 V    \;,
\end{align}
and the scalar potential \eqref{VD4} takes the form \cite{Wagemans:1990mv}
\begin{align}
    V  & = \ft 1 {16} \, \Im(\tau)^{-1} \,
      \sum_{i,j=1}^K \left( c_i c_j - 2 \Re(\tau) c_i s_j + |\tau|^2 s_i s_j \right)
      f^{(i)}_{\Ma\Mb\Mc} f^{(j)}_{\Md\Me\Mf} \nonumber \\ & \qquad \qquad \qquad \qquad \qquad \qquad
      \times \Big[\ft 1 3 \, M^{\Ma\Md} M^{\Mb\Me} M^{\Mc\Mf} 
        + ( \ft 2 3 \, \eta^{\Ma\Md} -  M^{\Ma\Md} ) \eta^{\Mb\Me} \eta^{\Mc\Mf}   \Big]
	\nonumber \\ & \qquad 
	        - \ft 1 {18} \, \sum_{i,j=1}^K \, c_i s_j
		             f^{(i)}_{\Ma\Mb\Mc} f^{(j)}_{\Md\Me\Mf} M^{\Ma\Mb\Mc\Md\Me\Mf} \; .
\end{align}
The kinetic term of the vector fields involves the field strength
\begin{align}
   \tilde {\cal F}{}_{\mua\mub}{}^{\Ma+} &= 2 \partial_{[\mua} \tilde A{}_{\mub]}{}^{\Ma+} 
            - g \, f{}_{\Mb\Mc}{}^\Ma \tilde A{}_{[\mua}{}^{\Mb+} \tilde A{}_{\mub]}{}^{\Mc+}  \; ,
\end{align}
and the scalar dependent matrices ${\cal I}_{\Ma\Mb}$ and ${\cal R}_{\Ma\Mb}$ which are defined by
\begin{align}
   ({\cal I}^{-1})^{\Ma\Mb} &= \frac 1 {\Im(\tau)} \sum_{i,j=1}^K 
        \left( c_i c_j - 2 \Re(\tau) c_i s_j + |\tau|^2 s_i s_j \right) 
	  \mathbbm{P}_i{}^{\Ma}{}_\Mc \mathbbm{P}_j{}^{\Mb}{}_\Md \, M^{\Mc\Md} \; , \nonumber \\
  {\cal R}_{\Ma\Mb} ({\cal I}^{-1})^{\Mb\Mc} &= 
    \frac 1 {\Im(\tau)} \sum_{i,j=1}^K 
        \left[ - c_i s_j + \Re(\tau) ( s_i s_j - c_i c_j) + |\tau|^2 s_i c_j \right]
	  \mathbbm{P}_{i\Ma\Mb} \mathbbm{P}_j{}^{\Mc}{}_\Me M^{\Mb\Me} \; .
\end{align}
In general when going to the electric frame for an arbitrary gauging there is still a topological term
for the electric fields of the form $AA\partial A + AAAA$ \cite{deWit:1984px}, but here this term is not present.

Comparing the scalar potential $V$ for non-vanishing phases $\alpha_i$ with that of the last subsection we find it to have 
a much more complicated $\tau$ dependence and one can indeed find gaugings where it possesses 
stationary points \cite{deRoo:2003rm,Wagemans:1990mv}.

\subsection{IIB flux compactifications}
\label{sec:D4IIB}

We now consider gaugings with an origin in type IIB supergravity.
$N=4$ supergravity can be obtained by an orientifold compactification of IIB \cite{Frey:2002hf,Kachru:2002he}
and in the simplest $T^6/\mathbbm{Z}_2$ case this yields the ungauged theory with $n=6$, i.e.\ the global symmetry group is
$G_0 = {\rm SL}(2) \times {\rm SO}(6,6)$. Here, the ${\rm SL}(2)$ factor is the symmetry that was already present
in ten dimensions and ${\rm SO}(6,6)$ contains the ${\rm GL}(6)$ symmetry group associated with the torus $T^6$.
The compactification thus
yields the theory in a symplectic frame in which ${\rm SL}(2) \times {\rm GL}(6)$ is realized off-shell. Turning on fluxes
results in gaugings of the theory that are purely electric in this particular symplectic frame.
This is the class of gaugings to be examined in this subsection.

An ${\rm SO}(6,6)$ vector decomposes under ${\rm GL}(6)={\rm U}(1) \times {\rm SL}(6)$ into
${\bf 6} \oplus {\bf \overline 6}$.
The vector fields $A_\mua{}^{\Ma\aa}$ split accordingly into electric
ones $A_\mua{}^{\LLa\aa}$ and magnetic ones $A_{\mua\,\LLa}{}^{\aa}$ where $\LLa=1 \ldots 6$ is a (dual) ${\rm SL}(6)$ vector index.
The ${\rm SO}(6,6)$ metric takes the form
\begin{align}
  \eta_{\Ma\Mb} &= \begin{pmatrix} \eta_{\LLa\LLb} & \eta_\LLa{}^\LLb \\ \eta^\LLa{}_\LLb & \eta^{\LLa\LLb} \end{pmatrix}
                 = \begin{pmatrix} 0 & \delta_\LLa^\LLb \\ \delta^\La_\LLb & 0 \end{pmatrix} \; .
\end{align}   
The gauge group generators \eqref{DefXD4} split as $X_{\Ma\aa}=(X_{\LLa\aa}, \, X^\LLa{}_\aa)$
and a purely electric gauging satisfies $X^\LLa{}_\aa=0$.
The tensors $\xi_{\aa\Ma}$ and $f_{\aa\Ma\Mb\Mc}$ decompose into the following
representations
\begin{align}
  ({\bf 2},{\bf 12}) \, & \rightarrow \,  ({\bf 2},{\bf 6}) \oplus ({\bf 2},{\bf \overline 6}) \, , \nonumber \\
  ({\bf 2},{\bf 220}) \, & \rightarrow \, 
     ({\bf 2},{\bf 6}) \oplus ({\bf 2},{\bf 20}) \oplus ({\bf 2},{\bf 84}) \oplus ({\bf 2},{\bf \overline{84}})
     \oplus ({\bf 2}, {\bf \overline{20}}) \oplus ({\bf 2},{\bf \overline 6}) \; . 
\end{align}     
From \eqref{DefXD4} one finds that the condition $X^\LLa{}_\aa=0$ demands most of these components to vanish,
only the $({\bf 2},{\bf 20})$ and
a particular combinations of the two $({\bf 2},{\bf 6})$'s are allowed to be non-zero.
Explicitly we find for the general electric gaugings in this frame
\begin{align}
   \xi_{\aa\Ma} &= (\xi_{\aa\LLa},\, \xi_\aa{}^\LLa) = (\xi_{\aa\LLa},\,0) \, , \nonumber \\
   f_{\aa\Ma\Mb\Mc} &= ( f_{\aa\LLa\LLb\LLc}, \, f_{\aa\LLa\LLb}{}^\LLc, \, f_{\aa\LLa}{}^{\LLb\LLc}, \, f_\aa{}^{\LLa\LLb\LLc} )
                    = ( f_{\aa\LLa\LLb\LLc}, \, \xi^{\phantom{\LLc}}_{\aa[\LLa} \delta_{\LLb]}^\LLc, \, 0, \, 0 ) \; .
   \label{IIBgaugings}		    
\end{align}
This Ansatz automatically satisfies most of the quadratic constraints \eqref{QConD4}, the only consistency constraint
left is
\begin{align}
   f_{(\aa[\LLa\LLb\LLc} \, \xi_{\ab)\LLd]} &= 0 \; .
   \label{fxi}
\end{align}
Thus for $\xi_{\aa\LLa} = 0$ we find $f_{\aa\LLa\LLb\LLc}$ to be unconstrained, i.e.\ every choice of $f_{\aa\LLa\LLb\LLc}$
gives a valid gauged theory. It turns out that $f_{\aa\LLa\LLb\LLc}$ corresponds to the possible three-form fluxes that can
be switched on. These theories and extensions of them
were already described and analyzed in \cite{D'Auria:2002tc,D'Auria:2003jk}.
It was noted in \cite{deWit:2003hq} that not all $N=4$ models that come from $T^6/\mathbbm{Z}_2$ orientifold
compactifications can be embedded
into the $N=8$ models from torus reduction of IIB, since for the latter the fluxes have to satisfy the constraint
$f_{\aa[\LLa\LLb\LLc} f_{\ab\LLd\Le\LLf]} = 0$.

Searching for solutions to the constraint \eqref{fxi} with $\xi_{\aa\LLa}$ non-vanishing one finds that
the possible solutions have the form
\begin{align}
   f_{\aa\LLa\LLb\LLc} &= \xi_{\aa[\LLa} \, A_{\LLb\LLc]} \; , && \text{or} &
   f_{\aa\LLa\LLb\LLc} &= \epsilon^{\ab\ac} \,  B_{\aa[\LLa} \, \xi_{\ab\LLb} \,  \xi_{\ac\LLc]} \; ,
\end{align}
with unconstraint $\xi_{\aa\LLa}$, $A_{\LLa\LLb}=A_{[\LLa\LLb]}$ and $B_{\aa\LLa}$, respectively.

Theories
with both $f_{\aa\Ma\Mb\Mc}$ and $\xi_{\aa\Ma}$ non-zero were not yet considered in the literature.
For $f_{\aa\Ma\Mb\Mc}=0$ the remaining quadratic constraints on $\xi_{\aa\Ma}$ demands it to be of the form
$\xi_{\aa\Ma} = v_{\aa} \, w_\Ma$, with $v_\aa$ arbitrary and $w_\Ma$ lightlike, i.e.\ $w_\Ma w^\Ma = 0$.
Thus for vanishing $f_{\aa\Ma\Mb\Mc}$ the solution for $\xi_{\aa\Ma}$ is unique up to ${\rm SL}(2) \times {\rm SO}(6,n)$
transformations. This solution corresponds to the gauging that can be obtained from Scherk-Schwarz reduction from $d=5$ with
a non-compact ${\rm SO}(1,1)$ twist, which was constructed in \cite{Villadoro:2004ci} for the case of one vector multiplet.
This suggests that in certain cases non-vanishing $\xi_{\aa\Ma}$ corresponds to torsion on the internal manifold. But this does
not apply to the IIB reductions here since $\xi_{\aa\LLa}$ is a doublet under the global ${\rm SL}(2)$ symmetry of IIB,
while a torsion parameter should be a singlet.
We have shown that these theories with non-vanishing $\xi_{\aa\LLa}$ are consistent $N=4$ supergravities,
but their higher-dimensional origin remains to be elucidated.

The list of gauged $N=4$ supergravities that were presented in this section is, of course, far from complete.
One could, for example, discuss other orientifold compactifications of IIA and IIB supergravity, for all of which turning
on fluxes yields gauged theories in four dimensions \cite{Angelantonj:2003rq,Angelantonj:2003up}.
However, the examples discussed were hopefully representative enough to show that indeed all the various gaugings
appearing in the literature can be embedded in the universal formulation presented above. 
New classes of gaugings are those with both $f_{\aa\Ma\Mb\Mc}$ and $\xi_{\aa\Ma}$ non-vanishing.
Every solution of the quadratic constraints \eqref{QConD4}
yields a consistent gauging . For additional examples see \cite{Jonas:Thesis}.

\chapter{The $N=4$ supergravities in $d=5$} \label{ch:D5}

Analogous to the presentation of the four-dimensional theory in the last chapter
we now describe the general five-dimensional gauged $N=4$ supergravity\footnote{We denote
by $N=4$ the half-maximal supergravity, although in five spacetime dimensions this theory is sometimes referred to as $N=2$.}
The first account of the ungauged $N=4$ supergravity in $d=5$ was given in \cite{Awada:1985ep},
where also the first gauging of the theory was already considered.
Those gaugings for which the gauge group is a product of a semi-simple and an Abelian factor
were already presented in \cite{Dall'Agata:2001vb}, examples of this type were already known for a while
\cite{Romans:1985ps,Pernici:1985ju,Gunaydin:1985cu,Gunaydin:1999zx,Andrianopoli:2000fi,Bergshoeff:2004kh}.
Also some non-semi-simple gaugings
were already constructed \cite{Villadoro:2004ci}. Our presentation incorporates all these known gaugings
and also includes new ones.
The construction of the general gaugings in this chapter follows closely the one in \cite{deWit:2004nw} for
the $d=5$ maximal supergravities.

\section{Embedding tensor and gauge fields}
\label{sec:D5qu}

\subsection{Linear and quadratic constraint}

The global symmetry group of ungauged $d=5$, $N=4$ supergravity is $G_0={\rm SO}(1,1) \times {\rm SO}(5,n)$,
where $n$ counts the number of vector multiplets. The theory contains Abelian vector gauge fields
that form one vector $A_\mua^\Ma$ and one scalar $A_\mua^\Nv$ under ${\rm SO}(5,n)$.
Note that the index $\Ma=1 \ldots 5+n$ now is a vector index of ${\rm SO}(5,n)$ while in the last
chapter we used it for ${\rm SO}(6,n)$. The vector fields carry ${\rm SO}(1,1)$ charges $1/2$ and $-1$, respectively, i.e.\
\begin{align}
   \delta_\Na A_\mua^\Ma &= \frac 1 2 A_\mua^\Ma \; , &
   \delta_\Na A_\mua^\Nv &= - A_\mua^\Nv \; ,
\end{align}
where $\delta_\Na$ denotes the ${\rm SO}(1,1)$ action. The corresponding algebra generator is denoted $t_\Na$ while
the ${\rm SO}(5,n)$ generators are $t_{\Ma\Mb}=t_{[\Ma\Mb]}$. For the representations of the vector gauge fields these
generators explicitly read
\begin{align}
   {t_{\Ma\Mb \, \Mc}}^\Md &= \delta^\Md_{[\Ma} \eta^{\phantom{\Md}}_{\Mb]\Mc} \; , &
   {t_{\Na\Ma}}^\Mb &= - \frac 1 2 \delta_\Ma^\Mb \; , &
   {t_{\Ma\Mb \, \Nv}}^\Nv &= 0 \; , &
   {t_{\Na\Nv}}^\Nv &= 1 \; .
   \label{tvf}
\end{align}
The covariant derivative \eqref{GenCovDiv} reads
\begin{align}
   D_\mua &= \partial_\mua - g \, A_\mua^\Ma \, {\Theta_\Ma}^{\Mb\Mc} t_{\Mb\Mc} - g \, A_\mua^\Ma \, {\Theta_\Ma}^\Na \, t_\Na
                           - g \, A_\mua^\Nv \, {\Theta_\Nv}^{\Mb\Mc} t_{\Mb\Mc} - g \, A_\mua^\Nv \, {\Theta_\Nv}^\Na \, t_\Na \; .
\end{align}
According to table \ref{LinCon2} there are only three irreducible components of the embedding tensor allowed in the present case.
These three components are parameterized by tensors $f_{\Ma\Mb\Mc}=f_{[\Ma\Mb\Mc]}$, $\xi_{\Ma\Mb}=\xi_{[\Ma\Mb]}$
and $\xi_\Ma$. In terms of these tensors the embedding tensor reads
\begin{align}
   {\Theta_\Ma}^{\Mb\Mc} &= f_\Ma{}^{\Mb\Mc} + \delta_\Ma^{[\Mb} \xi^{\Mc]} \; , &
   {\Theta_\Ma}^\Na &= \xi_\Ma  \; , &
   {\Theta_\Nv}^{\Ma\Mb} &= \xi^{\Ma\Mb} \; , &
   {\Theta_\Nv}^\Na &= 0 \; .
   \label{D5LinCon}
\end{align}
The covariant derivative becomes
\begin{align}
   D_\mua &= \nabla_\mua - g \, A_\mua^\Ma \, f_\Ma{}^{\Mb\Mc} \, t_{\Mb\Mc}
			 - g \, A_\mua^\Nv \, \xi^{\Mb\Mc} \, t_{\Mb\Mc}
                         - g \, A_\mua^\Ma \, \xi^\Mb \, t_{\Ma\Mb}  
                         - g \, A_\mua^\Ma \, \xi_\Ma \, t_\Na \; ,
  \label{CovDivD5}			 
\end{align}
where the indices are raised and lowered by using the ${\rm SO}(5,n)$ metric $\eta_{\Ma\Mb}$.
In order that the above expression is $G_0$ invariant we need $f_{\Ma\Mb\Mc}$ and $\xi_\Ma$ to carry ${\rm SO}(1,1)$
charge $-1/2$ and $\xi_{\Ma\Mb}$ to have charge $1$. By $G_0$ invariance we again mean the formal invariance treating the
$f_{\Ma\Mb\Mc}$, $\xi_{\Ma\Mb}$ and $\xi_\Ma$ as spurionic objects.

The quadratic constraints \eqref{QconGen1} on $\Theta$ yields the following constraint
on $f_{\Ma\Mb\Mc}$, $\xi_{\Ma\Mb}$ and $\xi_\Ma$:
\begin{align}
   \xi_\Ma \xi^\Ma &= 0 \; , &
   \xi_{\Ma\Mb} \xi^\Mb &= 0 \; , &
   f_{\Ma\Mb\Mc} \xi^\Mc &= 0 \; , \nonumber \\
   3 f_{\Me[\Ma\Mb} \, f_{\Mc\Md]}{}^\Me &= 2 f_{[\Ma\Mb\Mc} \, \xi_{\Md]} \; ,  &
   {\xi_{\Ma}}^\Md \, f_{\Md\Mb\Mc} &= \xi_\Ma \, \xi_{\Mb\Mc} - \xi_{[\Mb} \, \xi_{\Mc]\Ma} \; .
   \label{XiRel}
\end{align}
This implies for example that $\xi_\Ma$ has to vanish for $n=0$ since for an Euclidean metric $\eta_{\Ma\Mb}$ one 
has no lightlike vectors. In general, however, all three tensors may be non-zero at the same time.

It is convenient to introduce a composite index $\cMa=\{ \Nv, \, \Ma\}$ that combines all vector gauge fields
$A_\mua^\cMa = (A_\mua^\Nv, \, A_\mua^\Ma)$. The covariant derivate acts on an object in the vector field
representation as
\begin{align}
   D_\mua \, \Lambda^\cMa &= \nabla_\mua \, \Lambda^\cMa + g \, A_\mua^\cMb \, X_{\cMb\cMc}{}^\cMa  \, \Lambda^\cMc \; .
\end{align}
We already introduced the gauge group generators $X_{\cMa\cMb}{}^\cMc = (X_{\cMa})_\cMb{}^\cMc$ 
in section \ref{sec:gen:emb}. In the present case they explicitly read
\begin{align}
   {X_{\Ma\Mb}}^\Mc &= - f_{\Ma\Mb}{}^\Mc - \frac 1 2 \eta_{\Ma\Mb} \xi^\Mc + \delta_{[\Ma}^\Mc \xi_{\Mb]} \; , &
   {X_{\Ma\Nv}}^\Nv &= \xi_\Ma \; , &
   {X_{\Nv\Ma}}^\Mb &= - {\xi_\Ma}^\Mb  \; , &
   \label{GenD5}
\end{align}
and all other components vanish.  
The quadratic constraint ensures that the $X_{\cMa\cMb}{}^\cMc$ satisfy the condition \eqref{QconGen2}
that guarantees the closure of the gauge group and
identifies the $X_{\cMa\cMb}{}^\cMc$ themselves as generalized structure constants of the gauge group.
For gaugings with only $f_{\Ma\Mb\Mc}$ non-zero we see that this tensor is a structure constant for a subgroup $G$
of ${\rm SO}(5,n)$ that is gauged by using $A_\mua^\Ma$ as vector gauge fields.
If only $\xi_{\Ma\Mb}$ is non-zero we find a one-dimensional subgroup of ${\rm SO}(5,n)$
to be gauged with gauge field $A_\mua^\Nv$. And for gaugings with only $\xi_\Ma$ non-zero one finds a $4+n$ dimensional
gauge group ${\rm SO}(1,1) \ltimes {\rm SO}(1,1)^{3+n}$ where the first factor involves the ${\rm SO}(1,1)$ of $G_0$.

\subsection{Vector and tensor gauge fields}

We have already introduced the vector fields $A_\mua^\cMa = (A_\mua^\Nv, \, A_\mua^\Ma)$.
In $d=5$ the two-form fields are introduced as dual to the vector fields, i.e.\ we have
$B_{\mua\mub\,\cMa} = (B_{\mua\mub \, \Ma}, B_{\mua\mub \, \Nv})$. 
They also transform dual to the vector gauge field under $G_0$, i.e.\ $B_{\mua\mub \, \Ma}$ is
a vector with ${\rm SO}(1,1)$ charge $-1/2$ and $B_{\mua\mub \, \Nv}$ is a singlet carrying charge $1$. 
In the gauged theory we use both vector and two-form fields as free fields in the Lagrangian. However,
the latter do not have a kinetic term but couple to the vector fields via a topological term and
via St\"uckelberg type couplings in the vector field strengths. The two-forms then turn out to be dual
to the vectors fields due to their own equations of motion \cite{deWit:2004nw}. This is analogous to the four dimensional case
where the two-forms turned out to be dual to scalars via the equations of motion.

To translate the general formulas of section \ref{sec:GenNonAVecTen} to the particular case of half-maximal
$d=5$ supergravity we first need to give the tensors $d_{I\Ma\Mb}$ and $Z^{\Ma I}$ which in the index conventions
of the present chapter read $d_{\cMa\cMb\cMd}=d_{(\cMa\cMb\cMd)}$ and $Z^{\cMa\cMb}=Z^{[\cMa\cMb]}$. The complete
symmetry of $d_{\cMa\cMb\cMd}$ and the antisymmetry of $Z^{\cMa\cMb}$ was found in section \ref{subsec:TopTermsOdd}
to be crucial for the existence of an gauge invariant Lagrangian in $d=5$.
For the present case these tensors are defined by
\begin{align}
   d_{\Nv\Ma\Mb} &= d_{\Ma\Nv\Mb} = d_{\Ma\Mb\Nv} = \eta_{\Ma\Mb} \; , \qquad \text{all other components zero,}
   \label{Defdin5}
\end{align}
and
\begin{align}
   Z^{\Ma\Mb} &= \, \ft 1 2 \, \xi^{\Ma\Mb} \; , &
   Z^{\Nv\Ma} &= - Z^{\Ma\Nv} = \, \ft 1 2 \, \xi^\Ma \; .
   \label{ZD5}
\end{align}
From these definitions one finds \eqref{DefZfromX} to be satisfied, i.e.\ in our present notation
\begin{align}
   {X_{(\cMa\cMb)}}^\cMc &= d_{\cMa\cMb\cMd} Z^{\cMc\cMd} \; .
   \label{linConD5}
\end{align}
This relation is the general formulation of the five-dimensional linear constraint. One can show that the existence of
$Z^{\cMc\cMd}$ such that \eqref{linConD5} is satisfied is equivalent to the linear constraint \eqref{D5LinCon} on
the embedding tensor. 

With the above definitions at hand we can now read of the covariant field strengths of the vector and two-form
gauge fields from equation \eqref{DefHHH}. We find
\begin{align}
   {\cal H}_{\mua\mub}^\cMa \, &= \, 2 \partial_{[\mua} A^\cMa_{\mub]} + g X_{\cMb\cMc}{}^\cMa A^\cMb_\mua A^\cMc_\mub
                               + g Z^{\cMa\cMb} B_{\mua\mub\,\cMb} \; .,
   \nonumber \\
   Z^{\cMa\cMb} {\cal H}_{\mua\mub\muc \, \cMb} 
     &=  Z^{\cMa\cMb} \left[ 3 \, D_{[\mua} \, B_{\mub\muc] \cMb} 
             + 6 \, d_{\cMb\cMc\cMd} \, A^\cMc_{[\mua}  \left( \partial^{\phantom{\cMd}}_\mub A_{\muc]}^\cMd 
	                      + \ft 1 3 \, g \, {X_{\cMe\cMf}}^\cMd \, A_\mub^\cMe \, A_{\muc]}^\cMf \right) \right] \; .
\end{align}
These field strengths transform covariantly under the gauge transformations \eqref{GenGauge}, which
in our particular case read
\begin{align}
   \delta A_\mua^\cMa &= D_\mua \Lambda^\cMa - g Z^{\cMa\cMb} \Xi_{\mua \, \cMb} \; , \nonumber \\
   \Delta B_{\mua\mub \, \cMa} 
      &= \left( 2 D_{[\mua} \Xi_{\mub]\cMa} - 2 d_{\cMa\cMb\cMc} {\cal H}_{\mua\mub}^\cMb \Lambda^\cMc \right)  \;,
    \label{D5gaugetrafo}
\end{align}
where $\Lambda^\cMa$ and $\Xi_{\mua \, \cMb}$ are the gauge parameters and we use th the covariant variation
\begin{align}
  \Delta B_{\mua\mub \, \cMa} 
      &= \left( \delta B_{\mua\mub \, \cMa} 
         - 2 d_{\cMa\cMb\cMc} A_{[\mua}^\cMb \delta A_{\mub]}^\cMc \right) \; .
\end{align}
The two-forms appear in the Lagrangian only projected with $Z^{\cMa\cMb}$ and thus we also define their field strengths only
under this projection. The two-forms thus decouple from the
theory in the ungauged limit $g \rightarrow 0$. Also for the gauged theory there are never all two-forms entering the
Lagrangian. For example, for gaugings with only $f_{\Ma\Mb\Mc}$ non-zero we have $Z^{\cMa\cMb}=0$ and thus no two-forms are needed
at all.

\section{The general Lagrangian}
\label{sec:D5lag}

We have already introduced the vector fields $A_\mua^\cMa$ and the two-form fields
$B_{\mua\mub\,\cMa}$ in the last section. In addition the bosonic field content
consists of the metric and of scalars that form the coset
${\rm SO}(1,1) \times {\rm SO}(5,n)/{\rm SO}(5) \times {\rm SO}(n)$.
The ${\rm SO}(1,1)$ part of the scalar manifold is simply described by one real field $\Sigma$ that is a singlet under
${\rm SO}(5,n)$ and carries ${\rm SO}(1,1)$ charge $-1/2$. In addition we have the coset
${\rm SO}(5,n)/{\rm SO}(5) \times {\rm SO}(n)$ which is parameterized by a coset representative
${\cal V}=({\cal V}_\Ma{}^\ya, \, {\cal V}_\Ma{}^\xa)$, where $\ya=1\ldots 5$ and $\xa=1\ldots n$ are
${\rm SO}(5)$ and ${\rm SO}(n)$ vector indices. Our conventions for ${\cal V}$ here are the same as for the 
${\rm SO}(6,n)/{\rm SO}(6) \times {\rm SO}(n)$ coset representative we had in four dimensions, see equations
\eqref{DefEta}, \eqref{CosetSO6n} and \eqref{DefMVV} of the last chapter.
In addition to the symmetric matrix $M_{\Ma\Mb}={\cal V} {\cal V}^T$
and its inverse $M^{\Ma\Mb}$ we need the completely antisymmetric scalar tensor
\begin{align}
   M_{\Ma\Mb\Mc\Md\Me} &= \epsilon_{\ya\yb\yc\yd\ye}
                             {{\cal V}_\Ma}^{\ya} {{\cal V}_\Mb}^{\yb} {{\cal V}_\Mc}^{\yc} 
                             {{\cal V}_\Md}^{\yd} {{\cal V}_\Me}^{\ye} \; .
\end{align}

We now have all objects to give the bosonic Lagrangian of the general gauged $N=4$ supergravity in five dimensions
\begin{align}
   {\cal L}_{\text{bos}} &= {\cal L}_{\text{kin}} + {\cal L}_{\text{top}} + {\cal L}_{\text{pot}} \; .   
\end{align}
It consists of a kinetic part
\begin{align}
   e^{-1} {\cal L}_{\text{kin}} &= \ft 1 2 \, R 
               - \, \ft 1 4 \, \Sigma^2 \, M_{\Ma\Mb} \, {\cal H}^\Ma_{\mua\mub} \, {\cal H}^{\Mb\, \mua\mub}
               - \, \ft 1 4 \, \Sigma^{-4} \, {\cal H}^\Nv_{\mua\mub} \, {\cal H}^{\Nv\,\mua\mub}
	       \nonumber \\ & \qquad
	       - \, \ft 3 2 \, \Sigma^{-2} \, (D_\mua \Sigma)^2 
	       + \, \ft 1 {16} \, (D_\mua M_{\Ma\Mb}) (D^\mua M^{\Ma\Mb}) \; ,
\end{align}
a topological part \cite{deWit:2004nw}
\begin{align}
   {\cal L}_{\text{top}} &= - \frac e {8 \sqrt{2}} \epsilon^{\mua\mub\muc\mud\mue} 
    \bigg\{ g Z^{\cMa\cMb} B_{\mua\mub \, \cMa} \left[ D_\muc B_{\mud\mue \, \cMb}
                   + 4 d_{\cMb\cMc\cMd} A^\cMc_{[\muc} \Big( \partial_\mud A_{\mue]}^\cMd 
	                      + \ft 1 3 g {X_{\cMe\cMf}}^\cMc A_\mud^\cMe A_{\mue]}^\cMf \Big) \right]
			      \nonumber \\ & \qquad \qquad \qquad \quad
      - \ft 8 3 \, d_{\cMa\cMb\cMc} \, A_\mua^\cMa \, \partial_\mub A_\muc^\cMb \, \partial_\mud A_\mue^\cMc
      - 2 \, g \, d_{\cMa\cMb\cMc} \, {X_{\cMd\cMe}}^\cMa \, A_\mua^\cMb \, A_\mub^\cMd \, A_\muc^\cMe \, \partial_\mud A_\mue^\cMc
		      \nonumber \\ & \qquad \qquad \qquad \qquad
      - \ft 2 5 \, g^2 \, d_{\cMa\cMb\cMc} \, {X_{\cMd\cMe}}^\cMa \, {X_{\cMf\cMg}}^\cMc  \,
              A_\mua^\cMb \, A_\mub^\cMd \, A_\muc^\cMe \, A_\mud^\cMf \, A_\mue^\cMg \bigg\} \; ,
\end{align}
and a scalar potential
\begin{align}
   e^{-1} {\cal L}_{\text{pot}} &= -g^2 V \nonumber \\
    =& - \frac {g^2} {4} \Big[
       \xi_{\Ma\Mb\Mc} \xi_{\Md\Me\Mf} \Sigma^{-2} \left(
       \ft 1 {12} M^{\Ma\Md} M^{\Mb\Me} M^{\Mc\Mf} 
      -\ft 1 {4} M^{\Ma\Md} \eta^{\Mb\Me} \eta^{\Mc\Mf} 
      +\ft 1 {6} \eta^{\Ma\Md} \eta^{\Mb\Me} \eta^{\Mc\Mf} \right)
      \nonumber \\ & \qquad \qquad
      +\ft 1 4 \xi_{\Ma\Mb} \xi_{\Mc\Md} \Sigma^4 \left( M^{\Ma\Mc} M^{\Mb\Md} - \eta^{\Ma\Mc} \eta^{\Mb\Md} \right)
      +\xi_\Ma \xi_\Mb \Sigma^{-2} M^{\Ma\Mb}
      \nonumber \\ & \qquad \qquad \qquad
      +\ft 1 3 \sqrt{2} \xi_{\Ma\Mb\Mc} \xi_{\Md\Me} \Sigma M^{\Ma\Mb\Mc\Md\Me} \Big] \; .
   \label{PotD5}      
\end{align}
For $\xi_{\Ma}=0$ this scalar potential agrees with the one given in \cite{Dall'Agata:2001vb}.
The topological term ${\cal L}_{\text{top}}$ is a special case of equation \eqref{GenTopD5} which gave
${\cal L}_{\text{top}}$ for a general five-dimensional theory.
This topological term seems complicated, but its variation with respect to the vector and tensor gauge fields
takes a simple and covariant form
\begin{align}
   \delta {\cal L}_{\text{top}} &= \frac{e} {4 \sqrt{2}} \epsilon^{\mua\mub\muc\mud\mue}
   \left( \ft 1 3 \, g \, Z^{\cMa\cMb} \,  {\cal H}^{(3)}_{\mua\mub\muc \, \cMa} \, \Delta B_{\mud\mue \, \cMb}
          + d_{\cMa\cMb\cMc} \, {\cal H}_{\mua\mub}^\cMa \, {\cal H}_{\muc\mud}^\cMb \, \delta A_\mue^\cMc \right)
   + \text{tot. deriv.}	   \; ,
   \label{VaryLtop}   
\end{align}
Under gauge transformations \eqref{D5gaugetrafo} the Lagrangian is invariant up to a total derivative.

Varying the two-forms in the Lagrangian yields the equation of motion
\begin{align}
   Z^{\cMa\cMb} \left( \frac{1} {6 \sqrt{2}} \epsilon_{\mua\mub\muc\mud\mue} \,  {\cal H}^{\muc\mud\mue}_{\cMb} 
                      - {\cal M}_{\cMb\cMc} {\cal H}_{\mua\mub}^{\cMc} \right) &= 0 \; ,
  \label{Dual23}		      
\end{align}
where we have used
\begin{align}
   {\cal M}_{\cMa\cMb} &\equiv \begin{pmatrix} \Sigma^{-4} & 0 \\
                                          0 & \Sigma^2 M_{\Ma\Mb} \end{pmatrix} \; .
\end{align}
Due to equation \eqref{Dual23} the two-forms become dual to the vector gauge fields as was announced above.

\section{Killing spinor equations}
\label{sec:D5kill}

We now turn to the fermions of the five dimensional theory in order to give the Killing spinor equations.
The fermions come in representations of the maximal compact subgroup $H={\rm USp}(4) \times {\rm SO}(n)$ of $G_0$,
where ${\rm USp}(4)$ is the covering group of ${\rm SO}(5)$.
In the gravity multiplet there are four gravitini $\psi_{\mua\jja}$ and four spin $1/2$ fermions $\chi_\jja$, both
vectors under ${\rm USp}(4)$ and singlets under ${\rm SO}(n)$, $i=1\ldots 4$. In the $n$ vector multiplets there are $4n$
spin $1/2$ fermions $\lambda^\xa_\jja$ which form a vector under both ${\rm USp}(4)$ and ${\rm SO}(n)$, $\xa=1\ldots n$.
All fermions are pseudo-Majorana, i.e.\ they satisfy a pseudo-reality constraint of the form
$\xi_\jja = \Omega_{\jja\jjb} C ({\bar \xi}^\jjb)^T$, where $\Omega_{\jja\jjb}$ is the ${\rm USp}(4)$ invariant symplectic form
and $C$ is the charge conjugation matrix.

The coset representative ${\cal V}_\Ma{}^\ya$ transforms as a ${\bf 5}$ under ${\rm USp}(4)$ and can alternatively
be expressed as ${\cal V}_\Ma{}^{\jja\jjb}={\cal V}_\Ma{}^{[\jja\jjb]}$ subject to
\begin{align}
   {\cal V}_\Ma{}^{\jja\jjb} \Omega_{\jja\jjb} &= 0 \; , &
   ( {\cal V}_\Ma{}^{\jja\jjb} )^* &= \Omega_{\jja\jjc} \Omega_{\jjb\jjd} {\cal V}_\Ma{}^{\jjc\jjd} \; .
\end{align}
Under supersymmetry transformations parameterized by $\epsilon_\jja=\epsilon_\jja(x)$ we have
\begin{align}
   \delta \psi_{\mua\jja}  &= D_\mua \epsilon_\jja 
      - \frac i 6 \left( \Omega_{\jja\jjb} \Sigma {\cal V}_{\Ma}{}^{\jjb\jjc} {\cal H}_{\mub\muc}^\Ma
                          - \ft 1 4 \sqrt{2} \, \delta_\jja^\jjc \Sigma^{-2} {\cal H}_{\mub\muc}^\Nv \right)
		  \left( \Gamma_\mua{}^{\mub\muc} - 4 \delta_\mua^\mub \Gamma^\muc \right) \epsilon_\jjc
       \nonumber \\ & \qquad		  
      + \frac {i g} {\sqrt{6}} \, \Omega_{\jja\jjb} \, A_1^{\jjb\jjc} \, \Gamma_\mua  \, \epsilon_\jjc \; ,
   \nonumber \\
   \delta \chi_\jja &= - \ft 1 2 \, \sqrt{3} \, i \, (\Sigma^{-1} D_\mua \Sigma) \, \Gamma^\mua \epsilon_\jja
                      - \ft 1 6 \, \sqrt{3} \, 
		      \left( \Sigma \, \Omega_{\jja\jjb} \, {\cal V}_{\Ma}{}^{\jjb\jjc} {\cal H}_{\mua\mub}^\Ma
            + \ft 1 2 \sqrt{2} \, \Sigma^{-2} \, \delta_\jja^\jjc \, {\cal H}_{\mua\mub}^\Nv \right) \Gamma^{\mua\mub} \epsilon_\jjc
        \nonumber \\ & \qquad			     
		     + \sqrt{2} \, g \, \Omega_{\jja\jjb} \, A_2^{\jjc\jjb} \, \epsilon_\jjc \; ,
   \nonumber \\
  \delta \lambda_\jja^\xa &= i \, \Omega^{\jjb\jjc} \, ( {\cal V}_{\Ma}{}^{\xa} D_\mua {\cal V}_{\jja\jjb}{}^\Ma ) \Gamma^\mua \epsilon_\jjc
                       - \ft 1 4 \, \Sigma \, {\cal V}_{\Ma}{}^\xa \, {\cal H}_{\mua\mub}^\Ma \, \Gamma^{\mua\mub} \, \epsilon_\jja
		       + \sqrt{2} \, g \, \Omega_{\jja\jjb} \, A_2^{\xa\jjc\jjb} \, \epsilon_\jjc	      	      \; .
\end{align}
Here we have neglected higher order fermion terms. These fermion variations could formally be
read off from \cite{Dall'Agata:2001vb}. But the fermion shift matrices
$A_{1 \jja\jjb}$, $A_{2 \jja\jjb}$ and $A_{2 \jja\jjb}^\xa$ which are defined below now include contributions
from the vector $\xi_\Ma$.

Using ${\cal V}_\Ma{}^\xa$ and ${\cal V}_\Ma{}^{\jja\jjb}$ we can define 
from $f_{\Ma\Mb\Mc}$, $\xi_{\Ma\Mb}$ and $\xi_\Ma$
scalar dependent tensors that transform under $H$. The vector $\xi_\Ma$ gives
\begin{align}
   \tau^{\jja\jjb} &= \Sigma^{-1} {{\cal V}_\Ma}^{\jja\jjb} \, \xi^\Ma  \; ,&
   \tau^\xa &= \Sigma^{-1} {{\cal V}_\Ma}^\xa \, \xi^\Ma \; ,
\end{align}
from the 2-form $\xi_{\Ma\Mb}$ one gets
\begin{align}   
   \zeta^{\jja\jjb} &= \sqrt{2} \, \Sigma^2 \Omega_{\jjc\jjd} \, {{\cal V}_\Ma}^{\jja\jjc} {{\cal V}_\Mb}^{\jjb\jjd} \, \xi^{\Ma\Mb}  \; ,&
   \zeta^{\xa\jja\jjb} &= \Sigma^2 {{\cal V}_\Ma}^{\xa} {{\cal V}_\Mb}^{\jja\jjb} \, \xi^{\Ma\Mb} \; ,
\end{align}
and the 3-form $f_{\Ma\Mb\Mc}$ yields
\begin{align}   
   \rho^{\jja\jjb} &= - \ft 2 3 \, \Sigma^{-1} {{\cal V}^{\jja\jjc}}_{\Ma} {{\cal V}^{\jjb\jjd}}_{\Mb} {{\cal V}^\Mc}_{\jjc\jjd} \, {f^{\Ma\Mb}}_\Mc \; ,
   &
   \rho^{\xa\jja\jjb} &= \sqrt{2} \, \Sigma^{-1} \, \Omega_{\jjc\jjd} \,
                          {{\cal V}_\Ma}^{\xa} {{\cal V}_\Mb}^{\jja\jjc} {{\cal V}_\Mc}^{\jjb\jjd} \, f^{\Ma\Mb\Mc}  \; ,
\end{align}
where $\lambda^{\jja\jjb} = \lambda^{[\jja\jjb]}$, $\zeta^{\jja\jjb} = \zeta^{(\jja\jjb)}$, $\zeta^{\xa\jja\jjb} = \zeta^{\xa[\jja\jjb]}$,
$\rho^{\jja\jjb} = \rho^{(\jja\jjb)}$, $\rho^{\xa\jja\jjb} = \rho^{\xa(\jja\jjb)}$.
\footnote{
Our notation translates into that of \cite{Dall'Agata:2001vb} as follows:
$a_\mua = A_\mua^\Nv$, $\Lambda^\Ma_\Mb = \frac {g} {g_A}  {\xi^\Ma}_\Mb$, $f_{\Ma\Mb}^\Mc = - \frac g {g_S}  f_{\Ma\Mb}{}^\Mc$,
$U_{\jja\jjb} = - \frac {g} {6 g_A} \zeta_{\jja\jjb}$, $V_{\jja\jjb}^\xa = - \frac {g} {\sqrt{2} g_A} \zeta_{\jja\jjb}^{\xa}$,
$S_{\jja\jjb} = \frac g {3 g_S}  \rho_{\jja\jjb}$,
$T^\xa_{\jja\jjb} = \frac g {\sqrt{2} g_S}  \rho^\xa_{\jja\jjb}$.
}
The above tensors are the irreducible components of the $T$-tensor introduced for the general case in equation
\eqref{GenSplitT}. As explained in section \ref{sec:PresSUSY} these irreducible components are used to define the
the fermion shift matrices. In our case one finds
\begin{align}
   A_1^{\jja\jjb} &= \frac 1 {\sqrt{6}} \left( - \zeta^{\jja\jjb} 
                                        + 2 \rho^{\jja\jjb} \right) \, , \nonumber \\
   A_2^{\jja\jjb} &= \frac 1 {\sqrt 6} \left( \zeta^{\jja\jjb} 
                                     + \rho^{\jja\jjb} 
				     + \ft 3 {2} \, \tau^{\jja\jjb} \right) \, , \nonumber \\
   A_2^{\xa\jja\jjb} &= \frac 1 2 \left( - \zeta^{\xa\jja\jjb}
                                      + \rho^{\xa\jja\jjb}
				      - \, \ft 1 4 \, \sqrt{2} \, \tau^{\xa} \, \Omega^{\jja\jjb} \right) \; .
\end{align}
According to section \ref{sec:PresSUSY}
these matrices do not only appear in the fermion variations but also in the fermion mass terms that have to appear
in the Lagrangian of the gauged theory
\begin{align}
   e^{-1} {\cal L}_{\text{f.mass}} \, &= \, \frac {\sqrt{6} \, i \, g} 4 \, \Omega_{\jjc\jja} \, A_1^{\jja\jjb}  \, 
                 \bar \psi^\jjc_\mua \Gamma^{\mua\mub} \psi_{\mub \, \jjb}
                   + \sqrt{2} \, g \, \Omega_{\jjc\jjb} \, A_2^{\jjb\jja} \, \bar \psi^\jjc_\mua \Gamma^{\mua} \chi_{\jja}
                   + \sqrt{2} \, g \, \Omega_{\jjc\jjb} \, A_2^{\jjb\jja\xa} \, \bar \psi^\jjc_\mua \Gamma^{\mua} \lambda_{\jja}^{\xa}		   
    \; .		   
\end{align}
Note that we have only given those terms that involve the gravitini.
Supersymmetry imposes the condition \eqref{GenAA} on the fermion shift matrices, which here reads
\begin{align}
   \Omega_{\jjc\jjd} \left( A_1^{\jja\jjc} A_1^{\jjb\jjd} - A_2^{\jja\jjc} A_2^{\jjb\jjd} - A_2^{\xa\jja\jjc} A_2^{\xa\jjb\jjd} \right)
                 &= - \frac 1 4 \Omega^{\jja\jjb} V  \; ,
   \label{WardD5}		 
\end{align}
where the scalar potential appears on the right hand side. Again, this condition is satisfied as a consequence of the
quadratic constraint \eqref{XiRel}.

\section{Dimensional reduction from $d=5$ to $d=4$}
\label{sec:reduction}

Starting from a five dimensional supergravity one can perform a (twisted) circle reduction
to get a four dimensional supergravity.
For the maximal gauged supergravities this was discussed in \cite{Trigiante:2007ki}
for Scherk-Schwarz reductions from $d=5$.
We consider $N=4$ supergravities here and restrict our attention to simple circle reductions
starting with a five-dimensional theory that is already gauged.
Any five dimensional gauging described by $f_{\Ma\Mb\Mc}$,
$\xi_{\Ma\Mb}$ and $\xi_\Ma$ must give rise to a particular four dimensional gauging characterized by
$f_{\aa\Ma\Mb\Mc}$ and $\xi_{\aa\Ma}$. In other words the set of five dimensional gaugings is embedded into the
set of four dimensional gaugings and we now want to make this embedding explicit.
This yields additional examples of four dimensional gaugings, but it is also interesting
in the context of string dualities in presence of fluxes since the two tensors $f_{\Ma\Mb\Mc}$ and $\xi_{\Ma\Mb}$ in $d=5$
turn out to be parts of the single tensor $f_{\aa\Ma\Mb\Mc}$ under the larger duality group
in $d=4$. Thus, as usual, one gets a more
unified description of gaugings with different higher dimensional origin when compactifying the supergravity
theory further. With all the group structure at hand it is not necessary to
explicitly perform the dimensional reduction but we can read off the connection from the formulas for the
covariant derivatives \eqref{D4CovDivTh} and \eqref{CovDivD5} (that is from the embedding tensor).

A five dimensional theory with $n$ vector multiplets yields a four dimensional theory with $n+1$ vector multiplets.
One way to understand that is by counting scalar fields. There are $5n+1$ scalars already present in five dimensions and
in addition one gets one scalar from the metric and $6+n$ scalars form the vector fields which gives $6n+8$ in total and agrees
with the number of scalars in the coset ${\rm SL}(2) \times {\rm SO}(6,n+1)/{\rm SO}(2) \times {\rm SO}(6) \times {\rm SO}(n+1)$.
When breaking the ${\rm SO}(6,n+1)$ into ${\rm SO}(1,1)_A \times {\rm SO}(5,n)$
the vector representation splits into an ${\rm SO}(5,n)$ vector $v^\Ma$ and two scalars $v^\oplus$ and $v^\ominus$
with charges $0$, $1/2$ and $-1/2$, respectively,
under ${\rm SO}(1,1)_A$. When breaking the ${\rm SL}(2)$ into ${\rm SO}(1,1)_B$ the vector
splits into two scalars $v^+$ and $v^-$ with charges $1/2$ and $-1/2$ under ${\rm SO}(1,1)_B$. The four dimensional
vector fields therefore split into $A_\mua^{\Ma+}$, $A_\mua^{\Ma-}$, $A_\mua^{\oplus+}$, $A_\mua^{\oplus-}$,
$A_\mua^{\ominus+}$ and $A_\mua^{\ominus-}$. We can now identifying the five dimensional vector fields as
\begin{align}
   A_\mua^{\Ma} &= A_\mua^{\Ma+} \, , &
   A_\mua^\Nv &= A_\mua^{\ominus-} \, ,
   \label{IdentA}
\end{align}
and these fields carry charges $1/2$ and $-1$ under the diagonal of ${\rm SO}(1,1)_A$ and ${\rm SO}(1,1)_B$ and
the five dimensional ${\rm SO}(1,1)$ therefore has to be this diagonal. Thus the five dimensional
global symmetry generators are given in terms of the four dimensional ones as follows
\begin{align}
   t_{\Na} &= t^{{\rm SL}(2)}_{+-} + t^{{\rm SO}(6,n+1)}_{\ominus\oplus} \; , &
   t_{\Ma\Mb} &= t^{{\rm SO}(6,n+1)}_{[\Ma\Mb]} \; .
   \label{IdentT}
\end{align}
The vector fields $A_\mua^{\Ma-}$, $A_\mua^{\oplus+}$ are the four dimensional duals of
$A_\mua^{\Ma+}$ and $A_\mua^{\ominus-}$, they come from the two-form gauge fields in five dimensions.
The vector fields $A_\mua^{\oplus-}$ and $A_\mua^{\ominus+}$ are uncharged under the five dimensional
${\rm SO}(1,1)$, they are the Kaluza-Klein vector coming from the metric and its dual field.

Now, if a four dimensional vector field that was already a vector field in five dimensions \eqref{IdentA}
gauges a four dimensional symmetry that was already a symmetry in five dimensions \eqref{IdentT} the corresponding
gauge coupling in the covariant derivative in $d=4$ has to be the same as in $d=5$.
For the four dimensional covariant derivative \eqref{D4CovDivTh} one finds
\begin{align}
   D_\mua &= \nabla_\mua - g \, A_\mua{}^{\Ma+} 
                          \left(  \Theta_{+\Ma}{}^{\Mb\Mc} t_{\Mb\Mc}
                           + 2 f_{+\Ma}{}^{\ominus\oplus} t_{\ominus\oplus} 
                           + \xi_{+\Ma} t_{+-} \right)
			    \nonumber \\ & \qquad \quad
                           - g \, A_\mua{}^{\ominus-} 
			  \left(  f_{-\ominus}{}^{\Mb\Mc} t_{\Mb\Mc}
                           + \xi_{-\ominus} t_{\ominus\oplus}
                           - \xi_{-\ominus} t_{+-} \right) \, + D^{\text{add}}_{\mua} \, ,
\end{align}
where $\Theta_{\aa\Ma\Mb\Mc}$ is defined in \eqref{DefThetaF}\footnote{
Note that what we called $n$ in section \ref{ch:D4} 
is now $n+1$ and the index $\Ma$ now is an ${\rm SO}(5,n)$ vector index rather than
a ${\rm SO}(6,n+1)$ index.} and $D^{\text{add}}_{\mua}$ denotes exclusively four dimensional contributions to the covariant derivative.
By comparing with the known covariant derivative in five dimensions \eqref{CovDivD5} one gets
\begin{align}
   \xi_{+\Ma} &= \xi_\Ma \; , &
   f_{+\Ma\oplus\ominus} &= \ft 1 2 \, \xi_\Ma \; , &
   f_{-\ominus\Ma\Mb} &= \xi_{\Ma\Mb} \; , &
   f_{+\Ma\Mb\Mc} &= f_{\Ma\Mb\Mc}  \; . &
\end{align}
For a simple circle reduction it is natural to demand furthermore
$f_{\pm\Ma\Mb\oplus} = 0$, $f_{+\Ma\Mb\ominus}=0$, $f_{-\Ma\Mb\Mc} = 0$,
$f_{-\Ma\oplus\ominus} = 0$, $\xi_{-\Ma}=0$, $\xi_{\pm\oplus}=0$ and $\xi_{\pm\ominus}=0$.
Some of the last quantities, however, may be non-zero for more complicated dimensional reductions
and may then for example correspond to Scherk-Schwarz generators \cite{Villadoro:2004ci}.
But for the ordinary circle reduction we have just given the embedding of the five dimensional gaugings
into the four dimensional ones. In addition to the above equations we have to make sure that
$f_{\aa{\tilde \Ma}{\tilde \Mb}{\tilde \Mc}}$ is totally antisymmetric 
in the last three indices ($\tilde \Ma=\{\Ma,\oplus,\ominus\}$). One can then show that for these
tensors $f_{\aa{\tilde \Ma}{\tilde \Mb}{\tilde \Mc}}$ and $\xi_{\aa{\tilde \Ma}}$ the four dimensional
quadratic constraint \eqref{QConD4}
becomes precisely the five dimensional one \eqref{XiRel} for $f_{\Ma\Mb\Mc}$, $\xi_{\Ma\Mb}$
and $\xi_\Ma$. Also the four and the five dimensional scalar potentials \eqref{VD4}, \eqref{PotD5}
become the same if all scalars that are not yet present in $d=5$ are set to the origin\footnote{
The equality of the scalar potentials is most easily checked at the origin $M=\mathbbm{1}$. If the potentials
do agree there for all possible gaugings the statement is already proven due to the ${\rm SO}(1,1) \times {\rm SO}(5,n)$
covariance of the construction.}.

Due to the antisymmetry of $f_{\aa{\tilde \Ma}{\tilde \Mb}{\tilde \Mc}}$ one finds the following additional terms 
in the $d=4$ covariant derivative:
\begin{align}
   D^{\text{add}}_{\mua}
     &= - g \, A_\mua{}^{\Ma-} \left( 2 {\xi_{\Ma}}^{\Mb} \, t_{\Mb\ominus} + \xi_\Ma \, t_{--} \right)
                \nonumber \\ & \qquad \qquad
                   + g \, A_\mua{}^{\ominus+} \, \xi^{\Mb} \, \left( t_{\Mb\ominus} -  t_{\Mb\oplus} \right)
		   + g \, A_\mua{}^{\oplus+} \, \xi^{\Mb} \left( t_{\Mb\ominus} + t_{\Mb\oplus}   \right) \; .
\end{align}
These are couplings of vector fields to symmetry generators that both only occur in four dimensions.
If one explicitly performs the dimensional reduction by hand these gauge couplings originate from
the dualization of the various fields.

Thus, we showed how the gaugings of $N=4$ supergravity in five dimensions are naturally embedded into the
four dimensional ones by dimensional reduction.
Noteworthy, the five dimensional gaugings are parameterized in
terms of three tensors $f_{\Ma\Mb\Mc}$, $\xi_{\Ma\Mb}$ and $\xi_{\Ma}$ while the four dimensional ones are parameterized
in terms of two tensors $f_{\aa\Ma\Mb\Mc}$ and $\xi_{\aa\Ma}$ only. Thus with decreasing spacetime dimension one finds not
only a larger duality group but also a more uniform
description of the deformations. This is the typical picture of dualities in string theory where dimensional
reduction relates theories with different higher-dimensional origin.

\chapter{The maximal supergravities in $d=7$} \label{ch:D7}

In this chapter the general gauging of seven-dimensional maximal supergravity is presented.
Examples of these theories can be obtained by sphere reductions of M-theory or of type IIA or IIB supergravity
which lead to  gauge groups ${\rm SO}(5)$, ${\rm CSO}(4,1)$, and ${\rm SO}(4)$, respectively.
All the known gaugings as well as a number of new examples are incorporated in our formulation.
In particular, we obtain the theory with gauge group ${\rm SO}(4)$ that originates from a (warped)
$S^{3}$ reduction of type IIB supergravity.

\section{Embedding tensor and gauge fields}
\label{sec:embedding}

\subsection{Linear and quadratic constraint}

The global symmetry group of the ungauged seven-dimensional theory 
is $G_0=E_{4(4)}={\rm SL}(5)$. Its 24 generators $t^{\Ja}{}\!_{\Jb}$ are labeled
by indices $\Ja, \Jb=1, \dots, 5$ with $t^{\Ja}{}\!_{\Ja}=0$ and satisfy
the algebra
\bea
\Big[\,t^{\Ja}{}\!_{\Jb}\,,\;t^{\Jc}{}\!_{\Jd}\,\Big] &=&
\delta^{\Jc}_{\Jb}\,t^{\Ja}{}\!^{\phantom{\Jc}}_{\Jd}
-
\delta^{\Ja}_{\Jd}\,t^{\Jc}{}\!^{\phantom{\Jc}}_{\Jb} 
\;.
\label{sl5}
\eea
The vector fields $A_\mu^{\vphantom{[]}\Ja\Jb}=A_\mu^{[\Ja\Jb]}$ of the ungauged theory transform in the representation 
$\overline{\bf 10}$ of ${\rm SL}(5)$, so that
$\delta A_\mu^{\vphantom{[]}\Ja\Jb}= 2\Lambda_{\Jc}{}_{\vphantom{[]}}^{[\Ja}A_{\mu}^{\vphantom{[]}\Jb]\Jc}$.
The covariant derivatives \eqref{GenCovDiv} takes the form
\bea
D_{\mu} &=& \nabla_{\mu} 
- g A_{\mu}^{\vphantom{[]}\Ja\Jb} \Theta_{\Ja\Jb,\Jc}{}^{\Jd} \,t^{\Jc}{}\!_{\Jd}
\;,
\eea
We already discussed the linear constraint on the embedding tensor $\Theta_{\Ja\Jb,\Jc}{}^{\Jd}$
in section \ref{sec:gen:emb}. According to table \ref{LinCon1} only two of the four irreducible components of
$\Theta_{\Ja\Jb,\Jc}{}^{\Jd}$ are allowed to be non-zero. These two components are a ${\bf 15}$, 
described by a symmetric matrix $Y_{\Ja\Jb}=Y_{(\Ja\Jb)}$, and a ${\bf \overline{40}}$, 
described by a tensor $Z^{\Ja\Jb,\Jc}=Z^{[\Ja\Jb],\Jc}$ with $Z^{[\Ja\Jb,\Jc]}=0$.
In terms of these tensors the embedding tensor is given by
\bea
\Theta_{\Ja\Jb,\Jc}{}^{\Jd}&=&
\delta^{\Jd}_{[\Ja}\,Y^{\phantom{\Jd}}_{\Jb]\Jc}
-2\epsilon_{\Ja\Jb\Jc\Je\Jf}\,Z^{\Je\Jf,\Jd}
\;.
\label{linear}
\eea
The quadratic constraint \eqref{QconGen1} on the embedding tensor
reduces to the following condition on $Y_{\Ja\Jb}$ and $Z^{\Ja\Jb,\Jc}$:
\bea
Y_{\Ja\Jd}\,Z^{\Jd\Jb,\Jc} 
+2\epsilon_{\Ja\Je\Jf\Jg\Jh}\,Z^{\Je\Jf,\Jb}Z^{\Jg\Jh,\Jc}
&=& 0
\;,
\label{quadratic}
\eea
In terms of ${\rm SL}(5)$ representations this quadratic constraint
has different irreducible parts in the 
${\bf \overline{5}}$, the ${\bf \overline{45}}$, and the ${\bf \overline{70}}$ representation.
In particular, they give rise to the relations
\bea
Z^{\Ja\Jb,\Jc}\,Y_{\Jc\Jd}&=&0\;,
\qquad
Z^{\Ja\Jb,\Jc}\,X_{\Ja\Jb} ~=~0  \;.
\label{Q2}
\eea
The second equation of these equations already carries the full content
of the quadratic constraint. The gauge group generators $X_{\Ja\Jb}=X_{[\Ja\Jb]}$ are given by
\bea
\label{X-theta}
X_{\Ja\Jb} &=& \Theta_{\Ja\Jb,\Jc}{}^{\Jd}\;
t^{\Jc}{}\!_{\Jd}\;.
\eea
They can be taken in an a an arbitrary representation.
Acting on the ${\bf 5}$ and ${\bf 10}$ of ${\rm SL}(5)$ they read
\begin{align}
   (X_{\Ja\Jb}){}_{\Jc}{}^{\Jd} &= \Theta_{\Ja\Jb,\Jc}{}^{\Jd}~=~
                                       \delta^{\Jd}_{[\Ja}\,Y^{\phantom{\Jd}}_{\Jb]\Jc} 
                                         -2\epsilon_{\Ja\Jb\Jc\Je\Jf}\,Z^{\Je\Jf,\Jd}    \; ,
\nonumber \\
      (X_{\Ja\Jb}){}_{\Jc\Jd}{}^{\Je\Jf} 
                     &=  2 (X_{\Ja\Jb}){}^{\vphantom{\Jf]}}_{[\Jc}{}^{[\Je}_{\vphantom{\Jd]}}\delta_{\Jd]}^{\Jf]} \; .
\label{XP}
\end{align}
Summarizing, a consistent gauging of the seven-dimensional theory is defined by an embedding tensor
$\Theta_{\Ja\Jb,\Jc}{}^{\Jd}$ satisfying a linear 
and a quadratic ${\rm SL}(5)$ representation constraint
which schematically read
\bea
  \Big({\mathbb{P}}_{\bf 10}  +{\mathbb{P}}_{{\bf 175}}\Big)
  \,\Theta\phantom{\,   \Theta }  &=&   0\,,   \nn\\ 
\Big( {\mathbb{P}}_{\bf \overline{5}} + {\mathbb{P}}_{\bf \overline{45}}
+ {\mathbb{P}}_{\bf \overline{70}}
  \Big)\,\Theta\,   \Theta &=&  0\,.  
  \label{sum}
\eea
The first of these equations can be explicitly solved
in terms of two tensors $Y_{\Ja\Jb}$ and $Z^{\Ja\Jb,\Jc}$ leading to~(\ref{linear});
the quadratic constraint then translates into the 
conditions~(\ref{quadratic}) on these tensors. 
In the rest of this chapter we will demonstrate that
an embedding tensor~$\Theta$ solving equations~(\ref{sum})
defines a consistent gauging in seven dimensions. 

\subsection{Vector and tensor gauge fields}
\label{SecVecTen}

In the Lagrangian of ungauged $d=7$ maximal supergravity one in addition to the vector fields $A_\mua^{\Ja\Jb}$
has two-form fields fields $B_{\mua\mub\,\Ja}$ that transform in the ${\bf 5}$ of ${\rm SL}(5)$ \cite{Sezgin:1982gi}. 
On-shell one can introduce the dual gauge fields. Dual to the two-forms there are three forms
$S_{\mua\mub\muc\,\Ja}$ in the ${\bf \overline{5}}$ representation. In the gauged theory these
three-forms are present at the level of the Lagrangian. They appear via St\"uckelberg type couplings
in the field strengths of the two-form gauge fields and are necessary for the gauge invariance 
of these field strengths. They will always
appear projected under $Y_{\Ja\Jb}$, i.e.\ for $\Theta_{\Ja\Jb,\Jc}{}^{\Jd}=0$ they will decouple and
the ungauged theory is recovered.
For the general case this system of vector and tensor gauge fields
was already introduced in section \ref{sec:GenNonAVecTen}.
The formulas given there shall now be specialized to the present context.

The tensors $d_{I \Ma\Mb}$ and $c^{AI}_\Ma$ of section \ref{sec:GenNonAVecTen} are now given by
\begin{align}
   d_{\Jg,[\Ja\Jb][\Jc\Jd]} &= \epsilon_{\Jg\Ja\Jb\Jc\Jd} \; , &
   c^{\Ja\Jb}_{[\Jc\Jd]} &= - \, \delta^{[\Ja}_{[\Jc} \, \delta^{\Jb]}_{\Jd]} \; .
\end{align}
Comparing equations \eqref{DefGenY} and \eqref{linear} shows that $Y_{IA}$ and $Z^{MI}$ of
section \ref{sec:GenNonAVecTen} are identified with the tensors $Y_{\Ja\Jb}$ and $Z^{\Ja\Jb,\Jc}$
introduced above. The relation \eqref{DefZfromX} then translates into
\bea
(X_{\Ja\Jb}){}_{\Jc\Jd}{}^{\Je\Jf}+(X_{\Jc\Jd}){}_{\Ja\Jb}{}^{\Je\Jf}
~=~ 2\,Z_{\phantom{\Ja}}^{\Je\Jf,\Jg}\,d_{\Jg,[\Ja\Jb][\Jc\Jd]}
\;,
\label{Csym}
\eea
With these identifications the covariant field strengths \eqref{DefHHH} for the gauge fields are given by
\begin{align}
   {\cal H}^{(2)\Ja\Jb}_{\mua\mub} &= 
              2 \partial_{[\mua} A_{\mub]}^{\Ja\Jb} 
            + g {(X_{\Jc\Jd})_{\Je\Jf}}_{\vphantom{[\mua}}^{\Ja\Jb} 
                 A_{[\mua}^{\Jc\Jd} A_{\vphantom{[]}\mub]}^{\vphantom{\Jd}\Je\Jf}
            + g Z^{\Ja\Jb,\Jc} B_{\mua\mub\Jc} \;,
   \nonumber \\
   {\cal H}^{(3)}_{\mua\mub\muc\,\Ja} &= 3 D_{[\mua} B_{\mub\muc]\Ja} 
                                        + 6 \epsilon_{\Ja\Jb\Jc\Jd\Je} A^{\Jb\Jc}_{[\mua} 
                \Big( \partial^{\vphantom{\Jd}}_{\vphantom{[]}\mub} A^{\Jd\Je}_{\muc]} 
              + \ft 2 3 g {X_{\Jf\Jg,\Jh}}^\Jd A^{\Je\Jh}_\mub A^{\Jf\Jg}_{\muc]} \Big)
	     \nonumber \\ & \qquad \qquad \qquad \qquad \qquad \qquad \qquad \qquad \qquad \qquad \qquad \qquad
              + g Y_{\Ja\Jb} S^\Jb_{\mua\mub\muc}   \;,
   \nonumber \\	      
      Y_{\Ja\Jb}\, {\cal H}^{(4)\,\Jb}_{\mua\mub\muc\mud} &= 
                       Y_{\Ja\Jb}  \Big( 4D^{\phantom{\Ja}}_{[\mua} S_{\mub\muc\mud]}^\Jb
                      + 6 {\cal F}_{[\mua\mub}^{\Jb\Jc} B^{\phantom{\Ja}}_{\muc\mud]\Jc}  
                       + 3 g Z^{\Jb\Jc,\Jd} B_{[\mua\mub\,\Jc} B_{\muc\mud]\,\Jd}
             \nonumber \\  & \quad
                + 8 \epsilon_{\Jc\Jd\Je\Jf\Jg} A_{[\mua}^{\Jb\Jc} 
                        A_{\mub}^{\Jd\Je} \partial^{\vphantom{\Ja}}_{\muc} A_{\mud]}^{\Jf\Jg}
                    +4g  \epsilon_{\Jc\Jd\Je\Ji\Jj} {X_{\Jf\Jg,\Jh}}^{\Ji} 
                 A_{[\mua}^{\Jb\Jc} A_\mub^{\Jd\Je} A_\muc^{\Jf\Jg} A_{\mud]}^{\Jh\Jj}   \Big)          \;.
  \label{H234}
\end{align}
These field strengths transform covariantly under vector and tensor gauge transformations \eqref{GenGauge}
which read for the present case
\begin{align}
   \Delta A_{\mua}^{\Ja\Jb} &= D_\mua \Lambda^{\Ja\Jb} 
   - g Z^{\Ja\Jb,\Jc} \Xi_{\mua\,\Jc} 
   \;,
   \nonumber\\[1ex]
   \Delta B_{\mua\mub\,\Ja} &= 2 D_{[\mua} \Xi_{\mub]\,\Ja} 
             - 2 \epsilon_{\Ja\Jb\Jc\Jd\Je}\, {\cal H}_{\mua\mub}^{(2)\Jb\Jc} \Lambda^{\Jd\Je} 
             - g Y_{\Ja\Jb} \Phi_{\mua\mub}^\Jb 
   \;,
   \nonumber\\[1ex]
   Y_{\Ja\Jb}\,\,\Delta S_{\mua\mub\muc}^\Jb &= 
   Y_{\Ja\Jb}\,\Big(
   3 D^{\phantom{\Jb}}_{[\mua} \Phi_{\mub\muc]}^\Jb
                   - 3 {\cal H}_{[\mua\mub}^{(2)\Jb\Jc} \Xi_{\muc]\,\Jc} 
                   + {\cal H}^{(3)}_{\mua\mub\muc\,\Jc} \Lambda^{\Jc\Jb} 
\Big)\;,
\label{gauge}
\end{align}
with gauge parameters $\Lambda^{\Ja\Jb}$, $\Xi_{\mua\,\Ja}$,
and  $\Phi_{\mua\mub}^\Ja$, corresponding to vector and tensor gauge transformations, respectively.
The covariant variations \eqref{CovD1} take the form
\begin{align}
  \Delta A_{\mua}^{\Ja\Jb} &\equiv \delta A_{\mua}^{\Ja\Jb} 
  \;,
  \nonumber\\[1ex]
  \Delta B_{\mua\mub\,\Ja} &\equiv \delta B_{\mua\mub\,\Ja} 
  - 2 \epsilon_{\Ja\Jb\Jc\Jd\Je} A_{[\mua}^{\Jb\Jc} \,\delta A_{\mub]}^{\Jd\Je} 
  \;,
  \nonumber\\[1ex]
Y_{\Ja\Jb}\,
  \Delta S_{\mua\mub\muc}^\Jb &\equiv 
  Y_{\Ja\Jb}\,\Big(
  \delta S_{\mua\mub\muc}^\Jb
- 3 B^{\phantom{\Jb}}_{[\mua\mub\,\Jc}\, \delta A_{\muc]}^{\Jc\Jb} 
+ 2 \epsilon_{\Jc\Jd\Je\Jf\Jg} A_{[\mua}^{\Jb\Jc} A_{\mub}^{\Jd\Je} \,\delta A_{\muc]}^{\Jf\Jg}
\Big)
\;.
\label{covV}
\end{align}
The gauge transformations~(\ref{gauge}) consistently close into the algebra \eqref{GenGaAlg}.
For $g\rightarrow0$ one recovers from~(\ref{gauge}) the vector and tensor gauge
transformations of the ungauged theory~\cite{Sezgin:1982gi}.
The action of the tensor gauge transformations eventually allows to 
eliminate some of the vector and tensor gauge fields by fixing part of the gauge symmetry. 
We will discuss this in more detail in section~\ref{sec:gfixing}.

The deformed Bianchi identities \eqref{GenBianchi} read in the present context
\begin{align}
   D^{\phantom{\Ja}}_{[\mua} {\cal H}_{\mub\muc]}^{(2)\Ja\Jb} &= 
   \frac13 g Z^{\Ja\Jb,\Jc} {\cal H}^{(3)}_{\mua\mub\muc\,\Jc} 
   \;,\nonumber\\[1ex]
   D^{\phantom{\Ja}}_{[\mua} {\cal H}^{(3)}_{\mub\muc\mud]\,\Ja} 
        &= \frac 3 2 \epsilon_{\Ja\Jb\Jc\Jd\Je}\, 
        {\cal H}_{[\mua\mub}^{(2)\Jb\Jc} {\cal H}_{\muc\mud]}^{(2)\Jd\Je}
           + \frac 14g Y_{\Ja\Jb} \, {\cal H}^{(4)\,\Jb}_{\mua\mub\muc\mud}
           \;.
           \label{Bianchi}
\end{align}

There is a unique gauge invariant topological Lagrangian in seven dimensions that combines vector and tensor fields in such a way
that it is invariant under the full set of non-Abelian vector and tensor gauge transformations \eqref{gauge}
up to total derivatives. The leading terms of this Lagrangian were already given in \eqref{GenTopD7},
Completely it reads
\begin{align}
   {\cal L}_{\text{VT}} &= - \frac 1 {9}\, 
   \epsilon^{\mua\mub\muc\mud\mue\muf\mug} \nonumber \\ &
     \Bigg[ g Y_{\Ja\Jb} S^\Ja_{\mua\mub\muc} \Big( D^{\vphantom{\Ja}}_\mud S_{\mue\muf\mug}^\Jb
+ \frac 3 2 g Z^{\Jb\Jc,\Jd} B_{\mud\mue\,\Jc} B_{\muf\mug\,\Jd}
+ 3 {\cal F}^{\Jb\Jc}_{\mud\mue} B^{\phantom{\Ja}}_{\muf\mug\,\Jc}
\nonumber \\ & \qquad \qquad
                 + 4 \epsilon_{\Jc\Jd\Je\Jf\Jg} A^{\Jb\Jc}_{\mud} A^{\Jd\Je}_{\mue} \partial^{\vphantom{\Ja}}_{\muf} A_\mug^{\Jf\Jg}
        + g \epsilon_{\Jc\Jd\Je\Jj\Jk} {X_{\Jf\Jg,\Jh\Ji}}^{\Jj\Jk} A_\mud^{\Jb\Jc} A_\mue^{\Jd\Je} A_\muf^{\Jf\Jg} A_\mug^{\Jh\Ji} \Big) 
  \nonumber \\  &  \quad
      + 3 g Z^{\Ja\Jb,\Jc} (D_\mua B_{\mub\muc\,\Ja}) B_{\mud\mue\,\Jb} B_{\muf\mug\,\Jc} 
    - \frac 9 2 {\cal F}_{\mua\mub}^{\Ja\Jb} B_{\muc\mud\,\Ja} D_{\mue} B_{\muf\mug\,\Jb}
  \nonumber \\ & \quad
    + 18 \epsilon_{\Ja\Jb\Jc\Jd\Je} {\cal F}_{\mua\mub}^{\Ja\Ji} A^{\Jb\Jc}_{\muc} 
            \Big( \partial^{\vphantom{\Ja}}_\mud A^{\Jd\Je}_{\mue} 
                                       + \frac 2 3 g {X_{\Jf\Jg,\Jh}}^\Jd A^{\Je\Jh}_\mud A^{\Jf\Jg}_{\mue} \Big)\, B_{\muf \mug\,\Ji}
  \nonumber \\ & \quad
    + 9 g \epsilon_{\Ja\Jb\Jc\Jd\Je} Z^{\Ja\Ji,\Jj} A^{\Jb\Jc}_{\mua} \Big( \partial^{\vphantom{\Ja}}_\mub A^{\Jd\Je}_{\muc} 
          + \frac 2 3 g {X_{\Jf\Jg,\Jh}}^\Jd A^{\Je\Jh}_\mub A^{\Jf\Jg}_{\muc} \Big)\, B_{\mud\mue\,\Ji} B_{\muf\mug\,\Jj} \nonumber \\
  & + \frac {36} 5 \epsilon_{\Ja\Jc\Jd\Jg\Jh} \; \epsilon_{\Jb\Je\Jf\Ji\Jj} 
       A_\mua^{\Ja\Jb} A_\mub^{\Jc\Jd} A_\muc^{\Je\Jf} 
             (\partial_\mud^{\vphantom{\Ja}} A_\mue^{\Jg\Jh}) (\partial_\muf^{\vphantom{\Ja}} A_\mug^{\Ji\Jj})
     \nonumber \\ &   
    + 8 g \epsilon_{\Ja\Jc\Jd\Je\Jf} \; \epsilon_{\Jb\Jg\Jh\Jm\Jn} {X_{\Ji\Jj,\Jk\Jl}}^{\Jm\Jn}
          A_\mua^{\Ja\Jb} A_\mub^{\Jc\Jd} A_\muc^{\Jg\Jh} A_\mud^{\Ji\Jj} A_\mue^{\Jk\Jl} 
\partial_\muf^{\vphantom{\Ja}} A_\mug^{\Je\Jf} \nonumber \\
  & - \frac 4 7 g^2 \epsilon_{\Ja\Jc\Jd\Jo\Jp} \; \epsilon_{\Jb\Ji\Jj\Jq\Jr} {X_{\Je\Jf,\Jg\Jh}}^{\Jo\Jp} {X_{\Jk\Jl,\Jm\Jn}}^{\Jq\Jr}
        A_\mua^{\Ja\Jb} A_\mub^{\Jc\Jd} A_\muc^{\Je\Jf} A_\mud^{\Jg\Jh} A_\mue^{\Ji\Jj} A_\muf^{\Jk\Jl} A_\mug^{\Jm\Jn}
\Bigg]
\;.
\label{VT}
\end{align}
As $g\rightarrow0$ this topological term reduces to the ${\rm SL}(5)$ invariant
Chern-Simons term of the ungauged theory~\cite{Sezgin:1982gi}.
Under variation of the vector and tensor fields, the topological Lagrangian ${\cal L}_{\text{VT}}$ transforms as
\begin{align}
   \delta {\cal L}_{\text{VT}} = &
      - \frac 1 {18}\, \epsilon^{\mua\mub\muc\mud\mue\muf\mug} \Bigg[
         Y_{\Ja\Jb} \, {\cal H}^{(4)\Ja}_{\mua\mub\muc\mud} \, \Delta S^\Jb_{\mue\muf\mug} 
                + 6 {\cal H}^{(2)\Ja\Jb}_{\mua\mub} {\cal H}^{(3)}_{\muc\mud\mue\Ja} \,\Delta B_{\muf\mug\Jb}
\nonumber \\ & \qquad \qquad \qquad \qquad 
                - 2 {\cal H}^{(3)}_{\mua\mub\muc\Ja} {\cal H}^{(3)}_{\mud\mue\muf\Jb} \,\Delta A^{\Ja\Jb}_\mug \Bigg]
\; +{\rm total~derivatives} \;,
\label{varCS}
\end{align}
in terms of the covariant variations~(\ref{covV}). With~(\ref{gauge}) one explicitly verifies that this variation 
reduces to a total derivative. To show this one needs the deformed Bianchi identities \eqref{Bianchi} 
as well as the ${\rm SL}(5)$ relation
\begin{align}
     R_1^{[\Ja\Jb} R_2^{\vphantom{[]}\Jc\Jd}\, R_3^{\Je]\Jf} + 
     R_2^{[\Ja\Jb} R_3^{\vphantom{[]}\Jc\Jd}\, R_1^{\Je]\Jf} + 
     R_3^{[\Ja\Jb} R_1^{\vphantom{[]}\Jc\Jd}\, R_2^{\Je]\Jf}
      &= 0  \;,
   \label{RelSL5AAA}
\end{align}
for arbitrary tensors $R^{\Ja\Jb}_{1,2,3} = R^{[\Ja\Jb]}_{1,2,3}$.
\footnote{In terms of representations, this is the 
statement that the threefold symmetric product of three ${\bf 10}$
representations of ${\rm SL}(5)$ does not contain a ${\bf 5}$.}

\section{Coset space structure and the $T$-tensor}
\label{sec:CT}

In this section we introduce the scalar sector of maximal seven-dimensional
supergravity, which is described in terms of the scalar coset space
${\rm SL}(5)/{\rm SO}(5)$. This allows to manifestly realize the 
global ${\rm SL}(5)$ symmetry of the ungauged theory
while the local ${\rm SO}(5)\sim{\rm USp}(4)$ symmetry coincides
with the $R$-symmetry of the theory. 
For the gauged theory we further introduce the $T$-tensor 
as the ${\rm USp}(4)$ covariant analog 
of the embedding tensor~$\Theta$. 

\subsection{The ${\rm SL}(5)/{\rm SO}(5)$ coset space}
\label{sec:coset}

The scalar fields in seven dimensions parameterize the
coset space ${\rm SL}(5)/{\rm SO}(5)$. 
They are most conveniently 
described by a matrix ${\cal V}\in {\rm SL}(5)$
which transforms according to
\begin{align}
   {\cal V} \; &\rightarrow \; G \, {\cal V} \, H(x) &
   G \in {\rm SL}(5), \quad
   H(x) \in {\rm SO}(5) \; ,
   \label{TrafoCalV}
\end{align}
under global ${\rm SL}(5)$ and local ${\rm SO}(5)$ 
transformations, respectively 
(see~\cite{deWit:2002vz} for an introduction to 
the coset space structures in supergravity theories).
The local ${\rm SO}(5)$ symmetry
reflects the coset space structure of the scalar target space,
the corresponding connection is a composite field.
One can impose a gauge condition with respect to the 
local ${\rm SO}(5)$ invariance which amounts to 
fixing a coset representative,
i.e.\ a minimal parameterization of the coset space
in terms of the $14=24-10$ physical scalars. 
This induces a nonlinear realization of the 
global ${\rm SL}(5)$ symmetry obscuring the
group theoretical structure and complicating the calculations.
It is therefore most convenient to postpone this gauge fixing 
till the end.

In particular, 
the formulation~(\ref{TrafoCalV}) is indispensable
to describe the coupling to fermions 
with the group ${\rm SO}(5)\sim{\rm USp}(4)$
acting as the $R$-symmetry group of the theory. 
For ${\rm USp}(4)$ we use indices $\ja, \, \jb, \, \ldots = 1, \ldots, 4$
to label its fundamental representation.
The ${\rm USp}(4)$ invariant symplectic form $\Omega_{\ja\jb}$ has the properties
\begin{align}
   \Omega_{\ja\jb} &= \Omega_{[\ja\jb]} &
   (\Omega_{\ja\jb})^* &= \Omega^{\ja\jb} &
   \Omega_{\ja\jb} \Omega^{\jc\jb} &= \delta_\ja^\jc \; .
\end{align}
The lowest ``bosonic" ${\rm USp}(4)$ representations are defined in terms of the
fundamental representation (\!$\tinyyoung{\cr}$\,) with index structures according to
  \begin{align}
     {\bf 1}:  & ~~\cdot  & V_{\bf 1} \nonumber\\[.5ex]
     {\bf 5}:  & ~\tinyyoung{ \cr \cr} &   {V_{\bf 5}}^{\ja\jb} &= {V_{\bf 5}}^{[\ja\jb]} \;,
                                  &  \Omega_{\ja\jb } {V_{\bf 5}}^{\ja\jb} &= 0\;, \nonumber\\[.5ex]
     {\bf 10}: & ~\tinyyoung{&  \cr }
     &  {V_{\bf 10}}^{\ja\jb} &= {V_{\bf 10}}^{(\ja\jb)}\;, \nonumber\\[.5ex]
     {\bf 14}: &~\tinyyoung{&  \cr & \cr}
     & {{V_{\bf 14}}^{\ja\jb}}_{\jc\jd} &= {{V_{\bf 14}}^{[\ja\jb]}}_{[\jc\jd]} \;,
      &
                  {{V_{\bf 14}}^{\ja\jb}}_{\jc\jb} &= 0 \;,&
\Omega_{\ja\jb} {{V_{\bf 14}}^{\ja\jb}}_{\jc\jd} &= 0 = 
\Omega^{\jc\jd} {{V_{\bf 14}}^{\ja\jb}}_{\jc\jd} \;,\nonumber\\[.5ex]
     {\bf 35}: &~
     \tinyyoung{& & \cr \cr}
& {{V_{\bf 35}}^{\ja\jb}}_{\jc\jd} &= {{V_{\bf 35}}^{[\ja\jb]}}_{(\jc\jd)}\;, &
                 {{V_{\bf 35}}^{\ja\jb}}_{\jc\jb} &= 0 \;,&
\Omega_{\ja\jb} {{V_{\bf 35}}^{\ja\jb}}_{\jc\jd} &= 0 \;.
\label{USp4Reps}
  \end{align}
All objects in these representations are pseudo-real, 
i.e.~they satisfy reality constraints
\begin{align}
   (V_{\bf 1})^* &= V_{\bf 1}\;,\quad
   ({V_{\bf 5}}^{\ja\jb})^* = \Omega_{\ja\jc} \Omega_{\jb\jd} {V_{\bf 5}}^{\jc\jd}\;,\quad
   ({{V_{\bf 14}}^{\ja\jb}}_{\jc\jd})^* = 
   \Omega_{\ja\je} \Omega_{\jb\jf} \Omega^{\jc\jg} \Omega^{\jd\jh} \,
                                                    {{V_{\bf 14}}^{\je\jf}}_{\jg\jh} 
                                                    \;,
   \label{DefReality}    
\end{align}
etc. We use complex conjugation to raise and lower ${\rm USp}(4)$ indices.
According to \eqref{DefReality} pseudo-real objects are defined such that 
their indices are equivalently raised and lowered  using $\Omega_{\ja\jb}$ and $\Omega^{\ja\jb}$.

Under its sub-algebra $\mathfrak{usp}(4)$ the algebra $\mathfrak{sl}(5)$ splits 
as ${\bf 24}\rightarrow {\bf 10}+{\bf 14}$
into its compact and non-compact part, respectively. 
The elements $L=L_\Ja{}^\Jb t^{\Ja}{}_{\Jb}$ accordingly decompose as
\begin{align}
   {L_{\ja\jb}}^{\jc\jd} &~=~  
   2 \Lambda^{\vphantom{[\jc]}}_{[\ja}{}_{\vphantom{[\jc]}}^{[\jc} \delta^{\jd]}_{\jb]} 
   + {\Sigma^{\jc\jd}}_{\ja\jb}\;.
   \label{AlgebraSplit1}
\end{align}
The ${\rm SL}(5)$ vector indices $M$ are now represented as antisymmetric,
symplectic traceless index pairs $[\ja\jb]$ of ${\rm USp}(4)$.
In accordance with (\ref{USp4Reps}), $\Lambda$ and $\Sigma$ satisfy
$\Lambda_{[\ja}{}^{\jc}\,\Omega_{\jb]\jc}=0$, $\Sigma^{\ja\jb}{}_{\jc\jb}=0$,
${\Sigma^{\ja\jb}}_{\jc\jd} \, \Omega^{\jc\jd} = 0=\Omega_{\ja\jb} \, {\Sigma^{\ja\jb}}_{\jc\jd}$.
Note that this in particular implies the relation
\begin{align}
   \Omega_{\ja\je}\Omega_{\jb\jf}\Sigma^{\je\jf}{}_{\jc\jd} &= 
   \Omega_{\jc\je}\Omega_{\jd\jf}\Sigma^{\je\jf}{}_{\ja\jb} \; ,
\end{align}
i.e.\ viewed as a $5 \times 5$ matrix $\Sigma$ is symmetric.
In the split~(\ref{AlgebraSplit1}), the commutator (\ref{sl5}) 
between two elements $L_{1}=(\Lambda_{1},\Sigma_{1})$, $L_{2}=(\Lambda_{2},\Sigma_{2})$
takes the form
\begin{align}
   [L_1,L_2] &= L\;,
\end{align}
with $L=(\Lambda,\Sigma)$ according to
\begin{align}
{\Lambda_{\ja}}^\jb =&~ 
{{\Sigma_{1}}^{\jd\je}}_{\ja\jc} \, {{\Sigma_{2}}^{\jb\jc}}_{\jd\je} 
-{{\Sigma_{2}}^{\jd\je}}_{\ja\jc} \, {{\Sigma_{1}}^{\jb\jc}}_{\jd\je} 
+ {\Lambda_{1\;\ja}}^\jc \, {\Lambda_{2\;\jc}}^\jb - {\Lambda_{2\;\ja}}^\jc \, {\Lambda_{1\;\jc}}^\jb
\;,
   \nonumber \\[1ex]
   {{\Sigma}^{\jc\jd}}_{\ja\jb} =&~ 
                             - 2 {{{\Sigma_{1}}^{\je[\jc}}_{\ja\jb}} \, {\Lambda_{2\;\je}}^{\jd]} 
                             + 2 {{{\Sigma_{1}}^{\jc\jd}}_{\je[\ja}} \, {\Lambda_{2\;\jb]}}^\je
                             + 2 {{{\Sigma_{2}}^{\je[\jc}}_{\ja\jb}} \, {\Lambda_{1\;\je}}^{\jd]} 
                             - 2 {{{\Sigma_{2}}^{\jc\jd}}_{\je[\ja}} \, {\Lambda_{1\;\jb]}}^\je
                             \;.
   \label{SplitCom}    
\end{align}
The scalars of the supergravity multiplet parameterize the coset space ${\rm SL}(5)/{\rm SO}(5)$.
They are described by an ${\rm SL}(5)$ valued matrix 
${\cal V}_{\Ja}{}^{\ja\jb}={\cal V}_{\Ja}{}^{[\ja\jb]}$ with 
${\cal V}_{M}{}^{ab}\,\Omega_{ab}=0$.
Infinitesimally, the transformations~(\ref{TrafoCalV}) take the form
\begin{align}
   \delta {{\cal V}_{\Ja}}^{\ja\jb} &= {L_\Ja}^\Jb {{\cal V}_\Jb}^{\ja\jb} 
   +2 {{\cal V}_\Ja}^{\jc[\ja} {\Lambda_\jc}^{\jb]}(x) \;,&
    L \in \mathfrak{sl}(5) \;,\quad  \Lambda(x) \in  \mathfrak{usp}(4) \;.
    \label{TrafoCalV2}
\end{align}
The gauged theory is formally invariant under ${\rm SL}(5)$ transformations
only if the embedding tensor (\ref{X-theta}) is treated as a
spurionic object that simultaneously transforms under ${\rm SL}(5)$. Once $\Theta$ is frozen
to a constant, the theory remains invariant under local 
$G_{0}\times{\rm USp}(4)$ transformations
\begin{align}
   \delta {{\cal V}_{\Ja}}^{\ja\jb} &= 
   g\Lambda^{\Jc\Jd}(x)\,X_{\Jc\Jd,\Ja}{}^{\Jb}\, {{\cal V}_\Jb}^{\ja\jb} 
   +2 {{\cal V}_\Ja}^{\jc[\ja} {\Lambda_\jc}^{\jb]}(x)  \;,
    \label{TrafoCalV3}
\end{align}
parameterized by matrices $\Lambda^{\Ja\Jb}(x)$ and ${\Lambda_\ja}^{\jb}(x)$, respectively.

The inverse of ${\cal V}_{\Ja}{}^{\ja\jb}$ is denoted by ${{\cal V}_{\ja\jb}}^\Ja$, i.e.\
\begin{align}
   {{\cal V}_\Ja}^{\ja\jb}  {{\cal V}_{\ja\jb}}^\Jb &= \delta_\Ja^\Jb\;, &
   {{\cal V}_{\ja\jb}}^\Ja  {{\cal V}_\Ja}^{\jc\jd} 
        &= \delta_{\ja\jb}^{\jc\jd}  - \frac 1 4 \Omega_{\ja\jb} \Omega^{\jc\jd} \; .
\end{align}
Later on we need to consider the variation of ${\cal V}$, 
for example in order to derive field equations from the Lagrangian or to minimize the
scalar potential.
Since ${\cal V}$ is a group element, 
an arbitrary variation can be expressed as a right multiplication 
with an algebra element of ${\rm SL}(5)$
\begin{align*}
   \delta {{\cal V}_\Ja}^{\ja\jb} &= {{\cal V}_\Ja}^{\jc\jd} {L_{\jc\jd}}^{\ja\jb}(x)
   ~=~ {{\cal V}_\Ja}^{\jc\jd}\,{\Sigma^{\ja\jb}}_{\jc\jd}(x)
   - 2{{\cal V}_\Ja}^{\jc[\ja}\,
    \Lambda_{\jc}{}^{\jb]}(x) 
\; .
\end{align*}
Since the last term simply describes a
${\rm USp}(4)$ gauge transformation which 
leaves the Lagrangian invariant
it will be sufficient to consider general variations of the type
\begin{align}
   \delta_{\Sigma} {{\cal V}_\Ja}^{\ja\jb} &= 
    {{\cal V}_\Ja}^{\jc\jd}\,{\Sigma^{\ja\jb}}_{\jc\jd}(x)
   \; .
   \label{SigmaVariations}
\end{align}
The $14$ parameters of $\Sigma$ correspond to variation
along the manifold ${\rm SL}(5)/{\rm SO}(5)$.

Finally, we introduce the scalar currents $P_\mua$ and $Q_\mua$
that describe the gauge covariant space-time derivative of the scalar fields.
Taking values in the Lie algebra $\mathfrak{sl}(5)$ they are defined as
\begin{align}
   {{\cal V}_{\ja\jb}}^\Ja \left( \partial_\mua {{\cal V}_{\Ja}}^{\jc\jd} 
                             - g A_\mua^{\Jc\Jd} {X_{\Jc\Jd,\Ja}}^\Jb {{\cal V}_\Jb}^{\jc\jd} \right)
             &~\equiv~ {P_{\mua\,\ja\jb}}^{\jc\jd} 
             + 2 {Q^{\vphantom{\jd]}}_{\mua\,[\ja}}{}_{\vphantom{\jb]}}^{[\jc} \delta^{\jd]}_{\jb]} 
             \;,
   \label{DefPQ}      
\end{align}
in accordance with the split \eqref{AlgebraSplit1}.
The transformation behavior of these currents is derived directly
from~\eqref{TrafoCalV3} and shows that they
are invariant under local $G_{0}$ transformations.
Under local ${\rm USp}(4)$ transformations \eqref{TrafoCalV2},
${P_{\mua\,\ja\jb}}^{\jc\jd}$ transforms in the ${\bf 14}$, 
while ${Q_{\mua\,\ja}}^{\jb}$
transforms like a ${\rm USp}(4)$ gauge connection 
\begin{align}
   \delta {Q_{\mua\,\ja}}^\jb &= D_\mua {\Lambda_\ja}^\jb 
                             = \nabla_\mua {\Lambda_\ja}^\jb + {Q_{\mua\,\ja}}^\jc {\Lambda_\jc}^\ja
- {Q_{\mua\,\jc}}^\jb {\Lambda_\ja}^\jc
\;.
\end{align}
Thus $Q_\mua$ takes the role of a composite 
gauge field for the local ${\rm USp}(4)$ symmetry
and as such it appears in the covariant derivatives of all objects that transform under ${\rm USp}(4)$,
for example
\begin{align}
   D_\mua \psi^\ja &~=~ \nabla_\mua \psi^\ja - {Q_{\mua\,\jb}}^\ja \psi^\jb \nonumber \\[.5ex]
   D_\mua {P_{\mub\,\ja\jb}}^{\jc\jd}  &~=~ \nabla_\mua {P_{\mub\,\ja\jb}}^{\jc\jd}
                                        + 2 {Q_{\mua\,\je}}^{[\jc} {P_{\mub\,\ja\jb}}^{\jd]\je}
- 2 {Q_{\mua\,[\ja}}^\je {P_{\mub\jb]\je}}^{\jc\jd}  \nonumber \\[.5ex]
   D_\mua {{\cal V}_{\Ja}}^{\ja\jb} &~=~ \nabla_\mua {{\cal V}_{\Ja}}^{\ja\jb} 
+ 2 {Q_{\mua\,\jc}}^{[\ja} {\cal V}_\Ja^{\jb]\jc} 
- g A_\mua^{\Jc\Jd} {X_{\Jc\Jd\Ja}}^\Jb {{\cal V}_\Jb}^{\ja\jb}
~=~{{\cal V}_{\Ja}}^{\jc\jd}\,{P_{\mua\,\jc\jd}}^{\ja\jb}\;,
\end{align}
where $\psi^\ja$ is an arbitrary object in the fundamental representation of ${\rm USp}(4)$.

\subsection{The $T$-tensor}

All bosonic fields of the theory come in representations of 
${\rm SL}(5)$ while all fermionic fields come in representations of ${\rm USp}(4)$.
The object mediating between them is the scalar matrix ${\cal V}_{\Ja}{}^{\ja\jb}$.
e.g.\ it is convenient to define the ${\rm USp}(4)$ covariant
field strengths
\begin{align}
   {\cal H}^{(2)\ja\jb}_{\mua\mub} &\equiv \sqrt{2} \, \Omega_{\jc\jd} \, 
   {\cal V}_{\Ja}{}^{\ja\jc} {\cal V}_{\Jb}{}^{\jb\jd} \,
                                              {\cal H}^{(2)\Ja\Jb}_{\mua\mub} \;, &
   {\cal H}^{(3)}_{\mua\mub\muc\,\ja\jb} &\equiv  {{\cal V}_{\ja\jb}}^\Ja \, 
   {\cal H}^{(3)}_{\mua\mub\muc\,\Ja} \, ,
\end{align}
which naturally couple to the fermion fields.
More generally, the scalar matrix ${\cal V}_{\Ja}{}^{\ja\jb}$
maps tensors $R_{\Ja}$ and $S^{\Ja}$ in the 
${\rm SL}(5)$ representations ${\bf 5}$ and ${\bf\overline{5}}$,
respectively, into (scalar field dependent) tensors 
$R_{[\ja\jb]}$, $S^{[\ja\jb]}$ in the ${\bf 5}$ of ${\rm USp}(4)$ as
\begin{align}
   R_{[\ja\jb]} &=  {\cal V}_{\ja\jb}{}^{\Ja} R_{\Ja}\;,
   \qquad
      S^{[\ja\jb]} =  {\cal V}_{\Ja}{}^{\ja\jb} \,S^{\Ja}\;.
\end{align}
Similarly, tensors $R_{\Ja\Jb}$, $S^{\Ja\Jb}$ in the 
${\rm SL}(5)$ representations  ${\bf 10}$
and ${\bf \overline{10}}$, respectively, give rise to 
(scalar field dependent) tensors $R_{(\ja\jb)}$,
$S^{(\ja\jb)}$
in the ${\bf 10}$ of ${\rm USp}(4)$ as follows
\begin{align}
   R_{\ja\jb} &= \sqrt{2} \, \Omega^{\jc\jd} \, 
   {\cal V}_{\ja\jc}{}^{\Ja} {\cal V}_{\jb\jd}{}^{\Jb} \, R_{\Ja\Jb} &
   \Leftrightarrow &&
   R_{\Ja\Jb} &= - \sqrt{2} \, {\cal V}_{\Ja}{}^{\ja\jb} {\cal V}_{\Jb}{}^{\jc\jd} \, 
                            \delta^\je_{[\ja} \Omega^{\phantom{\ja}}_{\jb][\jc} \delta^\jf_{\jd]} \, R_{\je\jf} 
\;,
\nonumber \\[1ex]
   S^{\ja\jb} &= \sqrt{2} \Omega_{\jc\jd} 
   {\cal V}_{\Ja}{}^{\ja\jc} {\cal V}_{\Jb}{}^{\jb\jd} S^{\Ja\Jb} &
   \Leftrightarrow &&
   S^{\Ja\Jb} &= - \sqrt{2} \, {\cal V}_{\ja\jb}{}^{\Ja} {\cal V}_{\jc\jd}{}^{\Jb} \, 
                            \delta_\je^{[\ja} \Omega_{\phantom{\ja}}^{\jb][\jc} \delta_\jf^{\jd]} \, S^{\je\jf} \;,
\end{align}
where the normalization is chosen such that $R_{\ja\jb} S^{\ja\jb} = R_{\Ja\Jb} S^{\Ja\Jb}$.

Applying the analogous map to the embedding tensor $\Theta_{\Ja\Jb,\Jc}{}^{\Jd}$
(\ref{linear}) leads to the $T$-tensor~\cite{deWit:2002vt}
\begin{align}
  T_{(\je\jf)\,[\ja\jb]}{}^{[\jc\jd]} &\equiv   \sqrt{2}\,
  {\cal V}^{\Ja}{}_{\je\jg}{\cal V}^{\Jb}{}_{\jf\jh}\,\Omega^{\jg\jh}\,
                                              {\cal V}^{\Jc}{}_{\ja\jb} \,
                                              \Theta_{\Ja\Jb,\Jc}{}^{\Jd}\,{\cal V}_{\Jd}{}^{\jc\jd} 
\nonumber \\[1ex]
                          & =   \sqrt{2}\,\Omega_{\phantom{(\je}}^{\jh[\jc}\,{\delta}^{\jd]}_{(\je}
\,{\cal V}^{\Ja}{}^{\phantom{\jd]}}_{\jf)\jh}\,
                                {\cal V}^{\Jb}{}_{\ja\jb}\,Y^{\phantom{\ji}}_{\Ja\Jb} 
\nonumber \\ &\qquad \qquad \qquad
                           -2\sqrt{2}\,\epsilon_{\Ja\Jb\Jc\Jd\Je}\,Z^{\Jc\Jd,\Jf}\,
                           {\cal V}^{\Ja}{}_{\je\jg}{\cal V}^{\Jb}{}_{\jf\jh}\,
                             {\cal V}^{\Je}{}_{ab}\,{\cal V}_{\Jf}{}^{cd}\,\Omega^{gh}  \; .
\label{TT}
\end{align}
We shall see in the next section, that this tensor 
encodes the fermionic mass matrices as well as the scalar potential
of the Lagrangian.
This has first been observed for the $T$-tensor in the maximal $D=4$ 
supergravity~\cite{deWit:1982ig}.

Recall that the components  $Y_{\Ja\Jb}$ and $Z^{\Ja\Jb,\Jc}$ of $\Theta$
transform in the ${\bf 15}$ and the $\overline{\bf 40}$ of ${\rm SL}(5)$, respectively.
Under ${\rm USp}(4)$ they decompose as
\begin{align}
{\bf 15}+{\bf\overline{40}} & ~\rightarrow~ ({\bf 1}+{\bf 14})+  ({\bf 5}+{\bf 35}) \;.
\end{align}
Accordingly, the $T$-tensor can be decomposed into its four ${\rm USp}(4)$ irreducible components
that we denote by $B$, ${B^{[\ja\jb]}}_{[\jc\jd]}$, $C_{[\ja\jb]}$, and ${C^{[\ja\jb]}}_{(\jc\jd)}$,
respectively, with index structures according to~(\ref{USp4Reps}). This yields
\begin{align}
  T_{(\je\jf)\,\ja\jb}{}^{\jc\jd}
  =&~
  \ft12 B\,\Omega^{\vphantom{\jb}}_{\ja(\je}\,\delta_{\jf)}^{[\jc} \delta^{\jd]}_{\jb}
  - \ft12 B\,\Omega^{\vphantom{\jb}}_{\jb(\je}\,\delta_{\jf)}^{[\jc} \delta^{\jd]}_{\ja}
  + \delta_{(\je}^{[\jc}\,\Omega^{\vphantom{\jb}}_{\jf)\jg}\,B^{\jd]\jg}{}_{\ja\jb}
  \nonumber\\[1ex]
  &
  + \ft12 \, C^{\vphantom{\jb}}_{\ja(\je}\,\delta_{\jf)}^{[\jc} \delta^{\jd]}_{\jb}
  - \ft12 \,C^{\vphantom{\jb}}_{\jb(\je}\,\delta_{\jf)}^{[\jc} \delta^{\jd]}_{\ja}
  -\ft18 \Omega^{\jc\jd}\,C^{\vphantom{\jb}}_{\ja(\je}\,\Omega_{\jf)\jb}
  +\ft18 \Omega^{\jc\jd}\,C^{\vphantom{\jb}}_{\jb(\je}\,\Omega_{\jf)\ja}
  \nonumber\\[1ex]
  &
  + \ft14\Omega_{\ja\jb}\,C^{\vphantom{\jb}}_{\jg(\je}\,\delta_{\jf)}^{[\jc}\,
  \Omega^{\jd]\jg}_{\phantom{\jf}}
  + \ft12 \Omega_{\je[\ja}\,C^{\jc\jd}{}_{\jb]\jf}
  + \ft12 \Omega_{\jf[\ja}\,C^{\jc\jd}{}_{\jb]\je}
  + \ft14 \Omega_{\ja\jb}\,C^{\jc\jd}{}_{\je\jf}
  \;.
  \label{TABCD}
\end{align}
In appendices~\ref{app:tensors}, \ref{app:TQC} we present a more systematic account to these
decompositions in terms of ${\rm USp}(4)$ projection operators
which simplify the calculations.
In particular, the parameterization~(\ref{TABCD})
takes the compact form~$\eqref{TTwithTaus}$. 

For the components $Y_{\Ja\Jb}$ and $Z^{\Ja\Jb,\Jc}$ the parameterization~(\ref{TABCD}) 
yields explicitly
\begin{align}
   Y_{\Ja\Jb} &={{\cal V}_\Ja}^{\ja\jb} {{\cal V}_\Jb}^{\jc\jd}\,Y_{\ja\jb,\jc\jd}\;,
   \qquad 
   Z^{\Ja\Jb,\Jc} = 
   \sqrt{2} {{\cal V}_{\ja\jb}}^\Ja {{\cal V}_{\jc\jd}}^\Jb {{\cal V}_{\je\jf}}^\Jc \Omega^{\jb\jd}
                                      Z^{(\ja\jc)[\je\jf]}\;, \nonumber\\[2ex]
   & \begin{array}{lrl}
     \text{with} &  \qquad 
 Y_{\ja\jb,\jc\jd}&=  \frac 1 {\sqrt{2}}  
                 \Big[  ( \Omega_{\ja\jc} \Omega_{\jb\jd} - \ft 1 4 \Omega_{\ja\jb} \Omega_{\jc\jd} )\,B
+ \Omega_{\ja\je} \Omega_{\jb\jf} {B^{[\je\jf]}}_{[\jc\jd]} \Big]\;, \\[0.4cm]
 &  Z^{(\ja\jb)[\jc\jd]} &=\frac 1 {16} \Omega^{\ja[\jc} C^{\jd]\jb} 
                              + \frac 1 {16} \Omega^{\jb[\jc} C^{\jd]\ja} 
                     - \frac 1 8 \Omega^{\ja\je} \Omega^{\jb\jf} {C^{\jc\jd}}_{\je\jf}
                     \;,
   \end{array}	
  \nonumber \\[-0.65cm]
   \phantom{A}	     
\label{YZABCD}
\end{align}
where $C^{\ja\jb} = \Omega^{\ja\jc} \Omega^{\jb\jd} C_{\jc\jd}$.
Note that $\Theta$ and thus $Y_{\Ja\Jb}$ and $Z^{\Ja\Jb,\Jc}$ are constant matrices.
In contrast, the $T$-tensor and thus the tensors $B$, $C$ are functions of the scalar fields. 
It is useful to give also the inverse relations
\begin{align}
   B &~= ~
   \frac {\sqrt{2}} 5 \Omega^{\ja\jc} \Omega^{\jb\jd}  Y_{\ja\jb,\jc\jd}
       \;,
\nonumber\\[1ex]
      {B^{\ja\jb}}_{\jc\jd} &~=~ 
      \sqrt{2}\, \Big[
    \Omega^{\ja\je} \Omega^{\jb\jf} \delta^{\jg\jh}_{\jc\jd}
            -\ft15 \left( \delta^{\ja\jb}_{\jc\jd} - \ft 1 4 \Omega^{\ja\jb} \Omega_{\jc\jd} \right)
\Omega^{\je\jg} \Omega^{\jf\jh}   \Big]  \,
Y_{\je\jf,\jg\jh}
             \;,
       \nonumber \\[1ex]
   C^{\ja\jb} &~=~ 8 \,\Omega_{\jc\jd}\,Z^{(\ja\jc)[\jb\jd]}\;, \nonumber \\[1ex]
   {C^{\ja\jb}}_{\jc\jd} 
        &~=~   8 \left( - \Omega_{\jc\je}\Omega_{\jd\jf} \delta^{\ja\jb}_{\jg\jh}  \, 
               + \, \Omega^{\phantom{\ja}}_{\jg(\jc} \delta^{\ja\jb}_{\jd)\je} \Omega_{\jf\jh}  \right)
Z^{(\je\jf)[\jg\jh]}
\;.   
\label{BC}
\end{align}
Under the variation \eqref{SigmaVariations} 
of the scalar fields, these tensors transform as
\bea
\delta_\Sigma \, B &=& -\ft25\, \Sigma^{\ja\jb}{}_{\jc\jd}\,B^{\jc\jd}{}_{\ja\jb} 
\;,
\nonumber\\[1ex]
\delta_\Sigma \, B^{\ja\jb}{}_{\jc\jd} &=&
-2\,B\,\Sigma^{\ja\jb}{}_{\jc\jd}
-
\Sigma^{\ja\jb}{}_{\jg\jh}\,B^{\jg\jh}{}_{\jc\jd}
-\Sigma^{\jg\jh}{}_{\jc\jd}\,B^{\ja\jb}{}_{\jg\jh}+
\ft25
(\delta^{\ja\jb}_{\jc\jd}-\ft14\Omega^{\ja\jb}\Omega_{\jc\jd}) 
             \Sigma^{\je\jf}{}_{\jg\jh}\,B^{\jg\jh}{}_{\je\jf}
\;,
\nonumber\\[1ex]
\delta_\Sigma \, C^{\ja\jb} &=& 
   \ft12\, \Sigma^{\ja\jb}{}_{\jc\jd}\,C^{\jc\jd} 
    +2\, \Omega^{\je[\ja} \Sigma^{\jb]\jf}{}_{\jc\jd}\,C^{\jc\jd}{}_{\je\jf} 
\;,
\nonumber\\[1ex]
\delta_\Sigma \, C^{\ja\jb}{}_{\jc\jd} &=&
    4\,\Omega^{\jg[\ja}\,\Sigma^{\jb]\jh}{}_{\jg(\jc}\, C_{\jd)\jh}
   + \Omega^{\jg[\ja}_{\phantom{(\jc}}\,\delta^{\jb]}_{(\jc}\,\Sigma^{\jk\jh}{}_{\jd)\jg}\, C_{\jk\jh}
    +\Omega^{\jg\jk}\,\delta^{[\ja}_{(\jc}\,\Sigma^{\jb]\jh}{}_{\jd)\jg}\, C_{\jk\jh}
\nonumber\\[.5ex]
&&{}
+
\Sigma^{\ja\jb}{}_{\jg\jh}\,C^{\jg\jh}{}_{\jc\jd}
+\Sigma^{\jk[\ja}{}_{\jg\jh}\,\delta^{\jb]}_{(\jc}\,C^{\jg\jh}{}_{\jd)\jk}
\nonumber\\[.5ex]
&&
{}
+4\,\Sigma^{\jk\jm}{}_{\jl(\jc}\,\Omega_{\jd)\jk}\,\Omega^{\jn[\ja}\,C^{\jb]\jl}{}_{\jm\jn}
- \delta^{[\ja}_{(\jc}\,\Omega^{\phantom{\ja}}_{\jd)\jk}\,\Omega^{\jb]\jn}\,
\Sigma^{\jk\jm}{}_{\jl\jg}\,C^{\jg\jl}{}_{\jm\jn}
\;.
\label{varABCD}
\eea
These variations will be relevant in the next section, 
since  in the Lagrangian the tensors $B$, $C$ appear
in the fermionic mass matrices and in the scalar potential.
Furthermore, one derives from~\eqref{varABCD} the expressions for
the ${\rm USp}(4)$ covariant derivatives of these tensors
\bea
D_{\mu}  B &=& -\ft25\, P_{\mu\,\jc\jd}{}^{\ja\jb} B^{\jc\jd}{}_{\ja\jb} 
\;,
\nonumber\\[1ex]
D_{\mu}  B^{\ja\jb}{}_{\jc\jd} &=&
-2BP_{\mu\,\jc\jd}{}^{\ja\jb}
-
P_{\mu\,\jg\jh}{}^{\ja\jb}\!B^{\jg\jh}{}_{\jc\jd}
-P_{\mu\,\jc\jd}{}^{\jg\jh}\!B^{\ja\jb}{}_{\jg\jh}+
\ft25
(\delta^{\ja\jb}_{\jc\jd} -\ft14\Omega^{\ja\jb}\Omega_{\jc\jd} ) 
             P_{\mu\,\jg\jh}{}^{\je\jf}\!B^{\jg\jh}{}_{\je\jf}
\;,
\nonumber\\[1ex]
D_{\mu}  C^{\ja\jb} &=& 
   \ft12\, P_{\mu\,\jc\jd}{}^{\ja\jb}A^{\jc\jd} 
    +2\, \Omega^{\je[\ja} P_{\mu\,\jc\jd}{}^{\jb]\jf}
 C^{\jc\jd}{}_{\je\jf} 
\;,
\nonumber\\[1ex]
D_{\mu} C^{\ja\jb}{}_{\jc\jd} &=&
    4\,\Omega^{\jg[\ja}\, P_{\mu\,\jg(\jc}{}^{\jb]\jh} C_{\jd)\jh}
   + \Omega^{\jg[\ja}_{\phantom{(\jc}}\,\delta^{\jb]}_{(\jc}\,
   P_{\mu\,\jd)\jg}{}^{\jk\jh}  C_{\jk\jh}
    +\Omega^{\jg\jk}\,\delta^{[\ja}_{(\jc}\,P_{\mu\,\jd)\jg}{}^{\jb]\jh} C_{\jk\jh}
\nonumber\\[.5ex]
&&{}
+
P_{\mu\,\jg\jh}{}^{\ja\jb} C^{\jg\jh}{}_{\jc\jd}
+
P_{\mu\,\jg\jh}{}^{\jk[\ja}\delta^{\jb]}_{(\jc}\,C^{\jg\jh}{}_{\jd)\jk}
\nonumber\\[.5ex]
&&
{}
+4\,P_{\mu\,\jl(\jc}{}^{\jk\jm}\Omega_{\jd)\jk}\,\Omega^{\jn[\ja}\,C^{\jb]\jl}{}_{\jm\jn}
- \delta^{[\ja}_{(\jc}\,\Omega^{\phantom{\ja}}_{\jd)\jk}\,\Omega^{\jb]\jn}\,
P_{\mu\,\jl\jg}{}^{\jk\jm}C^{\jg\jl}{}_{\jm\jn}
\;.
\eea
Since the $T$-tensor~(\ref{TT}) is obtained
by a finite ${\rm SL}(5)$-transformation
from the embedding tensor~(\ref{linear}),
the ${\rm SL}(5)$-covariant quadratic constraints~(\ref{quadratic})
directly translate into quadratic relations 
among the tensors $B$, $C$.
e.g.\ the first equation of~(\ref{Q2})
gives rise to
\bea
Z^{(\ja\jb)[\je\jf]}\,
\Big[ 
\Omega_{\jc\je} \Omega_{\jd\jf} \,B
+ \Omega_{\je\jg} \Omega_{\jf\jh} {B^{[\jg\jh]}}_{[\jc\jd]} \Big]
&=& 0
\;,
\label{QZB1}
\eea
while the second equation yields
\bea
Z^{(\ja\jb)[\jc\jd]}\,T_{(\ja\jb)\,\je\jf}{}^{\jg\jh} &=& 0
\;.
\label{QZB2}
\eea
These equations can be further expanded into explicit
quadratic relations among the tensors $B$, $C$.
We give the explicit formulas in terms of ${\rm USp}(4)$ projectors
in appendix~\ref{app:TQC}. They are crucial to verify the
invariance of the Lagrangian~\eqref{L} presented in the next section.

Let us close this section by noting that the $T$-tensor~(\ref{TT})
naturally appears in the deformation of the Cartan-Maurer equations
induced by the gauging. Namely, the definition of the 
currents~$P_\mua$ and $Q_\mua$~\eqref{DefPQ} together with the 
algebra structure~(\ref{SplitCom}) gives rise to the following
integrability relations
\begin{align}
 2 \partial_{[\mua} {Q_{\mub]\ja}}^\jb + 2 {Q_{\ja[\mua}}^\jc {Q_{\mub]\jc}}^\jb
                             &= - 2 {P_{\ja\jc[\mua}}^{\jd\je} {P_{\mub]\jd\je}}^{\jb\jc}
                             - g \, {\cal H}_{\mua\mub}^{(2)\,\jc\jd} \, {T_{(\jc\jd)[\ja\je]}}^{[\jb\je]} 
                             \;,\label{Cartan-Maurer}
\\[1ex]
   D_{[\mua} {P_{\mub]\ja\jb}{}^{\jc\jd}} &= 
   - \; \frac 1 4 g \, {\cal H}_{\mua\mub}^{(2)\,\je\jf} 
                           \left( {T_{(\je\jf)[\ja\jb]}}^{[\jc\jd]} + 
        \Omega^{\jc\jg} \Omega^{\jd\jh} \Omega_{\ja\ji} \Omega_{\jb\jj} \; 
        {T_{(\je\jf)[\jg\jh]}}^{[\ji\jj]} \right) \; .
\nonumber
\end{align}
The terms in order $g$ occur proportional to the 
$T$-tensor. They will play an important role in the check of
supersymmetry of the Lagrangian that we present in the next section.
The fact that these equations appear manifestly
covariant with the full modified field strength ${\cal H}_{\mua\mub}^{(2)\,\jc\jd}$
on the r.h.s.\ is a consequence of 
the quadratic constraint~(\ref{QZB2}).

\section{Lagrangian and supersymmetry}
\label{SecLagrSUSY}

In this section we present the main results of this chapter. After establishing
our spinor conventions, we derive the supersymmetry transformations
of the seven-dimensional theory by requiring closure of
the supersymmetry algebra into the generalized
vector/tensor gauge transformations introduced in section~\ref{SecVecTen}.
We then present the universal Lagrangian of the maximal seven-dimensional theory
which is completely encoded in the embedding tensor~$\Theta$.

\subsection{Spinor conventions}
\label{SubSecSpinor}

Seven-dimensional world and tangent-space indices are denoted by
$\mu,\nu,\ldots$ and $m,n,\ldots$, respectively, and take the values
$1,2,\ldots,7$. Our conventions for the $\Gamma$-matrices in seven dimensions are
\begin{align}
   \left\{  \Gamma^{\ma}, \Gamma^{\mb}  \right\} &= 2 \eta^{\ma\mb} &
   (\Gamma^\ma)^\dag &= \Gamma_\ma \; ,&(\Gamma^\ma)^T  &=-C \Gamma^\ma C^{-1} 
\end{align}
with metric of signature $\eta=\text{diag}(-1,1,1,1,1,1,1) $ 
and the charge conjugation matrix $C$ obeying   
\begin{align}   
   C &= C^T = - C^{-1} = - C^\dag \; .
\end{align}
We use symplectic Majorana spinors, i.e.\
spinors carry  a fermionic representation of the $R$-symmetry group ${\rm USp}(4)$
and for instance a spinor  $\psi^\ja$ ($\ja=1, \ldots, 4$) in the fundamental representation 
of ${\rm USp}(4)$ satisfies a reality constraint of the form
\begin{align}
   {\bar \psi_{\ja}}^T &= \Omega_{\ja\jb} C \,\psi^\jb \;, 
   \label{symM}  
\end{align}
where $\bar \psi\equiv\psi^\dag \Gamma^0$.
The following formula is useful as it captures the symmetry property of spinor 
products\footnote{Note that our conventions differ from those of \cite{Sezgin:1982gi}
in that they use ${\phi}_\ja=\Omega_{\ja\jb}{\phi}^\jb$,  while 
in our conventions raising and lowering of indices is
effected by complex conjugation ${\phi}_\ja = ({\phi}^\ja)^*$.}
\begin{align}
   \bar \phi_\ja \Gamma^{(k)} \psi^\jb &= 
   \Omega_{\ja\jc} \Omega^{\jb\jd} \bar \psi_\jd (C^{-1})^T (\Gamma^{(k)})^T C \phi^\jc 
   ~=~ (-1)^{\frac12k(k+1)}\,\Omega_{\ja\jc} \Omega^{\jb\jd} \bar \psi_\jd\Gamma^{(k)} \phi^\jc  \; .
\end{align}
Products of symplectic Majorana spinors yield real tensors
\begin{align}
   \bar \phi_\ja \psi^\ja &&
   \bar \phi_\ja \Gamma^\mua \psi^\ja &&
   \bar \phi_\ja \Gamma^{\mua\mub} \psi^\ja &&
   \bar \phi_\ja \Gamma^{\mua\mub\muc} \psi^\ja &&
   \text{etc.}
\end{align}     
Finally, the epsilon tensor is defined by
\begin{align}
   e\,\Gamma^{\mu\nu\rho\sigma\tau\kappa\lambda} &\equiv \mathbbm{1} \; 
   \epsilon^{\mu\nu\rho\sigma\tau\kappa\lambda} \; .
\end{align}

\subsection{Supersymmetry transformations and algebra}

The field content of the ungauged maximal supergravity multiplet in seven dimensions is
given by the vielbein ${e_\mua}^\ma$, the gravitino $\psi^\ja_\mua$,
vector fields $A_\mua^{\Ja\Jb}$, two-form fields $B_{\Ja \, \mua\mub}$, 
matter fermions $\chi^{\ja\jb\jc}$, and scalar fields parameterizing 
${{\cal V}_\Ja}^{\ja\jb}$.
Their on-shell degrees of freedom are summarized in Table~\ref{TabMultiplet}.
Note the symmetry in the distribution of degrees of freedom
due to the accidental coincidence of the $R$-symmetry group 
${\rm USp}(4)$ and the little group ${\rm SO}(5)$.

\begin{table}[tb]
   \begin{center}
     \begin{tabular}{r|cccccc}
        fields & ${e_\mua}^\ma$ & $\psi^\ja_\mua$ & $A_\mua^{\Ja\Jb}$ & $B_{\mua\mub\,\Ja}$ 
& $\chi^{\ja\jb\jc}$ & ${{\cal V}_\Ja}^{\ja\jb}$
\\ \hline
little group ${\rm SO}(5)$    & {\bf 14} & {\bf 16} & {\bf 5}  & {\bf 10} & {\bf 4} & {\bf 1} \\
$R$-symmetry ${\rm USp}(4)$ & {\bf 1}  & {\bf 4}  
& {\bf 1} & {\bf 1}  & {\bf 16} & {\bf 5}
\\
global ${\rm SL}(5)$ & {\bf 1}  & {\bf 1}  
& ${\bf \overline{10}}$ & {\bf 5}  & {\bf 1} & {\bf 5}
        \\ \hline
        \# degrees of freedom & 14 & 64 & 50 & 50 & 64 & 14 
     \end{tabular}
     \caption{\label{TabMultiplet}{ \small The ungauged $D=7$ maximal super-multiplet.}}
   \end{center}     
\end{table}

Under the $R$-symmetry group ${\rm USp}(4)$ 
the gravitinos $\psi_{\mu}^{\ja}$ 
transform in the fundamental representation ${\bf 4}$
while the matter spinors $\chi^{\ja\jb\jc}$ 
transform in the ${\bf 16}$ representation, i.e.\
\begin{align}
   \chi^{\ja\jb\jc} &= \chi^{[\ja\jb]\jc}\;, &
   \Omega_{\ja\jb} \chi^{\ja\jb\jc} &= 0\;, &
   \chi^{[\ja\jb\jc]} &= 0 \; .
\end{align}
All spinors are symplectic Majorana, that is they satisfy
\begin{align}   
   {\bar \chi_{\ja\jb\jc}}^T &= \Omega_{\ja\jd} \Omega_{\jb\je} \Omega_{\jc\jf} C \chi^{\jd\je\jf}\;, &
   {\bar \psi_{\mua\ja}}^T &= \Omega_{\ja\jb} C \psi_\mua^\jb \; ,
\end{align}
in accordance with~(\ref{symM}).

We are now in position to derive the supersymmetry transformations.
Parameterizing them by $\epsilon^\ja=\epsilon^\ja(x)$ the final result takes the form
\begin{align}
   \delta {e_\mua}^\ma &= \frac 1 2 \bar \epsilon_\ja \Gamma^\ma \psi^\ja_\mua 
   \;,
   \nonumber \\[1ex]
   \delta {{\cal V}_\Ja}^{\ja\jb} &= \frac{1}{4} {{\cal V}_\Ja}^{\jc\jd} 
      \Big( \Omega_{\je[\jc} \bar \epsilon_{\jd]} \chi^{\ja\jb\je} 
            + \frac 1 4 \Omega_{\jc\jd} \bar \epsilon_\je \chi^{\ja\jb\je} 
+ \Omega_{\jc\je} \Omega_{\jd\jf} \bar \epsilon_\jg \chi^{\je\jf[\ja} \Omega^{\jb]\jg}
+ \frac 1 4 \Omega_{\jc\je} \Omega_{\jd\jf} \Omega^{\ja\jb} \bar \epsilon_\jg \chi^{\je\jf\jg} \Big) 
\;,
\nonumber \\[1ex]
   \Delta A_\mua^{\Ja\Jb} &= - {{\cal V}_{\ja\jb}}^{[\Ja} {{\cal V}_{\jc\jd}}^{\Jb]} \Omega^{\jb\jd} 
                             \Big( \frac{1}{2} \Omega^{\ja\je} \bar \epsilon_\je \psi_\mua^\jc
+\frac{1}{4} \bar \epsilon_\je \Gamma_\mua \chi^{\je\ja\jc}  \Big)
\;,
\nonumber  \\[1ex]
   \Delta B_{\mua\mub\,\Ja} &= {{\cal V}_{\Ja}}^{\ja\jb}
                             \Big( - \Omega_{\ja\jc} \bar \epsilon_\jb \Gamma_{[\mua} \psi^\jc_{\mub]}
+\frac{1}{8}  \Omega_{\ja\jc} \Omega_{\jb\jd} \bar \epsilon_\je \Gamma_{\mua\mub} \chi^{\jc\jd\je} \Big) 
\;,
\nonumber  \\[1ex]
   \Delta S_{\mua\mub\muc}^\Ja &= {{\cal V}_{\ja\jb}}^{\Ja} 
                           \Big( -\frac{3}{8} \Omega^{\ja\jc} \bar \epsilon_\jc \Gamma_{[\mua\mub} \psi^\jb_{\muc]}
-\frac{1}{32} \bar \epsilon_\je \Gamma_{\mua\mub\muc} \chi^{\ja\jb\je} \Big)  
\;,
\nonumber \\[1ex]
   \delta \psi_\mua^\ja &= D_\mua \epsilon^\ja 
            -\frac{1}{5 {\sqrt{2}}} {\cal H}^{(2)(\ja\jb)}_{\mub\muc} \Omega_{\jb\jc} 
\Big( {\Gamma^{\mub\muc}}_\mua + 8 \Gamma^\mub \delta^\muc_\mua \Big) \epsilon^\jc
            \nonumber \\ & \qquad 
            -\frac{1}{15} {\cal H}^{(3)}_{\mub\muc\mud[\jb\jc]} \Omega^{\ja\jb}     
\Big( {\Gamma^{\mub\muc\mud}}_\mua + \frac 9 2 \Gamma^{\mub\muc} \delta^\mud_\mua \Big) \epsilon^\jc
            - g \Gamma_\mua A_1^{\ja\jb} \Omega_{\jb\jc} \epsilon^\jc 
\;,
\nonumber \\[1ex]
   \delta \chi^{\ja\jb\jc} &= 2 \Omega^{\jc\jd} {P_{\mua\jd\je}}^{\ja\jb} \Gamma^\mua \epsilon^\je
               -\sqrt{2} \Big( {\cal H}_{\mua\mub}^{(2)\jc[\ja} \Gamma^{\mua\mub} \epsilon^{\jb]}
                         - \frac 1 5 \, 
                         (\Omega^{\ja\jb}\delta_{\jg}^{\jc}- \Omega^{\jc[\ja}\delta_{\jg}^{\jb]})\,
                         \Omega_{\jd\je}{\cal H}_{\mua\mub}^{(2)\jg\jd} \Gamma^{\mua\mub} \epsilon^\je
                         \Big)
            \nonumber \\ & \qquad 
           -\frac{1}{6} \Big( \Omega^{\ja\jd} \Omega^{\jb\je} {\cal H}^{(3)}_{\mua\mub\muc[\jd\je]} \Gamma^{\mua\mub\muc} \epsilon^{\jc}
                - \frac 1 5 (\Omega^{\ja\jb} \Omega^{\jc\jf}+4\Omega^{\jc[\ja} \Omega^{\jb]\jf})\,
                 {\cal H}^{(3)}_{\mua\mub\muc[\jf\je]} \Gamma^{\mua\mub\muc} \epsilon^\je
                 \Big)
      \nonumber  \\[1ex]
  &\qquad   + g A_2^{\jd,\ja\jb\jc} \Omega_{\jd\je} \epsilon^\je
   \;,
   \label{SUSYrules}
\end{align}
up to higher order fermion terms.
We have given the result in terms of the covariant variations $\Delta({\epsilon})$ of the vector
and tensor fields introduced in \eqref{covV}, from which the bare 
transformations $\delta({\epsilon})$ are readily deduced.
In the limit $g \rightarrow 0$ the above supersymmetry transformations reduce
to those of the ungauged theory~\cite{Sezgin:1982gi}.
Upon switching on the gauging, the formulas are covariantized
and the fermion transformations are modified by
the fermion shift matrices $A_1$ and $A_2$ defined by
\begin{align}
   A_1^{\ja\jb} &\equiv - \frac{1}{4 {\sqrt{2}}} 
   \Big( \frac 1 4 B \Omega^{\ja\jb} + \frac{1}{5} C^{\ja\jb} \Big) 
\;,
     \nonumber \\
   A_2^{\jd,\ja\jb\jc} &\equiv \frac{1}{2 {\sqrt{2}}} 
   \left[\Omega^{\je\jc} \Omega^{\jf\jd}\,({C^{\ja\jb}}_{\je\jf}- {B^{\ja\jb}}_{\je\jf})
+ \frac{1}{4} \Big( C^{\ja\jb} \Omega^{\jc\jd}
+ \frac 1 5 \Omega^{\ja\jb} C^{\jc\jd} 
+ \frac 4 5 \Omega^{\jc[\ja} C^{\jb]\jd} \Big) \right]   \; ,
   \label{FermionShifts}       
\end{align}
in terms of the components of the $T$-tensor~(\ref{TABCD}).
These will further enter the fermionic mass matrices
and the scalar potential of the full Lagrangian~(\ref{L}) below.
The coefficients in~(\ref{SUSYrules})
are uniquely fixed by requiring the closure of the
supersymmetry algebra
into diffeomorphisms, local Lorentz and ${\rm USp}(4)$-transformations,
and vector/tensor gauge transformations~(\ref{gauge}).
In particular, the fermion shifts~\eqref{FermionShifts}
are uniquely determined such that the 
commutator of two supersymmetry transformations
reproduces the correct order $g$ shift terms 
in the resulting vector/tensor gauge transformations~(\ref{gauge}).
Specifically, one finds for the commutator of two supersymmetry transformations
\begin{align}
   [ \delta(\epsilon_1), \delta(\epsilon_2) ] &=  \xi^\mua D_\mua  
                                           + \delta_{\text{Lorentz}}  \left(\epsilon^{\ma\mb} \right)
+ \delta_{{\rm USp}(4)} \left( {\kappa_\ja}^\jb \right)
+ \delta_{\text{gauge}}   \Big( \Lambda^{\Ja\Jb}, \Xi_{\Ja\mua}, \Phi^\Ja_{\mua\mub} \Big) \; .
   \label{ComSUSYalgebra}      
\end{align}
Here, we denote by $\xi^\mua D_\mua$ a covariant general coordinate
transformation with parameter $\xi^\mua$, i.e.\
\begin{align}
   \xi^\mua D_\mua &= {\cal L}_\xi + \delta_{\text{Lorentz}}  \left( \hat \epsilon{}^{\;\ma\mb} \right)
+ \delta_{{\rm USp}(4)} \left( {{\hat \kappa}_\ja}{}^\jb \right)
+ \delta_{\text{gauge}}   \left( \hat \Lambda^{\Ja\Jb}, \hat \Xi_{\Ja\mua}, \hat \Phi^\Ja_{\mua\mub} \right) 
\; ,
\end{align}
with the induced parameters
\begin{align}
   \hat \epsilon{}^{\;\ma\mb} &= - \xi^\mua {\omega_\mua}^{\ma\mb} \;,\nonumber \\
   {{\hat \kappa}_\ja}{}^\jb &= - \xi^\mua {Q_{\mua\ja}}^\jb\;, \nonumber \\
   \hat \Lambda^{\Ja\Jb} &= - \xi^\mua A_\mua^{\Ja\Jb} \;,\nonumber \\
   \hat \Xi_{\Ja\mua} &= - \xi^\mub B_{\Ja\,\mub\mua}
   -  \epsilon_{\Ja\Jb\Jc\Jd\Je}\xi^\mub A_\mub^{\Jb\Jc}A_\mua^{\Jd\Je}\;,\nonumber \\
   \hat \Phi^\Ja_{\mua\mub} &= - \xi^\muc S^\Ja_{\muc\mua\mub}
   -\xi^{\muc} A_{\muc}^{\Ja\Jb} B_{\mua\mub\,\Jb}
   -\frac23\epsilon_{\Jb\Jc\Jd\Je\Jf}\,\xi^{\muc} A_{\muc}^{\Jb\Jc}
   A_{[\mua}^{\Ja\Jd}A_{\mub]}^{\vphantom{\Jd}\Je\Jf}
     \; .
\end{align}
In addition to these transformations the right hand side of \eqref{ComSUSYalgebra}
consists of general coordinate, Lorentz, ${\rm USp}(4)$, and
vector/tensor gauge transformations with parameters given by
\begin{align}
   \xi^\mua &= \frac 1 2 \bar \epsilon_{2\ja} \Gamma^\mua \epsilon_1^\ja \;,
       \nonumber \\
   \epsilon^{\ma\mb} &= -\frac{1}{5 {\sqrt{2}}} {\cal H}^{(2)(\ja\jb)}_{\mc\md} \Omega_{\jb\jc} 
                    \bar \epsilon_{2\ja} \left( \Gamma^{\ma\mb\mc\md} + 
                    8 \eta^{\ma\mc} \eta^{\mb\md} \right) \epsilon_1^\jc 
                    + \frac{g}{20 {\sqrt{2}}} A^{\ja\jb} \Omega_{\jb\jc} \bar \epsilon_{2\ja} \Gamma^{\ma\mb} \epsilon_1^\jc 
\nonumber \\ & \qquad 
                  + \frac{1}{15} {\cal H}^{(3)}_{\mc\md\me[\ja\jb]} \Omega^{\jb\jc}
\bar \epsilon_{2\jc} \left( \Gamma^{\ma\mb\mc\md\me} + 
9 \eta^{\ma\mc} \eta^{\mb\md} \Gamma^\me \right) \epsilon_1^\ja
- \frac{g}{16 {\sqrt{2}}} D \bar \epsilon_{2\ja} \Gamma^{\ma\mb} \epsilon_1^\ja   
       \;,\nonumber \\
   {\kappa_\ja}^\jb &= \frac 1 4 \Lambda^{\jd\je} {T_{(\jd\je)[\ja\jc]}}^{[\jb\jc]} \;,
       \nonumber \\
   \Lambda^{\Ja\Jb} &= \sqrt{2} {{\cal V}_{\ja\jb}}^\Ja {{\cal V}_{\jc\jd}}^\Jb \Omega^{\jb\jd} \Lambda^{\ja\jc} \; ,
       \qquad \qquad 
   \text{with} \qquad  \Lambda^{\ja\jb}  = \frac{1}{2\sqrt{2}} \Omega^{\jc(\ja} \bar \epsilon_{2\jc} \epsilon_1^{\jb)} 
     \;,   \nonumber \\
   \Xi_{\Ja\mua} 
      &= \frac {1} 2 {{\cal V}_\Ja}^{\ja\jb} \bar \epsilon_{2\ja} \Gamma_\mua \epsilon_1^\jc \Omega_{\jc\jb} 
   \;,     \nonumber \\
   \Phi^\Ja_{\mua\mub} &= - \frac{1} 8 {{\cal V}_{\ja\jb}}^\Ja \Omega^{\ja\jc} 
                               \bar \epsilon_{2\jc} \Gamma_{\mua\mub} \epsilon_1^\jb
                               \;.
\end{align}
To this order in the fermion fields the fermionic field equations are not
yet required for verifying the closure~(\ref{ComSUSYalgebra}) of the algebra.
Closure on the three-form tensor fields 
$S^\Ja_{\mua\mub\muc}$ however makes use of the (projected) duality equation
\bea
   e^{-1} \, \epsilon^{\mua\mub\muc\mud\mue\muf\mug} Y_{\Ja\Jb} 
   {\cal H}^{(4)\Jb}_{\mud\mue\muf\mug}
   &=& 6 \, Y_{\Ja\Jb} \, \Omega^{\ja\jc} \Omega^{\jb\jd} \, 
   {{\cal V}_{\ja\jb}}^\Jb {\cal H}^{(3)}_{\jc\jd \, \mua\mub\muc}
     + \text{fermionic terms}\;,
   \label{Duality1}
\eea
This equation 
will arise as a first order equation of motion from
the full Lagrangian upon varying w.r.t.\ the $S^{\Ja}_{\mua\mub\muc}$.
We will confirm this in the next section.
Note that also this duality equation appears only under 
projection with $Y_{\Ja\Jb}$.

\subsection{The universal Lagrangian}

We can now present the universal Lagrangian of gauged maximal supergravity 
in seven dimensions up to higher order fermion terms:
\begin{align}
   e^{-1} {\cal L} =& 
               - \frac 1 2 R 
-  \Omega_{\ja\jc} \Omega_{\jb\jd} {\cal H}^{(2)\ja\jb}_{\mua\mub} {\cal H}^{(2)\jc\jd\mua\mub}
- \frac 1 {6} \Omega^{\ja\jc} \Omega^{\jb\jd} {\cal H}^{(3)}_{\mua\mub\muc\,\ja\jb} 
{\cal H}^{(3)}{}_{\jc\jd}^{\mua\mub\muc}
- \frac 1 2 {P_{\mua\ja\jb}}^{\jc\jd} {P^\mua}_{\jc\jd}{}^{\ja\jb}
\nonumber \\ & 
- \frac 1 2 \bar \psi_{\mua\ja} \Gamma^{\mua\mub\muc} D_\mub \psi_{\muc}^\ja
- \frac 1 {8} \bar \chi_{\ja\jb\jc} \slashchar D \chi^{\ja\jb\jc}
- \frac 1 2 {P_{\mua\ja\jb}}^{\jc\jd} \Omega_{\jc\je} \bar \psi_{\mub\jd} \Gamma^{\mua} \Gamma^{\mub} \chi^{\ja\jb\je}
\nonumber \\ & 
+ \frac{\sqrt{2}} {4} {\cal H}^{(2)\ja\jb}_{\mua\mub} \,
\Big( - \bar \psi^\muc_{\ja} \Gamma_{[\muc} \Gamma^{\mua\mub} \Gamma_{\mud]} \psi^{\mud\jc} \Omega_{\jc\jb} 
                         + \bar \psi_{\muc\jc} \Gamma^{\mua\mub} \Gamma^{\muc} \chi^{\jc\jd\je} \Omega_{\ja\jd} \Omega_{\jb\je}
+\frac 1 2 \bar \chi_{\ja\jc\jd} \Gamma^{\mua\mub} \chi^{\je\jd\jc} \Omega_{\je\jb}  \Big)
\nonumber \\ & 
               + \frac 1 {12} {\cal H}^{(3)}_{\ja\jb\mua\mub\muc} \,
\Big( -\Omega^{\ja\jc} \bar \psi^\mud_{\jc} \Gamma_{[\mud} \Gamma^{\mua\mub\muc} \Gamma_{\mue]} \psi^{\mue\jb}
                         +\frac 1 2 \bar \psi_{\mud\jc} \Gamma^{\mua\mub\muc} \Gamma^{\mud} \chi^{\ja\jb\jc}
+\frac 1 4 \Omega^{\ja\je} \bar \chi_{\jc\jd\je} \Gamma^{\mua\mub\muc} \chi^{\jc\jd\jb} \Big)
\nonumber \\ &  
               - \frac 5 2 g A_1^{\ja\jb} \Omega_{\jb\jc} \bar \psi_{\mua\ja} \Gamma^{\mua\mub} \psi^{\jc}_\mub
+ \frac 1 4 g A_2^{\jd,\ja\jb\jc} \Omega_{\jd\je} \bar \chi_{\ja\jb\jc} \Gamma^{\mua} \psi_\mua^{\je}
\nonumber \\ &  
+ \frac{g} {4 \sqrt{2}} \, \Big( \frac 3 {32} \delta^\jb_\jd \delta^\jc_\je B
+\frac 1 8 \delta^\jb_\jd \Omega_{\je\jf} C^{\jf\jc}
+ {B^{\jb\jc}}_{\jd\je} - {C^{\jb\jc}}_{\jd\je} \Big) 
\bar \chi_{\ja\jb\jc} \, \chi^{\ja\jd\je}
\nonumber \\ & 
+ \frac {g^2} {128} \left(15 B^2+2 C^{\ja\jb}C_{\ja\jb} - 
2 B^{\ja\jb}{}_{\jc\jd}B^{\jc\jd}{}_{\ja\jb} - 
2 C^{[\ja\jb]}{}_{(\jc\jd)}C_{[\ja\jb]}{}^{(\jc\jd)}
 \right) 
\nonumber\\[1ex]
&+ e^{-1}{\cal L}_{\rm VT}
\;,
\label{L}
\end{align}
with the tensors $A_{1}$, $A_{2}$ from~(\ref{FermionShifts}) and the
topological vector-tensor Lagrangian from (\ref{VT}): 
\begin{align*}
{\cal L}_{\rm VT} &= - \frac 1 {9}
\epsilon^{\mua\mub\muc\mud\mue\muf\mug}\;\times
\nonumber\\
&
\times     
\Big[ g Y_{\Ja\Jb} S^\Ja_{\mua\mub\muc} \Big( D^{\vphantom{\Ja}}_\mud S_{\mue\muf\mug}^\Jb
+ \frac 3 2 g Z^{\Jb\Jc,\Jd} B_{\mud\mue\,\Jc} B_{\muf\mug\,\Jd}
+ 3 {\cal F}^{\Jb\Jc}_{\mud\mue} B^{\vphantom{\Ja}}_{\muf\mug\,\Jc}
\nonumber \\ & \qquad \qquad
                 + 4 \epsilon_{\Jc\Jd\Je\Jf\Jg} A^{\Jb\Jc}_{\mud} A^{\Jd\Je}_{\mue} \partial^{\vphantom{\Ja}}_{\muf} A_\mug^{\Jf\Jg}
        + g \epsilon_{\Jc\Jd\Je\Jj\Jk} {X_{\Jf\Jg,\Jh\Ji}}^{\Jj\Jk} A_\mud^{\Jb\Jc} A_\mue^{\Jd\Je} A_\muf^{\Jf\Jg} A_\mug^{\Jh\Ji} \Big) 
  \nonumber \\  &  \quad
      + 3 g Z^{\Ja\Jb,\Jc} (D_\mua B_{\mub\muc\,\Ja}) B_{\mud\mue\,\Jb} B_{\muf\mug\,\Jc} 
    - \frac 9 2 {\cal F}_{\mua\mub}^{\Ja\Jb} B_{\muc\mud\,\Ja} D_{\mue} B_{\muf\mug\,\Jb}
  \nonumber \\ & \quad
    + 18 \epsilon_{\Ja\Jb\Jc\Jd\Je} {\cal F}_{\mua\mub}^{\Ja\Ji} A^{\Jb\Jc}_{\muc} 
            \Big( \partial^{\vphantom{\Ja}}_\mud A^{\Jd\Je}_{\mue} 
                                       + \frac 2 3 g {X_{\Jf\Jg,\Jh}}^\Jd A^{\Je\Jh}_\mud A^{\Jf\Jg}_{\mue} \Big) B_{\muf \mug\,\Ji}
  \nonumber \\ & \quad
    + 9 g \epsilon_{\Ja\Jb\Jc\Jd\Je} Z^{\Ja\Ji,\Jj} A^{\Jb\Jc}_{\mua} \Big( \partial^{\vphantom{\Ja}}_\mub A^{\Jd\Je}_{\muc} 
          + \frac 2 3 g {X_{\Jf\Jg,\Jh}}^\Jd A^{\Je\Jh}_\mub A^{\Jf\Jg}_{\muc} \Big) B_{\mud\mue\,\Ji} B_{\muf\mug\,\Jj} \nonumber \\
  & + \frac {36} 5 \epsilon_{\Ja\Jc\Jd\Jg\Jh} \; \epsilon_{\Jb\Je\Jf\Ji\Jj} 
       A_\mua^{\Ja\Jb} A_\mub^{\Jc\Jd} A_\muc^{\Je\Jf} 
             (\partial_\mud^{\vphantom{\Ja}} A_\mue^{\Jg\Jh}) (\partial_\muf^{\vphantom{\Ja}} A_\mug^{\Ji\Jj})
     \nonumber \\ &   
    + 8 g \epsilon_{\Ja\Jc\Jd\Je\Jf} \; \epsilon_{\Jb\Jg\Jh\Jm\Jn} {X_{\Ji\Jj,\Jk\Jl}}^{\Jm\Jn}
          A_\mua^{\Ja\Jb} A_\mub^{\Jc\Jd} A_\muc^{\Jg\Jh} A_\mud^{\Ji\Jj} A_\mue^{\Jk\Jl} 
\partial_\muf^{\vphantom{\Ja}} A_\mug^{\Je\Jf} \nonumber \\
  & - \frac 4 7 g^2 \epsilon_{\Ja\Jc\Jd\Jo\Jp} \; \epsilon_{\Jb\Ji\Jj\Jq\Jr} {X_{\Je\Jf,\Jg\Jh}}^{\Jo\Jp} {X_{\Jk\Jl,\Jm\Jn}}^{\Jq\Jr}
        A_\mua^{\Ja\Jb} A_\mub^{\Jc\Jd} A_\muc^{\Je\Jf} A_\mud^{\Jg\Jh} A_\mue^{\Ji\Jj} A_\muf^{\Jk\Jl} A_\mug^{\Jm\Jn}
\Big]
\;.
\end{align*}
This Lagrangian is the unique one invariant under the full set of 
non-Abelian vector/tensor gauge 
transformations~\eqref{gauge} and under local 
supersymmetry 
transformations~\eqref{SUSYrules}.
Furthermore it possesses the local ${\rm USp}(4)$ invariance 
introduced in~\eqref{TrafoCalV}, and is formally
invariant under global ${\rm SL}(5)$ transformations
if the embedding tensor~$\Theta$ is treated as a spurionic
object that simultaneously transforms. 
With fixed~$\Theta$, the global ${\rm SL}(5)$
is broken down to the gauge group.

In the limit $g \rightarrow 0$ the three-form fields $S^{\Ja}_{\mua\mub\muc}$
decouple from the Lagrangian, and~(\ref{L}) consistently 
reduces to the  ungauged theory of~\cite{Sezgin:1982gi} with global ${\rm SL}(5)$ 
symmetry.
Upon effecting the deformation by switching on $g$,
derivatives are covariantized $\partial_\mua \rightarrow D_\mua$ 
and the former Abelian field strengths are replaced by the full covariant
combinations ${\cal H}^{(2)}$ and ${\cal H}^{(3)}$ from~(\ref{H234}).
The extended gauge invariance~\eqref{gauge} 
moreover requires a unique extension of the former Abelian topological
term which in particular includes a first order kinetic term for the 
three-form fields $S^{\Ja}_{\mua\mub\muc}$.
As a consequence, the duality equation \eqref{Duality1} 
between the two-form and the three-form
tensor fields arises directly as a field equation of this Lagrangian. 
This ensures that the total number of degrees of freedom
is not altered by switching on the deformation
and does not depend on the explicit form of 
the embedding tensor.

In order to maintain supersymmetry under the extended 
transformations~\eqref{SUSYrules}, 
and in presence of the deformed Bianchi 
and Cartan-Maurer equations~\eqref{Bianchi}, 
\eqref{Cartan-Maurer}, the Lagrangian
finally needs to be augmented by the bilinear fermionic mass terms 
in order $g$ and a scalar potential in order $g^{2}$. 
These are expressed in terms of the scalar field dependent 
${\rm USp}(4)$-components $B$, $C$ of the $T$-tensor.
Cancellation of the terms in order $g^{2}$ in particular
requires the quadratic identities~\eqref{QZB1}, \eqref{QZB2},
expanded in components in~\eqref{Con5}, \eqref{Con35}. 
In particular, these identities give rise to
\bea
\frac18 A_2^{\ja,\jc\jd\je}A^{\vphantom{\jc}}_{2\,\jb,\jc\jd\je}-15 
A_{1}^{\ja\jc}A^{\vphantom{\jc}}_{1\,\jb\jc} &=&
\frac14\delta_{\jb}^{\ja}\,
\Big(
\frac18 A_2^{\jf,\jc\jd\je}A^{\vphantom{\jc}}_{2\,\jf,\jc\jd\je}-15 
A_{1}^{\jc\jd}A^{\vphantom{\jc}}_{1\,\jc\jd}
\Big)\;,
   \label{A1A1A2A2}
\eea
featuring the scalar potential on the r.h.s.\ and needed
for cancellation of the supersymmetry contributions from the scalar potential.
Indeed, the scalar potential which contributes to the 
Lagrangian~\eqref{L} in order $g^2$ may be written in the equivalent forms
\begin{align}
   V &= - \frac {1} {128} \left(15 B^2+2 C^{\ja\jb}C_{\ja\jb} - 
2 B^{\ja\jb}{}_{\jc\jd}B^{\jc\jd}{}_{\ja\jb} - 
2 C^{[\ja\jb]}{}_{(\jc\jd)}C_{[\ja\jb]}{}^{(\jc\jd)}
 \right) 
 \nonumber\\[1ex] & 
 = \frac 1 8 |A_2|^2 - 15 |A_1|^2 
 \;.
   \label{SPotential}
\end{align}
Under variation of the scalar fields given by 
$\delta_{\Sigma} {\cal V}_{M}{}^{ab}=\Sigma^{ab}{}_{cd}\,{\cal V}_{M}{}^{cd}$ 
the potential varies according to
\bea
  \delta_{\Sigma} V &= &
          -\frac {1}{16} {B^{[\ja\jb]}}_{[\jc\jd]} 
          {B^{[\jc\jd]}}_{[\je\jf]} {\Sigma^{[\je\jf]}}_{[\ja\jb]}
+\frac {1}{32} B {B^{[\ja\jb]}}_{[\jc\jd]} {\Sigma^{[\jc\jd]}}_{[\ja\jb]}
          -\frac {1}{64} C^{[\ja\jb]} C_{[\jc\jd]} {\Sigma^{[\jc\jd]}}_{[\ja\jb]}
\nonumber \\[1ex]  && 
{}
+\frac {1}{32} {C^{[\ja\jb]}}_{(\je\jf)} {C_{[\jc\jd]}}^{(\je\jf)} {\Sigma^{[\jc\jd]}}_{[\ja\jb]}
-\frac{1}8
{C^{[\jc\je]}}_{(\ja\jf)} {C^{[\jd\jf]}}_{(\jb\je)} 
{\Sigma^{[\ja\jb]}}_{[\jc\jd]} \;,
   \label{VaryV}
\eea
which in particular yields the contribution of the potential
under supersymmetry transformations.
Moreover, equation~\eqref{VaryV} 
is important when analyzing the ground states of the theory since
$\delta_{\Sigma} V = 0$ is a necessary condition for a stationary point
of the potential.
The residual supersymmetry of the corresponding solution
(assuming maximally symmetric space-times) is parameterized by
spinors~$\epsilon^\ja$ satisfying the condition
\begin{equation}
A_{2\,\ja,\jb\jc\jd}\, \epsilon^\ja = 0 \;.
\label{susy_groundstate}
\end{equation}
The gravitino variation imposes an extra condition
\bea
2A_{1\,\ja\jb}\,\epsilon^{\jb} &=&
\pm \sqrt{-V/15}\,\Omega_{\ja\jb}\,\epsilon^{\jb}
\;,
   \label{susy_groundstate2}
\eea
but the two conditions~(\ref{susy_groundstate})
and~(\ref{susy_groundstate2})
are in fact equivalent by virtue of~(\ref{A1A1A2A2}).\footnote{
More precisely, a solution of~\eqref{susy_groundstate}, (\ref{susy_groundstate2}) 
tensored with a Killing spinor of AdS$_{7}$
(or seven-dimensional Minkowski space, respectively,
depending on the value of $V$)
solves the Killing spinor equations $\delta\psi^{a}_{\mu}=0$,
$\delta\chi^{abc}=0$ obtained from~\eqref{SUSYrules}.
}

The full check of invariance of the Lagrangian~\eqref{L}
under the supersymmetry transformations~\eqref{SUSYrules}
is rather lengthy and makes heavy use
of the quadratic constraints~\eqref{quadratic}
on the embedding tensor and their
consequences collected in appendix~\ref{app:TQC}
as well as of the properties of the ${\rm SL}(5)/{\rm USp}(4)$ coset
space discussed in the previous section.
We have given the Lagrangian and transformation rules only
up to higher order fermion terms; however one does not 
expect any order $g$ corrections to these higher order
fermion terms, i.e.\ they remain unchanged w.r.t.\ those of 
the ungauged theory. 

Let us finally note that the bosonic part of the Lagrangian~(\ref{L})
can be cast into a somewhat simpler form in which the scalar fields parameterize
the ${\rm USp}(4)$-invariant symmetric unimodular matrix ${\cal M}_{\Ja\Jb}$
\begin{align}
   {\cal M}_{\Ja\Jb} &\equiv 
   {{\cal V}_\Ja}^{\ja\jb} {{\cal V}_\Jb}^{\jc\jd} \,\Omega_{\ja\jc} \Omega_{\jb\jd}
   \; ,
   \label{DefCalM}
\end{align}
with the inverse ${\cal M}^{\Ja\Jb}= ({\cal M}_{\Ja\Jb})^{-1}=
{{\cal V}_{\ja\jb}}^\Ja {{\cal V}_{\jc\jd}}^\Jb \,\Omega^{\ja\jc} \Omega^{\jb\jd}$. 
The bosonic part of the Lagrangian~(\ref{L}) can then
be expressed exclusively in terms of ${\rm USp}(4)$-invariant quantities
and takes the form
\bea
e^{-1} {\cal L}_{\text{bosonic}} &=&
- \frac 1 2 R 
- {\cal M}_{\Ja\Jc} {\cal M}_{\Jb\Jd} {\cal H}^{(2)\Ja\Jb}_{\mua\mub} {\cal H}^{(2)\mua\mub\,\Jc\Jd}
- \frac 1 6 {\cal M}^{\Ja\Jb} {\cal H}^{(3)}_{\mua\mub\muc\,\Ja} {\cal H}^{(3)}{}^{\mua\mub\muc}_{\Jb}
\nonumber \\ &&
+ \frac 1 8 (\partial_\mua {\cal M}_{\Ja\Jb}) (\partial^\mua {\cal M}^{\Ja\Jb})
 + e^{-1}{\cal L}_{\text{VT}}- g^{2}\,V\;,
 \label{LM}
 \eea
 with the scalar potential 
\bea
V&=&
\frac {1} {64} \Big( 
                         3 {X_{\Ja\Jb,\Je}}^\Jf {X_{\Jc\Jd,\Jf}}^\Je {\cal M}^{\Ja\Jc} {\cal M}^{\Jb\Jd}
-  {X_{\Ja\Jc,\Jd}}^\Jb {X_{\Jb\Je,\Jf}}^\Ja {\cal M}^{\Jc\Je} {\cal M}^{\Jd\Jf}
\Big)
\nonumber\\ && 
+\frac {1} {96} \Big(
 {X_{\Ja\Jb,\Je}}^\Jf {X_{\Jc\Jd,\Jg}}^\Jh 
{\cal M}^{\Ja\Jc} {\cal M}^{\Jb\Jd} {\cal M}^{\Je\Jg} {\cal M}_{\Jf\Jh}
 + {X_{\Ja\Jc,\Jd}}^\Jb {X_{\Jb\Je,\Jf}}^\Ja {\cal M}^{\Jc\Jd} {\cal M}^{\Je\Jf} \Big)
 \nonumber\\[2ex]
 &=&
 \frac1{64}
\Big(2{\cal M}^{\Ja\Jb}Y_{\Jb\Jc}{\cal M}^{\Jc\Jd}Y_{\Jd\Ja}-
({\cal M}^{\Ja\Jb}Y_{\Ja\Jb})^{2}\Big)
\nonumber\\
&&{}
+Z^{\Ja\Jb,\Jc}Z^{\Jd\Je,\Jf}\,\Big(
{\cal M}_{\Ja\Jd}{\cal M}_{\Jb\Je}{\cal M}_{\Jc\Jf}
-{\cal M}_{\Ja\Jd}{\cal M}_{\Jb\Jc}{\cal M}_{\Je\Jf} \Big)
\;.
 \label{LMP}
\eea
This is in analogy to the fact that 
the gravitational degrees of freedom can be described alternatively
in terms of the vielbein or in terms of the metric.
In particular, the scalar potential here is directly expressed 
in terms of the embedding tensor~(\ref{X-theta})
properly contracted with the scalar matrix ${\cal M}$
without having to first pass to the ${\rm USp}(4)$ tensors $B$, $C$.
In concrete examples this may simplify the computation
and the analysis of the scalar potential.
Of course, in order to describe the coupling to fermions
it is necessary to reintroduce ${\cal V}$, the tensors $B$, $C$,  
and to exhibit the local ${\rm USp}(4)$ symmetry.

\section{Examples}
\label{SecExpamples}

In this section, we will illustrate the general formalism 
with several examples. 
In particular, these include the maximally supersymmetric 
theories resulting from M-theory compactification 
on $S^{4}$~\cite{Pernici:1984xx,Pilch:1984xy,Nastase:1999cb,Nastase:1999kf},
as well as the (warped) type IIA/IIB compactifications on $S^{3}$
which so far have only partially been constructed in the literature.

In order to connect to previous results in the literature, we first discuss 
the possible gauge fixing of tensor gauge transformations
depending on the specific form of the embedding tensor.
In sections~\ref{SecExNoZ} and \ref{SecExNoY} we consider particular
classes of examples in which the embedding tensor is restricted 
to components in either the ${\bf 15}$ or the ${\bf \overline{40}}$
representation. Finally, we sketch in section~\ref{sec:YZ}
a more systematic approach towards classifying the solutions of the 
quadratic constraint~\eqref{quadratic} with both $Y_{\Ja\Jb}$ and $Z^{\Ja\Jb,\Jc}$
non-vanishing. Our findings are collected in Table~\ref{TabSumEx}.

\subsection{Gauge fixing}
\label{sec:gfixing}

We have already noted in section~\ref{SecVecTen}
that the extended local gauge transformations~\eqref{gauge}
allow to eliminate a number of vector and tensor fields
depending on the specific form of the components 
$Y_{\Ja\Jb}$ and $Z^{\Ja\Jb,\Jc}$ of the embedding tensor.
More precisely, $s\equiv{\rm rank}\,Z$ vector fields can be set to 
zero by means of tensor gauge transformations 
$\delta_\Xi$ of~\eqref{gauge},
rendering $s$ of the two-forms massive.
Here, $Z^{\Ja\Jb,\Jc}$ is understood as a rectangular $10\times5$ matrix.
Furthermore, $t\equiv{\rm rank}\,Y$ of the two-forms can be 
set to zero by means of tensor gauge transformations 
$\delta_\Phi$. The $t$ three-forms
that appear in the Lagrangian~\eqref{L} then turn into 
self-dual massive forms.
The quadratic constraint~\eqref{Q2} ensures that $s+t\leq5$.
Before gauge fixing, the degrees of freedom in the Lagrangian~\eqref{L}
are carried by the vector and two-form fields just as in the ungauged theory
(Table~\ref{TabMultiplet}) while the three-forms
appear topologically coupled. After gauge fixing the distribution
of these 100 degrees of freedom is summarized in Table~\ref{dof}.
In a particular ground state, in addition some of the vectors
may become massive by a conventional Brout-Englert-Higgs mechanism.

\begin{table}[bt]
\begin{center}
\begin{tabular}{r|c|c}
fields & $\#$ & $\#$ dof \\
\hline
massless vectors & $10-s$ & 5 \\
massless 2-forms& $5-s-t$ & 10\\
massive 2-forms& $s$ & 15 \\
massive sd.\ 3-forms & $t$ & 10
\end{tabular}
\caption{{\small Distribution of degrees of freedom after gauge fixing.}\label{dof}}
\end{center}
\end{table}

Let us make this a little more explicit. To this end, we employ 
for the two-forms a special basis $B_{\Ja}=(B_{x},B_{\alpha})$,
$x=1,\dots, t$; $\alpha=t\!+\!1, \dots, 5$,
such that the symmetric matrix $Y_{\Ja\Jb}$ takes block diagonal form,
$Y_{xy}$ is invertible (with inverse $Y^{xy}$), 
and all entries $Y_{x\alpha}$, $Y_{\alpha\beta}$ vanish.
For the tensor $Z$ the quadratic constraint~\eqref{quadratic} then implies
that only its components
\bea
Z^{\alpha\beta,\gamma}\;,\qquad 
Z^{x\alpha,\beta}=Z^{x(\alpha,\beta)}
\;,
\label{nZ}
\eea
are non-vanishing and need to satisfy
\bea
Y_{xy}\,Z^{y\alpha,\beta} 
+2\epsilon_{x\Ja\Jb\Jc\Jd}\,Z^{\Ja\Jb,\alpha}Z^{\Jc\Jd,\beta}
&=& 0
\;.
\label{quadraticZ}
\eea
Gauge fixing eliminates the two-forms $B_{x}$ which explicitly breaks the 
${\rm SL}(5)$ covariance.
Supersymmetry transformations thus need to be amended by a compensating term
$\delta^{\rm new}(\epsilon)=\delta^{\rm old}(\epsilon)+
\delta(\Phi_{\mua\mub}^{x})$.
It is convenient to define the modified three-forms
\bea
{\cal S}^{x}_{\mu\nu\rho} &\equiv& 
g^{-1} Y^{xy}\,{\cal H}^{(3)}_{\mu\nu\rho\,y}
~=~  S^{x}_{\mua\mub\muc} + 6 g^{-1} Y^{xy}
\epsilon_{y\Ja\Jb\Jc\Jd} 
A^{\Ja\Jb}_{[\mu}\partial^{\vphantom{\Jc}}_{\nu}A^{\Jc\Jd}_{\rho]}+\dots \;,
\label{newS}
\eea
which are by construction invariant under tensor gauge transformations
and will appear in the Lagrangian as massive fields.
Their transformation under local gauge and supersymmetry 
is given by
\bea
\delta({\Lambda})\,{\cal S}^{x}_{\mu\nu\rho}&=&
-gY_{yz}\Lambda^{xy} {\cal S}^{z}_{\mua\mub\muc} 
-\Lambda^{x\alpha} {\cal H}^{(3)}_{\mua\mub\muc\,\alpha} 
-2 Y^{xy}Z^{\Jb\Jc,\alpha}\,\epsilon_{y\Jb\Jc\Jd\Je}\, 
\Lambda^{\Jd\Je} {\cal H}_{\mua\mub\muc\,\alpha}^{(3)} 
\;,
\nonumber\\[1.5ex]
\delta({\epsilon})\,{\cal S}^{x}_{\mu\nu\rho}&=&
-{{\cal V}_{\ja\jb}}^{x} 
                          ( \ft{3}{8} \Omega^{\ja\jc} \bar \epsilon_\jc \Gamma_{[\mua\mub} \psi^\jb_{\muc]}
+\ft{1}{32} \bar \epsilon_\je \Gamma_{\mua\mub\muc} \chi^{\ja\jb\je})  
\nonumber\\[.5ex]
&&
- 3 g^{-1} Y^{xy}\,\epsilon_{y\Jb\Jc\Jd\Je} {\cal H}_{[\mua\mub}^{(2)\Jb\Jc} 
{{\cal V}_{\ja\jb}}^{[\Jd} {{\cal V}_{\jc\jd}}^{\Je]} \Omega^{\jb\jd} 
                             ( \Omega^{\ja\je} \bar \epsilon_\je \psi_{\muc]}^\jc
+\ft{1}{2} \bar \epsilon_\je \Gamma_{\muc]} \chi^{\je\ja\jc}  )
\nonumber\\[.5ex]
&&
-3g^{-1} Y^{xy}\,  D_{[\mua} \Big(
                             ( \Omega_{\ja\jc} \bar \epsilon_\jb \Gamma^{{\vphantom{]}}}_{\mub} \psi^\jc_{\muc]}
-\ft{1}{8}  \Omega_{\ja\jc} \Omega_{\jb\jd} \bar \epsilon_\je \Gamma_{\mub\muc]} 
\chi^{\jc\jd\je}) {{\cal V}_{y}}^{\ja\jb}\Big) 
\;.
\label{gf1}
\eea
In the Lagrangian these fields appear with a mass term descending from the kinetic
term of the modified field strength tensor
${\cal H}_{\mua\mub\muc\,\ja\jb}=
{\cal V}_{\ja\jb}{}^{\alpha}{\cal H}_{\mua\mub\muc\,\alpha}^{(3)} +
gY_{xy}{\cal V}_{\ja\jb}{}^{x}{\cal S}_{\mua\mub\muc}^{y}$
and a first order kinetic term from the Chern-Simons term
\bea
{\cal L}_{\rm VT} &= - \frac 1 {9}g
\epsilon^{\mua\mub\muc\mud\mue\muf\mug}
 Y_{xy}\, {\cal S}^x_{\mua\mub\muc} D^{\vphantom{\Ja}}_\mud {\cal S}_{\mue\muf\mug}^y
+\dots \;.
\label{gf2}
\eea
The remaining terms in the expansion~\eqref{newS} in particular lead to terms 
$A\,\partial\!A\,\partial\!A\,\partial\!A$
of order $g^{-1}$ in the topological term which obstruct a smooth limit back
to the ungauged theory. Indeed these terms have been observed in the 
original construction of the ${\rm SO}(p,q)$ gaugings~\cite{Pernici:1984xx}.
Generically the gauge fixing procedure described above leads to 
many more interaction terms between vector and tensor fields than those that 
are known from the particular case of the ${\rm SO}(p,q)$ theories.

\subsection{Gaugings in the ${\bf 15}$ representation:~ \\
${\rm SO}(p,5\!-\!p)$ and
${\rm CSO}(p,q,5\!-\!p\!-\!q)$}
\label{SecExNoZ}

As a first class of examples let us analyze those gaugings
for which the embedding tensor~$\Theta$ lives entirely in the ${\bf 15}$
representation of ${\rm SL}(5)$, i.e.\ $Z^{\Ja\Jb,\Jc}=0$,
and the gauge group generators~\eqref{XP} take the form
\bea
(X_{\Ja\Jb}){}_{\Jc}{}^{\Jd} &=&
\delta^{\Jd}_{[\Ja}\,Y^{\phantom{\Jd}}_{\Jb]\Jc} 
\;.
\label{CSO}
\eea
In this case, the quadratic constraint \eqref{quadratic} is automatically satisfied,
thus every symmetric matrix $Y_{\Ja\Jb}$ defines a viable gauging.
Fixing the ${\rm SL}(5)$ symmetry (and possibly rescaling the gauge coupling constant), 
this matrix can be brought into the form
\begin{align}
  Y_{\Ja\Jb} &= 
  \diag(\,\underbrace{1, \dots,}_{p}\underbrace{-1,\dots,}_{q} \underbrace{0, \dots}_{r}\,) \;,
  \label{YinCSO}
\end{align}
with $p+q+r=5$. The corresponding gauge group is
\begin{align}
   G_0={\rm CSO}(p,q,r)={\rm SO}(p,q) \ltimes \mathbbm{R}^{(p+q)\cdot r} \; ,
\end{align}
where the Abelian part combines $r$ vectors under ${\rm SO}(p,q)$. 
This completely classifies the gaugings in this sector.
The scalar potential~\eqref{LMP} reduces to
\bea
V&=& \ft1{64}
\Big(2{\cal M}^{\Ja\Jb}Y_{\Jb\Jc}{\cal M}^{\Jc\Jd}Y_{\Jd\Ja}-
({\cal M}^{\Ja\Jb}Y_{\Ja\Jb})^{2}\Big)\;.
\label{potCSO}
\eea
{}From Table~\ref{dof} one reads off the spectrum of these theories ($s=0$, $t=5-r$):
after gauge fixing
it consists of 10 vectors together with $r$ massless two-forms and
$5-r$ self-dual massive three-forms.
In particular, a nondegenerate $Y_{\Ja\Jb}$ ($r=0$)
corresponds to the semi-simple gauge groups ${\rm SO}(5)$, 
${\rm SO}(4,1)$ and ${\rm SO}(3,2)$
that have originally been constructed 
exclusively in terms of vector and three-form 
fields~\cite{Pernici:1984xx,Pernici:1984zw}. 

The ${\rm SO}(5)$ gauged theory has a higher-dimensional interpretation
as reduction of $D=11$ supergravity on the sphere 
$S^{4}$~\cite{Pilch:1984xy,Nastase:1999cb,Nastase:1999kf}.
Accordingly, its potential~(\ref{potCSO})
admits a maximally supersymmetric AdS$_{7}$ ground state.
The theories with ${\rm CSO}(p,q,r)$ gauge groups
are related to the compactifications on the (non-compact) manifolds $H^{p,q} \circ T^r$
\cite{Hull:1988jw}.
These are the four-dimensional hyper-surfaces of $\mathbb{R}^5$ defined by
\begin{align}
   Y_{\Ja\Jb} \, v^\Ja v^\Jb &= 1 \; , & &   v^\Ja \in \mathbb{R}^5 \; .
\end{align}
A particularly interesting example is the ${\rm CSO}(4,0,1)$ theory
which corresponds to the $S^{3}$ compactification of the ten-dimensional 
type IIA theory. The bosonic part of this theory has previously been 
constructed in~\cite{Cvetic:2000ah}. In order to derive its scalar potential
from~\eqref{potCSO} it is useful to parameterize the 
coset representative ${\cal V}$ as
\bea
{\cal V} &=& e^{b_{m}t^{m}}\,V_{4}\,e^{\phi\, t_{0}}\;, 
\label{M4}
\eea
where $V_{4}$ is an ${\rm SL}(4)/{\rm SO}(4)$ matrix and $t_{0}$, $t^{m}$
denote the ${\rm SO}(1,1)$ and four nilpotent generators, respectively,
in the decomposition ${\rm SL}(5)\rightarrow {\rm SL}(4)\times{\rm SO}(1,1)$. 
For the matrix ${\cal M}$ this yields a block decomposition into
\bea
{\cal M}_{\Ja\Jb} &=& \left(
\begin{array}{cc}
e^{-2\phi}\,M_{mn} + e^{8\phi}\, b_{m} b_{n} & e^{8\phi}\, b_{m}\\
e^{8\phi}\, b_{n} & e^{8\phi}
\end{array}
\right)
\eea
with $M=V^{{\rm \vphantom{T}}}_{4}\,V_{4}^{{\rm T}}$. 
Plugging this into~\eqref{potCSO} 
with $Y_{\Ja\Jb}={\rm diag}(1,1,1,1,0)$
yields the potential
\bea
V &=& \ft1{64}e^{4\phi}\,
\Big(2\,M^{mn}\delta_{nk}M^{kl}\delta_{lm}-(M^{mn}\delta_{mn})^{2}\Big)\;,
\label{potIIA}
\eea
(where $M_{mk}M^{kn}=\delta_{m}^{n}$)
in agreement with~\cite{Cvetic:2000ah}. The presence of the dilaton pre-factor
$e^{4\phi}$ shows that this potential does not admit any stationary points,
rather the ground state of this theory is given by a domain wall solution
corresponding to the (warped) $S^{3}$ reduction of the 
type IIA theory~\cite{Cvetic:2000ah,Bergshoeff:2004nq}.

We can finally determine all the stationary points
of the scalar potentials~\eqref{potCSO} in this sector of gaugings.
The variation of the potential has been given in~\eqref{VaryV}.
Since $Z^{\Ja\Jb,\Jc}=0$, the tensors $C^{ab}$, $C^{[ab]}{}_{(cd)}$
vanish such that requiring $\delta_{\Sigma} V=0$ reduces to the matrix equation
\bea
2{\bf B}^{2}- B\, {\bf B} 
&=& \ft15\, {\rm Tr}(2{\bf B}^{2}- B\, {\bf B} )\,\mathbb{I}_{5}\;,
\label{matrix}
\eea
for the traceless symmetric matrix ${\bf B}=B^{[ab]}{}_{[cd]}$, where
$\mathbb{I}_{5}$ denotes the $5\times5$ unit matrix. 
According to~\eqref{BC} ${\bf B}$ is related by 
$\sqrt{2}{\bf Y}={\bf B}+B\,\mathbb{I}_{5}$ to the matrix ${\bf Y}=Y_{[\ja\jb],[\jc\jd]}$.
Fixing the local ${\rm USp}(4)$-invariance
the matrix ${\bf B}$ can be brought into diagonal form. Equation~\eqref{matrix}
then has only three inequivalent solutions
\bea
{\bf B}\propto {\rm diag}(0,0,0,0,0) &\Longrightarrow&
{\bf Y}= {\rm diag}(1,1,1,1,1) \;,
\nonumber\\
{\bf B}\propto {\rm diag}(1,1,1,1,-4)&\Longrightarrow&
{\bf Y}= 2^{-1/5}\,{\rm diag}(1,1,1,1,2) \;,
\nonumber\\
{\bf B}\propto {\rm diag}(1,1,1,-3/2,-3/2) &\Longrightarrow&
{\bf Y}= {\rm diag}(0,0,0,1,1) \;.
\label{statpoints}
\eea
The first two solutions correspond to the ${\rm SO}(5)$ and the ${\rm SO}(4)$
invariant stationary points of the theory with gauge group 
${\rm SO}(5)$~\cite{Pernici:1984xx,Pernici:1984zw}.
The third solution is a stationary point
in the ${\rm CSO}(2,0,3)$ gauged theory. We will come back to this 
in section~\ref{sec:YZ} and show that it
gives rise to a Minkowski vacuum related to a 
Scherk-Schwarz reduction from eight dimensions.

Analyzing the remaining supersymmetry of these vacua we note 
that in this sector of theories $A_{1}^{\ja\jb}\propto\Omega^{\ja\jb}$.
According to~(\ref{susy_groundstate2}) thus supersymmetry is
either completely preserved (${\cal N}=4$) or completely broken 
(${\cal N}=0$). Only the first stationary point in~\eqref{statpoints}
preserves all supersymmetries: this is the maximally supersymmetric
AdS$_{7}$ vacuum mentioned above.

\subsection{Gaugings in the ${\bf \overline{40}}$ representation:~ \\ 
${\rm SO}(p,4\!-\!p)$ and
${\rm CSO}(p,q,4\!-\!p\!-\!q)$}
\label{SecExNoY}

Another sector of gaugings is characterized by restricting
the embedding tensor to the ${\bf \overline{40}}$ representation
of ${\rm SL}(5)$, i.e.\ setting $Y_{\Ja\Jb}=0$. 
These gaugings are parameterized by a tensor $Z^{\Ja\Jb,\Jc}$
for which the quadratic constraint~\eqref{quadratic} reduces to
\bea
\epsilon_{\Ja\Je\Jf\Jg\Jh}\,Z^{\Je\Jf,\Jb}Z^{\Jg\Jh,\Jc} &=& 0 \;.
\label{qZ}
\eea
Rather than attempting a 
complete classification of these theories we will present a representative
class of examples.
Specifically, we consider gaugings with the tensor $Z^{\Ja\Jb,\Jc}$ given by
\begin{align}
   Z^{\Ja\Jb,\Jc} \, &= \, v^{[\Ja} \, w^{\Jb]\Jc}
   \;,
   \label{ZeqVW}
\end{align}
in terms of a vector $v^\Ja$ and a symmetric matrix $w^{\Ja\Jb}=w^{(\Ja\Jb)}$. 
This Ansatz automatically solves the quadratic constraint~\eqref{qZ} and thus
defines a class of viable gaugings.
The ${\rm SL}(5)$ symmetry can be used to further bring $v^{\Ja}$ into the form 
$v^{\Ja}=\delta^{\Ja}_{5}$ introducing the index split 
${\Ja}=({i},{5})$, $i=1, \dots, 4$.
The remaining ${\rm SL}(4)$ freedom can be fixed
by diagonalizing the corresponding $4\times4$ block $w^{ij}$
\begin{align}
  w^{ij} &= 
  \diag(\,\underbrace{1, \dots,}_{p}\underbrace{-1,\dots,}_{q} \underbrace{0, \dots}_{r}\,) \; .
\end{align}
For simplicity we restrict to cases with $w^{i5}=w^{55}=0$.
The gauge group generators then take the form
\bea
(X_{ij}){}_{k}{}^{l} &=&
2\epsilon_{ijkm}\, w^{ml}
\;,
\label{cso4}
\eea
and generate the group~${\rm CSO}(p,q,r)$ with $p+q+r=4$.
According to Table~\ref{dof}, these theories contain only
vector and two-forms, $4\!-\!r$ of which become massive after gauge 
fixing. The scalar potential is obtained from~\eqref{LMP}
and in the parameterization of~\eqref{M4} takes the form
\begin{align}
V &= \ft14e^{14\phi}\,b_{m}w^{mk}M_{kl}\,w^{ln}\,b_{n}
+\ft14e^{4\phi}\,\Big(2\,M_{mn}w^{nk}M_{kl}w^{lm}-(M_{mn}w^{mn})^{2}\Big)\;.
\label{potIIB}
\end{align}
A particularly interesting case  is the theory with $r=0$ and
compact gauge group ${\rm SO}(4)$.
The existence of this maximal supergravity in seven dimensions was 
anticipated already in \cite{Boonstra:1998mp} in the context of holography to 
six-dimensional super Yang-Mills theory. 
Indeed, its spectrum should consist of vector and two-form 
tensor fields only (cf.\ Table~IV in~\cite{Morales:2004xc}).
Its higher-dimensional origin is a (warped) $S^3$ reduction of type IIB supergravity.
Again, this is consistent with the fact that 
due to the presence of the dilaton pre-factor
the potential~\eqref{potIIB} in this case
does not admit any stationary points 
but only a domain wall solution.
So far, only the ${\cal N}=2$ truncation of this theory had been 
constructed~\cite{Salam:1983fa,Cvetic:2000dm}, in which the scalar manifold
truncates to an ${\rm GL}(4)/{\rm SO}(4)$ coset space and only a single (massless)
two-form is retained in the spectrum.

In analogy to the discussion of the last section it seems natural that the 
other ${\rm CSO}(p,q,r)$ gaugings in this sector
are related to reductions of the type IIB theory over the 
non-compact manifolds $H^{p,q} \circ T^r$. In particular, the 
potential~\eqref{potIIB} of the ${\rm CSO}(2,0,2)$ theory admits a 
stationary point with vanishing potential.
This is related to the Minkowski vacuum obtained by
Scherk-Schwarz reduction from eight dimensions as we will discuss in the next section.

\subsection{Further examples}
\label{sec:YZ}

We will finally indicate a more systematic approach towards 
classifying the general gaugings with an embedding tensor combining
parts in the ${\bf 15}$ and the ${\bf \overline{40}}$ representation.
To this end, we go to the special basis introduced in section~\ref{sec:gfixing},
in which the only non-vanishing components of the embedding tensor are given by
\bea
Y_{xy}\;,\qquad Z^{x(\alpha,\beta)}\;,\qquad Z^{\alpha\beta,\gamma}
\;,
\label{nonv}
\eea
with ${\rm rank}\, Y\equiv t$, and the range of indices  $x, y=1, \dots, t$ and 
$\alpha, \beta=t\!+\!1, \dots, 5$ . 
Further fixing (part of) the global ${\rm SL}(5)$ symmetry, 
the tensor $Y_{xy}$ can always be brought into the standard form 
\begin{align}
  Y_{xy} &= 
  \diag(\,\underbrace{1, \dots,}_{p}\underbrace{-1,\dots}_{q}\,) \;.
  \label{Yxy}
\end{align}
The possible gaugings can then systematically be found by 
scanning the different values of $t$, $p$, and $q$,
and determining the real 
solutions of the quadratic constraint~\eqref{quadraticZ}.
We will in the following discuss a (representative rather than complete) 
number of examples for the different values of $t$. 
A list of our findings is collected in Table~\ref{TabSumEx}.

\mathversion{bold}
\subsection*{$t=5$}
\mathversion{normal}

{}From~\eqref{nonv} one reads off that a nondegenerate matrix $Y_{\Ja\Jb}$ implies
a vanishing tensor $Z^{\Ja\Jb,\Jc}$. 
Thus we are back to the situation discussed 
in section~\ref{SecExNoZ}. The possible gauge groups are 
${\rm SO}(5)$,  ${\rm SO}(4,1)$, and  ${\rm SO}(3,2)$.

\mathversion{bold}
\subsection*{$t=4$}
\mathversion{normal}

{}The quadratic constraint~\eqref{quadraticZ} implies that also in this case 
the tensor $Z^{\Ja\Jb,\Jc}$ entirely vanishes. These gaugings
are again completely covered by the discussion of section~\ref{SecExNoZ}, 
with possible gauge groups 
${\rm CSO}(4,0,1)$,  ${\rm CSO}(3,1,1)$, and  ${\rm CSO}(2,2,1)$.

\mathversion{bold}
\subsection*{$t=3$}
\mathversion{normal}

Now we consider the cases $Y_{\Ja\Jb}=\diag(1,1,\pm 1,0,0)$. 
In this case the tensor $Z$ may have non-vanishing components
for which the quadratic constraint~\eqref{quadraticZ} 
imposes
\bea
\epsilon_{xyz}\,Z^{y\alpha,\gamma}\,\epsilon_{\gamma\delta}\,Z^{z\delta,\beta}
&=& \ft18\,Y_{xu}Z^{u\alpha,\beta} 
\;.
\label{quadraticZ23}
\eea
For $Z=0$, these gaugings have been discussed in section~\ref{SecExNoZ}, 
with possible gauge groups ${\rm CSO}(3,0,2)$ and  ${\rm CSO}(2,1,2)$.
There, gauge group generators take the form
\bea
L_\Ja{}^\Jb  &=&
\left( \begin{array}{cc}  \lambda^{z} (t^{z})_{x}{}^{y} & Q_{x\alpha} \\
0_{2 \times 3} & 0_{2\times 2} \end{array} \right) \;,\qquad
\lambda^{z}\in\mathbb{R}\;,\quad Q_{x\alpha}\in\mathbb{R}\;,
   \label{cso302}      
\eea
where $(t^{z})_{x}{}^{y}=\epsilon^{zyu}Y_{ux}$ 
denote the generators of the adjoint representation 
of the semi-simple part $\mathfrak{so}(p,3\!-\!p)$ and 
the $Q_{x\alpha}$ parameterize the 6 nilpotent
generators transforming as a couple of ${\bf 3}$ vectors 
under $\mathfrak{so}(p,3\!-\!p)$.
The components~$Z^{\ka\kb,\kc}$ are not constrained by~\eqref{quadraticZ23}
and may be set to arbitrary values $Z^{\ka\kb,\kc} = \epsilon^{\ka\kb} v^\kc$ 
parameterized by a two-component vector~$v^{\ka}$
without altering the form~\eqref{cso302} of the gauge group.
For the remaining components $Z^{\la\alpha,\beta}$, 
equation~\eqref{quadraticZ23} 
shows that the $2\times2$ matrices 
$(\Sigma^{\la}){}_\alpha{}^\beta \equiv -16 \epsilon_{\alpha\gamma}Z^{\la\gamma,\beta}$
satisfy the algebra
\bea
   [\Sigma^x,\Sigma^y] &= 2 \epsilon^{xyu}Y_{uz}\, \Sigma^z \; , 
\eea
i.e.\ yield a representation of the algebra $\mathfrak{so}(3)$ or 
$\mathfrak{so}(2,1)$, respectively, depending on the signature of $Y_{uz}$.
A real non-vanishing solution of~\eqref{quadraticZ23} thus can only exist
in the $\mathfrak{so}(2,1)$ sector, i.e.\
for $Y_{\Ja\Jb}=\diag(1,1,-1,0,0)$. It is given by 
$Z^{\la\ka,\kb} = - \, \frac 1 {16} \, \epsilon^{\ka\kc} \, (\Sigma^{\la})_\kc{}^\kb$
with the $\Sigma^{x}$ expressed in terms of the Pauli matrices as
\begin{align}
   \Sigma^1 &= \sigma_{1} \; ,\quad
   \Sigma^2 = \sigma_{3}  \; ,\quad
   \Sigma^3 = i\sigma_{2}  \;,
\end{align}
and providing a real representation of $\mathfrak{so}(2,1)$.
In this case, the gauge group generators 
schematically take the form
\bea
{{L_\Ja}}^\Jb  &= &
\left( \begin{array}{cc} \lambda^{z} (t^{z})_{x}{}^{y} & Q^{(4)}_{\vphantom{[]}x\alpha} \\
0_{2 \times 3} & \ft12\lambda^{z}\,(\Sigma^{z})_{\alpha}{}^{\beta} \end{array} \right) \;,
   \label{notcso302}      
\eea
such that the semi-simple part $\mathfrak{so}(2,1)$ is 
embedded into the diagonal. The nilpotent generators $Q_{x\alpha}$
now transform in the tensor product ${\bf 3}\otimes{\bf 2}={\bf 2}+{\bf 4}$ 
of $\mathfrak{so}(2,1)$ and moreover
turn out to be projected onto the irreducible ${\bf 4}$ representation. 
Compared to~\eqref{cso302},
the gauge group thus shrinks to
\bea
\mathfrak{so}(2,1) \ltimes \mathbb{R}^{4}\;.
\eea
Again, further switching on $Z^{\ka\kb,\kc}$ does not 
change the form of the algebra.
None of the theories in this sector possesses a stationary 
point in its scalar potential.

\mathversion{bold}
\subsection*{$t=2$}
\mathversion{normal}

In the case $Y_{\Ja\Jb}=(1,\pm 1,0,0,0)$ only
the $Z^{\ka\kb,\kc}$ components are allowed to be nonzero 
in order to fulfill the quadratic constraint \eqref{quadraticZ}.
These components can be parameterized by a traceless 
matrix ${Z_{\ka}}^{\kb}$ as
\begin{align}
   Z^{\ka\kb,\kc} &= \ft18\epsilon^{\ka\kb\kd} {Z_{\kd}}^\kb \;.
\end{align}
For this solution the gauge generators take the form
\begin{align}
    {{L_\Ja}}^\Jb  &= \left( \begin{array}{cc}  \lambda \, t_{2\,x}{}^{y} & Q_{x}{}^{\alpha} \\
0_{3 \times 2} &  \lambda \, Z_{\alpha}{}^{\beta} \end{array} \right) \;,
\qquad
\lambda\in\mathbb{R}\;,\quad Q_{x}{}^{\alpha}\in\mathbb{R}\;,
   \label{GenYZr2}
\end{align}
where $t_{2}=\Big(\!\!\!{\footnotesize\begin{array}{rc} 0 & \!\!\!\!\!1 
   \\[-1ex] \mp 1 & \!\!\!\!\!0 \end{array} }\!\!\Big)$ denotes
a generator of $\mathfrak{so}(2)$ or  $\mathfrak{so}(1,1)$, respectively,
and $Q_{x}{}^{\alpha}$ parameterizes  a generically unconstrained block  of six translations.
Thus, generically the gauge group $G_0$ in this case is seven-dimensional, 
namely either $G_0={\rm SO}(2) \ltimes \mathbb{R}^6$ or $G_0={\rm SO}(1,1) \ltimes \mathbb{R}^6$.
The number of independent translations is 
reduced in case the equation
\bea
t_{2}\, Q - Q Z &=&0  \;,
\label{redu}
\eea
has nontrivial solutions $Q$. In this case, the gauge group shrinks to 
$G_0={\rm SO}(2) \ltimes \mathbb{R}^s$ or $G_0={\rm SO}(1,1) \ltimes \mathbb{R}^s$, 
with $s=4,5$.
The scalar potential in this sector can be computed from~\eqref{LMP}
and takes the form
\bea
V &=& \ft1{64}\,
\Big( 2 {\rm Tr}\,[{\hat Y}{}^{2}] - ({\rm Tr}\,\hat Y)^{2} + 
2 \,({\rm det}{\cal M}_{\alpha\beta})\, {\rm Tr}\,[\hat Z^{2}] \Big)
\;,
\label{potSS}
\eea
in terms of the 
matrices $\hat Y_{x}{}^{y}=Y_{xz}{\cal M}^{zy}$
and 
$\hat Z_{\alpha}{}^{\beta}=
Z_{(\alpha}{}^{\gamma}{\cal M}_{\delta)\gamma}{\cal M}^{\delta\beta}$.
Here, ${\cal M}^{xy}$ and ${\cal M}_{\alpha\beta}$ denote the
diagonal blocks of the symmetric unimodular matrix 
defined in~\eqref{DefCalM}, and 
${\cal M}_{\alpha\gamma}{\cal M}^{\gamma\beta}=\delta_{\alpha}^{\beta}$.
Since the matrix  $\hat Y_{x}{}^{y}$ has only two non-vanishing eigenvalues,
this potential is positive definite. In particular, this
implies that $V=0$ is a sufficient condition for a stationary point.
It further follows from~\eqref{potSS} that $V$ only vanishes
for  $\hat Y_{x}{}^{y}\propto \delta_{x}{}^{y}$ and 
$Z_{(\alpha}{}^{\gamma}{\cal M}_{\delta)\gamma}=0$, i.e.\ for
compact choice of $t_{2}$ and $Z$. 
With vanishing $Z$ or vanishing $t_{2}$ one recovers the
Minkowski vacua in the ${\rm CSO}(2,0,3)$ and the ${\rm CSO}(2,0,2)$
theory, respectively, discussed in 
sections~\ref{SecExNoZ} and \ref{SecExNoY} above.

In turn, every compact choice of $t_{2}$ and $Z$ defines a theory with 
a Minkowski vacuum in the potential. The gravitino masses and thereby the 
remaining supersymmetries at this ground state are determined from 
the eigenvalues of $A_{1\,ab}$~\eqref{susy_groundstate2}
according to
\bea
m_{\pm}^{2} &=&   
\frac1{1600}\,\bigg(1\pm \sqrt{-\frac12{\rm Tr}Z^{2}}\;\bigg)^{2} \;.
\eea
Half of the supersymmetry (${\cal N}=2$) is thus preserved iff ${\rm Tr}Z^{2}=-2$.
With~\eqref{redu} one finds that precisely at this value the dimension of the 
gauge group decreases from 7 down to~5; the group then is ${\rm CSO}(2,0,2)$.

All the gaugings in this sector have a well defined higher-dimensional origin,
namely they descend by Scherk-Schwarz reduction~\cite{Scherk:1979zr} 
from the maximal theory in eight dimensions. Indeed, Scherk-Schwarz reduction 
singles out one generator from the ${\rm SL}(2)\times {\rm SL}(3)$
global symmetry group of the eight-dimensional theory~\cite{Salam:1984ft}. 
With the seven-dimensional embedding tensor branching as
\bea
Y:\quad{\bf 15}&\rightarrow& ({\bf 3},{\bf 1})+({\bf 2},{\bf 3})+({\bf 1},{\bf 6}) \;,
\nonumber\\
Z:\quad{\bf \overline{40}}&\rightarrow& ({\bf 1},{\bf \overline{3}})+({\bf 1},{\bf 8})+
({\bf 2},{\bf 1}) +
({\bf 2},{\bf 3})+ ({\bf 2},{\bf \overline{6}})+({\bf 3},{\bf \overline{3}})\;,
\eea
a Scherk-Schwarz gauging corresponds to
switching on components $({\bf 3},{\bf 1})+({\bf 1},{\bf 8})$
in the adjoint representation of~${\rm SL}(2)\times {\rm SL}(3)$.
This precisely amounts to the parameterization in terms of 
matrices $Y_{xy}$, $Z_{\alpha}{}^{\beta}$
introduced above. 
We have seen that for compact choice of $t_{2}$ and $Z$,
the potential~\eqref{potSS} admits a Minkowski ground state
as expected from the Scherk-Schwarz origin.
Moreover, we have shown that for a particular ratio between
the norms of $t_{2}$ and $Z$, this ground state preserves $1/2$
of the supersymmetries.

\mathversion{bold}
\subsection*{$t=1,0$}
\mathversion{normal}
As $t$ becomes smaller, the consequences of the quadratic constraint~\eqref{quadraticZ}
become more involved. We refrain from attempting a complete classification in this sector
and refer to the examples that we have discussed 
in sections~\ref{SecExNoZ} and \ref{SecExNoY} above.

\begin{table}[tb]
   \begin{center}
      \begin{tabular}{c|c|c|c|c|c|c}
      $t$ & $Y_{\Ja\Jb}$ &  $Z^{\ka\kb,\kc}$  & $Z^{x\ka,\kb}$  &\quad gauge group 
      & stat.\ point & susy \\[0.1cm]
\hline
 $5$&$(+\!+\!+\!+\!+)$ &  &  & ${\rm SO}(5)$ & $\times$\,,\,$\times$ & 4\,,\,0 \\[0.2cm]
 $5$&$(+\!+\!+\!+\!-)$ &  &  & ${\rm SO}(4,1)$ & $-$ &  \\[0.2cm]
 $5$&$(+\!+\!+\!-\!-)$ &  &  & ${\rm SO}(3,2)$ & $-$ &  \\[0.2cm]
 \hline
 $4$&$(+\!+\!+\!+\,0)$ &  & & ${\rm CSO}(4,0,1)$ & $-$ &  \\[0.2cm]
 $4$&$(+\!+\!+\!-\,0)$ &  &  & ${\rm CSO}(3,1,1)$ & $-$ &  \\[0.2cm]
 $4$&$(+\!+\!-\!-\,0)$ &  &  & ${\rm CSO}(2,2,1)$ & $-$ &  \\[0.2cm]
\hline
 $3$&$(+\!+\!+\,0\,0)$ &$\epsilon^{\ka\kb}v^{\kc}$ & & ${\rm CSO}(3,0,2)$ & $-$ &  \\[0.2cm]
 $3$&$(+\!+\!-\,0\,0)$ & $\epsilon^{\ka\kb}v^{\kc}$ &  & ${\rm CSO}(2,1,2)$ & $-$ &  \\[0.2cm]
 $3$&$(+\!+\!-\,0\,0)$ & $\epsilon^{\ka\kb}v^{\kc}$ &
\!\! $\frac 1{16}\epsilon^{\kc\ka}(\Sigma^{\la})_\kc{}^\kb$\!
  & ${\rm SO}(2,1)\!\ltimes\! \mathbb{R}^4$ & $-$ &  \\[0.2cm]
 \hline
 $2$&$(+\!+\,0\,0\,0)$ & $\ft18\epsilon^{\ka\kb\kd} {Z_{\kd}}^\kc$ &
  & ${\rm SO}(2) \!\ltimes\! \mathbb{R}^s$ & $\times$ &
 $2\rightarrow0$  \\[0.2cm]
 $2$&$(+\!-\,0\,0\,0)$ & $\ft18\epsilon^{\ka\kb\kd} {Z_{\kd}}^\kc$ &
   & ${\rm SO}(1,1) \!\ltimes\! \mathbb{R}^s$ & $-$ &  \\[0.2cm]
\hline
 $1$&$(+\,0\,0\,0\,0)$ &&& ${\rm CSO}(1,0,4)$ & $-$ &  \\[0.2cm]
\hline
       $0$ &$(0\,0\,0\,0\,0)$  &  $v^{[\alpha} \, w^{\beta]\gamma}$ &
     & ${\rm SO}(p,4\!-\!p)$ 
       & $-$      & \\[0.2cm] 
       $0$ &$(0\,0\,0\,0\,0)$  &  $v^{[\alpha} \, w^{\beta]\gamma}$ &
       & $\begin{array}{c}
       {\rm CSO}(p,q,r)\\[-.5ex]
       (p\!+\!q\!+\!r=4) \end{array}$ 
       & $\begin{array}{c} \times\\[-.5ex](p\!=\!2\!=\!r) \end{array}$      
        & 0
      \end{tabular}
   \end{center}
   \caption{\label{TabSumEx}{\small Examples for gaugings of $D=7$ maximal supergravity.}}
\end{table}

\chapter{The maximal supergravities in $d=2$}
\label{ch:D2}

In this chapter we present the embedding tensor and the bosonic Lagrangian (up to the scalar potential)
of gauged maximal supergravity in two dimensions. Dimensional reduction of gravity and supergravity to two dimensions yields
an effective theory which is an integrable classical theory and whose symmetry group
is infinite dimensional \cite{Belinsky:1971nt,Pohlmeyer:1975nb,Maison:1978es}
(for an introductory presentation we refer to \cite{Nicolai:1991tt}).
For the particular case of $d=2$ maximal ($N=16$)
supergravity the integrability and the symmetry structure are well known \cite{Nicolai:1987kz,Nicolai:1988jb,Nicolai:1998gi}.
The global symmetry group is the affine Lie group $G_0={\rm E}_{9(9)}$ and the infinite tower of dual scalars that
can be introduced onshell arrange
in the coset space ${\rm E}_{9(9)}/K({\rm E}_9)$, where $K({\rm E}_9)$ is the maximal compact
subgroup of ${\rm E}_{9(9)}$. In the next section we apply the general method of the embedding tensor to this particular situation.
We find the embedding tensor and the vector gauge fields to transform in the (dual) basic representation of ${\rm E}_{9(9)}$ and
we work out the quadratic constraint on the embedding tensor. In the second section of this chapter we first introduce the ungauged
maximal $d=2$ supergravity, explain its integrability structure and finally give the bosonic Lagrangian of the gauged theory.
Examples of gaugings include those that originate from torus reduction of higher dimensional supergravity and
the ${\rm SO}(9)$ gauging that descend from a warped sphere reduction of IIA supergravity.

\section{The embedding tensor}

\subsection{Symmetry algebra and basic representation}

The global onshell symmetry group of ungauged $d=2$ maximal supergravity is $G_0 = {\rm E}_{9(9)}$.
The corresponding algebra ${\mathfrak e}_{9(9)}$ is an infinite dimensional
affine Lie algebra or Kac-Moody algebra \cite{Goddard:1986bp,Kac:1990gs}.
In this subsection we give a description of ${\mathfrak e}_{9(9)}$ starting from the finite Lie algebra $\mathfrak{e}_{8(8)}$.
The $\mathfrak{e}_{8(8)}$ generators $t_\alpha$ ($\alpha=1 \ldots 248$) obey 
\begin{align}
   [t_\alpha ,t_\beta ] \, &= \, f_{\alpha\beta}{}^\gamma \, t_\gamma \; ,
\end{align}
with structure constants $f_{\alpha\beta}{}^\gamma$. 
To lower and raise algebra indices $\alpha$ we use the Cartan Killing form 
\begin{align}
   \eta_{\alpha\beta} &= \ft 1 {60} \, f_{\alpha\delta}{}^\gamma \, f_{\beta\gamma}{}^\delta \; , &
   \eta^{\alpha\beta} &= (\eta_{\alpha\beta})^{-1} \; .
\end{align}
We now consider the loop group of ${\rm E}_{8(8)}$.
Its algebra generators are $T^m_\alpha$, $m \in \mathbbm{Z}$, and the commutator reads
\begin{align}
   [ T^m_\alpha , T^n_\beta ] &= f_{\alpha\beta}{}^\gamma \, T^{m+n}_\gamma \; .
\end{align}
This commutator is naturally obtained by introducing a complex spectral parameter~$y$ and
identifying $T^m_\alpha$ with the formal product of the $\mathfrak{e}_{8(8)}$ generators and a power of $y$,
namely $T^m_\alpha \, = \, y^m \, t_\alpha$.
The spectral parameter will be essential later for the description of the linear system
of $d=2$ maximal supergravity. The Lie algebra of ${\rm E}_{9(9)}$ is given by the unique central extension of the above loop algebra
\cite{Goddard:1986bp,Kac:1990gs}. The central element is denoted by $k$. The algebra reads
\begin{align}
   [ T^m_\alpha , T^n_\beta ] &= f_{\alpha\beta}{}^\gamma \, T^{m+n}_\gamma \, + \, k \, m \, \eta_{\alpha\beta} \, \delta^{0,m+n} \; ,
   &
   [ k , T^m_\alpha ] &= 0 \; .
   \label{E9algebra}
\end{align}
We will refer to the $\mathfrak{e}_{8(8)}$ subalgebra spanned by the generators
$T^0_\alpha$ as the zero-mode algebra.
There is a natural action of the Witt-Virasoro algebra on the generators $T^m_\alpha$. 
The Witt-Virasoro algebra has generators $L_{m}$, $m\in\mathbbm{Z}$, and is given by
\begin{align}
   [ L_m, L_n ] &= (m-n) \, L_{m+n} \; .
\end{align}
In terms of the spectral parameter we can identify $L_m = - y^{m+1} \partial_y$ which yields the commutators
\begin{align}
   [ L_m , T^n_\alpha ] &= \, - \, n \,T^{m+n}_\alpha \; , &
   [ L_m , k ] &= 0 \; .
   \label{Lalgebra}
\end{align}
In order to define an invariant non-degenerate inner product on ${\mathfrak e}_{9(9)}$ one needs to pick one of the
generators $L_m$ in addition to $T^m_\alpha$ and $k$. The inner product is then defined on these generators and is
invariant under their action, but it is not invariant under the action of the remaining Witt-Virasoro generators.
Usually one chooses $L_0$, but it is crucial for our construction
to choose $L_{1}$. The reason is that for a generic gauging we are going to discuss $L_{1}$ becomes a generator of the gauge group 
while $L_0$ remains ungauged. For gauge invariance we therefore need an inner product which is invariant under $L_{1}$.
It is given by
\begin{align}
   (T^m_\alpha , T^n_\beta) &= \eta_{\alpha\beta} \delta^{m+n,1} \; , &
   (L_{1} , k ) &= - 1 \; , && \text{all others zero.} 
   \label{DefInvForm}
\end{align}
One can easily check that this inner product is indeed invariant under $T^m_\alpha$, $k$ and $L_{1}$ action\footnote{
The value of $(L_1,k)$ is fixed by the invariance condition
           $([T^m_\alpha,L_1],T^n_\beta)+(L_1,[T^m_\alpha,T^n_\beta])=0$.}.

Having thus defined the global symmetry algebra we now turn to its representation theory.
An irreducible representation of ${\mathfrak e}_{9(9)}$ is first of all characterized by
its level, i.e.\ by its value of $k$.
For the adjoint representation defined by the commutators \eqref{E9algebra} we have $k=0$.\footnote{Note that the
adjoint and the co-adjoint representation are not equivalent here.} 
The adjoint representation is not a highest weight representation,
in contrast to the case of finite dimensional Lie algebras. Highest weight representations
of affine Lie algebras only exist for $k>0$, and in fact the possible values for $k$ are quantized
\cite{Goddard:1986bp,Kac:1990gs}, in our case $k\in\mathbbm{N}$.
For ${\rm E}_{9(9)}$ there is a unique highest weight representation with $k=1$ called the basic
representation\footnote{For larger values of $k$ there more highest-weight representation.}.
The embedding tensor and the vector fields of $d=2$ maximal supergravity transform in the (dual) basic representation.
It is therefore this particular highest weight representation
which we need to understand in detail.

In addition to the value of $k$ one needs to specify an irreducible vacuum representation of the zero-mode algebra
in order to completely determine the highest weight representation of an affine Lie algebra.
The vacuum representation is annihilated 
by $T^m_\alpha$, $m<0$, and all other states can be constructed via the action of $T^m_\alpha$, $m>0$. 
In case of the basic representation the vacuum state is a singlet of ${\rm E}_{8(8)}$. 
We therefore denote the basic representation by $Q^1_{\bf 1}$, where the superscript refers to the value of $k$
and the subscript indicates the vacuum representation.
The $L_0$ grading of the algebra generators $T^m_\alpha$ ($m=0$ zero modes, $m=1$ first level, etc.) 
carries over to the basic representation. We choose $L_0=0$ for the vacuum singlet.
Acting with $T^1_\alpha$ yields a ${\bf 248}$ representation of ${\rm E}_{8(8)}$ at level $L_0=1$.
Acting again with $T^1_\alpha$ yields several irreducible components at level $L_0=2$, namely
a singlet ${\bf 1}$, a ${\bf 248}$ and a ${\bf 3875}$.
The decomposition of $Q^1_{\bf 1}$ under ${\rm E}_{8(8)}$ yields an infinite number of irreducible
${\rm E}_{8(8)}$ representations, but finitely many for each $L_0$ level. For $L_0 \leq 6$ these ${\rm E}_{8(8)}$ irreducible
components and their multiplicities are listed in table~\ref{BasicComp}.

\begin{table}[tb]
   \begin{center}
     \begin{tabular}{c|l@{$\quad$}l@{$\quad$}l@{$\quad$}l@{$\quad$}l@{$\quad$}l@{$\quad$}l@{$\quad$}l}
        $L_0$ &  \multicolumn{8}{c}{${\rm E}_{8(8)}$ representations} \\ \hline
         $0$   & ${\bf 1}$ \\
	 $1$   & & ${\bf 248}$ \\
	 $2$   & ${\bf 1}$ & ${\bf 248}$ & ${\bf 3875}$ \\
	 $3$   & ${\bf 1}$ & ${\bf 248}_2$ & ${\bf 3875}$ & ${\bf 30380}$ \\
	 $4$   & ${\bf 1}_2$ & ${\bf 248}_3$ & ${\bf 3875}_2$ & ${\bf 30380}$ & ${\bf 27000}$ & ${\bf 147250}$  \\
	 $5$   & ${\bf 1}_2$ & ${\bf 248}_5$ & ${\bf 3875}_3$ & ${\bf 30380}_3$ & ${\bf 27000}$ & ${\bf 147250}$ & ${\bf 779247}$ \\
	 $6$   & ${\bf 1}_4$ & ${\bf 248}_7$ & ${\bf 3875}_6$ & ${\bf 30380}_4$ & ${\bf 27000}_3$ & ${\bf 147250}_2$ & ${\bf 779247}_2$
	                & ${\bf 2450240}$ 
     \end{tabular}
     \caption{\label{BasicComp} \small Decomposition of the ${\rm E}_{9(9)}$ basic representation 
                     into irreducible ${\rm E}_{8(8)}$ components. These ${\rm E}_{8(8)}$ representations
		     have a natural $L_0$ grading. The total number of components is infinite,
		     here we only list those with $L_0\leq 6$. The subscripts indicate the multiplicity of
		     the representations. }
   \end{center}     
\end{table}

For the lowest $L_0$ levels we now explicitly give the action of $\mathfrak{e}_{9(9)}$ on the basic representation.
We denote by $X^l$, $Y^l_\alpha$ and $Z^l_{\alpha\beta}$
the ${\bf 1}$, ${\bf 248}$ and ${\bf 3875}$ component,
at level $L_0=l=0,1,2$.
Since $Z^l_{\alpha\beta}$ only contains a ${\bf 3875}$ representation it obeys
\begin{align}
   Z^l_{\alpha\beta}&=Z^l_{(\alpha\beta)} \; , &
   \eta^{\alpha\beta} \, Z^l_{\alpha\beta} &= 0 \; , &
   f_{\epsilon\gamma}{}^{\alpha} \, f_{\delta}{}^{\beta\epsilon} \, Z^l_{\alpha\beta} &= 2 \, Z^l_{\gamma\delta} \; .
\end{align}
The compact version of these equations is $({\cal P}_{3875})_{\alpha\beta}{}^{\gamma\delta} Z^l_{\gamma\delta} = Z^l_{\gamma\delta}$,
where the projector is given by \cite{Koepsell:1999uj}
\begin{align}
   ({\cal P}_{3875})_{\alpha\beta}{}^{\gamma\delta}
      &= \ft 1 7 \, \delta_{(\alpha}^{\gamma} \delta_{\beta)}^{\delta}
         - \ft 1 {56} \eta_{\alpha\beta} \eta^{\gamma\delta} - \ft 1 4 f_{\epsilon(\alpha}{}^{\gamma} f_{\beta)}{}^{\delta\epsilon}
	  \; .
\end{align}
We denote the action of $T^m_\alpha$ and $L_{m}$ by $\delta^m_\alpha$ and $\delta^m_L$, respectively\footnote{
Note the sign in the general relation $[\delta_X,\delta_Y]=-\delta_{[X,Y]}$.}. We have the zero mode action
\begin{align}
   \delta^0_\alpha X^l &= 0  \; ,&
   \delta^0_\alpha Y^l_\beta &= - f_{\alpha\beta}{}^\gamma Y^l_\gamma \; ,&
   &\text{etc.}
   \label{ActionBasic1}
\end{align}
The action of the positive level generators reads
\begin{align}
   \delta^1_\alpha X^0 &= Y^1_\alpha \; , &
   \delta^1_\alpha Y^1_\beta &= \eta_{\alpha\beta} \, X^2 
                              + f_{\alpha\beta}{}^\gamma \, Y^2_\gamma
			      + Z^2_{\alpha\beta} \; ,  \nonumber \\
   \delta^2_\alpha X^0 &= - 2 Y^2_\alpha \; , &
   &\text{etc.}
   \label{ActionBasic2}
\end{align}
The negative level generators act as
\begin{align}
   \delta_\alpha^{-1} X^2 &= \ft 1 4 \, Y^1_\alpha \; , &
   \delta_\alpha^{-1} Y^1_\beta &= \eta_{\alpha\beta} \, X^0 \; ,
   \nonumber \\
   \delta_\alpha^{-1} Y^2_\beta &= \ft 1 2 \, f_{\alpha\beta}{}^\gamma \, Y^1_\gamma \; , &
   \delta_\alpha^{-1} Z^2_{\beta\gamma} &= 14 \, ({\cal P}_{3875})_{\beta\gamma\alpha}{}^\delta \, Y_\delta^{1} \; ,
   \nonumber \\
   \delta_\alpha^{-2} Y^2_\beta &= - \eta_{\alpha\beta} \, X^0 \, , &   
   &\text{etc.}
   \label{ActionBasic3}
\end{align}
Finally, the action of the Witt-Virasoro generators is given by
\begin{align}
   \delta^0_L X^l &= l \, X^l \; , &
   \delta^2_L X^0 &= 4 X^2 \; , &
   \delta^{-2}_L X^2 &= 0 \; , \nonumber \\
   \delta^0_L Y^l_\alpha &= l \, Y^l_\alpha \; , &
   \delta^1_L Y^1_\alpha &= -2 Y^2_\alpha \; , &  
   \delta^{-1}_L Y^2_\alpha &= - Y^1_\alpha \; , & 
   &\text{etc.}
   \label{ActionBasic4}
\end{align}
We thus explicitly gave the symmetry action for levels $l \leq 2$ of $Q^1_{\bf 1}$.
These levels will be found to be already sufficient to explain the gaugings that originate from torus reductions
of higher-dimensional gauged maximal supergravity.

\subsection{Vector gauge fields}
\label{sec:VecFieldsD2}

In order to apply the general formalism of chapter \ref{ch:GenGauged} to the description of gauged maximal 
supergravities in $d=2$ we first need to introduce vector gauge fields in a particular representation of $G_0={\rm E}_{9(9)}$.
But in contrast to all higher dimensions in $d=2$ it is a priori not clear which is the appropriate $G_0$-representation.
For $d>4$ the vector fields
already appear in the Lagrangian ungauged theory. 
For $d=4$ only the electric vector fields appear in the Lagrangian, but only together with their dual
magnetic vector fields they form a representation under the global symmetry group $G_0$. Therefore, for the gauged theory
also the magnetic vector fields are introduced in the Lagrangian.
For $d=3$ the ungauged theory can be formulated entirely in
terms of scalars, but in the gauged theory also their dual vector fields are introduced
as gauge fields in the Lagrangian \cite{Nicolai:2001sv,Nicolai:2001ac,deWit:2003ja}. 
In $d=2$ the reasoning for a particular vector field representation is less direct, but it is the
basic representation of ${\rm E}_{9(9)}$ which is the appropriate choice. More precisely, in
our conventions it is the dual of the basic representation which we choose as vector field representation.

Eventually, it is the consistency of the general gauged theory which justifies this choice.
But this choice is well motivated from extrapolation of the representation theory of the higher
dimensional maximal supergravities. In figure \ref{Fig:Dynkin} we gave the Dynkin diagrams for the
global symmetry groups $G_0={\rm E}_{11-d(11-d)}$ of maximal supergravities in $2\leq d \leq 8$.
For $d=8$ the Dynkin diagram has three knots and with each decreasing dimension one additional knot appears,
i.e.\ there is one additional ``new'' simple root in the root lattice of $G_0$. A highest weight representation of
$G_0$ is uniquely characterized by the inner products of the corresponding highest weight with the simple roots.
These inner products need to be non-negative integer numbers.
It turns out that the vector field representations in dimensions $3\leq d \leq 7$ are always those highest weight
representations whose highest weight has vanishing inner product with all simple roots that were already present
for the respective higher dimensions and whose inner product with the ``new'' simple root equals one. Extrapolating this rule
to $d=2$ yields $Q^1_{\bf 1}$ as the vector field representation.

The basic representation is infinite dimensional and it may seem strange to introduce an infinite number of vector gauge fields
in the Lagrangian of the gauged theory. But these vector gauge fields  will only appear
projected with the embedding tensor
and for any particular gauging, i.e.\ for any particular valid embedding tensor, only a finite subset of the
gauge fields will enter the Lagrangian. 

When considering dimensional reductions to $d=2$ maximal supergravity one can identify the higher dimensional
origin of some of the vector fields in the basic representation.
Let us consider as a
first example the circle reduction from $d=3$ gauged supergravity. The three dimensional
global symmetry group is embedded as the zero mode ${\rm E}_{8(8)}$ into the $d=2$ symmetry group ${\rm E}_{9(9)}$.
The decomposition of the basic representation under this zero mode ${\rm E}_{8(8)}$ was given in table \ref{BasicComp}.
We identify the highest singlet, i.e.\ the vacuum representation, with the Kaluza-Klein vector field and 
the ${\bf 248}$ at level $L_0=1$ with the vector fields that were already present in $d=3$.
The other irreducible ${\rm E}_{8(8)}$ components of $Q^1_{\bf 1}$ do not have an
immediate interpretation in this context. However, it is also natural to consider
the vector fields in the ${\bf 1}$ and ${\bf 3875}$ at level $L_0=2$ of $Q^1_{\bf 1}$
as originating from two-form gauge fields in $d=3$.

As a second example let us consider the torus reduction from $d=11$ supergravity. To identify the higher dimensional
origin of the gauge fields we should first identify the ${\rm SL}(9)$ symmetry group 
of the internal torus as a subgroup of the global symmetry group ${\rm E}_{9(9)}$. 
One can embed ${\rm SL}(9)$ already in the zero-mode ${\rm E}_{8(8)}$ but this is not the embedding we are searching for
since the zero-mode ${\rm E}_{8(8)}$ is the symmetry group of $d=3$ maximal supergravity and the torus ${\rm SL}(9)$
can not be realized in $d=3$ already.
However, it is possible to embed ${\rm SL}(9)$ into ${\rm E}_{9(9)}$ such that it is
not contained in
any ${\rm E}_{8(8)}$ subgroup of ${\rm E}_{9(9)}$. In order to construct this
embedding we first decompose the algebra $\mathfrak{e}_{9(9)}$ under the zero mode ${\rm E}_{8(8)}$. This yields
one ${\bf 248}$ representation for each  $L_0$ level as depicted in figure \ref{Fig:AlgD2b}.
We indicate the level, i.e.\ the charge under $L_0$, as a subscript,
i.e.\ we have ${\bf 248}_l$, $l\in\mathbbm{Z}$.
We decompose these ${\rm E}_{8(8)}$ representations further under a $\mathbbm{R}^+ \times {\rm SL}(8)$
subgroup of ${\rm E}_{8(8)}$. They branch as (see also table \ref{Fig:AlgD3})
\begin{align}
   {\bf 248}_l \, & \rightarrow \, \overline {\bf 8}_{+3,l} \,\oplus\, {\bf 28}_{+2,l} \,\oplus\, \overline {\bf 56}_{+1,l}
                             \,\oplus\, {\bf 1}_{0,l} \,\oplus\, {\bf 63}_{0,l} \,\oplus\, {\bf 56}_{-1,l} \,\oplus\, 
			     \overline {\bf 28}_{-2,l} \,\oplus\, {\bf 8}_{-3,l} \; ,
\end{align}
where the first subscript indicates the $\mathbbm{R}^+$ charges which in the following is denoted by $q$.
The ${\rm SL}(9)$ algebra which is contained
in the zero modes is composed out of
$\overline {\bf 8}_{+3,0} \,\oplus\, {\bf 1}_{0,0} \,\oplus\, {\bf 63}_{0,0}\,\oplus\, {\bf 8}_{-3,0}$.
The ${\rm SL}(9)$ algebra which corresponds to the torus symmetry is given by
\begin{align}
   \mathfrak{sl}(9) &= \overline {\bf 8}_{+3,-1} \,\oplus\, {\bf 1}_{0,0} \,\oplus\, {\bf 63}_{0,0}\,\oplus\, {\bf 8}_{-3,+1} \;.
\end{align}
We have chosen those ${\rm SL}(8)$ representations for which $q+3l=0$.
This guarantees that the generators form a closed subalgebra of $\mathfrak{e}_{9(9)}$. The linear combination $q+3l$
also gives a natural grading for all other components of the decomposition.
At each level in this new grading the above components form an irreducible ${\rm SL}(9)$ representation, namely
\begin{align}
   \overline {\bf 84}_{n} &= \overline {\bf 56}_{+1,(n+2)/3} \,\oplus\, \overline {\bf 28}_{-2,(n+5)/3} \; , \nonumber \\
   {\bf 80}_n &= \overline {\bf 8}_{+3,(n-3)/3} \,\oplus\, {\bf 1}_{0,n/3} \,\oplus\, {\bf 63}_{0,n/3}\,\oplus\, {\bf 8}_{-3,(n+3)/3} \; , \nonumber \\
   {\bf 84}_{n} &=  {\bf 28}_{+2,(n-2)/3} \,\oplus\, {\bf 56}_{-1,(n+1)/3} \; ,
\end{align}
where $n=q+3l$.
We have thus described the decomposition of $\mathfrak{e}_{9(9)}$ under the torus ${\rm SL}(9)$ which we already
anticipated in figure \ref{Fig:AlgD2a}. The tower of ${\bf 80}$ representations constitutes an
infinite dimensional subgroup of ${\rm E}_{9(9)}$,
namely the affine extension $\widehat{\rm SL}(9)$ of ${\rm SL}(9)$.

In order to decompose the basic representation under this torus ${\rm SL}(9)$ it is convenient to first decompose 
under $\widehat{\rm SL}(9)$ because one then still remains with a finite number of irreducible components.
Concretely, the basic representation of ${\rm E}_{9(9)}$
decomposes into three highest weight representations of $\widehat{\rm SL}(9)$ (this can be inferred, for example,
from the decompositions given in \cite{Kac:1988iu}).
They all have $k=1$ and the vacuum ${\rm SL}(9)$ representations are ${\bf 9}$, $\overline {\bf 36}$
and ${\bf 126}$ with $q+3l$ charges $0$, $1$ and $2$. The decomposition of these three irreducible $\widehat {\rm SL}(9)$
representation under ${\rm SL}(9)$ is given in table \ref{SL9hatRep1} for the first few $q+3l$ levels. 

\begin{table}[tb]
   \small
     \begin{tabular}{c|l@{$\quad$}l@{$\quad$}l@{$\quad$}l@{$\quad$}l@{$\quad$}l}
        $q+3l$ &  \multicolumn{6}{c}{${\rm SL}(9)$ representations} \\ \hline  & \\[-0.4cm] 
         $0$   & ${\bf 9}$ \\
	 $3$   & ${\bf 9}$   & $\overline{\bf 315}$ \\
	 $6$   & ${\bf 9}_2$ & $\overline{\bf 315}_2$ & ${\bf 396}$ & $\overline {\bf 2700}$ \\ 
	 $12$   & ${\bf 9}_4$ & $\overline{\bf 315}_5$ & ${\bf 396}_2$ & $\overline {\bf 2700}_2$ & $\overline {\bf 3465}$
	                    & $\overline {\bf 7560}$ 
     \end{tabular}
     \\[0.3cm]
     \begin{tabular}{c|l@{$\quad$}l@{$\quad$}l@{$\quad$}l@{$\quad$}l@{$\quad$}l@{$\quad$}l@{$\quad$}l}
        $q+3l$ &  \multicolumn{8}{c}{${\rm SL}(9)$ representations} \\ \hline & \\[-0.4cm] 
         $1$   & $\overline {\bf 36}$ \\
         $4$   & $\overline {\bf 36}$ & $\overline {\bf 45}$ & ${\bf 720}$ \\
         $7$   & $\overline {\bf 36}_3$ & $\overline {\bf 45}$ & ${\bf 720}_2$ 
	              & $\overline {\bf 2079}$ & ${\bf 3780}$  \\
         $13$   & $\overline {\bf 36}_5$ & $\overline {\bf 45}_3$ & ${\bf 720}_5$ 
	              & $\overline {\bf 2079}_3$ & ${\bf 3780}_2$ & ${\bf 4950}$ & ${\bf 6048}$ & ${\bf 19800}$ \\
     \end{tabular}
     \\[0.3cm]
     \begin{tabular}{c|l@{$\quad$}l@{$\quad$}l@{$\quad$}l@{$\quad$}l@{$\quad$}l@{$\quad$}l@{$\quad$}l@{$\quad$}l@{$\quad$}l}
        $q+3l$ &  \multicolumn{10}{c}{${\rm SL}(9)$ representations} \\ \hline & \\[-0.4cm] 
         $2$   & ${\bf 126}$ \\
         $5$   & ${\bf 126}$ & ${\bf 630}$ & $\overline {\bf 1008}$ \\
         $8$   & ${\bf 126}_3$ & ${\bf 630}_2$ & $\overline {\bf 1008}_2$ & ${\bf 540}$ & $\overline {\bf 1890}$
	         & ${\bf 8316}$ \\
         $14$   & ${\bf 126}_6$ & ${\bf 630}_5$ & $\overline {\bf 1008}_5$ & ${\bf 540}$ & $\overline {\bf 1890}_2$
	         & ${\bf 8316}_3$ & ${\bf 990}$ & $\overline {\bf 2772}$ & $\overline {\bf 15840}$ 
		 & ${\bf 27720}$\\
     \end{tabular}
   \begin{center}
     \caption{\label{SL9hatRep1} \small The three $k=1$ irreducible $\widehat {\rm SL}(9)$
                     representation that are contained in the basic representation
		     are further decomposed under ${\rm SL}(9)$.
		     The ${\rm SL}(9)$ representations
		     have a natural $q+3l$ grading (see the main text).
		     We only list those with $q+3l\leq 14$. The subscripts indicate the multiplicity of
		     the representations. }
   \end{center}     
\end{table}

The three vacuum representations are naturally identified as the Kaluza-Klein vector (the ${\bf 9}$),
the vector fields that originate from the three-form (the $\overline {\bf 36}$) and 
the vector fields coming from the dual six-form (the ${\bf 126}$) in $d=11$. For the remaining ${\rm SL}(9)$
representations there is no natural interpretation in the context of a torus reduction from $d=11$. But we have
identified a higher dimensional origin for different vector fields than in the reduction from
$d=3$. On the way from table \ref{BasicComp} to table \ref{SL9hatRep1} a highly non-trivial re-ordering of the states
of the basic representation takes place. For example the vacuum $\overline {\bf 36}$ and ${\bf 126}$ representation
in table \ref{SL9hatRep1} are composed out of states in the ${\bf 248}$ representation
at $L_0=1$ and the ${\bf 3875}$ representation at $L_0=2$ in table \ref{BasicComp}.\footnote{One needs to decompose
all ${\rm SL}(9)$ and ${\rm E}_{8(8)}$ representations under the common subgroup ${\rm SL}(8)$
in order to make the mapping between table \ref{BasicComp} and \ref{SL9hatRep1} explicit.}
Consideration of more complicated dimensional reductions might disclose a higher dimensional
origin of other vector fields within the basic representation. 

\subsection{Linear and quadratic constraint}

We now apply the general methods of section \ref{sec:gen:emb} to the case of $d=2$ maximal supergravity.
First, we introduce indices $\cAa$ and $\cMa$ for the adjoint and for the basic representation
of ${\rm E}_{9(9)}$, respectively. The vector fields in the dual basic representation are thus denoted by $A_\mua^{\cMa}$.
Since the gauge algebra may generically contain the $k$ and $L_{1}$ generators we want the
index $\cAa$ to also take corresponding values, i.e.\ we have symmetry generator
\begin{align}
  T_\cAa &= \{ T_{(m)\alpha}, T_{(0)k} , T_{(1)L} \} = \{ T^m_\alpha, k , L_{1} \} \; .
  \label{DefcAa}
\end{align}  
In equation \eqref{DefInvForm} we defined an invariant non-degenerate inner product 
$\eta_{\cAa\cAb} = (T_\cAa,T_\cAb)$ on these generators.
This also induces an inner product $\eta^{\cAa\cAb}$ on the dual adjoint representation.
In addition, we have structure constants $f_{\cAa\cAb}{}^\cAc$ and
generators $T_{\cAa{\cMa}}{}^{\cMb}$ that describe the symmetry action on the basic representation.

The general covariant derivative \eqref{GenCovDiv} takes the form
\begin{align}
   D_\mua &=  \partial_\mua - g \, A_\mua^\cMa \, \Theta_{\cMa}{}^\cAa \, T_\cAa \; .
\end{align}
The embedding tensor $\Theta_{\cMa}{}^\cAa$ transforms in the tensor product of the basic ($k=1$) and
the adjoint ($k=0$) representation. Since the latter is not a highest weight representation also 
the irreducible components of $\Theta_{\cMa}{}^\cAa$ are in general not highest weight. However,
by the linear constraint we demand $\Theta_{\cMa}{}^\cAa$ to only contain a single irreducible component
$\Theta_{\cMa}$ which transforms in the basic representation, explicitly
\begin{align}
   \Theta_{\cMa}{}^\cAa  &= \eta^{\cAa\cAb} \, T_{\cAb\cMa}{}^{\cMb} \, \Theta_{\cMb} \; .
   \label{LconD2}
\end{align}
This linear constraint is justified by the fact that all known gaugings
are incorporated in $\Theta_{\cMb}$, as will be explained below.
In terms of $\Theta_{\cMa}$ the covariant derivative reads
\begin{align}
   D_\mua &=  \partial_\mua - g \,  \eta^{\cAa\cAb} \, T_{\cAa\cMa}{}^{\cMb} \, A_\mua^\cMa  \, \Theta_{\cMb} \, T_\cAb \; .
   \label{CovDivD2}
\end{align}
The quadratic constraint \eqref{QconGen1} then takes the form
\begin{align}
   \eta^{\cAa\cAb} \, T_{\cAa\cMa}{}^{\cMc} \, T_{\cAb\cMb}{}^{\cMd} \, \Theta_\cMc \, \Theta_{\cMd} &= 0 \; .
   \label{QConD2}
\end{align}
We now give this quadratic constraint for the case that only the first $L_0$-components of $\Theta_\cMc$ are
non-vanishing.
The symmetry action on the components $X^l$, $Y^{l}_\alpha$ and $Z^{l}_{\alpha\beta}$ of the basic representation were given
in equation \eqref{ActionBasic1} to \eqref{ActionBasic4} for levels $l\leq 2$. These tensors are now
regarded as components of $\Theta_\cMc$ and we consider the case
\begin{align}
   \Theta_\cMc &= \{  X^0, \; Y^1_\alpha , \; X^2 , \; Z^2_{\alpha\beta} \} \; ,  &&
   \text{all other components zero.}   
\end{align}
The quadratic constraint \eqref{QConD2} then takes the form\footnote{
From $\delta_\cAa \Theta_\cMa = T_{\cAa\cMa}{}^{\cMb} \Theta_\cMb$ and relation \eqref{ActionBasic1} to \eqref{ActionBasic4}
we find $T_{(1)\alpha,(0)X}{}^{(1)Y\beta} = \delta_\alpha^\beta$, $T_{(1)\alpha,(1)Y\beta}{}^{(2)X} = \eta_{\alpha\beta}$,
etc. In addition, we find from \eqref{DefInvForm} that
$\eta^{(m)\alpha,(n)\beta}=\delta^{m+n,1} \eta^{\alpha\beta}$, $\eta^{(1)L,(0)k}=-1$.
Thus, we can evaluate the quadratic constraint \eqref{QConD2} explicitly on the lowest components of $\Theta_\cMa$.}
\begin{align}
   Y^1_{\alpha} \, f^\alpha{}_{(\beta}{}^\delta \, Z^2_{\gamma)\delta} &= 0 \; , &
   (\eta_{\alpha\beta} X^2 + Z^2_{\alpha\beta}) f^\beta{}_{(\gamma}{}^\rho \, Z^2_{\delta)\rho} &= 0  \; .
   \label{QConD2Expl}
\end{align}
For $Y^1_{\alpha} = 0$ the last equation coincides with the quadratic constraint of $d=3$ maximal supergravity
\cite{Nicolai:2000sc,Nicolai:2001sv} if one identifies the $d=3$ embedding tensor as 
\begin{align}
   \Theta^{\text{d=3}}_{\alpha\beta}  = \eta_{\alpha\beta} X^2 + Z^2_{\alpha\beta} \; .
\end{align}
This means that a simple torus reduction of gauged $d=3$ maximal supergravity yields a gauged $d=2$ maximal supergravity
with non-vanishing components $X^2$ and $Z^2_{\alpha\beta}$ of $\Theta_\cMa$ and all other components zero.
Such a torus reduction can also be twisted, i.e.\ a generator of the $d=3$ symmetry group ${\rm E}_{8(8)}$ can be
chosen as Scherk-Schwarz generator. This leads to an additional gauging in $d=2$ and the Scherk-Schwarz generator 
is described by the
component $Y^1_\alpha$ of $\Theta_\cMa$. If the $d=3$ theory is already gauged before the Scherk-Schwarz reduction
there is a consistency condition between the $d=3$ embedding tensor and the Scherk-Schwarz generator given by the
first equation of \eqref{QConD2Expl}. If the $d=3$ theory is ungauged, i.e.\ $\Theta^{\text{d=3}}_{\alpha\beta}=0$,
an arbitrary Scherk-Schwarz generator $Y^1_\alpha$ can be chosen. In the next section we will explain that there
is an additional freedom in a $d=3$ to $d=2$ torus reduction, namely the field strength of the Kaluza-Klein vector field
can be chosen to be non-vanishing (see equation \eqref{AbackgroundD2} below).
This corresponds to the component $X^0$ of $\Theta_\cMa$ to be switched on and
according to \eqref{QConD2Expl} there is no consistency condition on the choice of $X^0$.
Thus, we have motivated the linear constraint \eqref{LconD2} by identifying
the higher-dimensional origin of the components
$X^0$, $Y^1_\alpha$, $X^2$ and $Z^{2}_{\alpha\beta}$ of $\Theta_\cMa$.
The analogous consideration for torus reductions from $d+1$ to $d\geq3$
dimensions can be found in appendix \ref{app:DimRedEmb}.
The only non-generic feature in the $d=3$ to $d=2$ reduction
is the additional freedom of choosing $X^0$.

\begin{table}[tbp]
   \begin{center}
     \begin{small}
     \begin{tabular}{l c||c|c|c|c|c@{\quad}c}
         & \multicolumn{6}{c}{\normalsize ~\qquad\qquad~vector fields}
	 \\[0.2cm]
	 && ${\bf 1}_{0}$ & ${\bf 248}_{+1}$ & ${\bf 1}_{+2}$ & ${\bf 248}_{+2}$ & ${\bf 3875}_{+2}$ & \ldots  \\ 
	      \cline{2-8} &&&\\[-0.41cm] \cline{2-8}
	 &&&&& \\[-0.2cm]
	 \multirow{8}{0.4cm}{\begin{sideways} \normalsize symmetry generators ~~~~~~~~~~~~ \end{sideways}} 
	 & $\vdots$ & &&&
        \\[0.3cm] 
	 & ${\bf 248}_{+3}$ & --- & --- &  --- &  ${\bf 1}_{0}$ &  --- & \ldots  
        \\[0.2cm] \cline{2-8} &&&&&& \\[-0.3cm]
	 & ${\bf 248}_{+2}$ & --- & ${\bf 1}_{0}$ &  ${\bf 248}_{+1}$ &  ${\bf 248}_{+1}$ &  ${\bf 248}_{+1}$ & \ldots  
        \\[0.2cm] \cline{2-8} &&&&&& \\[-0.3cm]
	 & ${\bf 248}_{+1}$ & --- & ${\bf 248}_{+1}$ & ${\bf 248}_{+2}$ & ${\bf 1}_{+2}\oplus{\bf 248}_{+2}$
	                          & ${\bf 248}_{+2}\oplus{\bf 3875}_{+2}$ & \ldots  \\
	 &  & & & & $\oplus{\bf 3875}_{+2}$	                          
	 \\[0.4cm]
	 & $L_{+1}$ &  ${\bf 1}_0$ & ${\bf 248}_{+1}$ & ${\bf 1}_{+2}$ & ${\bf 248}_{+2}$ & ${\bf 3875}_{+2}$ & \ldots 
        \\[0.2cm] \cline{2-8} &&&&&& \\[-0.3cm]
	 & $k_{0}$ & --- & ${\bf 248}_{+2}$ & ${\bf 1}_{+3}$ & ${\bf 248}^b_{+3}$ & ${\bf 3875}_{+2}$ & \ldots 
	\\[0.4cm]
	 & ${\bf 248}_{0}$ & ${\bf 248}_{+1}$ & ${\bf 1}_{+2}\oplus{\bf 248}_{+2}$ & ${\bf 248}^a_{+3}$
	     & ${\bf 1}_{+3}\oplus{\bf 248}^{a+b}_{+3}$ & ${\bf 248}^{a}_{+3} \oplus {\bf 3875}_{+3}$ & \ldots  \\
	 &  &  & $\oplus{\bf 3875}_{+2}$ && $\oplus{\bf 3875}_{+3}$ & $\oplus{\bf 30380}_{+3}$\\
	 &  &  &  & & $\oplus{\bf 30380}_{+3}$ 
        \\[0.2cm] \cline{2-8} &&&&&& \\[-0.3cm]
	 & ${\bf 248}_{-1}$ &  ${\bf 248}_{+2}$ & ${\bf 1}_{+3}\oplus{\bf 248}^{a+b}_{+3} $
	                                             & \ldots & \ldots & \ldots & \ldots 
	 \\
	 &  &  & $\oplus {\bf 3875}_{+3}$  &&
	 \\
	 &  &  & $\oplus{\bf 30380}_{+3}$  &&
	\\[0.3cm]
	 & $\vdots$ & &&&
     \end{tabular}
     \end{small}
     \caption{\label{ThetaD2D3}\small The $d=2$ couplings of the vector fields $A_\mua^\cMa$ to the symmetry generators $T_\cAa$                                      
                                      are decomposed under the $d=3$ symmetry group ${\rm E}_{8(8)}$.
				      The representations inside the table denote the components
				      of the embedding tensor $\Theta_\cMa$.
				      $A_\mua^\cMa$ and $\Theta_\cMa$ transform in the
				      (dual) basic representation which branches according to table \ref{BasicComp}.
				      The adjoint representation branches into
				      an infinite tower of ${\bf 248}$ representations, see table \ref{Fig:AlgD2b}.
				      In addition we have
				      the symmetry generators $k$ and $L_1$. The subscripts in the table denote the
				      charges under $L_0$, which corresponds to the rescalings of the internal circle.
				      For each entry the $L_0$ charges of $\Theta_\cMa$ and
				      $T_\cAa$ add up to $1$ plus the $L_0$ charge of $A_\mua^\cMa$
				      (this charge convention slightly differs from
				      the one used in appendix \ref{app:DimRedEmb}). The two ${\bf 248}$
				      representation at level $L_0=3$ of the basic representation are
				      distinguished by superscripts $a$ and $b$.}
   \end{center}
\end{table}   				      

We already explained in section \ref{sec:MaxTorus} that the positive mode generators $T^m_\alpha$, $m>0$, of the 
Kac-Moody symmetry algebra correspond to the shift-symmetries of an infinite tower of dual scalars that can be introduced
in $d=2$ maximal supergravity. In contrast, the existence of the negative mode generators $T^m_\alpha$, $m<0$, corresponds to a
non-trivial symmetry enhancement and the generators act non-linearly on the infinite tower of scalars. We will make this explicit
in the next section when we introduce a linear system in order to give the complete symmetry action. In gauged $d=2$
theories that originate from dimensional reduction of $d=3$ maximal supergravity
one expects scalar shift symmetries $T^m_\alpha$, $m>0$, to be gauged, but the extended symmetry generators $T^m_\alpha$, $m<0$,
should not contribute to the gauge algebra.

Which vector fields $A_\mua^\cMa$ are coupled to which symmetry generators $T_\cAa$ by the various components of
$\Theta_{\cMa}$ is shown in table \ref{ThetaD2D3}.
We find that the Kaluza Klein vector field ${\bf 1}_0$ is coupled
to the zero-mode generators ${\bf 248}_0$ by the ${\bf 248}_{+1}$ (i.e.\ $Y^1_\alpha$) component of $\Theta_\cMa$.
The vector fields ${\bf 248}_{+1}$ that were already present in $d=2$ are coupled to the zero-mode generators
${\bf 248}_0$ by the ${\bf 1}_{+2}$ and ${\bf 3875}_{+2}$ (i.e.\ $X^2$ and $Z^2_{\alpha\beta}$) component of $\Theta_\cMa$.
This agrees with the above identification of $Y^1_\alpha$,  $X^2$ and $Z^2_{\alpha\beta}$ as the Scherk-Schwarz
generator and the components of the $d=3$ embedding tensor, respectively.
As expected, we also find from table \ref{ThetaD2D3} that no
negative mode generators $T^m_\alpha$, $m<0$, are excited by these components of $\Theta_\cMa$.
In contrast, the ${\bf 248}_{+2}$ (i.e.\ $Y^2_\alpha$) component of $\Theta_\cMa$ describes gaugings whose algebra also contains
the ${\bf 248}_{-1}$ (i.e.\ $T^{-1}_\alpha$) generators. Therefore, these gaugings can not originate from gauged
$d=3$ maximal supergravity. The $d=3$ embedding tensor does indeed not contain a ${\bf 248}$ component.

We now aim for a more general interpretation of the quadratic
constraint \eqref{QConD2}. It was explained in section \ref{sec:gen:emb} that the quadratic constraint can be understood
as a projector equation $\mathbbm{P}_2(\Theta\otimes\Theta)=0$. In the higher dimensional theories there is a finite number
of irreducible components in the twofold symmetric tensor product of $\Theta$ and $\mathbbm{P}_2$
can thus be defined by declaring which
representations are allowed and which are forbidden. In contrast, for the present case of $d=2$ maximal supergravity, there
is an infinite number of irreducible ${\rm E}_{9(9)}$ representations in the twofold symmetric tensor product 
$(Q^1_{\bf 1} \otimes Q^1_{\bf 1})_{\text{sym}}$ and the interpretation of the projector $\mathbbm{P}_2$ in terms of representation
theory is not clear immediately.

The basic representation is a highest weight representation with $k=1$.
Its twofold symmetric tensor product thus decomposes into highest weight representations with $k=2$.
There are only three of those $k=2$ representations. We denote them by
$Q^2_{\bf 1}$, $Q^2_{\bf 248}$ and $Q^2_{\bf 3875}$, because they have have vacuum ${\rm E}_{8(8)}$ representations
${\bf 1}$, ${\bf 248}$ and ${\bf 3875}$, respectively. 
The decomposition of these representation under ${\rm E}_{8(8)}$ is given in table \ref{Lev2Comp}
for the first $L_0$ levels. 
It turns out that $Q^2_{\bf 248}$ is not contained in the symmetric tensor product $(Q^1_{\bf 1} \otimes Q^1_{\bf 1})_{\text{sym}}$,
but $Q^2_{\bf 1}$ and $Q^2_{\bf 3875}$ both appear with infinite multiplicity.
There is natural $L_0$ grading on these $Q^2_{\bf 1}$ and $Q^2_{\bf 3875}$ components of the tensor product.
The twofold product of the vacuum singlet of $Q^1_{\bf 1}$ yields a singlet in the tensor product at $L_0=0$. This singlet
must be the highest weight of the first $Q^2_{\bf 1}$ component in the tensor product. For $L_0=1$ there is only one
${\bf 248}$ representation in $(Q^1_{\bf 1} \otimes Q^1_{\bf 1})_{\text{sym}}$ and this ${\bf 248}$ belongs to the $Q^2_{\bf 1}$
representation that starts at $L_0=0$, i.e.\ no irreducible ${\rm E}_{9(9)}$ representation starts at $L_0=1$. For $L_0=2$
one finds most of the ${\rm E}_{8(8)}$ representations that appear in the tensor product to be contained in
the $Q^2_{\bf 1}$ that starts at $L_0=0$, except for one ${\bf 1}$ and one ${\bf 3875}$ representation. These
representations thus are the vacuum representations of a $Q^2_{\bf 1}$ and a $Q^2_{\bf 3875}$ starting at $L_0=2$.
Continuing that analysis
one finds a finite number of $Q^2_{\bf 1}$ and $Q^2_{\bf 3875}$ representation to start at each $L_0$ level, as shown
in table \ref{Qdecomp} for the first few levels.

\begin{table}[tb]
     \begin{tabular}{l|l@{$\quad$}l@{$\quad$}l@{$\quad$}l@{$\quad$}l@{$\quad$}l@{$\quad$}l@{$\quad$}l}
        $L_0$ &  \multicolumn{8}{c}{${\rm E}_{8(8)}$ representations} \\ \hline
         $\,0$   & ${\bf 1}$ \\
	 $\,1$   & & ${\bf 248}$ \\
	 $\,2$   & ${\bf 1}$ & ${\bf 248}$ & ${\bf 3875}$ & ${\bf 27000}$ \\
	 $\,3$   & ${\bf 1}$ & ${\bf 248}_3$ & ${\bf 3875}$ & ${\bf 27000}$ & ${\bf 30380}_2$ & ${\bf 779247}$ \\
	 \multirow{2}{0.2cm}{$\,4$} 
	       & ${\bf 1}_3$ & ${\bf 248}_5$ & ${\bf 3875}_4$ & ${\bf 27000}_4$ & ${\bf 30380}_3$ & ${\bf 779247}_2$
	       & ${\bf 147250}_2$ & ${\bf 4096000}_2$ \\
	         &&&&&&& \multicolumn{2}{l}{\quad ${\bf 2450240}$ \quad ${\bf 4881384}$}
     \end{tabular}
     \\[0.5cm]
     \begin{tabular}{l|l@{$\quad$}l@{$\quad$}l@{$\quad$}l@{$\quad$}l@{$\quad$}l@{$\quad$}l}
        $L_0$ &  \multicolumn{7}{c}{${\rm E}_{8(8)}$ representations} \\ \hline
         $\,0$   & ${\bf 248}$ \\
	 $\,1$   & ${\bf 248}$ & ${\bf 1}$ & ${\bf 3875}$  & ${\bf 30380}$ \\
	 $\,2$   & ${\bf 248}_3$ & ${\bf 1}$ & ${\bf 3875}_2$  & ${\bf 30380}_2$ & ${\bf 27000}$ & ${\bf 147250}$ 
	                 & ${\bf 779247}$
     \end{tabular}
     \\[0.5cm]
     \begin{tabular}{l|l@{$\quad$}l@{$\quad$}l@{$\quad$}l@{$\quad$}l@{$\quad$}l@{$\quad$}l@{$\quad$}l}
        $L_0$ &  \multicolumn{8}{c}{${\rm E}_{8(8)}$ representations} \\ \hline
         $\,0$   & ${\bf 3875}$ \\
	 $\,1$   & ${\bf 3875}$ & ${\bf 248}$ & ${\bf 30380}$ & ${\bf 147250}$\\
	 $\,2$   & ${\bf 3875}_3$ & ${\bf 248}_2$ & ${\bf 30380}_2$ & ${\bf 147250}_2$
	         & ${\bf 1}$ & ${\bf 27000}$ & ${\bf 779247}$ & ${\bf 2450240}$
     \end{tabular}
     \caption{\label{Lev2Comp} \small Decomposition of the three level $k=2$ representations of
                              ${\rm E}_{9(9)}$ into irreducible ${\rm E}_{8(8)}$ components.
		     These ${\rm E}_{8(8)}$ representations
		     have a natural $L_0$ grading. The total number of components is infinite,
		     we only list the lowest $L_0$ levels. The subscripts indicate the multiplicities of
		     the representations. }
\end{table}

\begin{table}[tb]
     \begin{center}
     \begin{tabular}{c|c@{\qquad}c}
        $L_0$ & \# of $Q^2_{\bf 1}$ &  \# of $Q^2_{\bf 3875}$ \\ \hline
         $0$     & $1$  & ---\\
	 $1$    & ---   & ---\\
	 $2$    & $1$ & $1$ \\
	 $3$    & $1$ & $1$ \\
	 $4$    & $2$ & $1$ 
     \end{tabular}
     \caption{\label{Qdecomp} \small The decomposition of the two-fold symmetric tensor product
                                      $(Q^1_{\bf 1} \otimes Q^1_{\bf 1})_{\text{sym}}$ into irreducible
				      $Q^2_{\bf 1}$ and $Q^2_{\bf 1}$ representations is shown for the first
				      $L_0$ levels. We give the number of $Q^2_{\bf 1}$ and $Q^2_{\bf 1}$ 
				      representations whose highest weight is at the respective $L_0$ level.}
     \end{center}
\end{table}

In the language of conformal field theory what we consider here is a coset model\footnote{
We are grateful to Sakura Sch\"afer-Nameki for pointing out this relation to us.} \cite{Goddard:1984vk,Goddard:1986ee}
\begin{align}
   \frac {\mathfrak{G}} {\mathfrak{H}} &= 
      \frac {(\widehat {\rm E}_8)_1 \oplus (\widehat {\rm E}_8)_1} { (\widehat {\rm E}_8)_2 } \; ,
\end{align}
where the subscripts denote the level $k$ for the ${\rm E}_8$ representations under consideration.
The central charge of the Virasoro algebra on $({\rm E}_8)_k$ is given by
\begin{align}
   c_k=\frac{k \, \text{dim}({\rm E}_8)} {k+g^\vee} \; ,
\end{align}   
where $\text{dim}({\rm E}_8)=248$ is the dimension and $g^\vee=30$ is the dual Coxeter number of ${\rm E}_8$. This yields
$c_1=8$ and $c_2=31/2$ and thus the coset CFT has central charge $c_{\text{coset}}=2 c_1 - c_2 = 1/2$. Since
$c_{\text{coset}}$ lies between zero and one one is dealing with a minimal model which in this particular case
turns out to be the Ising model. For the CFT implications of this result we refer to \cite{DiFrancesco:1997nk}.
For our purpose it is important that the two infinite towers of $Q^2_{\bf 1}$ and $Q^2_{\bf 3875}$ representations
whose beginnings are shown in table \ref{Qdecomp} form the irreducible highest weight representations $V^{\text{Vir}}_{(1,1)}$ 
and $V^{\text{Vir}}_{(2,1)}$ of the Virasoro-Witt algebra (with central charge $1/2$), i.e.
the symmetric tensor product of the basic representation branches as
\begin{align}
   (Q^1_{\bf 1} \otimes Q^1_{\bf 1})_{\text{sym}} \; &= \;
    V^{\text{Vir}}_{(1,1)} \, \otimes \, Q^2_{\bf 1}  \; \oplus \; V^{\text{Vir}}_{(2,1)} \, \otimes \, Q^2_{\bf 3875} \; .
\end{align}
The Virasoro-Witt generators on $V^{\text{Vir}}_{(1,1)}$ and $V^{\text{Vir}}_{(2,1)}$ are given by
\begin{align}
   L^{\text{coset}}_m &= L^{\mathfrak{G}}_m - L^{\mathfrak{H}}_m \; ,
\end{align}
where $L^{\mathfrak{G}}_m$ and $L^{\mathfrak{H}}_m$ are the Virasoro-Witt generators induced by
$\mathfrak{G}=(\widehat {\rm E}_8)_1 \oplus (\widehat {\rm E}_8)_1$ (i.e.\ by $(Q^1_{\bf 1} \otimes Q^1_{\bf 1})_{\text{sym}}$) and
by $\mathfrak{H}=(\widehat {\rm E}_8)_2$ (i.e.\ by $Q^2_{\bf 1}$ and $Q^2_{\bf 3875}$). 
We are interested in $L^{\text{coset}}_1$. The Sugawara construction gives
an expressions for $L_1$ on the basic representation and for $L^{\mathfrak{G}}_1$ and $L^{\mathfrak{H}}_1$
in terms of the Kac-Moody generators $T_{\cAa\cMa}{}^{\cMb}$, namely
\begin{align}
   (L_1)_{\cMa}{}^{\cMb} 
      &= \frac 1 {k_1+g^\vee} 
          \, \sum_{n=0}^{\infty} \, \eta^{\alpha\beta} \, T_{(1+n)\alpha\,\cMa}{}^{\cMe} \, T_{(-n)\beta\,\cMe}{}^{\cMb}
                           \; , \nonumber \\
   (L^{\mathfrak{G}}_1)_{\cMa\cMb}{}^{\cMc\cMd} 
      &= 2 \, (L_1)_{(\cMa}{}^{(\cMc} \, \delta_{\cMb)}^{\cMd)}  \; , \nonumber \\
   (L^{\mathfrak{H}}_1)_{\cMa\cMb}{}^{\cMc\cMd} 
      &= \frac 4 {k_2+g^\vee} 
          \, \sum_{n=0}^{\infty} \, \eta^{\alpha\beta} \, T_{(1+n)\alpha\,(\cMa}{}^{(\cMe} \, \delta_{\cMb)}^{\cMf)}
                                                       \, T_{(-n)\beta\,(\cMe}{}^{(\cMc}  \, \delta_{\cMf)}^{\cMd)} \; ,
\end{align}
where $k_1=1$ and $k_2=2$. This yields
\begin{align}
   (L^{\text{coset}}_1)_{\cMa\cMb}{}^{\cMc\cMd}
     &= \frac{2} {k_2+g^\vee} \left(  (L_1)_{(\cMa}{}^{(\cMc} \, \delta_{\cMb)}^{\cMd)} 
                                    - \sum_{n=0}^{\infty} \, \eta^{\alpha\beta} \, 
        			          T_{(1+n)\alpha\,(\cMa}{}^{(\cMc} \, T_{(-n)\beta\,\cMb)}{}^{\cMd)} \right)
    \nonumber \\					  
     &= - \, \frac{1} {k_2+g^\vee} \, \eta^{\cAa\cAb} \, T_{\cAa(\cMa}{}^{(\cMc} \, T_{|\cAb|\cMb)}{}^{\cMd)} \;,
\end{align}
where the index $\cAa$ runs over the Kac-Moody generators and over $k$ and $L_1$, as introduced in \eqref{DefcAa}.
Comparing the last equation with \eqref{QConD2} we find that the quadratic constraint on $\Theta_\cMa$ can be written as
\begin{align}
   L^{\text{coset}}_1 ( \Theta \otimes \Theta) &= 0 \; , 
\end{align}
i.e.\ $L^{\text{coset}}_1$ is the projector $\mathbbm{P}_2$ we were searching for. The symmetric tensor product
$\Theta \otimes \Theta$ is only allowed to contain those representations that are annihilate by $L^{\text{coset}}_1$.
According to table \ref{Qdecomp} these are the $Q^2_{\bf 1}$ at $L_0=0$, the $Q^2_{\bf 1}$ and
$Q^2_{\bf 3875}$ at $L_0=2$, a particular linear combination of the two $Q^2_{\bf 1}$ representations at 
$L_0=4$, etc.

To summarize, we have argued that the embedding tensor~$\Theta$ of $d=2$ maximal supergravity transforms in the basic
representation of ${\rm E}_{9(9)}$. Due to this assumption we could embed the known gaugings that originate from $d=3$ torus reduction
into $\Theta$. We will show in the next section that this assumption also allows for the formulation of the general gauged theory
in terms of the embedding tensor.
As usual, one therefore needs the additional quadratic constraint \eqref{QConD2}.
Every embedding tensor in the basic representation that satisfies this quadratic constraint
defines a valid gauging.

\section{The gauged theory}

\subsection{Lagrangian of the ungauged theory}

The $d=2$ maximal supergravity is obtained from torus reduction of higher dimensional maximal supergravity.
This statement is of course also true for $d>2$, but it has a particular meaning for the two-dimensional case.
In $d=2$ the ordinary Einstein-Hilbert action describes a topological invariant (the Euler number)
and thus does not yield any equation of motion.
From a purely two-dimensional perspective it is thus not clear what a theory of gravity or supergravity should look like
and the higher dimensional origin is necessary to define it.

We thus start from the ungauged $d=3$ maximal supergravity, which has a global ${\rm E}_{8(8)}$ symmetry.
For the Lie algebra of ${\rm E}_{8(8)}$ we need the decomposition
$\mathfrak{e}_{8(8)}=\mathfrak{h} \oplus \mathfrak{k}$ into the compact part $\mathfrak{h}=\mathfrak{so}(16)$
and the non-compact part $\mathfrak{k}$. This is a symmetric space decomposition, i.e.\
we have the following commutators
\begin{align}
   [ \mathfrak{h} , \mathfrak{h} ] &= \mathfrak{h} \; , &
   [ \mathfrak{h} , \mathfrak{k} ] &= \mathfrak{k} \; , &
   [ \mathfrak{k} , \mathfrak{k} ] &= \mathfrak{h} \; .
\end{align}
We indicate the projection $\mathfrak{h}$ and $\mathfrak{k}$ by corresponding subscripts, i.e.\ for
$\Lambda \in \mathfrak{e}_{8(8)}$ we have
\begin{align}
   \Lambda &= \Lambda_\mathfrak{h} + \Lambda_\mathfrak{k} \; , \qquad \qquad
   \Lambda_\mathfrak{h} \in \mathfrak{h} \; , \qquad
   \Lambda_\mathfrak{k} \in \mathfrak{k} \; .
   \label{DefHK}
\end{align}
In addition we define the following involution on $\mathfrak{e}_{8(8)}$ elements
\begin{align}
   - \, \Lambda^T \, &= \, \Lambda_\mathfrak{h} \,  - \,  \Lambda_\mathfrak{k} \; .
\end{align}
The $128$ scalars of maximal $d=3$ supergravity arrange in an
${\rm E}_{8(8)}/{\rm SO}(16)$ coset and are the only bosonic degrees of freedom of the theory. They are described by a coset
representative ${\cal V}$ which is a group element of ${\rm E}_{8(8)}$ and transforms under global ${\rm E}_{8(8)}$
transformations from the left and local ${\rm SO}(16)$ transformations from the right, i.e.\
\begin{align}
  {\cal V} \, & \rightarrow \, g \, {\cal V} \, h(x) \; , \qquad \qquad
  g \in {\rm E}_{8(8)} \; , \quad h(x) \in {\rm SO}(16) \; .
  \label{E8coset}
\end{align}
The scalar currents are defined by
\begin{align}
   {\cal V}^{-1} \partial_{\hat \mua} {\cal V} &= Q_{\hat \mua} + P_{\hat \mua}  \; , \qquad \qquad
   Q_{\hat \mua} \in \mathfrak{h} \; , \qquad
   P_{\hat \mua} \in \mathfrak{k} \; .
   \label{DefE8currents}
\end{align}
The current $Q_{\hat \mua}$ is a
composite connection for the local ${\rm SO}(16)$ gauge invariance, i.e.\ it appears in covariant derivatives
of all quantities that transform under ${\rm SO}(16)$, in particular
\begin{align}
   D_{\hat \mua} P_{\hat \mub} &= \partial_{\hat \mua} P_{\hat \mub} + [Q_{\hat \mua},P_{\hat \mub}] \; .
\end{align}
The integrability conditions for \eqref{DefE8currents} are given by 
\begin{align}
   D_{[{\hat \mua}} P_{{\hat \mub}]} &= 0 \; , &
   Q_{{\hat \mua}{\hat \mub}} \, &\equiv \, 2 \partial_{[{\hat \mua}} Q_{{\hat \mub}]} + [Q_{\hat \mua},Q_{\hat \mub}] = - [P_{\hat \mua},P_{\hat \mub}] \; .
   \label{IntConPQ}   
\end{align}
The bosonic Lagrangian of maximal $d=3$ supergravity is given by the Einstein-Hilbert term coupled to the non-linear sigma model
of the scalar coset, i.e.\ \cite{Marcus:1983hb}
\begin{align}
   {\cal L}_{\text{d=3}} &= \ft 1 2 \hat e R^{(3)} - \ft 1 2 \, \hat e \, \tr(P_{\hat \mua} P^{\hat \mua}) \; ,
\end{align}
where $\hat e$ is the determinant of the three-dimensional vielbein.
The general Ansatz for a torus reduction and the resulting effective Lagrangian were given in \eqref{RedMetric}
and \eqref{Seff1} (compared to section \ref{sec:TorusPure} we rescaled the Lagrangian by a factor $1/2$).
For our particular case the effective two-dimensional Lagrangian reads
\begin{align}
   {\cal L}_{\text{d=2}} &= \ft 1 2 \, e \, \rho \, R^{(2)} \, - \, \ft 1 8 \, e \, \rho^3 \, A_{\mua\mub} \, A^{\mua\mub} \, 
            - \, \ft 1 2 \, e \, \rho \, \tr(P_\mua P^\mua) \;,
   \label{LeffD2}   
\end{align}
where $A_{\mua\mub}=2 \partial_{[\mua} A_{\mub]}$ is the field strength of the Kaluza-Klein vector field, $\rho$ is the
dilaton and $e$ is the determinant of the two-dimensional vielbein. The equations of motion for $A_\mua$ can be integrated to
an algebraic equation for the field strength
\begin{align}
   A_{\mua\mub} &= \xi \, e \, \rho^{-3} \, \epsilon_{\mua\mub} \; ,
   \label{AbackgroundD2}
\end{align}
where $\xi$ is an integration constant and $\epsilon_{\mua\mub}$ is the Levi-Civita tensor ($\epsilon_{01}=1$).
Due to the last equation the vector fields can be integrated out and their kinetic term becomes a scalar potential for $\rho$.
One thus obtains a deformation of the ungauged theory which is parameterized in terms of $\xi$. Therefore, $\xi$ has to
be a component of the embedding tensor and it turns out that it is the $L_0=0$ singlet of $\Theta_\cMa$, which was denoted
$X^1$ above\footnote{The factor between $X^1$ and $\xi$ still needs to be determined.}.
For the description of the ungauged theory we set $\xi=0$ and thus the vector field term in the Lagrangian
\eqref{LeffD2} vanishes.

In two-dimensions a Weyl rescaling is not possible.
As mentioned above the usual Einstein-Hilbert term is a total derivative anyway. What is possible in two dimensions is
a conformal rescaling, i.e.\ the metric $g_{\mua\mub}$ can be brought into diagonal form by fixing part of the general coordinate
invariance. We make us of this freedom by choosing the following conformal gauge
\begin{align}
   g_{\mu \nu} &= \eta_{\mu \nu} \exp{2 {\hat \sigma}} \; ,
   \label{ConfGauge}
\end{align}
where $\eta_{\mua\mub}$ is the flat space Minkowski metric and ${\hat \sigma}$ is the conformal factor.
The vielbein determinant and the curvature scalar then read
\begin{align}   
   e &= \sqrt{- \det(g_{\mua\mub})} = \exp{2 {\hat \sigma}} \; , \nonumber \\
   R^{(2)} &= - 2 g^{\mua \mub} \partial_\mua \partial_\mub {\hat \sigma} 
           = - 2 \exp(-2 {\hat \sigma}) \eta^{\mua \mub} \partial_\mua \partial_\mub {\hat \sigma} \; .
\end{align}
In the rest of this chapter we raise and lower indices no longer with $g_{\mua\mub}$ but with $\eta_{\mua\mub}$,
i.e.\ we use flat space conventions\footnote{Note that the connection on the two-dimensional manifold is non-zero
wrt $g_{\mua\mub}$. For example, we have a non-trivial covariant derivative in
\begin{align*}
   D_\mu \partial_\nu \rho &= \partial_\mu \partial_\nu \rho - (\partial_\mu {\hat \sigma}) (\partial_\nu \rho)
           - (\partial_\mu \rho) (\partial_\nu {\hat \sigma}) + \eta_{\mu \nu} (\partial_\lambda {\hat \sigma}) (\partial^\lambda \rho)  \;.
\end{align*}
But we never use covariant derivatives in the following, instead we write out the appropriate $\partial_\mua {\hat \sigma}$
corrections explicitly. See for example the left hand side of the conformal constraint \eqref{ConfConst1}, which
equals $-\ft 1 2 D_\pm \partial_\pm \rho$. Note also that $D_\mu \partial^\mu \rho = \partial_\mu \partial^\mu \rho$.}.
In conformal gauge the Lagrangian \eqref{LeffD2} for $X_1=0$ takes the form
\begin{align}
   {\cal L} &= (\partial_\mua {\hat \sigma})  (\partial^\mua \rho)
	       \, - \, \ft 1 2 \, \rho \, \tr(P_\mua P^\mua) \; .
   \label{EffLconfD2}	       
\end{align}
From this Lagrangian one gets the following equations of motion
\begin{align}
   \Box \rho &= 0 \; , &
   \Box {\hat \sigma} + \ft 1 2 \, \tr(P_\mua P^\mua) &= 0 \; , &
   D_\mua(\rho P^\mua) &= 0 \; .
   \label{UngaugedEOMd2}
\end{align}
However, the Lagrangian \eqref{EffLconfD2} does not reproduce all equations of motion of the theory.
Those equations that descend from variation of the gauged fixed metric components\footnote{
These are the off-diagonal and the traceless part of the components, i.e.\ $g_{01}$
and $g_{00}+g_{11}$.} are missing and have to be imposed by hand as a conformal constraint. 
It is convenient to introduce light-cone coordinates $x^\pm=1/\sqrt{2}(x^0 \pm x^2)$.\footnote{
In light-cone coordinates we have $\eta_{\pm\mp}=1$, $\eta_{\pm\pm}=0$, $\epsilon_{\pm\mp}=\pm$, $\epsilon_{\pm\pm}=0$.}
The conformal constraint then reads
\begin{align}
   (\partial_\pm {\hat \sigma}) (\partial_\pm \rho) - \ft 1 2 \partial_\pm \partial_\pm \rho 
      &= \ft 1 2 \, \rho \, \tr(P_\pm P_\pm)
   \label{ConfConst1}
\end{align}
Defining $\sigma=\hat \sigma - \ft 1 2 \ln((\partial_+ \rho)(\partial_- \rho))$ and using $\Box \rho = 0$ we can
write the conformal constraint as
\begin{align}
   (\partial_\pm \sigma) (\partial_\pm \rho) &=  \ft 1 2 \, \rho \, \tr(P_\pm P_\pm)
   \label{ConfConst2}
\end{align}
In contrast to $\hat \sigma$ we find $\sigma$ to transforms as a scalar (i.e.\ to stay invariant)
under those coordinate transformations that are compatible with the conformal
gauge \eqref{ConfGauge}, i.e.\ under $x^+ \mapsto {x'}^+=f(x^+)$ and $x^- \mapsto {x'}^-=g(x^-)$. 
The field equations \eqref{UngaugedEOMd2} are not modified when replacing $\hat \sigma$ by $\sigma$.
When can thus also make this replacement in the Lagrangian \eqref{EffLconfD2}.

\subsection{Linear system}

In the last subsection the bosonic Lagrangian and the field equations of maximal $d=2$ supergravity were presented.
We now show that this theory is invariant under the global symmetry  group $G_0 = {\rm E}_{9(9)}$. 
The Lagrangian \eqref{EffLconfD2} is manifestly invariant under the three-dimensional symmetry group ${\rm E}_{8(8)}$,
which is the subgroup of ${\rm E}_{9(9)}$ generated by the zero-mode generators $T^0_\alpha$.
The complete ${\rm E}_{9(9)}$ symmetry is not realized at the level of the Lagrangian but only onshell.
We already encountered such a situation for the four-dimensional theories in chapter \ref{ch:D4}.
There, the global symmetry group $G_0$ transformed electric and magnetic vector fields into each other,
but in the ungauged theory the latter were only introduced onshell as dual to the electric fields.
Things are similar in $d=2$, but with vector fields replaced by scalars.
There is an infinite tower of dual scalars that can be defined onshell out of the scalars ${\cal V}$ 
which appear in the Lagrangian. The ${\rm E}_{9(9)}$ symmetry is only defined on this infinite tower
of scalars and thus is an onshell symmetry.

In order to define dual scalars one needs to identify currents that are conserved due to the equations of motion.
The simplest example is the current of the dilaton $\partial_\mua \rho$ which is conserved due to $\Box \rho = 0$. 
The dual dilaton $\tilde \rho$ is thus defined by
\begin{align}
   \partial_\mua \tilde \rho &= \epsilon_{\mua\mub} \, \partial^\mub \rho \; , &
   \text{or} &&
   \partial_\pm \rho &= \pm \partial_\pm \rho \; .
   \label{DualDilaton}
\end{align}
In most of the following we find it convenient to give the duality equations in light-cone coordinates. The last two equations show
that those formulas can be easily translated into a coordinate independent notation by using the Levi-Civita tensor. 
Equation \eqref{DualDilaton} only defines $\tilde \rho$ up to a constant shift $\tilde \rho \mapsto \tilde \rho + \lambda$.
Such shift-symmetries appear for all of the dual scalars we are going to introduce.
Note that a further dualization of $\tilde \rho$ simply yields the dilaton $\rho$, i.e.\ there is no infinite tower
of dual fields in the case of the dilaton. This is different for the non-Abelian scalars ${\cal V}$.

The field equation $D_\mua(\rho P^\mua)$ of ${\cal V}$ can also be written as a conservation law
$\partial_\mua(\rho J^\mua)=0$ for the current $J_\mua = (\partial_\mua {\cal V}){\cal V}^{-1} = {\cal V} P_\mua {\cal V}^{-1}$.
While $P_\mua$ transforms under local ${\rm SO}(16)$ transformations $J_\mua$ transforms under global ${\rm E}_{8(8)}$
transformations. The dual scalars to ${\cal V}$ are called dual potentials. The dual potentials $Y_i$, $i\in \mathbbm{N}$,
are $\mathfrak{e}_{8(8)}$ algebra valued, i.e.\ they transforms in the adjoint representation of ${\rm E}_{8(8)}$.
The first dual potential $Y_1$ is defined by
\begin{align}
   \partial_\pm Y_1 &= \pm \rho J_\pm = \pm \rho {\cal V} P_\pm {\cal V}^{-1} \; .
   \label{DefY1}
\end{align}
This equation is consistent since $J_\mua$ is conserved. We can now use $Y_1$ to define another conserved current and
thus a second dual potential $Y_2$. Using $Y_1$ and $Y_2$ we can define the third dual potential $Y_3$, etc. 
To convey an impression of how that works we give the defining equations for $Y_2$ and $Y_3$ explicitly
\begin{align}
   \partial_\pm Y_2 &= - \left( \pm \rho \tilde \rho + \ft 1 2 \rho^2 \right) {\cal V} P_\pm {\cal V}^{-1} 
                       - \ft 1 2 [Y_1,\partial_\pm Y_1]  \; ,
		         \nonumber \\
   \partial_\pm Y_3 &= - \left( \mp \ft 1 2 \rho^3 \mp \rho \tilde \rho^2 - \rho^2 \tilde \rho \right) {\cal V} P_\pm {\cal V}^{-1}
                       - [Y_1,\partial_\pm Y_2] - \ft 1 6 [Y_1,[Y_1,\partial_\pm Y_1]]] \; .
   \label{DefY23}		       
\end{align}
Obviously, it is rather inconvenient to define each dual potential separately. Fortunately, there is a generating function
for all the dual potentials which is called the linear system \cite{Belinsky:1971nt,Pohlmeyer:1975nb,Maison:1978es}.
For the definition of the linear system
we need the spectral parameter $y$ and its inverse $w=1/y$. We first need to introduce the
complex valued function $\gamma(w)$ by
\begin{align}
   \gamma &= \frac 1 \rho \left( w + \tilde \rho - \sqrt{(w+\tilde \rho)^2 - \rho^2} \right)  \; , &
    w &= \frac \rho 2 \left( \gamma + \frac 1 \gamma \right) - \tilde \rho \; .
    \label{DefGamma}
\end{align}
The second equation is a consequence of the first one. Since a square root appears in the definition of
$\gamma$ it formally takes values on a Riemann surface which is two-fold covering of the complex $w$-plane.
For our purposed we simply consider $\gamma$ as formal series in negative powers of $w$, i.e.\
\begin{align}
   \gamma &= \frac 1 2 \rho w^{-1} - \frac 1 2 \rho \tilde \rho w^{-2} 
                + \frac 1 8 \left( \rho^3 + 4 \rho \tilde \rho^2 \right)  w^{-3} 
		- \frac 1 8 \left( 3 \rho^3 \tilde \rho + 4 \rho \tilde \rho^3 \right) w^{-4}
		+ {\cal O}(w^{-5}) 
   \label{ExpandGamma}		
\end{align}
We refer to $\gamma$ as the variable spectral parameter, in contrast to $w$ (or $y$) which is referred to the constant spectral
parameter. While $\partial_\pm w = 0$ we find the variable spectral parameter to obey the following differential equation
\begin{align}
   \frac{\partial_\pm \gamma } \gamma &= \frac{\partial_\pm \rho } \rho \frac{1 \mp \gamma} {1 \pm \gamma} \; .
\end{align}

We now introduce the ${\rm E}_{8(8)}$ valued function $\hV(w)$. It also has a formal expansion in negative powers of $w$.
Since $\hV$ it is group valued we use the following expansion
\begin{align}
   \hV &= \ldots e^{(-Y_4 w^{-4})} e^{(-Y_3 w^{-3})} e^{(-Y_2 w^{-2})} e^{(-Y_1 w^{-1})} {\cal V} \; ,
   \label{ExpandHV}
\end{align}
where the dual potentials $Y_i$ appear as coefficients in the expansion and we have $\hV(w=\infty)={\cal V}$.
The following linear system defines $\hV$ in terms
of the scalars ${\cal V}$ that appear in the Lagrangian
\begin{align}
   \hV^{-1} \partial_\pm \hV \, &= Q_\pm + \hat P_\pm \; , &
   \hat P_\pm &= \frac{1\mp \gamma} {1\pm \gamma} \, P_\pm \; .
   \label{LinearSystem}
\end{align}
For a flat space sigma model (with Lagrangian ${\cal L}=\ft 1 2 \, \tr P_\mua P^\mua$) one can define the linear system
by using the constant spectral parameter $w$. For a curved space sigma model one accounts for the dilaton
dependence in the equations of motion by replacing $w$ with the variable spectral parameter $\gamma$. 
The integrability conditions for \eqref{LinearSystem} read
\begin{align}
   Q_{\mua\mub} + [\hat P_\mua,\hat P_\mub] &= 0 \; ,  &
   D_{[\mua} \hat P_{\mub]} &= 0 \; .
\end{align}
We find
\begin{align}
   [\hat P_\mua,\hat P_\mub] &=[P_\mua,P_\mub] \; , &
   \epsilon^{\mu \nu} D_\mu \hat P_\nu &= \frac{1+\gamma^2} {1-\gamma^2} \epsilon^{\mu \nu} D_\mu P_\nu
                             - \frac{2 \gamma} {1-\gamma^2} \rho^{-1} D_\mu(\rho P^\mu) \; .
\end{align}
Thus, the integrability equations are satisfied due to the integrability equations \eqref{IntConPQ} for $P_\mua$
and $Q_\mua$ and due to the equations of motions \eqref{UngaugedEOMd2}. By expanding the linear system
in powers of $w$ according to \eqref{ExpandGamma} and \eqref{ExpandHV} we find the defining equations
for all dual potentials. In particular, we reproduce the defining equations
\eqref{DefY1} and \eqref{DefY23} for the first three dual potentials.

The whole tower of dual scalars is contained in $\hV$. One can consider $\hV$ as the
coset representative of the scalar coset $G_0/H = {\rm E}_{9(9)}/K({\rm E}_{9})$, where $K({\rm E}_{9})$
is the maximal compact subgroup of ${\rm E}_{9(9)}$. In order to define the Lie algebra $\mathfrak{ke}_9$ of $K({\rm E}_{9})$
we consider an algebra element $\Lambda \in \mathfrak{e}_{9(9)}$ as an $\mathfrak{e}_{8(8)}$ valued function
of $w$ and thus of $\gamma$. $\Lambda(\gamma)$ is contained in the subalgebra $\mathfrak{ke}_9$ if it obeys
\begin{align}
   \Lambda^T(\gamma) &= - \Lambda^T \left( \frac 1 \gamma \right) \; .
\end{align}

Analogous to \eqref{E8coset} we want the ${\rm E}_{9(9)}$ symmetry to act on $\hV$ form the left while $K({\rm E}_{9})$
shall act from the right. A generic ${\rm E}_{9(9)}$ action destroys the form \eqref{ExpandHV} of $\hV$, because
$\hV$ may not have an expansion in negative powers of $w$ anymore. We therefore demand for a compensating 
$K({\rm E}_{9})$ action that restores the form \eqref{ExpandHV}, i.e.\
\begin{align}
   \hV \; &\rightarrow  \; g \; \hV \; h(g,x) \; , \qquad g \in {\rm E}_{9(9)} \; , \quad h(g,x) \in K({\rm E}_{9}) \; .
   \label{CosetHV}
\end{align}
This defines our ${\rm E}_{9(9)}$ action.
Apart form the ${\rm SO}(16)$ transformations in the zero-mode ${\rm E}_{8(8)}$
there is no additional freedom of $K({\rm E}_{9})$ transformation. We are thus dealing with a gauged fixed version of
the scalar coset, known as Breitenlohner-Maison gauge \cite{Breitenlohner:1986um}.
Those symmetry transformations $g$
that are generated by the non-negative Kac-Moody generators $T^m_\alpha$, $m \geq 0$, do not destroy
the Breitenlohner-Maison gauge and thus do not need a compensating $K({\rm E}_{9})$ transformations.
These transformations correspond to the zero-mode ${\rm E}_{8(8)}$ and to the shift-symmetries of
the dual potentials $Y_i$. On the other hand, those ${\rm E}_{9(9)}$ transformations $g$ that are generated by
the negative modes $T^m_\alpha$, $m<0$, of the Kac-Moody algebra need a compensating $K({\rm E}_{9})$ transformation
that restores the gauge fixing. They are thus not realized linearly on the coset representative $\hV$. 
An explicit expression for the symmetry action on $\hV$ is given in the next subsection for
infinitesimal ${\rm E}_{9(9)}$ transformation. 

The gauge fixing of $\hV$ is crucial in order
to guarantee convergence of the expressions in which $\hV$ enters and in order to interprete $\hV$ as a tower of dual potentials.
However, in principle one is not restricted to the Breitenlohner-Maison gauge. 
For example, in table \ref{Fig:AlgD2a} we gave the decomposition
of $\mathfrak{e}_{9(9)}$ under the $\mathbbm{R}^+ \times {\rm SL}(9)$ that corresponds to the torus group in dimensional reduction
from $d=11$. A valid gauge fixing is given by demanding $\hV$ to be generated only by the non-negative generators
in this decomposition. The choice of the gauge fixing is analogous to the choice of the symplectic frame in $d=4$ supergravity. 
This analogy suggests that for every particular gauging of $d=2$ supergravity there should be a natural gauge fixing
that simplifies the form of the gauged Lagrangian. In our presentation we stick to the
Breitenlohner-Maison gauge which is induced by dimensional reductions from $d=3$ and
corresponds to the form \eqref{EffLconfD2} of the ungauged Lagrangian.

\subsection{Global symmetry action}

Equation \eqref{CosetHV} defines the symmetry action of ${\rm E}_{9(9)}$ on the scalars, but it does not give an
explicit expression for the necessary compensating $K({\rm E}_{9})$ transformations. In this subsection we give
explicit formulas for the action of an algebra element
\begin{align}
   L &= L^\cAa T_\cAa = \Lambda^{\alpha}(w) \, t_\alpha \, +  \, \lambda \, L_{1} \, + \, \kappa \, k 
   \label{SymParamD2}
\end{align}
on the scalars $\sigma$, $\rho$, $\tilde \rho$ and $\hV$. These formulas are needed in the next subsection to write down
the appropriate covariant derivatives for these scalars and to check gauge invariance of the Lagrangian.
According to \eqref{SymParamD2} we have symmetry parameters $\lambda$ and $\kappa$ which are reals numbers
and an parameter $\Lambda$(w) that describes the $\mathfrak{e}_{9(9)}$ transformations.
We treat $\Lambda(w)$ as an $\mathfrak{e}_{8(8)}$ valued function of $w$ (respectively of $y=1/w$). 

The action of the central extension $k$ on the scalars is given by\footnote{
In terms of our notation in \eqref{ActionBasic1} to \eqref{ActionBasic4} we now have
$\delta_\kappa = \kappa \delta^0_k$, $\delta_\lambda = \lambda \delta^1_L$ and 
$\delta_\Lambda = \sum_{m=-\infty}^{\infty} \Lambda^\alpha_m \delta^m_\alpha$, where
$\Lambda(w)= \sum_{m=-\infty}^{\infty} \Lambda^\alpha_m w^{-m}$.}
\begin{align}
   \delta_{\kappa} \rho &= 0 \; , &
   \delta_{\kappa} \tilde \rho &= 0 \; , &
   \delta_{\kappa} \sigma &= \, - \, \kappa \; , &
   \delta_{\kappa} \hV &= 0 \; , &
   \label{Kaction}
\end{align}
i.e.\ the central extension acts as a shift symmetry on the conformal factor $\sigma$ and leaves all other scalars invariant
\cite{Breitenlohner:1986um}.

The Witt-Virasoro generator $L_1$ is represented as a differential operator $L_1 = - y^2 \partial_y = \partial_w$. This 
representation defines
the action of $L_1$ on $\hV(w)$, $\gamma(w)$ and $\sigma$. 
According to \eqref{DefGamma} and \eqref{ExpandGamma} the action on $\rho$ and $\tilde \rho$ is induced by the action
on $\gamma(w)$. We find
\begin{align}
   \delta_{\kappa} \rho &= 0 \; , &
   \delta_{\kappa} \tilde \rho &= \lambda \; , &
   \delta_{\kappa} \sigma &= 0 \; , &
   \delta_{\kappa} \hV &= \lambda \, \partial_w  \, \hV \; , &
   \label{L1action}
\end{align}
i.e.\ $L_{1}$ acts as a shift symmetry on $\tilde \rho$ and it also acts non-trivially on the dual potentials $Y_i$, $i \geq 2$,
that are contained in $\hV$, but it leaves $\rho$, $\sigma$, ${\cal V}$ and $Y_1$ invariant.

In order to give the action of the Kac-Moody generators $T^m_\alpha=w^{-m} t_\alpha$ it is very convenient to introduce the following
notation. For an arbitrary function $f(w)$ of the spectral parameter $w$ we define
\begin{align}
   \langle f(w) \rangle_w  \equiv \oint_l \frac {dw} {2 \pi i} f(w) = - \text{Res}_{w=\infty} f(w) \; ,
\end{align}
The path $l$ is chosen such that only the residual at $w=\infty$ is picked up. For our purposes it is sufficient to
consider $f(w)$ as a formal expansion in $w$, i.e.\ $f(w)=\sum_{m=-\infty}^{\infty} f_m w^{m}$. We then have
$\langle f(w) \rangle_w  = f_{-1}$. Sometimes we also use variables $v$ or $u$ instead of $w$. An expression that appears
regularly is
\begin{align}
   \left \langle \frac{ f(v) } {v-w} \right \rangle_v 
          &=  \left \langle \sum_{m=0}^{\infty} \frac{ f(v) \, w^m } {v^{m+1}} \right \rangle_v
     = \sum_{m=0}^{\infty} f_m w^m \; .
\end{align}
Another useful relation is
\begin{align}
   \Big\langle \Big\langle \frac{f(w,v)} {v-w} \Big\rangle_v \Big\rangle_w 
    - \Big\langle \Big\langle \frac{f(w,v)} {v-w} \Big\rangle_w \Big\rangle_v
   &= \langle f(w,w) \rangle_w \; .
   \label{RelationIntInt}   
\end{align}
For $\Lambda = \Lambda^\alpha(w) t_\alpha \, \in \mathfrak{e}_{9(9)}$
we introduce the abbreviation $\tilde \Lambda=\hV^{-1} \Lambda \hV$. According
to \eqref{DefHK} the compact and non-compact part of $\tilde \Lambda$ are denoted 
\begin{align}
   \tilde \Lambda_{\mathfrak{h}} &= [\hV^{-1} \Lambda \hV]_{\mathfrak{h}} \; , &
   \tilde \Lambda_{\mathfrak{k}} &= [\hV^{-1} \Lambda \hV]_{\mathfrak{k}} \; . 
\end{align}
In terms of this notation the action of $T^m_\alpha=w^{-m} t_\alpha$ on the scalar fields is given by \cite{Nicolai:1998gi}
\begin{align}
     \delta_\Lambda \, \rho &= 0 \; ,
       \nonumber \\
     \delta_\Lambda \, \tilde \rho &= 0 \; ,
       \nonumber \\
     \delta_\Lambda \, \sigma &= \, - \, \tr \, \Big\langle \Lambda(w) \, \partial_w \, \hV(w) \, \hV^{-1}(w) \Big\rangle_{w} \; ,
       \nonumber \\ 
     \hV^{-1} \delta_\Lambda \, \hV(w) &= \tilde \Lambda (w) 
                               - \left\langle \frac 1 {v-w} \left( \tilde \Lambda_{\mathfrak h}(v)
                               + \frac{\gamma(v)\,(1-\gamma^2(w))}{\gamma(w)\,(1-\gamma^2(v))} \,
                                                   \tilde \Lambda_{\mathfrak k}(v) \right) \right\rangle_{v} \;.
    \label{E9action}						   
\end{align}
The variation of $\hV$ can also be written as $\delta_\Lambda \hV = \Lambda \hV + \hV h$.
The first term in this variation describes the left action of $\Lambda$ on $\hV$, the second term 
is the compensating $K({\rm E}_9)$ transformation from the right, i.e.\ $h \in \mathfrak{ke}_{9(9)}$. 
Equation \eqref{E9action} gives an explicit expression for $h$ in terms of $\Lambda$.

According to \eqref{ExpandHV} the ${\rm E}_{8(8)}/{\rm SO}(16)$ representative ${\cal V}$ is contained
in $\hV$. Equation \eqref{E9action} thus also gives the action of ${\rm E}_{9(9)}$ on ${\cal V}$.
Using \eqref{RelationIntInt} we find
\begin{align}
   {\cal V}^{-1}\,\delta_\Lambda {\cal V} &= 
       \left\langle w^{-1} \, \hV^{-1} \delta_\Lambda \, \hV(w) \right\rangle
       \, = \,
     \left\langle
     \frac{2 \gamma(w)} {\rho\,(1-\gamma(w)^2)} \, \tilde \Lambda_{{\mathfrak k}}(w) \; \right\rangle_{w}\;,
\end{align}
The action of  $k$, $L_1$ and $T^m_{\alpha}$ given in \eqref{Kaction}, \eqref{L1action} and \eqref{E9action}
satisfies the symmetry algebra  \eqref{E9algebra}, \eqref{Lalgebra}.

\subsection{Lagrangian of the gauged theory}

We now present the bosonic Lagrangian of gauged $d=2$ maximal supergravity.
The gaugings are parameterized by an embedding tensor $\Theta_{\cMa}$. The ungauged Lagrangian \eqref{EffLconfD2}
is formulated in terms of scalar fields $\sigma$, $\rho$ and ${\cal V}$. Onshell one can define the dual
dilaton $\tilde \rho$ and the ${\rm E}_{9(9)}/K({\rm E}_9)$ coset representative $\hV$ that contains the dual potentials $Y_{i}$.
We also introduced vector fields $A_\mua^\cMa$ in the dual basic representation of ${\rm E}_{9(9)}$. All these fields
can feature in the Lagrangian of the gauged supergravity. The dual scalars and the vector fields only appear projected with
the embedding tensor $\Theta_{\cMa}$. Thus, $\Theta_{\cMa}$ determines the field content of the theory and for
$\Theta_\cMa \rightarrow 0$ the ungauged supergravity is recovered.

The general covariant derivative \eqref{CovDivD2} can be written as
\begin{align}
   D_\mua &= \partial_\mua - g \, {\cal A}^\alpha_\mua(w) \, t_\alpha 
	                   - g \, {\cal B}_\mua \, L_{1} 
			   - g \, {\cal C}_\mua \, k \; ,
\end{align}
where we introduced the following $\Theta$-projections of the vector fields $A_\mua^\cMa$
\begin{align}
   {\cal B}_\mua &= \, - \, T_{(0)k,\,\cMa}{}^{\cMb} \, A_\mua^\cMa  \, \Theta_{\cMb} \, = \, - \,  A_\mua^\cMa  \, \Theta_{\cMa} \; , 
     \nonumber \\[0.1cm]
   {\cal C}_\mua &= \, - \, T_{(1)L,\,\cMa}{}^{\cMb} \, A_\mua^\cMa  \, \Theta_{\cMb}  \; ,
     \nonumber \\[0.1cm]
   {\cal A}^\alpha_\mua(w) &= \sum_{m=-\infty}^{m=\infty} 
                        w^{-m} \, \eta^{\alpha\beta} \, T_{(1-m)\beta,\,\cMa}{}^{\cMb} \, A_\mua^\cMa  \, \Theta_{\cMb} \; .
   \label{DefABC}			
\end{align}
These projections correspond to the symmetry parameters $\Lambda$, $\lambda$ and $\kappa$ introduced in \eqref{SymParamD2}.
The action of the covariant derivative on the various scalars reads
\begin{align}
    D_\mua \tilde \rho &= \partial_\mua \tilde \rho - {\cal B}_\mua
        \nonumber \\[0.1cm]
    D_\mua \sigma &= \partial_\mua \sigma + {\cal C}_\mua + \tr \, \Big\langle {\cal A}_\mua(w)\, \partial_w \hV(w) \hV^{-1}(w) \Big\rangle_{w} 
        \nonumber \\[0.1cm]
    {\cal V}^{-1} D_\mua {\cal V} & = {\cal V}^{-1} \partial_\mua {\cal V} - \Big\langle
     \frac{2 \gamma(w)} {\rho\,(1-\gamma(w)^2)} \, \tilde {\cal A}_\mua(w)_{\mathfrak{k}} \; \Big\rangle_{w}
        \; \equiv \;  {\cal P}_\mua + Q_\mua     \; ,
        \nonumber \\[0.1cm]
   \hV^{-1} D_\mua \hV(w) &= \hV^{-1} \partial_\mua \hV(w) - {\cal B}_\mua \hV^{-1} \partial_w \hV(w)
                             - \tilde {\cal A}_\mua (w)  \nonumber \\ & \qquad \qquad 
                             + \left\langle \frac 1 {v-w} \left( [\tilde {\cal A}_\mua(v)]_{\mathfrak h}  
                               + \frac{\gamma(v)\,(1-\gamma^2(w))}{\gamma(w)\,(1-\gamma^2(v))}
                                                   [\tilde {\cal A}_\mua(v)]_{\mathfrak k} \right) \right\rangle_{v} \; ,
\end{align}
where $\tilde {\cal A}_\mua = \hV^{-1} {\cal A}_\mua \hV$.
Note that we defined the covariant generalization ${\cal P}_\mua$ of the scalar current $P_\mua$. The current $Q_\mua$
remains unchanged compared to the ungauged theory\footnote{This is due to our particular ${\rm SO}(16)$ gauge
choice in equation \eqref{E9action}.}.

We can now present the bosonic part of the general gauged Lagrangian
\begin{align}
   {\cal L} &= {\cal L}_{\text{kin}} + {\cal L}_{\text{top}} \; .
   \label{LgaugedD2}
\end{align}
It consists of a kinetic term
\begin{align}
   {\cal L}_{\text{kin}} &= \partial^\mua \rho \, D_\mua \sigma - \ft 1 2 \, \rho \, \tr ( {\cal P}_\mua {\cal P}^\mua ) \; ,
\end{align}
and a topological term
\begin{align}
   {\cal L}_{\text{top}} &= \epsilon^{\mua\mub} \Bigg\{ g \, \tr \Big\langle {\cal A}_{\mua} \,
                                       \,(\partial_{\mub}\hV-\hV Q_{\nu})\,\hV^{-1} \,
                               - \frac{1+\gamma^{2}}{1-\gamma^{2}}\, {\cal A}_{\mua} \,  \hV P_{\mub}\hV^{-1}  \Big\rangle_{w}
     \nonumber \\ & \qquad \quad
               - g \, \Big( {\cal C}_{\mua} 
	                                 + \tr\Big\langle  {\cal A}_{\mua} \, \partial_w \hV \, \hV^{-1} \Big\rangle_{w}\Big)
                                           \partial_\mub \tilde \rho 
     \nonumber \\ & \qquad \quad
               + \ft 1 2 \, g^2 \, {\cal C}_\mua {\cal B}_\mub   + \ft 1 2 \, g^2 \,
	            \tr \Big\langle \Big \langle 
		          \frac 1 {v-w} [\tilde {\cal A}_\mua(w)]_{\mathfrak h} \, [\tilde {\cal A}_\mub(v)]_{\mathfrak h}
			\nonumber \\ & \qquad \qquad  \quad
			  + \frac {(\gamma(v)-\gamma(w))^2 + (1-\gamma(v)\gamma(w))^2} {(v-w) (1-\gamma(v))^2 (1-\gamma(w))^2} 
                 			   [\tilde {\cal A}_\mua(w)]_{\mathfrak k} \, [\tilde {\cal A}_\mub(v)]_{\mathfrak k}
                        \Big \rangle_v \Big \rangle_w  \Bigg\} \; .
\end{align}
In addition, supersymmetry demands a scalar potential
whose general form (in terms of $\Theta_\cMa$) still needs to be determined.
The scalar potential and the Lagrangian ${\cal L}_{\text{kin}} + {\cal L}_{\text{top}}$
which we are considering here are both gauge invariant on its own.

The variation of the Lagrangian with respect to the vector fields reads
\begin{align}
   \delta {\cal L} =& \, - \, g \, \delta {\cal C}_{\mua} \, {\cal X}^\mua
            \, + \, g \, \tr \langle \delta {\cal A}_\mua \, {\cal Y}^\mua  \rangle_w  \; ,
   \label{VarVec}			    
\end{align}
where
\begin{align}
   {\cal X}_\mua &= - \, \partial_\mua \rho + \epsilon_{\mua\mub} D^\mub \tilde \rho  \; ,
   \nonumber \\
   {\cal Y}_\mua &= \hV \left[ \frac {2 \gamma} {1-\gamma^2} {\cal P}^\mua 
	                    - \frac{ 1+\gamma^2 } {1 - \gamma^2} \epsilon^{\mua\mub} {\cal P}_\mub
	                    + \epsilon^{\mua\mub} \hV^{-1} (D_\mub \hV - \hV Q_\mub)  \right] \hV^{-1}
			    \, - \, \partial_w \hV \, \hV^{-1} \, {\cal X}^\mua   \; .
\end{align}
As field equations we thus find $\Theta$-projections of ${\cal X}_\mua=0$ and ${\cal Y}_\mua=0$. These are
the covariantized versions of the duality equations \eqref{DualDilaton} and \eqref{LinearSystem}
that render $\rho$ dual to $\tilde \rho$ and ${\cal V}$ dual to $\hV$. These duality equation transform
in the (gauged fixed) adjoint representation of $G_0$. They arrange in the algebra valued current
$Z_\mua = Z_\mua^\cAa \, T_\cAa$ as follows
\begin{align}
   Z_\mua &= {\cal Y}^\alpha_\mua(w) t_\alpha \, + \, {\cal X}_\mua L_{1} \; .
\end{align}
In the covariant formulation the variations \eqref{VarVec} give rise to the following field equations
\begin{align}
   T_{\cAa\cMa}{}^\cMb \, \Theta_\cMb \, Z^\cAa_\mua &= 0 \; .
   \label{ProjDuality}
\end{align}
For the derivation of the order $g^2$ terms
in \eqref{VarVec} from the above Lagrangian one needs the following constraint on the projections of the vector fields
\begin{align}
   \tr \Big\langle {\cal A}_\mua(w) \, \delta {\cal A}_\mub(w) \Big \rangle_w 
          - {\cal B}_\mua \, \delta {\cal C}_\mub - {\cal C}_\mua \, \delta {\cal B}_\mub \, &= \, 0 \; .
\end{align}   
By virtue of the definitions \eqref{DefABC} the last equation is equivalent to the quadratic constraint \eqref{QConD2}
on $\Theta_\cMa$. 

The parameter of gauge transformations $L^\cMa$ transforms in the dual basic representation.
Under gauge transformations the scalars transform according to \eqref{Kaction}, \eqref{L1action} and \eqref{E9action}
with $\Theta$-projected parameters $\Lambda$, $\lambda$ and $\kappa$ given by 
\begin{align}
   \eta^{\cAa\cAb} \, T_{\cAb\cMa}{}^{\cMb} \, \Theta_\cMb \, L^\cMa \, T_\cAa
      &= \Lambda^\alpha(w) \, t_\alpha \, + \, \lambda \, L_1 \, + \, \kappa \, k \; .
\end{align}
The Lagrangian \eqref{LgaugedD2} is invariant under gauge transformations if we define the gauge transformations of
the vector fields as follows\footnote{We have checked gauge invariance explicitly up to terms of order $g^1$ so far,
but the $g^2$ terms of the Lagrangian are already completely determined by demanding \eqref{VarVec}.}
\begin{align}
   \Delta A_\mua^\cMa &= D_\mua L^\cMa \, + \, \epsilon_{\mua\mub}  \, T_{\cAa\cMb}{}^{\cMa} \, L^\cMb \, Z^{\mub\,\cAa} \;  .
\end{align}
Thus, the vector fields transform into the duality equations of the scalars, just as the $d=4$ two-forms transform into the
duality equation between electric and magnetic vector fields. For the $\Theta$-projected vector fields one finds
\begin{align}
   \Delta {\cal A}_\mua &= D_\mua \Lambda 
            + \ft 1 2 \, \epsilon_{\mua\mub} \, [ \Lambda , {\cal Y}^\mub ]
	    - \ft 1 2 \, \epsilon_{\mua\mub} \, (\partial_w \Lambda) \, {\cal X}^\mub + \text{additional terms}  \; ,
    \nonumber \\
   \Delta {\cal C}_\mua 
          &= D_\mua \kappa + \ft 1 2 \, \epsilon_{\mua\mub}
	        \, \tr \langle \Lambda \, \partial_w \, {\cal Y}^\mub \rangle_w   + \text{additional terms} \; ,             
\end{align}
where the additional terms are such that they vanish under the contraction with ${\cal X}_{\mua}$ and
${\cal Y}_\mua$ in \eqref{VarVec}.

For the vector fields ${\cal A}_\mua$ we have the expansion
\begin{align}
   {\cal A}^\alpha_\mua &= \sum_{m=-\infty}^{\infty} \, {\cal A}_\mua^{\alpha(m)} \, T^m_{\alpha}
                         = \sum_{m=-\infty}^{\infty} \, w^{-m} \,{\cal A}_\mua^{\alpha(m)} \, t_{\alpha} \; .
\end{align}
Due to the gauge fixing \eqref{ExpandHV} of $\hV(w)$ we find that $\partial_w \hV \hV^{-1}$ has an expansion in negative
powers of $w$ that starts with $w^{-2}$ and $\hV {\cal Y}^\mua \hV^{-1}$ has an expansion in negative powers of $w$
that starts with $w^{-1}$. From the variation \eqref{VarVec} we thus find that the positive mode vector fields
${\cal A}_\mua^{\alpha(m)}$, $m>0$, do not enter the Lagrangian at all, i.e.\ a gauging of the shift symmetries of the dual
potentials is not visible in the Lagrangian. From the Lagrangian itself this fact is not obvious since
the quadratic constraint was used to derive \eqref{VarVec}. The gauge fixing of $\hV$ thus seems
to induce a truncation of the
gauge group in the Lagrangian. Nevertheless the Lagrangian is invariant under all gauge transformations. The shift symmetries
of the dual potentials $Y_i$ only do not seem to be gauged because no vector gauge field is coupled to them.

We have thus presented the bosonic Lagrangian of maximal gauged $d=2$ supergravity. What is missing in the description of
the complete gauged supergravity are the fermionic correction (the fermionic mass terms) and the scalar potential.
In order to work them out one needs to understand the irreducible components of the
$T$-tensor, i.e.\ the branching of
the basic representation of ${\rm E}_{9(9)}$ under $K({\rm E}_9)$. Corresponding to the finite number of fermions
there are finite $K({\rm E}_9)$ components in this decomposition that correspond the fermion mass matrices.
In addition, we already checked that the replacement $\partial_\mua \rightarrow D_\mua$ in the kinetic terms of
the fermions\footnote{Normally the fermions do not transform under $G_0$ but only under $H$, but since we have
a gauged fixed realization of $G_0$ we have induced $H$ that act on the fermions.}
yields fermionic contributions to the variations \eqref{VarVec} that agree with the fermionic corrections
to the linear system given in \cite{Nicolai:1988jb,Nicolai:1998gi}. Finally, we also still need to give the covariant version
of the conformal constraint \eqref{ConfConst2}.

\subsection{${\rm SO}(p,9-p)$ gaugings}

We want to finish this chapter with a short discussion of the ${\rm SO}(9)$ gaugings that originate from
a warped sphere reductions of $d=10$ IIA supergravity \cite{Boonstra:1998mp,Nicolai:2000zt,Bergshoeff:2004nq}.
Closely related are the compactifications on the non-compact manifolds $H^{p,8-p}$ that result in gauge
groups ${\rm SO}(p,9-p)$ (for details see section \ref{SecExNoZ}). We want to identify the embedding tensors
$\Theta_\cMa$ that defines these gaugings.

The embedding tensor $\Theta_\cMa$ transforms in the basic representation of ${\rm E}_{9(9)}$.
The appropriate starting point for our analysis is the decomposition of the basic representation under ${\rm SL}(9)$
which is given in table \ref{SL9hatRep1}. The corresponding decomposition of the symmetry algebra is given
in table \ref{Fig:AlgD2a}. We want to show that an embedding tensor for which only the
$\overline {\bf 45}$ component at $q+3l=4$ is non-zero always satisfied the quadratic constraint \eqref{QConD2}.
To do so we first need to specify into which other components this $\overline {\bf 45}$ can transform under the symmetry action of the
basic representation. Under the zero-mode $\mathfrak{sl}(9)_0$ it transform into itself.
Under the $\overline{\bf 84}_{+1}$ symmetry generator it transforms into the $\overline {\bf 315}$.
at $q+3l=3$. Under the ${\bf 80}_{+3}$ symmetry generator it transforms into the $\overline {\bf 36}$ at $q+3l=1$.
These are all non-negative symmetry generators under which the $\overline {\bf 45}$ transforms. With this information
we find the quadratic constraint to be a projector equation from the symmetric tensor product
$(\overline {\bf 45} \otimes \overline {\bf 45})_{\text{sym}}$ to the tensor product $\overline {\bf 36} \otimes \overline {\bf 45}$,
where the $\overline {\bf 36}$ is the one at $q+3l=1$. Asking for the irreducible components in these tensor products we
find
\begin{align}
   (\overline {\bf 45} \otimes \overline {\bf 45})_{\text{sym}} &= \overline {\bf 495} \oplus \overline {\bf 540} \; , &
   \overline {\bf 36} \otimes \overline {\bf 45} &= \overline {\bf 630} \oplus \overline {\bf 990} \; .
\end{align}
Since the two tensor products have no representation in common
there is no non-trivial ${\rm SL}(9)$ invariant projector between them and thus the quadratic constraint is satisfied for
every choice of the $\overline {\bf 45}$ component.

The $\overline {\bf 45}$ can be parameterized by a symmetric $9 \times 9$ matrix $Y$. By fixing part of the ${\rm SL}(9)$ symmetry
this matrix can be brought into the form
\begin{align}
  Y &= \diag(\,\underbrace{1, \dots,}_{p}\underbrace{-1,\dots,}_{q} \underbrace{0, \dots}_{r}\,) \;,
\end{align}
with $p+q+r=9$. Such an embedding tensor gauges a subgroup ${\rm CSO}(p,q,r)$ of the zero-mode ${\rm SL}(9)$.
The corresponding gauge fields are the one in the $\overline {\bf 36}$ at $q+3l=1$. For $r=0$ we have the ${\rm SO}(p,9-p)$
gaugings we were looking for. In addition, some of the positive generators of the symmetry algebra are excited.
The $\overline {\bf 84}_{+1}$ is not excited, the ${\bf 84}_{+2}$ is gauged completely, 44 generators of the ${\bf 80}_{+3}$
are gauged, etc.

The gauged Lagrangian \eqref{LgaugedD2} is formulated in a manifestly ${\rm E}_{8(8)}$ covariant way.
For the analysis in this subsection we used the ${\rm SL}(9)$ decomposition of the basic and of the adjoint representation.
The relation between the ${\rm E}_{8(8)}$ and the ${\rm SL}(9)$ was explained in section \ref{sec:VecFieldsD2}.
If we consider the ${\rm SO}(9)$ gauging in the ${\rm E}_{8(8)}$ picture we find the ${\rm SO}(9)$ algebra itself to be contained
in the zero-modes and in the first positive and the first negative algebra modes of ${\rm E}_{9(9)}$. At the level of the
Lagrangian all positive algebra modes seem to be cut (shift symmetries for the dual potential are not gauged in the Lagrangian)
and what is left of the ${\rm SO}(9)$ gauge group is a semi-direct product ${\rm SO}(8) \ltimes \mathbbm{R}^8$. Parts of 
the excited ${\bf 84}_{+2}$ and ${\bf 80}_{+3}$ generators are also contained in the zero-mode ${\rm E}_{8(8)}$. In total 
one finds a gauge group
\begin{align}
   G = (({\rm SO}(8) \ltimes \mathbbm{R}_+^8) \ltimes \mathbbm{R}_+^{28} ) \ltimes \mathbbm{R}_+^8
\end{align}
From this perspective it is not obvious that an ${\rm SO}(9)$ gauge group is realized.

\chapter{Conclusions and outlook}

We presented the general structure of gauged supergravities in various space-time dimensions.
We reviewed the concept of the embedding tensor~$\Theta$ which parameterizes the possible gaugings
of the respective theory. $\Theta$ is defined as a tensor under the global symmetry group which
couples the vector gauge fields to the symmetry generators in the covariant derivative.
Furthermore, $\Theta$ also determines all other couplings that have to
be introduced in the gauged theory in order to preserve gauge invariance and supersymmetry,
such as St\"uckelberg type couplings in the $p$-form field strengths, generalized topological terms and
fermionic mass terms. For consistency of the construction the embedding tensor has to obey a linear and a quadratic constraint
which can be formulated as representation constraints on the irreducible components of $\Theta$ and
on $\Theta \otimes \Theta$.

These general methods were then applied to particular extended supergravity theories.
For each case we identified the allowed irreducible components of $\Theta$
and in terms of those we gave the respective universal Lagrangian and the supersymmetry transformations of the gauged theory.
We also discussed particular examples of gaugings and whenever possible we identified
the origin of these gaugings in dimensional reduction.

The gaugings of $d=4$ half-maximal supergravity are parameterized by two ${\rm SL}(2) \times {\rm SO}(6,n)$ tensors
$f_{\aa\Ma\Mb\Mc}$ and $\xi_{\aa\Ma}$.
All previously known examples of gaugings can be described by turning on either $f_{\aa\Ma\Mb\Mc}$ or
$\xi_{\aa\Ma}$. E.g.\ in orientifold compactifications of IIB we found the flux parameters to
be contained in $f_{\aa\Ma\Mb\Mc}$. The phases that were introduced by de Roo-Wagemans
\cite{deRoo:1985jh,deRoo:1986yw,Wagemans:1990mv} in order to have
$AdS$ and Minkowski vacua in the gauged supergravities are also contained as parameters in $f_{\aa\Ma\Mb\Mc}$.
On the other hand, Scherk-Schwarz reduction from $d=5$ with a non-trivial ${\rm SO}(1,1)$ twist yields
a theory with
vanishing $f_{\aa\Ma\Mb\Mc}$ but non-vanishing $\xi_{\aa\Ma}$ \cite{Villadoro:2004ci}.
For a general gauging both tensors can be non-vanishing.
It would be interesting to further study these new theories
by classifying their ground states, computing the mass spectrum and analyzing stability.
Also the higher-dimensional origin of many of these theories
such as the ones with non-vanishing de Roo-Wagemans phases is still unknown.
The compactifications that yield these gaugings might be of unconventional type
\cite{Hull:2004in,Dabholkar:2005ve,Hull:2006tp}.

For the $d=5$ half-maximal supergravity the general gaugings
are parameterized by three ${\rm SO}(1,1) \times {\rm SO}(5,n)$ tensors
$f_{\Ma\Mb\Mc}$, $\xi_{\Ma\Mb}$ and $\xi_\Ma$. The gaugings with $\xi_\Ma=0$
were already described in~\cite{Dall'Agata:2001vb}, but it is necessary to incorporate $\xi_\Ma$
to also include non-semi-simple gaugings that result from Scherk-Schwarz dimensional reduction~\cite{Villadoro:2004ci}.
For a generic gauging all three tensors may be non-zero. We discussed the dimensional reduction of these
five-dimensional theories and showed how the parameters of the $d=5$ gaugings are contained in those of the $d=4$ gaugings.

For the maximal seven-dimensional supergravities the gaugings are described by two ${\rm SL}(5)$ tensors
$Y_{\Ja\Jb}$ and $Z^{\Ja\Jb,\Jc}$. In terms of these tensors we gave the universal Lagrangian
that combines vector, two-form and three-form tensor fields.
The Lagrangian is invariant under an extended set of non-Abelian gauge transformations as well as under maximal 
supersymmetry. The $p$-form gauge fields enter via St\"uckelberg type couplings
in the generalized field strengths of the respective lower rank gauge fields and they all couple through
a unique gauge invariant topological term.
This ensures that the total number of degrees of freedom is independent of $Y_{\Ja\Jb}$ and $Z^{\Ja\Jb,\Jc}$.
As particular examples we have recovered the known seven-dimensional 
gaugings as well as a number of new gaugings. Some of these theories have a definite
higher-dimensional origin, such as the Scherk-Schwarz reduction from $d=8$ dimensions and
the (warped) sphere reductions from string and M-theory.

Finally, we discussed the gaugings of $d=2$ maximal supergravity. In this case the global symmetry group of the
ungauged theory is the infinite dimensional affine Lie group ${\rm E}_{9(9)}$. We have shown that the vector fields and the
embedding tensor~$\Theta$ transform in the (dual) basic representation, i.e.\ in the unique level one representation of ${\rm E}_{9(9)}$.
The basic representation is infinite dimensional and thus the embedding tensor contains an infinite number of
gauge parameters. We worked out the quadratic constraint on $\Theta$ and gave its interpretation as
a projector equation on the infinite tower of level two representations in the tensor product $\Theta \otimes \Theta$.
We then presented the bosonic Lagrangian (up to the scalar potential) of the gauged theory and showed that the quadratic constraint on $\Theta$
ensures gauge invariance. We also identified the ${\rm SO}(9)$ gauging that originates from a warped sphere reduction
of IIA supergravity.

The construction of the gauged $d=2$ maximal supergravities is not yet complete. The fermionic correction
to the Lagrangian and the modified supersymmetry rules of the gauged theory still need to be calculated. In order to do so
one has to work out the irreducible representation of the $T$-tensor, i.e.\ the branching of the basic representation of $ {\rm E}_{9(9)}$ under
$K({\rm E}_{9})$. The fermionic mass matrices are contained in the $T$-tensor as finite dimensional irreducible $K({\rm E}_{9})$ components.
Once these are determined the scalar potential is fixed as well.

Another interesting question for these $d=2$ theories concerns the bosonic Lagrangian which we have presented.
Analogous to the choice of a
symplectic frame in $d=4$ there should be different $d=2$ Lagrangians for different ``scalar frames''.
We mentioned already that such a ``scalar frame'' can
be defined by the gauge fixing of the ${\rm E}_{9(9)}/K({\rm E}_{9})$ coset representative $\hV$. We always worked in the so-called Breitenlohner-Maison
gauge, but it would be interesting to identify the general data that define a particular gauge fixing and to work out the Lagrangian for the generic
case. Analogous to the $d=4$ theories we would then expect that for any particular gauging,
i.e.\ for any particular embedding tensor, there exists a natural
scalar gauge fixing which is closely related to the higher-dimensional origin of the respective theory. It would be of great interest to study these
questions in the future.


\begin{appendix}

\chapter[Dimensional reduction of the embedding tensor]
{Dimensional reduction of the embedding tensor of maximal supergravity} \label{app:DimRedEmb}

A torus reduction of a gauged supergravity yields a lower-dimensional gauged supergravity with the same number of supercharges.
Therefore the higher-dimensional embedding tensor has to be
contained in the lower-dimensional one. In this appendix we consider the maximal supergravities and show  
how the $d+1$-dimensional embedding tensor is contained in the $d$ dimensional one. 
Step by step this yields the higher-dimensional linear constraint as a consequence of the three-dimensional one.

We start with the reduction from $d=7$ to $d=6$.
From a circle reduction one naively expects the symmetry group $\mathbbm{R}^+ \times {\rm
SL}(5)$, i.e.\ the product of
the seven-dimensional symmetry group and the circle rescalings. The actual $d=6$ symmetry algebra decomposes under this
group as 
\begin{align}
  \mathfrak{so}(5,5) \rightarrow 
  {\bf 1}_0 \oplus \mathfrak{sl}(5)_{0} \oplus \overline {\bf 10}_{+4} \oplus {\bf 10}_{-4} \; ,
\end{align}
where the subscripts indicate
the $\mathbbm{R}^+$ charges and ${\bf 1}_0$ is the $\mathbbm{R}^+$ generator. We already gave this decomposition
in figure \ref{Fig:AlgD6}, but the $\mathbbm{R}^+ \times {\rm SL}(5)$ has a different physical meaning now\footnote{We also use
different conventions for the $\mathbbm{R}^+$ charges now.}.
The $\overline {\bf 10}_{+4}$ generator in the $d=6$ symmetry algebra originates from the gauge symmetry
of the $d=7$ vector fields, analogous to equation \eqref{ShiftChi} in section \ref{sec:MaxTorus}.
The additional dual generators ${\bf 10}_{-4}$ appear as an enhancement of the symmetry and have no ancestors in $d=7$. 
Similarly we can trace back the origin of the six-dimensional vector fields, which transform in the representation ${\bf 16}_s$
under ${\rm SO}(5,5)$ and branch under $\mathbbm{R}^+ \times {\rm SL}(5)$ as
\begin{align}
   {\bf 16}_s \rightarrow {\bf 1}_{-5} \oplus \overline {\bf 10}_{-1} \oplus {\bf 5}_{+3} \; . 
\end{align}   
The singlet in this decomposition is
the Kaluza-Klein vector field from the metric, the $\overline {\bf 10}_{-1}$ are the $d=7$ vector fields and
the ${\bf 5}_{+3}$ are vector fields that originate from the seven-dimensional two-forms. Therefore it is clear that the only 
six-dimensional couplings from the vector fields to the symmetry generators that can already be present in $d=7$ are those that
couple the vector fields $\overline {\bf 10}_{-1}$ and ${\bf 5}_{+3}$ to the symmetry generators $\mathfrak{sl}(5)_{0}$
and $\overline {\bf 10}_{+4}$, since all other couplings do involve vector fields or symmetry generators that are not yet present in
$d=7$.\footnote{Of course, the Kaluza Klein vector field and the circle rescalings do exist in $d=7$ as part of the metric and
of general coordinate transformations, but one could not couple them without breaking Lorentz invariance.
Note that also couplings from the $\overline {\bf 10}_{-1}$ vector fields to the $\overline {\bf 10}_{+4}$ symmetry
generators and from the ${\bf 5}_{+3}$ vector fields to the $\mathfrak{sl}(5)_{0}$ symmetry generators would break Lorenz invariance
in $d=7$, but we will find those couplings to be excluded anyway.}. 

Under $\mathbbm{R}^+ \times {\rm SL}(5)$ the embedding tensor decomposes as
\begin{align}
   {\bf 144}_s \rightarrow {\bf 24}_{+5} \oplus {\bf 10}_{+1} \oplus {\bf 15}_{+1}
				                         \oplus \overline {\bf 40}_{+1} \oplus \overline {\bf 5}_{-3} 
							 \oplus {\bf 45}_{-3} \oplus {\bf 5}_{-7} \;.
\end{align}							 
In table \ref{ThetaD6D7} we specify which of these irreducible components couple which vector fields
to which symmetry generators. A box is set around those couplings that could already be present in $d=7$. The table shows
that only the components ${\bf 15}_{+1}$ and $\overline {\bf 40}_{+1}$ of the embedding tensor are allowed in $d=7$,
since all other components also appear outside the box. Since the ${\bf 15}_{+1}$ and $\overline {\bf 40}_{+1}$ do couple
the $d=7$ vector fields ${\bf 10}_{-1}$ to the $d=7$ symmetry generators $\mathfrak{sl}(5)_{0} = {\bf 24}_0$, they constitute the
embedding tensor in $d=7$, as already stated in table \ref{LinCon1} of the last section. In addition, the $\overline {\bf 40}_{+1}$
component also couples the ${\bf 5}_{+3}$ vector fields to the $\overline {\bf 10}_{+4}$ symmetry generators, which in $d=7$
is a coupling from the two-form gauge fields to the gauge-symmetry of the vector-fields. We did indeed introduce couplings
of this kind in the section \ref{sec:GenNonAVecTen} and found them necessary for gauge invariance.

\begin{table}[h]
     \begin{tabular}{l c||@{\qquad}c@{\qquad}c@{\qquad}c@{\qquad}c@{}c|}
         & \multicolumn{5}{c}{~~~~~vector fields}
	 \\[0.2cm]
	 && ${\bf 1}_{-5}$ & & $\overline{\bf 10}_{-1}$ & ${\bf 5}_{+3}$  \\ \cline{2-7} &&&\\[-0.44cm] \cline{2-7}
	 &&& \vline &&& \\[-0.3cm]
	 \multirow{4}{1cm}{\begin{sideways}\parbox{2.5cm}{symmetry\\ generators}\end{sideways}} &
	 $\overline{\bf 10}_{+4}$ & ---
	                    & \vline & ${\bf 24}_{+5}$ & ${\bf 10}_{+1} \oplus \overline{\bf 40}_{+1}$ &
	 \\[0.4cm]	      
	 & ${\bf 24}_{0}$ & 
	    ${\bf 24}_{+5}$ & \vline & ${\bf 10}_{+1} \oplus {\bf 15}_{+1} \oplus \overline{\bf 40}_{+1}$ 
	                                       & $\overline {\bf 5}_{-3} \oplus {\bf 45}_{-3}$ &
         \\[0.1cm] \cline{4-7} && \\[-0.4cm]
	 & ${\bf 1}_{0}$ & ---
	               & & ${\bf 10}_{+1}$   &  $\overline {\bf 5}_{-3}$
         \\[0.4cm]
	 & ${\bf 10}_{-4}$ & 
	    ${\bf 10}_{+1}$  & & $\overline {\bf 5}_{-3} \oplus {\bf 45}_{-3}$ &   ${\bf 5}_{-7}$
     \end{tabular}
     \caption{\label{ThetaD6D7}\small The $d=6$ couplings of the vector fields to the symmetry generators
                                      are decomposed under the $d=7$ symmetry group ${\rm SL}(5)$.}
\end{table}

The same argument as we just used to find the $d=7$ linear constraint from the one in $d=6$ can be applied
to the other dimensions as well, and the necessary decompositions of the embedding tensors are given in table
\ref{ThetaD3D4} to \ref{ThetaD7D8} for $3 \leq d \leq 7$. The analogous reduction from $d=3$ to $d=2$ 
will be discussed in chapter \ref{ch:D2}. One finds that the vector fields in $d$ dimensions always decompose into
the Kaluza-Klein vector, vector fields from $d+1$ dimensions, two-form fields from $d+1$ dimensions and for $d \leq 4$ also
additional vector fields from dualization appear. The decomposition of the symmetry generators is always analogous to the
$d=6$ case we discussed, but for $d=3$ one has an additional singlet ${\bf 1}_{+2}$, which is the shift-symmetry of the scalar
dual to the Kaluza-Klein vector, and via symmetry enhancement also the dual generator ${\bf 1}_{-2}$ appears. 
In all cases we have a similar $2 \times 2$ box of couplings that could already be present in $d+1$ dimensions, and one
can easily check consistency with the linear constraints given in table \ref{LinCon1}.

\begin{table}[tbp]
   \begin{center}
     \begin{small}
     \begin{tabular}{l c||c@{\quad}c@{\;}c@{\quad}c@{\;}r@{\quad}c@{\quad}c@{\quad}c}
         & \multicolumn{8}{c}{\normalsize ~\qquad\qquad~vector fields}
	 \\[0.2cm]
	 && ${\bf 1}_{-2}$ && ${\bf 56}_{-1}$ & ${\bf 133}_0$ & & ${\bf 1}_{0}$ & ${\bf 56}_{+1}$ & ${\bf 1}_{+2}$  \\ 
	      \cline{2-10} &&&\\[-0.4cm] \cline{2-10}
	 &&& \\[-0.2cm]
	 \multirow{4}{0.4cm}{\begin{sideways} \normalsize symmetry generators ~~~~~ \end{sideways}} &
	 ${\bf 1}_{+2}$ & --- & & ---
	                     & ${\bf 133}_{+2}$ & & --- & ${\bf 56}_{+1}$  & ${\bf 1}_{0}$
         \\[0.65cm] \cline{4-6} &&&\vline&&&\vline \\[-0.3cm]
	 & ${\bf 56}_{+1}$ & --- & \vline & ${\bf 133}_{+2}$
	                     & ${\bf 56}_{+1} \oplus {\bf 912}_{+1}$ &\vline& ${\bf 56}_{+1}$ &
	             \parbox{1.6cm}{${\bf 1}_{0} \oplus {\bf 133}_{0}$ \\ $\oplus {\bf 1539}_{0}$} & ${\bf 56}_{-1}$  
	 \\[0.75cm]	      
	 & ${\bf 133}_{0}$ & 
	             ${\bf 133}_{+2}$ & \vline & ${\bf 56}_{+1} \oplus {\bf 912}_{+1}$ 
		      & \parbox{1.6cm}{${\bf 1}_{0} \oplus {\bf 133}_{0}$ \\ $\oplus {\bf 1539}_{0}$}
		              &\vline& ${\bf 133}_{0}$ & ${\bf 56}_{-1} \oplus {\bf 912}_{-1}$
		       & ${\bf 133}_{-2}$ 
         \\[0.25cm] \cline{4-6} && \\[-0.3cm]
	 & ${\bf 1}_{0}$ & 
	             --- & & ${\bf 56}_{+1}$  
		      & ${\bf 133}_{0}$ && ${\bf 1}_{0}$ & ${\bf 56}_{-1}$ & ---
         \\[0.75cm]
	 & ${\bf 56}_{-1}$ & 
	            ${\bf 56}_{+1}$ & &  \parbox{1.6cm}{${\bf 1}_{0} \oplus {\bf 133}_{0}$ \\ $\oplus {\bf 1539}_{0}$}   
		      & ${\bf 56}_{-1} \oplus {\bf 912}_{-1}$
		       && ${\bf 56}_{-1}$ & ${\bf 133}_{-2}$ & ---
         \\[0.75cm]
	 & ${\bf 1}_{-2}$ & ${\bf 1}_{0}$ & & ${\bf 56}_{-1}$ & ${\bf 133}_{-2}$ && ---  & --- & ---
     \end{tabular}
     \end{small}
     \caption{\label{ThetaD3D4}\small The $d=3$ couplings of the vector fields to the symmetry generators
                                      are decomposed under the $d=4$ symmetry group ${\rm E}_{7(7)}$.
				      The two components of the embedding tensor branch as ${\bf 1} \rightarrow {\bf 1}_0$
				      and
				      ${\bf 3875} \rightarrow {\bf 133}_{+2} \oplus {\bf 56}_{+1} \oplus {\bf 912}_{+1}
               	                         \oplus {\bf 1}_{0} \oplus \overline {\bf 133}_{0} 
		        		 \oplus {\bf 1539}_{0} \oplus {\bf 912}_{-1} \oplus {\bf 56}_{-1} \oplus {\bf 133}_{-2}$ and most of
				      these components appear several times in the table.
                                      The subscripts indicate the charges under rescalings of the internal circle,
				      i.e.\ under the symmetry generator ${\bf 1}_0$.
				      Only couplings in the box can originate from $d=4$ gaugings.}
     \vspace{2.5cm}
     \begin{tabular}{l c||@{\qquad}c@{\qquad}c@{\;}c@{\qquad}c@{\;}l@{\qquad}c}
         & \multicolumn{5}{c}{~~~~~~\qquad\qquad~~~~~~vector fields}
	 \\[0.2cm]
	 && ${\bf 1}_{-3}$ & & $\overline {\bf 27}_{-1}$ & ${\bf 27}_{+1}$ && ${\bf 1}_{+3}$  \\ 
	      \cline{2-8} &&&\\[-0.44cm] \cline{2-8}
	 &&& \vline &&& \vline \\[-0.3cm]
	 \multirow{5}{1cm}{\begin{sideways}\parbox{2.5cm}{symmetry\\ generators}\end{sideways}} &
	 $\overline {\bf 27}_{+2}$ & ---
                   & \vline & ${\bf 78}_{+3}$ & ${\bf 27}_{+1} \oplus \overline {\bf 351}_{+1}$ & \vline & $\overline {\bf 27}_{-1}$
	 \\[0.5cm]	      
	 & ${\bf 78}_{0}$ & 
	             ${\bf 78}_{+3}$ & \vline & ${\bf 27}_{+1} \oplus \overline {\bf 351}_{+1}$ 
		      & $\overline {\bf 27}_{-1} \oplus {\bf 351}_{-1}$ & \vline & ${\bf 78}_{-3}$
         \\[0.1cm] \cline{4-6} && \\[-0.4cm]
	 & ${\bf 1}_{0}$ & ---
	               & & ${\bf 27}_{+1}$   &  $\overline {\bf 27}_{-1}$ & & ---
         \\[0.5cm]
	 & ${\bf 27}_{-2}$ & 
	    ${\bf 27}_{+1}$  & & $\overline {\bf 27}_{-1} \oplus {\bf 351}_{-1}$ &  ${\bf 78}_{-3}$ & & ---
     \end{tabular}
     \caption{\label{ThetaD4D5}\small Like table \ref{ThetaD3D4}, but for the $d=4$ couplings under the
                                      $d=5$ symmetry group ${\rm E}_{6(6)}$.
              			      The embedding tensor branches as
				      ${\bf 912} \rightarrow {\bf 78}_{+3} \oplus {\bf 27}_{+1} \oplus \overline {\bf 351}_{+1}
				                         \oplus {\bf 351}_{-1} \oplus \overline {\bf 27}_{-1} 
							 \oplus {\bf 78}_{-3}$.}
   \end{center}     
\end{table}

\begin{table}[tbp]
   \begin{center}
     \begin{tabular}{l c||@{\qquad}c@{\qquad}c@{\qquad}c@{\qquad}c@{}c|}
         & \multicolumn{5}{c}{~~~~~vector fields}
	 \\[0.2cm]
	 && ${\bf 1}_{-4}$ & & ${\bf 16}_{s,-1}$ & ${\bf 10}_{+2}$  \\ \cline{2-7} &&&\\[-0.44cm] \cline{2-7}
	 &&& \vline &&& \\[-0.3cm]
	 \multirow{4}{1cm}{\begin{sideways}\parbox{2.5cm}{symmetry\\ generators}\end{sideways}} &
	 ${\bf 16}_{s,+3}$ & ---
	                    & \vline & ${\bf 45}_{+4}$ & ${\bf 16}_{c,+1} \oplus {\bf 144}_{+1}$ &
	 \\[0.4cm]	      
	 & ${\bf 45}_{0}$ & 
	             ${\bf 45}_{+4}$ & \vline & ${\bf 16}_{c,+1} \oplus {\bf 144}_{+1}$ 
	                                       & ${\bf 10}_{-2}\oplus{\bf 120}_{-2}$ &
         \\[0.1cm] \cline{4-7} && \\[-0.4cm]
	 & ${\bf 1}_{0}$ & ---
	               & & ${\bf 16}_{c,+1}$   &  ${\bf 10}_{-2}$
         \\[0.4cm]
	 & ${\bf 16}_{c,-3}$ & 
	    ${\bf 16}_{c,+1}$  & & ${\bf 10}_{-2}\oplus{\bf 120}_{-2}$ &  ${\bf 16}_{s,-5}$
     \end{tabular}
     \caption{\label{ThetaD5D6}\small Like table \ref{ThetaD3D4}, but for the $d=5$ couplings under the
                                      $d=6$ symmetry group ${\rm SO}(5,5)$.
              			      The embedding tensor branches as
				      $\overline {\bf 351} \rightarrow {\bf 45}_{+4} \oplus {\bf 16}_{c,+1} \oplus {\bf 144}_{+1}
				                         \oplus {\bf 120}_{-2} \oplus {\bf 10}_{-2} 
							 \oplus {\bf 16}_{s,-5}$.}
     \vspace{1cm}
     \begin{tabular}{l c||@{\qquad}c@{\qquad}c@{\qquad}c@{\qquad}c@{}c|}
         & \multicolumn{5}{c}{~~~~~vector fields}
	 \\[0.2cm]
	 && ${\bf (1,1)}_{-6}$ & & ${\bf (2,3)}_{-1}$ & ${\bf (1,\overline{3})}_{+4}$  
	          \\[0.1cm] \cline{2-7} &&&\\[-0.44cm] \cline{2-7}
	 &&& \vline &&& \\[-0.3cm]
	 \multirow{4}{1cm}{\begin{sideways}\parbox{3cm}{symmetry\\ generators}\end{sideways}} &
	 ${\bf (2,3)}_{+5}$ & ---
	                    & \vline & ${\bf (3,1)}_{+6}\oplus{\bf (1,8)}_{+6}$ 
			                    & ${\bf (2,\overline{3})}_{+1} \oplus {\bf (2,6)}_{+1}$ &
	 \\[0.4cm]	      
	 & ${\bf (3,1)}_{0}$ &
	               ${\bf (3,1)}_{+6}$ & \vline & ${\bf (2,\overline{3})}_{+1}$     & ${\bf (3,3)}_{-4}$ &
         \\[0.18cm]
	 & ${\bf (1,8)}_{0}$ &
	               ${\bf (1,8)}_{+6}$ & \vline & ${\bf (2,\overline{3})}_{+1} \oplus {\bf (2,6)}_{+1}$ 
	                                       & ${\bf (1,\overline{6})}_{-4} \oplus {\bf (1,3)}_{-4}$ &
         \\[0.1cm] \cline{4-7} && \\[-0.4cm]
	 & ${\bf (1,1)}_{0}$ & ---
	               & & ${\bf (2,\overline{3})}_{+1}$   &  ${\bf (1,3)}_{-4}$
         \\[0.4cm]
	 & ${\bf (2,\overline{3})}_{-5}$ & 
	    ${\bf (2,\overline{3})}_{+1}$  & & ${\bf (1,\overline{6})}_{-4} \oplus {\bf (3,3)}_{-4}$ &  ${\bf (2,1)}_{-9}$
     \end{tabular}
     \caption{\label{ThetaD7D8}\small Like table \ref{ThetaD3D4}, but for the $d=7$ couplings under the
                                      $d=8$ symmetry group ${\rm SL}(2) \times {\rm SL}(3)$.
              			      The two components of the embedding tensor branch as
				      ${\bf 15} \rightarrow {\bf (3,1)}_{+6} \oplus {\bf (2,\overline{3})}_{+1} 
				                          \oplus {\bf (1,\overline{6})}_{-4}$ 
				      and 			  
				      $\overline {\bf 40} \rightarrow {\bf (1,8)}_{+6} \oplus {\bf (2,\overline{3})}_{+1} 
    	                                     \oplus {\bf (2,6)}_{+1} \oplus {\bf (1,3)}_{-4} \oplus {\bf (3,3)}_{-4}
					       \oplus {\bf (2,1)}_{-9}$, and the
				      two ${\bf (2,\overline{3})}_{+1}$ components appear in different linear combinations.}
   \end{center}     
\end{table}

Those components in the decomposition of the $d$ dimensional embedding tensors
that do not originate from the $d+1$ dimensional embedding tensor
can still have a well defined origin in generalized dimensional reduction. For example a Scherk-Schwarz reduction  
is always possible if the
theory that is reduced possesses a global symmetry $G_0$, i.e.\ if all fields, and in particular the scalars ${\cal V}$
transform in some representation of $G_0$ \cite{Scherk:1979zr}.
One then chooses a symmetry generator $Z=Z^\sa t_\sa$ and demands all fields to have a
particular dependence on the internal coordinate $y$, namely
\begin{align}
   \frac{\partial} {\partial y} \, {\cal V} &= Z \, {\cal V} \; ,
\end{align}
and similarly for the non-scalar fields. In the lower dimensional theory the Kaluza-Klein vector then couples
as a gauge field to the symmetry generator $Z$. Looking at the table \ref{ThetaD3D4} to \ref{ThetaD7D8} one finds
the component of embedding tensor with the highest $\mathbbm{R}^+$ charge always to be in the adjoint representation of the
$d+1$ dimensional symmetry group $G_0$. It couples the Kaluza-Klein vector field to a symmetry generators of $G_0$.
Therefore, this component corresponds to the Scherk-Schwarz generator $Z$. Tables
\ref{ThetaD3D4} to \ref{ThetaD7D8} also give the remaining gauge couplings that result from this reduction.
Note that the quadratic constraint on the embedding tensor is satisfied automatically if only the Scherk-Schwarz generator $Z$
is switched on.

\chapter{Gauged half-maximal supergravities in $d=3$}
\label{app:D3}

The general gauged half-maximal supergravity in $D=3$ was given in \cite{deWit:2003ja,Nicolai:2001ac}.
Here we shortly describe the underlying group theory and the tensors that parameterize the gauging.
We then give the fermion shift matrices and the scalar potential in the same form as it was given
for the four and five dimensional theories in chapter \ref{ch:D4} and \ref{ch:D5}.
Finally we describe the embedding of the four dimensional gaugings into the 
three dimensional ones. This relation is necessary in order to
calculate the four and five dimensional scalar potentials from the known three dimensional one.

\section[Scalar potential and fermion shift matrices]{General gauging, scalar potential and fermion shift matrices}
\label{app:D3embedding}

The global symmetry group of the ungauged theory is $G={\rm SO}(8,n)$, where
$n$ again counts the number of vector multiplets. The vector fields  $A_\mua{}^{\Ma\Mb}=A_\mua{}^{[\Ma\Mb]}$
transform in the adjoint representation of $G$. Here $\Ma,\Mb=1,\ldots, 8+n$ are ${\rm SO}(8,n)$ vector indices.
The general gauging is parameterized by the two real tensors
$\lambda_{\Ma\Mb\Mc\Md}=\lambda_{[\Ma\Mb\Mc\Md]}$ and
$\lambda_{\Ma\Mb}=\lambda_{(\Ma\Mb)}$, with $\eta^{\Ma\Mb} \lambda_{\Ma\Mb} = 0$,
and one real scalar $\lambda$. Together they constitute the embedding tensor
\begin{align}
   \Theta_{\Ma\Mb\Mc\Md} &= \lambda_{\Ma\Mb\Mc\Md} + \lambda_{[\Mc[\Ma} \, \eta_{\Mb]\Md]} 
                             + \lambda \, \eta_{\Mc[\Ma} \, \eta_{\Mb]\Md} \; ,
\end{align}
which enters into the covariant derivative
\begin{align}
   D_\mua &= \partial_\mua - A_\mua{}^{\Ma\Mb} \Theta_{\Ma\Mb}{}^{\Mc\Md} t_{\Mc\Md} \;.
\end{align}
Due to the above definition the embedding tensor automatically satisfies the linear constraint
\begin{align}
   \Theta_{\Ma\Mb \, \Mc\Md} &= \Theta_{\Mc\Md \, \Ma\Mb} \; .
\end{align}
In addition it has to satisfy the quadratic constraint
\begin{align}
   \Theta_{\Ma\Mb\Mg}{}^{\Mi} \Theta_{\Mc\Md\Mi}{}^{\Mh} - \Theta_{\Mc\Md\Mg}{}^{\Mi} \Theta_{\Ma\Mb\Mi}{}^{\Mh}
      &= \Theta_{\Ma\Mb[\Mc}{}^{\Mi} \Theta_{\Md]\Mi\Mg}{}^{\Mh} \; ,
   \label{D3QC}      
\end{align}
which may be written as a constraint on $\lambda_{\Ma\Mb\Mc\Md}$, $\lambda_{\Ma\Mb}$ and $\lambda$.

The scalars of the theory form the coset ${\rm SO}(8,n)/{\rm SO}(8) \times {\rm SO}(n)$ and in the following we
use the same conventions and notations as for the ${\rm SO}(6,n)/{\rm SO}(6) \times {\rm SO}(n)$ coset in four dimension,
in particular we again have
\begin{align}
   M_{\Ma\Mb} &= {{\cal V}_\Ma}^\xa {{\cal V}_\Mb}^\xa + {{\cal V}_\Ma}^{\ya} {{\cal V}_\Mb}^{\ya} \; , &
   \eta_{\Ma\Mb} &= {{\cal V}_\Ma}^\xa {{\cal V}_\Mb}^\xa - {{\cal V}_\Ma}^{\ya} {{\cal V}_\Mb}^{\ya} \; ,
\end{align}
where now $\xa = 1,\ldots,n$ and $\ya = 1,\ldots, 8$.
In addition we need the scalar dependent object
\begin{align}
   M_{\Ma\Mb\Mc\Md\Me\Mf\Mg\Mh} &= \epsilon_{\ya\yb\yc\yd\ye\yf\yg\yh}
                             {{\cal V}_\Ma}^{\ya} {{\cal V}_\Mb}^{\yb} {{\cal V}_\Mc}^{\yc} 
                             {{\cal V}_\Md}^{\yd} {{\cal V}_\Me}^{\ye} {{\cal V}_\Mf}^{\yf} 
			     {{\cal V}_\Mg}^{\yg} {{\cal V}_\Mh}^{\yh} \; .
\end{align}
The scalar potential then takes the form
\begin{align}
   V &= - \, \frac 1 {24} \, \Bigg[ 
               \lambda_{\Ma\Mb\Mc\Md} \lambda_{\Me\Mf\Mg\Mh} 
             \Big( - \ft 1 2 M^{\Ma\Me} M^{\Mb\Mf} M^{\Mc\Mg} M^{\Md\Mh}
	           + 3 M^{\Ma\Me} M^{\Mb\Mf} \eta^{\Mc\Mg} \eta^{\Md\Mh}
	           \nonumber \\ & \qquad
		   - 4  M^{\Ma\Me} \eta^{\Mb\Mf} \eta^{\Mc\Mg} \eta^{\Md\Mh}
		   + \ft 3 2 M^{\Ma\Me} \eta^{\Mb\Mf} \eta^{\Mc\Mg} \eta^{\Md\Mh}
                   + \ft 1 3 M^{\Ma\Mb\Mc\Md\Me\Mf\Mg\Mh} 
	     \Big)
	\nonumber \\ & \qquad     
           + \lambda_{\Ma\Mb} \lambda_{\Mc\Md}
	     \left( - \ft 3 2 M^{\Ma\Mc} M^{\Mb\Md} + \ft 3 2 \eta^{\Ma\Mc} \eta^{\Mb\Md}
	            + \ft 3 4 M^{\Ma\Mb} M^{\Mc\Md} \right) 
        \nonumber \\ & \qquad		    
	   + 192 \lambda^2 - 24 \, \lambda \, \lambda_{\Ma\Mb} M^{\Ma\Mb} 
	   \Bigg] \; .
    \label{VD3}	   
\end{align}
Although written differently, this is the same potential as given in \cite{deWit:2003ja}.

The maximal compact subgroup of $G$ is $H={\rm SO}(8) \times {\rm SO}(n)$.
All the fermions and the fermion shift matrices $A_1$ and $A_2$ transform under $H$.
Let $A, \, \dot A = 1, \ldots, 8$ be (conjugate) ${\rm SO}(8)$ spinor indices. The Gamma-matrices of ${\rm SO}(8)$
satisfy
\begin{align}
   \Gamma^{(\ya}_{A \dot A} \Gamma^{\yb)}_{B \dot A} &= \delta^{\ya\yb} \delta_{AB} \, ,&
   \Gamma^{\ya\yb}_{AB} & \equiv \Gamma^{[\ya}_{A \dot A} \Gamma^{\yb]}_{B \dot A} \, .
\end{align}
Then the fermion shift matrices $A_1$ and $A_2$ are defined through the so-called $T$-tensor as follows \cite{deWit:2003ja}
\begin{align}
   T^{AB \, CD} &= \frac 1 {16} \Gamma^{AB}_{\ya\yb} \Gamma^{CD}_{\yc\yd}
                    {{\cal V}_\Ma}^\ya {{\cal V}_\Mb}^\yb {{\cal V}_\Mc}^\yc {{\cal V}_\Md}^\yd
                    \Theta^{\Ma\Mb \, \Mc\Md} \; , \nonumber \\
   T^{AB \, \ya\xa} &= \frac 1 4 \Gamma^{AB}_{\yc\yd}
                    {{\cal V}_\Ma}^\yc {{\cal V}_\Mb}^\yd {{\cal V}_\Mc}^\ya {{\cal V}_\Md}^\xa
                    \Theta^{\Ma\Mb \, \Mc\Md} \; , \nonumber \\
   A_1^{AB} &= - \frac 8 3 T^{AC \, BC} + \frac 4 {21} \delta^{AB} T^{CD \, CD}  \; , \nonumber \\
   {A_2^{AB}}_{\ya\xa} &= 2 {T^{AB}}_{\ya\xa} 
                        - \frac 2 3 \Gamma^{C(A}_{\ya\yb} {T^{B)C}}_{\yb\xa}
			- \frac 1 {21} \delta^{AB} \Gamma^{CD}_{\ya\yb} {T^{CD}}_{\yb\xa} \; .
\end{align}
The quadratic constraint \eqref{D3QC} guarantees that $A_1$ and $A_2$ satisfy
\begin{align}
    A_1^{AC} A_1^{BC} - {A_2^{AC}}_{\ya\xa} {A_2^{BC}}_{\ya\xa} &= - \frac 1 {128} \delta^{AB} V \; ,
\end{align}
with the scalar potential $V$ appearing on the right hand side.

\section{From $d=4$ to $d=3$}
\label{sec:reduction43}

Performing a circle reduction of four dimensional $N=4$ supergravity with $n$ vector multiplets
yields a three dimensional $N=8$ supergravity with $n+2$ vector multiplets. 
The embedding of the global symmetry groups is given by 
\begin{align}
   {\rm SO}(8,n+2) \, \supset \, {\rm SO}(2,2)  \times {\rm SO}(6,n)
                   \, \supset \, {\rm SL}(2) \times {\rm SO}(6,n) \; ,
\end{align}   
where the ${\rm SL}(2)$ is just one of the factors in ${\rm SO}(2,2) = {\rm SL}(2) \times {\rm SL}(2)$.
Accordingly we split the fundamental representation of ${\rm SO}(8,n+2)$ as 
$v^{\tilde \Ma} = (v^\Ma, v^{x \aa})$ where $\aa=1,2$ and $x=1,2$.
Note that the ${\rm SO}(8,n+2)$ vector index is denoted by $\tilde \Ma$, while $\Ma$ is an ${\rm SO}(6,n)$ vector index.
The ${\rm SO}(2,2)$ metric is given by
\begin{align}
   \eta_{x \aa \; y \ab} &= \epsilon_{xy} \epsilon_{\aa\ab} \, , &
   \text{which yields} &&
    \eta_{x \aa \; y \ab} \eta^{y \ab \; z \ac} &= \delta^{z \ac}_{x \aa} \; .
\end{align}
The ${\rm SL}(2)$ generators $t_{(\aa\ab)}$, $t_{(xy)}$ and the ${\rm SO}(2,2)$
generators $t_{x \aa \, y \ab}=t_{y \ab\, x\aa}$
are related as follows
\begin{align}
   t_{x \aa \, y \ab} = - \frac 1 2 \left( \epsilon_{\aa\ab} t_{xy} + \epsilon_{xy} t_{\aa\ab} \right) \; ,
\end{align}
where we use the conventions $(t_{\cMa\cMb})_\cMc{}^\cMd = \delta^\cMd_{[\cMa} \eta^{\phantom{\cMd}}_{\cMb]\cMc}$ for the
${\rm SO}(2,2)$ generators ($\cMa=x\aa$).
The embedding of the $D=4$ vector fields into the $D=3$ ones is then given by
\begin{align}
   A_\mua^{\Ma \alpha} \, = \, A_\mua^{\Ma \, 1 \alpha} \; ,
\end{align}
where $A_\mua^{\Ma \, 1 \alpha}$ denotes the corresponding components of the $D=3$ vector fields
$A_\mua^{\tilde \Ma \tilde \Mb} = A_\mua^{[\tilde \Ma \tilde \Mb]}$.
Analogous to the reduction from $D=5$ to $D=4$ described in section \ref{sec:reduction},
now the covariant derivatives in $D=4$ and $D=3$ have to agree for those terms already present in $D=4$, i.e.\
\begin{align}
   D_\mua  \, &\supset \, \partial_\mua 
       - 2 A_\mua^{\Ma \, 1 \alpha} {\Theta_{\Ma \, 1 \alpha}}^{\Mb\Mc} t_{\Mb\Mc}
       + A_\mua^{\Ma \, 1 \alpha} {\Theta_{\Ma \, 1 \alpha}}^{x \ab \, y \ac} \, \epsilon_{xy} \, t_{\ab\ac}
       \nonumber \\
        &=  \, \partial_\mua - A_\mua{}^{\Ma\aa} {\Theta_{\Ma\aa}}^{\Mb\Mc} t_{\Mb\Mc}
                             - A_\mua{}^{\Ma\aa} {\Theta_{\Ma\aa}}^{\ab\ac} t_{\ab\ac}  \; .    
\end{align}
This yields
\begin{align}
   \lambda_{1 \aa \, \Ma\Mb\Mc} &= - \, \ft 1 2 \, f_{\aa \Ma\Mb\Mc} \; , &
   \lambda_{\Ma \, 1\aa \, x\ab \, y\ac} \epsilon^{xy} &= \, \ft 1 2 \, \epsilon_{\aa(\ac} \xi_{\ab)\Ma} \; , &
   \lambda_{1 \aa \, \Ma} &= \xi_{\aa \Ma} \; ,
\end{align}
while we demand the other components of $\lambda_{\tilde \Ma \tilde \Mb \tilde \Mc \tilde \Md}$ and
$\lambda_{\tilde \Ma \tilde \Mb}$ to vanish and also $\lambda=0$. However,
the antisymmetry of $\lambda_{\tilde \Ma \tilde \Mb \tilde \Mc \tilde \Md}$ 
and the symmetry of $\lambda_{\tilde \Ma \tilde \Mb}$ has to be imposed, for example
\begin{align}
   \lambda_{\Ma \, z\aa \, x\ab \, y\ac} &= \tilde \lambda_{\Ma \, [\{z\aa\} \, \{x\ab\} \, \{y\ac\}] } \, , &
   \tilde \lambda_{\Ma \, z\aa \, x\ab \, y\ac} &= \, \ft 1 2 \, \delta_z^1 \epsilon_{xy} \epsilon_{\aa(\ac} \xi_{\ab)\Ma} \, .
\end{align}
We have thus defined the embedding of the four dimensional gaugings into the three dimensional ones.
The quadratic constraint \eqref{D3QC} in $D=3$ is satisfied iff the $D=4$ quadratic constraint \eqref{QConD4}
is satisfied. The $D=3$ scalar potential \eqref{VD3} reduces to the $D=4$ potential \eqref{VD4} when all $D=3$
extra scalars are set to the origin.

\chapter{The $T$-tensor of maximal $d=7$ supergravity}

\section{${\rm USp}(4)$ invariant tensors}
\label{app:tensors}

In this appendix we introduce a number of ${\rm USp}(4)$ invariant tensors 
which explicitly describe the projection of ${\rm USp}(4)$ tensor products
onto their irreducible components and derive some relations between them.
All of these tensors are constructed from the invariant 
symplectic form $\Omega_{\ja\jb}$ and the relations that they satisfy
can be straightforwardly derived form the properties of $\Omega_{\ja\jb}$. 
We have used these tensors extensively in the course of our calculations, 
while the final results in the main text are formulated 
explicitly in terms of $\Omega_{\ja\jb}$.

We label the fundamental representation of
${\rm USp}(4)$ by indices $\ja$, $\jb$, \ldots \ running from $1$ to $4$.
The lowest bosonic representations of
${\rm USp}(4)$ have been 
collected in~\eqref{USp4Reps} built in terms of the fundamental representation.
In particular, 
the ${\bf 5}$ representation is 
given by an antisymmetric symplectic traceless tensor ${V_{{\bf 5}}}^{[\ja\jb]}$,
objects in the ${\bf 10}$ are described by a symmetric tensor ${V_{{\bf 10}}}^{(\ja\jb)}$, etc.

On the ${\bf 5}$ and ${\bf 10}$ representation 
of ${\rm USp}(4)$ there are nondegenerate symmetric forms given by
\begin{align}
   \delta_{[\ja\jb][\jc\jd]} &= - \Omega_{\ja [\jc} \Omega_{\jd] \jb} - \frac 1 4 \Omega_{\ja\jb} \Omega_{\jc\jd}  \; , &
 \nonumber\\
   \delta^{[\ja\jb][\jc\jd]} &= \left( \delta_{[\ja\jb][\jc\jd]} \right)^* = - \Omega^{\ja [\jc} \Omega^{\jd] \jb}  
                                                                         - \frac 1 4 \Omega^{\ja\jb} \Omega^{\jc\jd} \; , &
 \nonumber\\
   \delta^{[\ja\jb]}_{[\jc\jd]} &= 
   \delta^{[\ja}_{[\jc} \delta^{\jb]}_{\jd]} - \frac 1 4 \Omega_{\jc\jd} \Omega^{\ja\jb} 
   = 
   \delta^{\ja\jb}_{\jc\jd}  - \frac 1 4 \Omega_{\jc\jd} \Omega^{\ja\jb} \; , &
\nonumber\\[3ex] 
   \delta^{(\ja\jb)(\jc\jd)} &= \left( \delta_{(\ja\jb)(\jc\jd)} \right)^* = - \Omega^{\ja (\jc} \Omega^{\jd) \jb}  \; , \nonumber \\
    \delta_{(\ja\jb)(\jc\jd)} &= - \Omega_{\ja (\jc} \Omega_{\jd) \jb} \; , \nonumber \\[.8ex]
    \delta^{(\ja\jb)}_{(\jc\jd)} &= \delta^{(\ja}_{(\jc} \delta^{\jb)}_{\jd)} \; .
\end{align}
Note that $\delta^{[\ja\jb][\jc\jd]}$ is the inverse of $\delta_{[\ja\jb][\jc\jd]}$, i.e.\
\begin{align}
   \delta_{[\ja\jb][\jc\jd]} \delta^{[\jc\jd][\je\jf]} &= \delta_{[\ja\jb]}^{[\je\jf]} \; ,
\end{align}
and the same is true for $\delta^{(\ja\jb)(\jc\jd)}$ and $\delta_{(\ja\jb)(\jc\jd)}$. Furthermore we have
\begin{align}
   \Omega^{\ja\jb} \delta_{[\ja\jc][\jb\jd]} &= \frac 5 4 \Omega_{\jc\jd} \; , &
   \delta^{[\jb\jc]}_{[\ja\jc]} &= \frac 5 4 \delta^\jb_\ja \; , &
   \delta^{[\ja\jb]}_{[\ja\jb]} &= 5 \; , \nonumber \\
   \Omega^{\ja\jb} \delta_{(\ja\jc)(\jb\jd)} &= \frac 5 2 \Omega_{\jc\jd} \; , &
   \delta^{(\jb\jc)}_{(\ja\jc)} &= \frac 5 2 \delta^\jb_\ja \; , &
   \delta^{(\ja\jb)}_{(\ja\jb)} &= 10 \; .
\end{align}   
We use the index pairs $(\ja\jb)$ and $[\ja\jb]$ as composite indices for the ${\bf 5}$ and ${\bf 10}$ representation;
they are raised and lowered using the above metrics and when having several of them we use the usual bracket notation 
for symmetrization and anti-symmetrization.

The following tensors represent some 
projections onto the irreducible components 
 of particular ${\rm USp}(4)$ representations:
\begin{align}
   \tau_{(\ja\jb)(\jc\jd)(\je\jf)} &= \Omega_{(\je(\ja} \Omega_{\jb)(\jc} \Omega_{\jd)\jf)} \; ,
         &&  \left[ ({\bf 10} \otimes {\bf 10})_{\text{asymm.}} \mapsto {\bf 10} \right] \;, \nonumber \\
   \tau_{(\ja\jb)[\jc\jd][\je\jf]} &= \Omega_{(\ja[\jc} \Omega_{\jd][\je} \Omega_{\jf]\jb)} \; , 
         &&   \left[ ({\bf 5} \otimes {\bf 5})_{\text{asymm.}} \mapsto {\bf 10} \right] \;,\nonumber \\
   \tau_{[\ja\jb](\jc\jd)(\je\jf)} &= \Omega_{[\ja(\jc} \Omega_{\jd)(\je} \Omega_{\jf)\jb]}  
                                      - \frac 1 4 \Omega_{\ja\jb} \delta_{(\jc\jd)(\je\jf)}  \; ,
         &&   \left[ ({\bf 10} \otimes {\bf 10})_{\text{symm.}} \mapsto {\bf 5} \right] \;,\nonumber \\
   \tau_{(\ja\jb)[\jc\jd][\je\jf][\jg\jh]} &= 
   \hat \tau_{(\ja\jb)\,\text{\bf [}[\jc\jd][\je\jf][\jg\jh]\text{\bf ]}}  \; ,
         &&   \left[ ({\bf 5} \otimes {\bf 5} \otimes {\bf 5})_{\text{asymm.}} \mapsto {\bf 10} \right]\;,
   \label{DefTaus}  
\end{align}
where
\begin{align}
   \hat \tau_{(\ja\jb)[\jc\jd][\je\jf][\jg\jh]} &= \Omega_{(\ja[\jc} \Omega_{\jd][\je} \Omega_{\jf][\jg} \Omega_{\jh]\jb)}
                                                   + \frac 1 4  \tau_{(\ja\jb)[\jc\jd][\je\jf]} \Omega_{\jg\jh}
                                                   \;.
\end{align}
The contractions of these $\tau$-tensors with $\Omega$ yield
\begin{align}
   \Omega^{\jd\jf} \tau_{(\ja\jb)(\jc\jd)(\je\jf)} &= - \frac 3 2 \delta_{(\ja\jb)(\jc\je)} \; , \nonumber  \\
   \Omega^{\jd\jf} \tau_{(\ja\jb)[\jc\jd][\je\jf]} &= \frac 1 2 \delta_{(\ja\jb)(\jc\je)} \; , &
   \Omega^{\jb\jd} \tau_{(\ja\jb)[\jc\jd][\je\jf]} &= - \delta_{[\ja\jc][\je\jf]} \; , \nonumber  \\
   \Omega^{\jd\jf} \tau_{[\ja\jb](\jc\jd)(\je\jf)} &= - \frac 3 2 \delta_{[\ja\jb][\jc\je]} \; , &
   \Omega^{\jb\jd} \tau_{[\ja\jb](\jc\jd)(\je\jf)} &= \frac 3 4 \delta_{(\ja\jc)(\je\jf)} \; , \nonumber  \\
   \Omega^{\jf\jh} \tau_{(\ja\jb)[\jc\jd][\je\jf][\jg\jh]} &= \frac 1 2 \tau_{[\jc\jd](\ja\jb)(\je\jf)} \; , &
   \Omega^{\jb\jd} \tau_{(\ja\jb)[\jc\jd][\je\jf][\jg\jh]} &= - \frac 3 4 \tau_{(\ja\jc)[\je\jf][\jg\jh]} \; , \nonumber  \\
   \Omega^{\jd\jg} \Omega^{\jf\jh} \tau_{(\ja\jb)[\jc\jd][\je\jf][\jg\jh]} &= \frac 3 8 \delta_{(\ja\jb)(\jc\je)} \; .
\end{align} 
Note that $\tau_{(\ja\jb)(\jc\jd)(\je\jf)}$ is totally antisymmetric in the three index pairs.
Since the ${\bf 10}$ is the adjoint representation, the structure constants of ${\rm USp}(4)$ are 
${\tau_{(\ja\jb)(\jc\jd)}}^{(\je\jf)}$. The ${\rm USp}(4)$ generators in the ${\bf 5}$ representation
are ${\tau_{(\ja\jb)[\jc\jd]}}^{[\je\jf]}$ and satisfy the algebra
\begin{align}
    {\tau_{(\ja\jb)[\je\jf]}}^{[\jg\jh]} {\tau_{(\jc\jd)[\jg\jh]}}^{[\ji\jj]}
  - {\tau_{(\jc\jd)[\je\jf]}}^{[\jg\jh]} {\tau_{(\ja\jb)[\jg\jh]}}^{[\ji\jj]}
              &= {\tau_{(\ja\jb)(\jc\jd)}}^{(\jg\jh)} {\tau_{(\jg\jh)[\je\jf]}}^{[\ji\jj]}  \; .
\end{align}
As defined above, $\tau_{(\ja\jb)[\jc\jd][\je\jf]}$ describes the mapping
$({\bf 5} \otimes {\bf 5})_{\text{asymm.}} \mapsto {\bf 10}$. However since
$({\bf 5} \otimes {\bf 5})_{\text{asymm.}} = {\bf 10}$ this must be a bijection. Indeed one finds
\begin{align} 
   x_{(\ja\jb)} &= \sqrt{2} \; {\tau_{(\ja\jb)}}^{[\jc\jd][\je\jf]} \, x_{[\jc\jd][\je\jf]} = \Omega^{\jc\jd} \, x_{[\ja\jc][\jb\jd]} 
   && \Leftrightarrow &
   x_{[\jc\jd][\je\jf]} &= \sqrt{2} \; {\tau^{(\ja\jb)}}_{[\jc\jd][\je\jf]} \, x_{(\ja\jb)} \; ,
   \label{SO5eqUSp4}
\end{align}
for tensors $x_{(\ja\jb)}$ and $x_{[\ja\jc][\jb\jd]}=-x_{[\jb\jd][\ja\jc]}$.
When regarding $(\bf{5} \otimes \bf{5})_{\text{asymm.}}$ as the adjoint representation of ${\rm SO}(5)$, 
formula \eqref{SO5eqUSp4} describes the isomorphism between the algebras of ${\rm USp}(4)$ and ${\rm SO}(5)$.
Some other useful relations in this context are
\begin{align}       
   \tau_{(\ja\jb)[\jc\jd][\je\jf]} \, {\tau^{(\ja\jb)}}_{[\jg\jh][\ji\jj]}
           &= \frac 1 2 \delta_{[\jc\jd]\,\text{\bf[}[\jg\jh]} \delta_{[\ji\jj]\text{\bf]}\,[\je\jf]} \; , &
   \Omega^{\jb\jd} \, \delta^{\phantom{[\jg}}_{[\je\jf]\,\text{\bf[}[\ja\jb]} \delta_{[\jc\jd]\text{\bf]}}^{[\jg\jh]}
                &= {\tau_{(\ja\jc)[\je\jf]}}^{[\jg\jh]} \; .
\end{align}
The last equation states that under the bijection~\eqref{SO5eqUSp4}
the generators of the ${\rm SO}(5)$ vector representation\footnote{
When denoting ${\rm SO}(5)$ vector indices by $\underline{\Ja}$, $\underline{\Jb}$, \ldots, \ 
the ${\rm SO}(5)$ generators 
in the vector representation 
are given by ${t_{\underline{\Ja\Jb},\underline{\Jc}}}^{\underline{\Jd}} = 
\delta^{\phantom{\Jd}}_{\underline{\Jc}[\underline{\Ja}} \delta_{\underline{\Jb}]}^{\underline{\Jd}}$.}
yield ${\tau_{(\ja\jb)[\jc\jd]}}^{[\je\jf]}$.

Also the five-dimensional $\epsilon$-tensor 
can be expressed in terms of $\Omega_{\ja\jb}$. A useful relation  is
\begin{align}
   \epsilon^{[\ja\jb][\jc\jd][\je\jf][\jg\jh][\ji\jj]} x_{[\jc\jd][\je\jf]} y_{[\jg\jh][\ji\jj]}
    &= 4 \tau^{[\ja\jb](\jc\jd)(\je\jf)} x_{(\jc\jd)} y_{(\je\jf)} \;,
\end{align}
where $x$ and $y$ in the $({\bf 5} \otimes {\bf 5})_{\text{asymm.}}={\bf 10}$
representation are related by \eqref{SO5eqUSp4}.

There is no singlet in the product of three ${\rm SO}(5)$ vectors and thus no invariant tensor of the
form $\tau_{[\ja\jb][\jc\jd][\je\jf]}$. This gives rise to the identity
\begin{align}        
   0 &= \delta^{[\ja}_{[\jc} \Omega_{\jd][\je} \delta_{\jf]}^{\jb]} - \text{traces} \nonumber \\
     &= \delta^{[\ja}_{[\jc} \Omega_{\jd][\je} \delta_{\jf]}^{\jb]}
     + \frac 1 4 \Omega^{\ja \jb} \eta_{[\jc \jd][\je \jf]} + \frac 1 4  \Omega_{\jc \jd} \delta^{[\ja \jb]}_{[\je \jf]}
     + \frac 1 4 \Omega_{\je \jf} \delta^{[\ja \jb]}_{[\jc \jd]} + \frac 1 {16} \Omega^{ij} \Omega_{kl} \Omega_{mn}  \; .
\end{align}
Using this equation one finds
\begin{align}
   \Omega^{\jc \jd} \lambda_{\jc [\ja} \mu_{\jb] \jd}  &= - \frac 1 4 \Omega_{\ja \jb} 
               \eta^{[\jc \jd][\je\jf]} \lambda_{[\jc\jd]}  \mu_{[\je\jf]}  \; , &
   \Omega^{\jc \jd} \lambda_{[\ja \jc]} \lambda_{[\jb \jd]} &= \frac 1 4 \Omega_{\ja \jb} 
               \eta^{[\jc \jd][\je\jf]} \lambda_{[\jc\jd]}  \lambda_{[\je\jf]}        
               \;,
\end{align}
for tensors $\lambda_{[\ja \jb]}$, $\mu_{[\ja \jb]}$ in the ${\bf 5}$ representation.

\section{$T$-tensor and quadratic constraints}
\label{app:TQC}

In terms of the tensors~\eqref{DefTaus} defined in the previous section
the decomposition of the $T$-tensor of $d=7$ maximal supergravity into its ${\rm USp}(4)$ irreducible components
can be stated in the systematic form
\begin{align}
   {T_{(\ja\jb)[\jc\jd]}}^{[\je\jf]} &= \sqrt{2} \; \Omega^{\jg\jh} \, {X_{[\ja\jg][\jb\jh][\jc\jd]}}^{[\je\jf]} \nonumber \\
                                     =
                                      - & B {\tau_{(\ja\jb)[\jc\jd]}}^{[\je\jf]}
                                      - {B^{[\jg\jh]}}_{[\jc\jd]} {\tau_{(\ja\jb)[\jg\jh]}}^{[\je\jf]}
+ C^{[\jg\jh]} {\tau_{(\ja\jb)[\jg\jh][\jc\jd]}}^{[\je\jf]}
+ {C^{[\je\jf]}}_{(\jg\jh)} {\tau_{[\jc\jd](\ja\jb)}}^{(\jg\jh)} \;,
   \label{TTwithTaus}      
\end{align}
from which~\eqref{TABCD} is recovered with the explicit definitions of~\eqref{DefTaus}.
Similarly, the variation of the scalar potential under
$\delta_{\Sigma} {\cal V}_{M}{}^{ab}=\Sigma^{ab}{}_{cd}\,{\cal V}_{M}{}^{cd}$
takes the more concise form
\begin{align}
  \delta_{\Sigma} V &= 
          -\frac {g^{2}}{16} {B^{[\ja\jb]}}_{[\jc\jd]} 
          {B^{[\jc\jd]}}_{[\je\jf]} {\Sigma^{[\je\jf]}}_{[\ja\jb]}
+\frac {g^{2}}{32} B {B^{[\ja\jb]}}_{[\jc\jd]} {\Sigma^{[\jc\jd]}}_{[\ja\jb]}
          -\frac {g^{2}}{64} C^{[\ja\jb]} C_{[\jc\jd]} {\Sigma^{[\jc\jd]}}_{[\ja\jb]}
\nonumber \\[1ex]  &
{}
+\frac {g^{2}}{16} {C^{[\ja\jb]}}_{(\je\jf)} {C_{[\jc\jd]}}^{(\je\jf)} {\Sigma^{[\jc\jd]}}_{[\ja\jb]}
+\frac{g^{2}}8
{\tau^{(\ja\jb)}}_{[\je\jf][\ji\jj]} {\tau^{(\jc\jd)}}_{[\jg\jh][\jk\jl]} 
{C^{[\jg\jh]}}_{(\ja\jb)} C^{[\je\jf](\jc\jd)} 
\Sigma^{[\ji\jj][\jk\jl]} \;,
\end{align}
from which \eqref{VaryV} is deduced.

The quadratic constraint \eqref{quadratic} 
on the components $Y_{\Ja\Jb}$ and $Z^{\Ja\Jb,\Jc}$ of the embedding tensor~$\Theta$
translates under the ${\rm USp}(4)$ split into quadratic constraints 
on the components
$B$, ${B^{\ja\jb}}_{\jc\jd}$, $C^{\ja\jb}$ and ${C^{\ja\jb}}_{\jc\jd}$ of the $T$-tensor. 
These constraints prove essential when checking the 
algebra of the supersymmetry transformation \eqref{SUSYrules}
and the invariance of the Lagrangian \eqref{L} under these transformations.
According to \eqref{sum} the quadratic constraint on $\Theta$ decomposes into 
a ${\bf \overline{5}}$, a ${\bf \overline{45}}$ and a ${\bf \overline{70}}$ under ${\rm SL}(5)$
which under ${\rm USp}(4)$ branch as
\begin{align}
   {\bf\overline 5} ~\rightarrow~ & {\bf 5} \, , &
   {\bf\overline{45}} ~\rightarrow~& {\bf10} \oplus {\bf 35} \, , &
   {\bf\overline{70}} ~\rightarrow~& {\bf 5} \oplus {\bf30} \oplus {\bf35} \, ,
\end{align}
In closed form, these constraints have been given in~\eqref{QZB2}. 
The check of supersymmetry of the Lagrangian however needs the explicit 
expansion of these equations in terms of $B$  and $C$.
The two ${\bf 5}$ parts and the ${\bf 10}$ part read
\begin{align}
     4 B C^{[\ja\jb]} 
       - {B^{[\ja\jb]}}_{[\jc\jd]}  C^{[\jc\jd]} 
       - 4 {B^{[\ji\jj]}}_{[\jc\jd]} {C^{[\jc\jd]}}_{(\jg\jh)} {\tau^{(\jg\jh)[\ja\jb]}}_{[\ji\jj]}  &= 0 \; , 
       \nonumber \\
     B B^{[\ja\jb]} 
       + {B^{[\ja\jb]}}_{[\jc\jd]}  C^{[\jc\jd]} 
        + \tau^{[\ja\jb](\jc\jd)(\je\jf)} {C^{[\jg\jh]}}_{(\jc\jd)} C_{[\jg\jh](\je\jf)} &= 0 \; , 
\nonumber \\
     {\tau_{[\jc\jd]}}^{(\ja\jb)(\jg\jh)} {B^{[\jc\jd]}}_{[\je\jf]} {C^{[\je\jf]}}_{(\jg\jh)} &= 0 \;,
\label{Con5}
\end{align}      
respectively.
In particular, a proper linear combination of the first two equations yields the quadratic
relation~\eqref{A1A1A2A2} cited in the main text.
The two ${\bf 35}$ parts of the quadratic constraint are
\begin{align}
     {\tau_{(\jc\jd)}}^{[\ja\jb][\je\jf]} B C_{[\je\jf]} + B {C^{[\ja\jb]}}_{(\jc\jd)}
     + {\tau_{(\jc\jd)}}^{[\je\jf][\jg\jh]} {B^{[\ja\jb]}}_{[\je\jf]} C_{[\jg\jh]}
     + {B^{[\ja\jb]}}_{[\je\jf]} {C^{[\je\jf]}}_{(\jg\jh)}  &= 0  \; , 
   \nonumber \\[1.5ex]
   \mathbb{P}_{{\bf 35}} \Big( 
     B {C^{[\ja\jb]}}_{(\jc\jd)} 
       - 4 \tau^{(\je\jf)[\jg\jh][\ja\jb]} \tau_{(\jc\jd)[\ji\jj][\jk\jl]} {C^{[\ji\jj]}}_{(\je\jf)} {B^{[\jk\jl]}}_{[\jg\jh]} 
       \qquad \qquad \qquad \qquad \phantom{a} \nonumber & \\
       - 3 {\tau_{[\je\jf](\jc\jd)}}^{(\jg\jh)} C^{[\je\jf]} {C^{[\ja\jb]}}_{(\jg\jh)}
    + 4  {\tau^{(\je\jf)[\ja\jb]}}_{[\jg\jh]} {\tau_{[\ji\jj](\jc\jd)}}^{(\jk\jl)} {C^{[\ji\jj]}}_{(\je\jf)} {C^{[\jg\jh]}}_{(\jk\jl)} 
     \Big)
       &= 0 \; ,
   \label{Con35}    
\end{align}
where the projector $\mathbb{P}_{\bf 35}$ is defined by
\begin{align}
   \mathbb{P}_{{\bf 35}}\left( {X^{[\ja\jb]}}_{(\jc\jd)} \right) &=
    \left( \delta^{[\ja\jb]}_{[\je\jf]} \delta_{(\jc\jd)}^{(\jg\jh)} 
      - {\tau_{(\jc\jd)}}^{[\ja\jb][\ji\jj]} {\tau^{(\jg\jh)}}_{[\je\jf][\ji\jj]}
      - \frac 4 3 {\tau^{[\ja\jb]}}_{(\jc\jd)(\ji\jj)} {\tau_{[\je\jf]}}^{(\jg\jh)(\ji\jj)} \right) {X^{[\je\jf]}}_{(\jg\jh)} \; .
\end{align}
Note that also the first equation of \eqref{Con35} has to be projected with $\mathbb{P}_{\bf 35}$
in order to reduce it to a single irreducible part. However this equation is satisfied 
also without the projection,
since it contains the above ${\bf 10}$ and one of the ${\bf 5}$ constraints as well.\footnote{
Indeed this first equation of \eqref{Con35} is equivalent to~\eqref{QZB1}.}
Finally the ${\bf 30}$ component of the quadratic constraint is 
obtained by completely symmetrizing \eqref{QZB2}
in the three free index pairs, i.e.\
\begin{align}
   Z^{(\jg\jh)\text{\bf(}[\ja\jb]} {T_{(\jg\jh)}}^{[\jc\jd][\je\jf]\text{\bf)}} &= 0 \;.
\end{align}

\end{appendix}


\cleardoublepage
\addcontentsline{toc}{chapter}{Bibliography}  

\providecommand{\href}[2]{#2}\begingroup\raggedright\endgroup

\end{document}